\documentclass[aps,prc,showpacs,twocolumn,superscriptaddress,nofootinbib,preprintnumbers,floatfix]{revtex4-1}

\usepackage{tikz}
\usepackage{longtable}
\usepackage{epstopdf}
\usepackage{graphicx}
\usepackage{slashed}
\usepackage{dcolumn}
\usepackage{bm}
\usepackage{amssymb,amsthm,amsmath,amstext,amsbsy,amsopn}
\usepackage{bbm}
\usepackage{nicefrac}
\usepackage{mathrsfs}
\usepackage{pstricks}
\usepackage{hyperref}
\usepackage{leftidx}
\usepackage{environ}
\usepackage{mathtools}
\usepackage{xspace}
\usepackage{overpic}
\usepackage{subfigure}
\usepackage{tensor}
\usepackage{gensymb} 
\usepackage{cancel}
\usepackage{float}
\usepackage{color}
\usepackage{calligra}
\usepackage{soul}
\usepackage{changes}
\usepackage{appendix}
\usepackage{comment}
\usepackage{multirow}

\newcommand{\bea}{\begin{eqnarray}}
\newcommand{\eea}{\end{eqnarray}}
\newcommand{\beq}{\begin{equation}}
\newcommand{\eeq}{\end{equation}}

\usepackage[normalem]{ulem}  

\renewcommand\sout{\bgroup \color{red} \ULdepth=-.5ex \ULset}

\begin{document}



\title{Chiral enhancement in soft-photon bremsstrahlung and charge asymmetry in low-energy lepton-proton scattering }
\author{Bhoomika Das}
     \email[]{bhoomika.das@iitg.ac.in}
     \affiliation{Department of Physics, Indian Institute of Technology Guwahati, 
                 Guwahati - 781039, Assam, India.}%
\author{Rakshanda Goswami}
     \email[]{r.goswami@iitg.ac.in}
     \affiliation{Department of Physics, Indian Institute of Technology Guwahati, 
                 Guwahati - 781039, Assam, India.}%
\author{Pulak Talukdar}
     \email[]{pulaktalukdar45@gmail.com}
    \affiliation{Department of Physics, S. B. Deorah College, Guwahati - 781007, Assam, India.}%
\author{Udit Raha}
     \email[]{udit.raha@iitg.ac.in}
     \affiliation{Department of Physics, Indian Institute of Technology Guwahati, 
                 Guwahati - 781039, Assam, India.}%
\author{Fred Myhrer}
     \email[]{myhrer@mailbox.sc.edu}
     \affiliation{Department of Physics and Astronomy, 
                 University of South Carolina, Columbia, SC 29208, USA.}%
\begin{abstract} 
We carry out a Lorentz gauge evaluation of the single soft-photon bremsstrahlung radiative corrections to 
the unpolarized elastic lepton-proton scattering cross section at low-energies using the framework of heavy 
baryon chiral effective field theory (HB$\chi$PT). We systematically incorporate all next-to-leading order 
[i.e., ${\mathcal O}(\alpha^3/M)$] contributions to the cross section arising from radiative and 
proton's low-energy recoil effects. An important component of the soft-photon bremsstrahlung corrections 
arises from the interference between lepton and proton bremsstrahlung processes, which are characterized as 
charge-odd. We identify a class of Feynman diagrams involving the \underline{final-state} radiating proton
that induces a chiral enhancement, thereby modifying the naive amplitude hierarchy based on the standard 
chiral power-counting scheme. The recently derived HB$\chi$PT result for the analytically computed exact 
two-photon exchange (TPE) counterpart at NLO is combined with our charge-odd soft-photon bremsstrahlung 
contribution to yield the complete charge-odd radiative correction. By incorporating the leading hadronic 
correction to the leading order Born one-photon exchange (OPE) process within the charge-even radiative 
correction, we obtain a prediction for the charge asymmetry observable, which can be compared with 
forthcoming MUSE data on elastic lepton–anti-lepton scattering off the proton. In all our calculations, we
use finite lepton masses. 
\end{abstract}

\maketitle

\section{Introduction}
\label{sec:I}

The higher-order Quantum Electrodynamic (QED) radiative corrections to the one-photon exchange (OPE) are 
crucial for the precision determination of the unpolarized elastic (anti-)lepton-proton ($\ell^\pm$p) 
scattering cross sections. In experiments, the undetected {\it soft} bremsstrahlung photons play a vital
role in both elastic and inelastic processes, namely, $\ell^\pm\,{\rm p}\to \ell^\pm\,{\rm p}$ and 
$\ell^\pm\,{\rm p} \to \ell^\pm\,{\rm p}\,\gamma^*$, respectively. By a {\it soft-photon} we mean that 
the energy of the emitted photon is tiny, i.e., $E_{\gamma^*}\to 0$,\footnote{The distinction between 
soft-photon and hard-photon bremsstrahlung can be defined in terms of an experimental resolution 
threshold $\Delta_{\gamma^*}$, which sets the minimum photon energy required for an event to be 
detectable. This threshold, determined by the detector's acceptance and design, renders the hard-photon
events observable only above $\Delta_{\gamma^*}$, while soft-photon emissions fall below this limit.} 
rendering both elastic and inelastic processes physically indistinguishable.\footnote{In this work, we 
will use the symbol $\gamma^*$ to denote a real (on-shell) bremsstrahlung photon, in contrast to a 
virtual (off-shell) photon, which will be denoted by $\gamma$.} The bremsstrahlung contributions to the
cross section are further categorized as being ``charge-even" and ``charge-odd". This is based on the 
type of interference matrix elements considered between products of Feynman diagrams with the external
lepton and proton legs radiating the photon. For the $\ell^\pm$-p scatterings, the ``charge-even" 
cross section is the sum of the $\ell^-$ and the $\ell^+$ contributions, while the ``charge-odd" 
counterpart is their difference. Furthermore, in this paper, the individual squared amplitudes and the
lepton-lepton or proton-proton interference bremsstrahlung matrix elements contributing to the 
charge-even cross section $\sigma^{(\rm even)}_{\gamma\gamma^*}$ will be referred to as the 
``direct-interference" terms. In contrast, the lepton-proton interference bremsstrahlung matrix 
elements contributing to the charge-odd cross section $\sigma^{(\rm odd)}_{\gamma\gamma^*}$, shall be
referred to as the ``exchange-interference" terms. Among the one-loop virtual photon-loop corrections
in QED, only the two-photon exchange (TPE) radiative correction contributes to the charge-odd cross 
section $\sigma^{(\rm odd)}_{\gamma\gamma}$ (for recent TPE analyses, see e.g., 
Refs.~\cite{Kivel:2012vs,Lorenz:2014yda,Tomalak:2014sva,Tomalak:2014dja,Tomalak:2015aoa,Tomalak:2015hva,Tomalak:2016vbf,Tomalak:2017npu,Koshchii:2017dzr,Tomalak:2018jak,Bucoveanu:2018soy,Talukdar:2019dko,Peset:2021iul,Talukdar:2020aui,Kaiser:2022pso,Guo:2022kfo,Engel:2023arz,Cao:2021nhm,Choudhary:2023rsz,Goswami:2025yky}; 
including the review of Ref.~\cite{Afanasev:2017gsk}). The remaining virtual photon-loop corrections,
such as the vertex, self-energy, and vacuum polarization corrections, contribute to the charge-even 
cross section $\sigma^{(\rm even)}_{\gamma\gamma}$. The total charge-even and -odd contributions can 
therefore be summarized as    
\begin{eqnarray}
\sigma^{(\rm even)}_{\rm tot} \!&=&\! \sigma^{(\rm even)}_{\gamma\gamma} 
+  \sigma^{(\rm even)}_{\gamma\gamma^*}\,,\quad \text{and}
\\
\sigma^{(\rm odd)}_{\rm tot} \!&=&\! \sigma^{(\rm odd)}_{\gamma\gamma} 
+  \sigma^{(\rm odd)}_{\gamma\gamma^*}\,,
\end{eqnarray}
respectively. It should be noted, however, that neither the virtual photon-loops nor the soft-photon 
bremsstrahlung cross sections are physical observables. Each of them is affected by infrared (IR) 
divergence -- an artifact that precludes direct comparison with measured cross section data. To 
facilitate a meaningful comparison, the unphysical divergences are eliminated {\it via} the concomitant
inclusion of a degenerate phase-space of inelastic real soft-photon emissions, constituting an 
unavoidable background to the basic elastic scattering 
process~\cite{Bloch:1937pw,Yennie:1961ad,Kinoshita:1962ur,Lee:1964is}. The IR divergences stemming from
the virtual photon-loop integrals are systematically canceled order by order in perturbation theory, 
by the corresponding divergences from the soft-photon bremsstrahlung phase-space integrals when the 
incoherent sum of the virtual photon-loops and soft bremsstrahlung cross sections is considered. 
Consequently, the virtual corrections to the elastic cross section receive finite modifications from 
the soft-photon bremsstrahlung.

One can experimentally measure the difference between the lepton and anti-lepton proton scattering cross 
sections {\it via} the {\it charge asymmetry} observable ${\mathcal A}_{\ell^\pm}$, defined by the 
following ratio of the differential cross sections:
\begin{equation}
{\mathcal A}_{\ell^\pm}= \frac{{\rm d}\sigma^{(\rm odd)}_{\rm tot}}{{\rm d}\sigma^{(\rm even)}_{\rm tot}}
= \frac{{\rm d}\sigma^{(-)}_{el}-{\rm d}\sigma^{(+)}_{el}}{{\rm d}\sigma^{(-)}_{el}+{\rm d}\sigma^{(+)}_{el}}\,. 
\label{eq:Asy}
\end{equation}
In particular, the measurement of ${\mathcal A}_{\ell^\pm}$ provides a rather ``clean" procedure to 
estimate the charge-odd TPE effects, with various systematic errors contributing to the radiative 
corrections dropping out in the ratio. Several ongoing experimental efforts plan to measure this 
asymmetry. The MUon Scattering Experiment (MUSE) at the Paul Scherrer Institute 
(PSI)~\cite{MUSE:2017dod,Cline:2021ehf,Strauch:2024imt} is an ongoing effort designed to simultaneously
scatter leptons and anti-leptons off a proton target at unprecedented low four-momentum transfers 
($|Q^2|\lesssim 0.08$~GeV${{}^2}/c$ for a maximum beam three-momentum $p_{\rm beam}\lesssim 210$~MeV/$c$).
A major goal of MUSE is to pin down the low-energy TPE corrections with a systematic uncertainty of 
$0.1\%$, thereby offering potential insights into resolving the long-standing discrepancies in the 
proton charge 
radius~\cite{Pohl:2010zza,Pohl:2013,Antognini:1900ns,Bernauer:2014,Carlson:2015,Bernauer:2020ont}
-- an issue that remains unsettled in experimental analyses of low-energy electron-proton scattering 
(see e.g., Ref.~\cite{Gao:2021sml} for a review of recent experimental efforts).    

To facilitate the confrontation of theoretical predictions with the future high-precision MUSE data on
$\ell$-p elastic cross section, an accurate estimation of the low-energy TPE contribution, along with 
its {\it exchange-interference} soft-photon bremsstrahlung counterpart, is needed. Furthermore, given 
that these two corrections are rather poorly constrained in comparison with the other radiative 
corrections, a careful assessment of the role of low-energy non-perturbative effects due to the proton's 
structure, as well as the chiral {\it recoil-radiative} corrections~\cite{Talukdar:2020aui}, is deemed
essential. While a substantial body of recent literature focuses on two-photon exchange (TPE) 
effects~\cite{Kivel:2012vs,Lorenz:2014yda,Tomalak:2014sva,Tomalak:2014dja,Tomalak:2015aoa,Tomalak:2015hva,Tomalak:2016vbf,Tomalak:2017npu,Koshchii:2017dzr,Tomalak:2018jak,Bucoveanu:2018soy,Talukdar:2019dko,Peset:2021iul,Talukdar:2020aui,Kaiser:2022pso,Guo:2022kfo,Engel:2023arz,Cao:2021nhm,Choudhary:2023rsz,Goswami:2025yky}, 
relatively few studies address the estimation of charge-odd soft-photon bremsstrahlung corrections and 
their impact on charge asymmetry at 
low-energies.~\cite{Koshchii:2017dzr,Talukdar:2020aui,Afanasev:2021nhy,Afanasev:2022uwm}. For recent 
developments in this direction within the framework of QED-inspired hadronic models, we refer the reader
to the works presented in Refs.~\cite{Koshchii:2017dzr,Afanasev:2021nhy,Afanasev:2022uwm}. However, a key
limitation of model-dependent approaches lies in their potential to obscure subtle off-shell, 
non-perturbative features of the proton’s low-energy structure through the use of {\it ad hoc} on-shell 
phenomenological form factors. To address this, a robust, model-independent analysis of soft-photon 
bremsstrahlung corrections at low-energies is essential -- one that systematically accounts for 
uncertainties arising from the perturbative truncation of higher-order effects. This objective can be 
effectively pursued within the framework of low-energy effective field theory (EFT).
 
In this paper, we present an EFT evaluation of the {\it leading order} (LO) and {\it next-to-leading order} (NLO) 
bremsstrahlung corrections within the well-known framework of the SU(2) formulation of Heavy Baryon Chiral 
Perturbation Theory (HB$\chi$PT)~\cite{Jenkins:1990jv,Bernard:1992qa,Ecker:1994pi,Bernard:1995dp,Fettes:2000gb}. 
Such an EFT is intended to capture the essential non-perturbative features of low-energy QCD manifested through 
the chiral dynamics of pions and nucleons in response to gauge-invariant hadronic and electroweak external 
currents below a pre-determined cut-off momentum scale, $\Lambda_\chi\simeq 4\pi f_\pi \approx 1$ GeV$/c$, where
$f_\pi$ is the pion decay constant. Low-energy symmetries, such as chiral symmetry (and its breakings), play a 
crucial role in shaping the spectrum and underlying dynamics. The short-distance/ultraviolet physics beyond the 
scale $\Lambda_\chi$ is tacitly subsumed into a well-defined set of low-energy constants (LECs). These LECs are 
universally determined either phenomenologically or {\it via} lattice-QCD calculations. The hadrons are 
effectively treated as elementary particles in this framework, and the formalism incorporates a simultaneous 
low-energy expansion in inverse powers of the nucleon mass $M$, i.e., ${\mathcal Q}/M$, along with the standard
chiral expansion in powers of ${\mathcal Q}/\Lambda_\chi\sim {\mathcal Q}/M$, where ${\mathcal Q}$ is a generic
momentum scale near the pion mass $m_\pi$. Such an EFT power-counting scheme ensures a systematic perturbative
improvement of the theoretical predictions to an intended level of accuracy.

In the work of Talukdar {\it et al}.~\cite{Talukdar:2020aui}, the soft-photon bremsstrahlung radiative 
corrections to the elastic cross section up-to-and-including NLO [i.e, ${\mathcal O}(\alpha^3/M)$], were 
evaluated in HB$\chi$PT at low-energies relevant to MUSE kinematics. The evaluations were performed by adopting 
the {\it Coulomb gauge}. In this work, we will adopt the {\it Lorentz gauge} and demonstrate that there exists a
certain class of chirally \underline{enhanced} bremsstrahlung diagrams from sub-leading orders, which effectively
contribute to the cross section at LO and NLO. These \underline{enhanced} contributions are generated from the 
\underline{final-state} proton-radiating amplitudes having insertions of ${\mathcal O}(M^0)$ and 
${\mathcal O}(1/M)$ \underline{final-state} proton propagator components. As we will demonstrate in the next 
section, this calls for the {\it promotion} of some of these amplitudes to lower orders, which are formally 
{\it next-to-next-to-leading order} (NNLO) and {\it next-to-next-to-next-to-leading order} (N${}^3$LO) by the 
chiral power-counting. Moreover, this also implies that the ``effectively promoted" LO and NLO corrections to 
the bremsstrahlung cross section are now likely to receive additional contributions from renormalized pion-loops.
This possibility is somewhat unexpected, standing in contrast to the conventional scenario in HB$\chi$PT, where 
hadronic structure-dependent radiative corrections from the proton are anticipated to contribute only at NNLO in
the cross section. In light of the recent {\it exact} analytical evaluation of the TPE up-to-and-including NLO 
in HB$\chi$PT by Choudhary {\it et al}.~\cite{Choudhary:2023rsz} performed in the Lorentz gauge, it is 
appropriate to adopt the same gauge for the charge-odd bremsstrahlung calculation to maintain consistency in the
cancellation of IR divergences. The same applies to the charge-even bremsstrahlung corrections, wherein the 
IR-divergent terms must be systematically canceled with those stemming from the charge-even virtual corrections. 

Our purpose of this paper is to present a consistent analytical evaluation of the soft-photon bremsstrahlung 
contributions to the elastic cross section up-to-and-including NLO. Together with the virtual photon-loop 
corrections, obtained earlier in Refs.~\cite{Talukdar:2020aui} and \cite{Choudhary:2023rsz}, we determine the 
full IR-finite radiative corrections to the elastic cross section of ${\mathcal O}(\alpha^3/M)$. Especially, by 
including the exact TPE results of Choudhary {\it et al.}, the accuracy of our radiative corrections is 
expected to supersede the previous predictions by Talukdar {\it et al.}, who used the 
{\it soft-photon approximation} (SPA) to analytically evaluate the intricate 4-point scalar TPE 
loop-integrals~\cite{Koshchii:2017dzr,Talukdar:2019dko,Talukdar:2020aui,Mo:1968cg,Maximon:1969nw,Maximon:2000hm,Tsai:1961zz,Vanderhaeghen:2000ws,Bucoveanu:2018soy}. 
The radiative corrections results are subsequently used to evaluate the charge asymmetry ${\mathcal A}_{\ell^\pm}$, 
which in turn assesses the extent to which the TPE contributions could dominate the observable over the 
ostensibly residual charge-odd bremsstrahlung contributions. Thus, an accurate measurement of the asymmetry is 
crucial to the precision with which the TPE effects can be extracted in future measurements. 

Our paper is organized as follows: In Sec.~\ref{sec:II}, we present the basic ingredients of the HB$\chi$PT 
framework that allows a consistent evaluation of the LO and NLO soft-photon bremsstrahlung corrections to the 
elastic lepton-proton cross section. We incorporate the low-energy recoil-radiative effects in a perturbative 
approach. In addition, we also consider the leading hadronic structure-dependent corrections to the LO Born or 
OPE cross section at NNLO accuracy. Section~\ref{sec:III} deals with our numerical results and their discussion.
Here we combine our bremsstrahlung and hadronic OPE corrections results with the charge-even virtual 
photon-loop corrections obtained by Talukdar {\it et al}.~\cite{Talukdar:2020aui} and the charge-odd exact TPE 
results of Choudhary {\it et al}.~\cite{Choudhary:2023rsz}. Our predictions pertain to the full radiative 
corrections and charge asymmetry, in addition to estimating the soft-photon bremsstrahlung corrections at NLO 
accuracy. Finally, we summarize and conclude in Sec.~\ref{sec:IV} with an outlook for subsequent future 
work. Appendices A and B contain a detailed collection of relevant Feynman amplitudes and soft-photon 
phase-space integrals, as well as a brief description of the so-called S-frame transformations needed for 
evaluating the bremsstrahlung contributions.  

\section{Theoretical framework of HB$\chi$PT}
\label{sec:II}
We deploy the EFT framework of HB$\chi$PT~\cite{Jenkins:1990jv,Bernard:1992qa,Ecker:1994pi,Bernard:1995dp}, 
where the ``heavy" proton is treated non-relativistically, while leptons, pions, and the photon are treated 
relativistically. The radiative corrections are evaluated up to $\mathcal{O}(\alpha/M)$, which is formally 
NLO in the power-counting. In the following, we will explicitly present only the relevant LO, NLO, and NNLO 
components of the pion-nucleon effective chiral Lagrangian $\mathcal{L}_{\pi N}$,  that are required for our
calculations.\footnote{While our analysis accounts for non-vanishing contributions from N$^3$LO Feynman 
amplitudes, all insertions from the N$^3$LO chiral Lagrangian$\mathcal{L}^{(3)}_{\pi N}$ [see 
Eq.~\eqref{eq:LpiN-3} in Appendix A] either yield vanishing contributions to the bremsstrahlung cross 
section or contribute only at orders beyond the target ${\mathcal O}(\alpha^3/M)$ NLO accuracy. Consequently,
$\mathcal{L}^{(3)}_{\pi N}$ will not be explicitly required in the analysis.} This includes 
parts of the chiral Lagrangian involving purely pionic, as well as the standard leptonic interactions in QED. 
For details, we refer to the exposition presented in 
Refs.~\cite{Talukdar:2020aui,Bernard:1995dp,Fettes:2000gb}. Thus, to NNLO accuracy  in this work
\begin{eqnarray}
\mathcal{L}_{\pi N}=\mathcal{L}^{(0)}_{\pi N}+\mathcal{L}^{(1)}_{\pi N}+\mathcal{L}^{(2)}_{\pi N}+ ...\,\,,
\label{eq:LpiN}
\end{eqnarray}
where the complete lowest chiral order parts of the $\pi N$ Lagrangian are given as
\begin{eqnarray}
\mathcal{L}_{\pi N}^{(0)} \!&=&\! {\bar N}_v (i v\cdot {\mathcal D} + g_A S \cdot u) N_v\,;  
\label{eq:LpiN-0}
\\
\mathcal{D}_\mu \!&=&\! \partial_\mu +\frac{1}{2}
\bigg[u^\dag \left({\mathbb I}\partial_\mu+i\tau^3\frac{e}{2}A_\mu\right)u 
\nonumber\\
&&\hspace{1cm} +\, u \left({\mathbb I}\partial_\mu+i\tau^3\frac{e}{2}A_\mu\right) u^\dag\bigg] 
+ i {\mathbb I}\frac{e}{2} A_\mu\,, \quad\,
\label{eq:chiral_del}
\end{eqnarray}
where $\mathcal{D}_\mu$ represents the chiral covariant derivative, including the terms with the photon 
vector field $A_\mu$. Furthermore, ${\mathbb I}$ denotes the $2\times2$ identity matrix, and $\tau^3$ is
the third component of the Pauli iso-spin matrix. The next chiral order Lagrangian is 
\begin{eqnarray}
\mathcal{L}_{\pi N}^{(1)} \!&=&\! {\bar N}_v\! \left[\frac{1}{2 M} (v \cdot {\mathcal{D}})^2
- \frac{1}{2 M} \mathcal{D} \cdot \mathcal{D} - \frac{i g_A}{2M}\big\{S\cdot {\mathcal D} , v\cdot u\big\} \right.
\nonumber\\
&&\hspace{0.3cm} \left. +\, c_1 {\rm Tr}(\chi_+) + \left( c_2 - \frac{g_A^2}{8M}\right) ( v \cdot u)^2 
+ c_3 u \cdot u  \right.
\nonumber\\
&&\hspace{0.3cm} \left. +\, \left( c_4 +\frac{1}{4M}\right) [S^\mu , S^\nu]u_\mu u_\nu 
+ c_5 {\rm Tr}(\tilde{\chi}_+) \right.
\nonumber\\
&&\hspace{0.3cm} \left. -\, \frac{i}{4M}[S^\mu , S^\nu] \Big\{(1 +c_6)f^+_{\mu \nu} 
+ c_7 {\rm Tr}(f^+_{\mu\nu}) \Big\} \right]\! N_v .
\nonumber\\
\label{eq:LpiN-1}
\end{eqnarray}
In the above expressions, $N_v \!=\! ({\rm p}_v , {\rm n}_v)^{\rm T}$ denotes the heavy nucleon 
iso-doublet spinor field, and $u=\sqrt{U}$ is the SU(2) iso-triplet fields, where the non-linear pion
field $\boldsymbol{\pi}=(\pi_1,\pi_2,\pi_3)$ is defined, e.g.,  {\it via} the {\it sigma gauge} 
parametrization, i.e.,  
\begin{equation}
U=u^2=\sqrt{1-\frac{\boldsymbol{\pi}^2}{f_\pi}}+\frac{i}{f_\pi}\boldsymbol{\tau}\cdot\boldsymbol{\pi}\,\,\, \text{and} \,\,\,
u_\mu = i u^\dag \partial_\mu U u^\dag\,. 
\end{equation} 
The $\boldsymbol{\tau}$ denotes the Pauli isospin matrices and $g_A=1.267$ is the nucleon axial coupling 
constant. The four-vectors $v^\mu$ and $S^\mu$ denote the velocity and spin of the nucleon, respectively, 
satisfying $S\cdot v =0$. We find it convenient to choose, $v^\mu=(1,{\bf 0})$ and 
$S^\mu=(0,\boldsymbol{\sigma}/2)$. In addition, $c_{1,\cdots,7}$ are the seven LECs with well-known 
finite values which have been determined in phenomenological analyses, e.g., Ref.~\cite{scherer2003}. In 
particular, the values, $c_6 = \kappa_v$, and $c_7=(\kappa_s - \kappa_v)/2$, are determined {\it via} the
nucleon iso-vector and iso-scalar anomalous magnetic moments, $\kappa_v = 3.71$ and $\kappa_s = -0.12$, 
respectively~\cite{Bernard1998}. The remaining terms in the NLO Lagrangian are as given below:
\begin{eqnarray}
\chi_+ \!&=&\! u^\dagger\chi u + u\chi^\dagger u\,, \quad \tilde{\chi}_+=\chi_+ -\frac{\mathbb I}{2}{\rm Tr} (\chi_+)\,,
\nonumber\\
f^{+}_{\mu\nu} \!&=&\! e\left(\partial_\mu A_\nu-\partial_\nu A_\mu\right)
(u \hat{\mathcal{Q}} u^\dagger + u^\dagger\hat{\mathcal{Q}} u)\, , \quad  {\rm where} 
\nonumber\\
\hat{\mathcal{Q}} \!&=&\! \frac{1}{2}({\mathbb I}+\tau^3)
= 
\begin{pmatrix}
1 & 0\\ 
0   & 0
\end{pmatrix}\,. 
\label{eq9}
\end{eqnarray}
In our analysis, we have ignored isospin violation except for considering the physical mass of the 
proton. Hence, we only consider the limit $m_d = m_u$, i.e., $\tilde{\chi}_+ \rightarrow 0$, meaning 
that the term proportional to $c_5$ vanishes. 

Finally, we display the relevant portion of the NNLO $\pi N$ Lagrangian which are needed in our analysis;
the ellipsis denotes the excluded parts of the complete NNLO Lagrangian, see e.g., 
Ref.~\cite{Fettes:2000gb}, which, for example, contains counterterm LECs used for renormalizing 
UV-divergent pion-loops:
\begin{eqnarray}
{\mathcal{L}}_{\pi N}^{(2)} \!&=&\! \bar{N}_v \left[  - \frac{i}{4M^2} (v \cdot \mathcal{D})^3 
+ \frac{i}{8 M^2} \Big \{ \mathcal{D}^2 (v \cdot \mathcal{D}) \right. 
\nonumber\\
&&\hspace{0.6cm} \left. -\,(v \cdot \mathcal{D}^\dag) {\mathcal{D}^\dag}^2 \Big \}+\cdots\right] N_v\,,
\label{eq:LpiN-2}
\end{eqnarray} 

Before we proceed with our analysis of the radiative corrections, we first discuss the kinematics associated with the 
unpolarized elastic lepton-proton scattering. In addition, we include the kinematics of the corresponding inelastic 
single soft-photon emission processes with an energy $E_{\gamma^*}\lesssim \Delta_{\gamma^*}$, which is lower than any
other energy scale in the problem of interest and defines the limit of an experimentally detectable photon. Our 
analysis is performed in the laboratory frame ({\it lab.}-frame) where the initial target proton is at rest. In this 
frame, we adopt the following four-momentum assignments: 
\begin{eqnarray*}
\ell^\pm(p)+{\rm p}(P_p) \!\!&\longrightarrow&\!\! \ell^\pm(p^\prime)+{\rm p}(P_p^\prime)\, \quad \text{and}  
\\
\ell^\pm(p)+{\rm p}(P_p) \!\!&\longrightarrow&\!\! \ell^\pm(p^{\prime})+{\rm p}(P^{\prime}_p) 
+ \gamma^* (k)\,; \quad \ell\equiv e,\mu\, .
\end{eqnarray*}
Here $p^\mu = (E, {\bf p})$ and $P^\mu_p = (M,{\bf 0})$ denote the initial  (anti-)lepton and proton four-momenta, 
respectively, and $p^{\prime\mu} = (E^\prime, {\bf p}^{\prime})$ and $P^{\prime\mu}_p=(E^\prime_p,{\bf P}^{\prime}_p)$ 
their corresponding outgoing four-momenta. Hence, $Q^\mu=(p-p^\prime)^\mu=(P^\prime_p-P_p)^\mu$ is the four-momentum 
transferred between the (anti-)lepton and proton in the above elastic process. 

Regarding the inelastic soft-photon bremsstrahlung process, $k^\mu=(E_{\gamma^*},{\bf k})$ is the four-momentum of the
emitted soft-photon. In this situation, we either have the four momentum transfer to be $Q^\mu$ or 
$q^{\prime\mu}=(Q-k)^\mu$, depending on whether the photon is radiated by the (anti-)lepton or proton, respectively. 
Since we are only concerned with the undetectable soft-photon emission, the real photon emission amplitudes are 
evaluated in the limit $k\to 0$. One crucial assumption is that the soft-photon emission does not alter the elastic 
kinematics. This implies that the four-momentum transfer for the soft-photon bremsstrahlung process $q^{\prime\mu}$ 
is practically indistinguishable from its elastic counterpart $Q^\mu$, e.g., 
Refs.~\cite{Yennie:1961ad,Mo:1968cg,Vanderhaeghen:2000ws,Maximon:1969nw,Maximon:2000hm,Tsai:1961zz}. 

It is notable that in HB$\chi$PT, by virtue of re-parametrization-invariance, the proton's initial- and final-state 
four-momenta can be expressed as $P_p^\mu = M v^\mu + p_p^\mu $ and ${P}^{\prime\mu}_p = M v^\mu + {p}^{\prime\mu}_p$, 
where $p^\mu_p$ and $p^{\prime \mu}_p$ are the corresponding {\it residual} four-momenta. In the {\it lab.}-frame we 
therefore have $p_p=(0,{\bf 0})$, which implies that $v \cdot p_p =0$ and 
$v\cdot p^\prime_p = v\cdot Q=E-E^\prime=-Q^2/(2M)>0$. (For a  $t$-channel process, we always have $Q^2<0$.)  Finally, 
we mention that the standard form of the heavy proton propagator used in HB$\chi$PT, accurate up to chiral order 
$\nu=2$ [i.e., ${\mathcal O}(1/M^2)$], is given by
\begin{eqnarray*} 
iS^{(p)}_{\rm full}(l) \!&=&\! iS^{(\nu=0)}(l) + iS^{(\nu=1)}(l) + iS^{(\nu=2)}(l)\,,
\end{eqnarray*} 
where
\begin{eqnarray}
iS^{(0)}(l) \!&=&\! \frac{i}{v\cdot l+i0}\,,
\nonumber\\
iS^{(1)}(l) \!&=&\! \frac{i}{2M}\!\left[1-\frac{l^2}{(v\cdot l +i0)^2}\right]\,, \quad \text{and}
\nonumber\\
iS^{(2)}(l) \!&=&\! \frac{i}{4M^2}\left[\frac{(v \cdot l)^3-l^2 (v \cdot l)}{(v\cdot l +i0)^2}\right]
+ {\mathcal O}\left(\frac{1}{M^{3}}\right)\,. \qquad\, 
\label{eq:p_prop}
\end{eqnarray} 
Here, $l^\mu$ denotes the generic residual off-shell proton momentum. As already pointed out in the introduction, an 
important subtlety arises with the choice of the proton's off-shell momentum as $l^\mu\to Q^\mu$. Introducing the term 
$v\cdot Q\sim {\mathcal O}(1/M)$ in the propagator denominators results in ${\mathcal O}(M)$ chiral enhancements for 
both the LO and NLO components. As shown in the following analysis, the use of the proton propagator of 
Eq.~\eqref{eq:p_prop} leads to a {\it modified ordering} of a certain class of Feynman amplitudes as prescribed by the 
standard HB$\chi$PT power-counting scheme. For an alternative heavy-baryon propagator description, we refer the reader
to, e.g., Ref.~\cite{Hanhart2007u}. 

\subsection{The Born or OPE process} 
\label{Sec:II-a)}
The {\it lab.}-frame unpolarized differential cross section for the elastic OPE/Born (anti-)lepton-proton scattering 
process is given by
\begin{eqnarray}
\left[{\rm d}\sigma_{el}\right]_\gamma \!&=&\! \frac{(2\pi)^4\delta^4\left(p+p_p-p^\prime-p^\prime_p\right)}{4M|{\bf p}|}
\nonumber\\
&&\times\,\frac{{\rm d}^3{\bf p}^{\prime}}{(2\pi)^3 2E^\prime}\frac{{\rm d}^3{\bf P}^{\prime}_p}{(2\pi)^3 2E^\prime_p}\, 
\frac{1}{4}\sum_{\rm spins}|\mathcal{M}_\gamma|^2\,,
\end{eqnarray} 
where the total OPE amplitude can be expressed as the following sum of Born amplitudes of different chiral powers, 
written as 
\begin{eqnarray}
{\mathcal M}_\gamma={\mathcal M}^{(0)}_\gamma + {\mathcal M}^{(1)}_\gamma + {\mathcal M}^{(2)}_\gamma + \cdots\,\,\,,
\end{eqnarray}
where the superscripts denote the different chiral orders. In particular, the amplitude ${\mathcal M}^{(2)}_\gamma$ 
includes the leading hadronic corrections arising from the proton's low-energy structure at NNLO in the chiral 
power-counting. The OPE amplitude arising from the LO chiral Lagrangian is given by 
\begin{eqnarray}
\mathcal{M}_{\gamma}^{(0)} = -\frac{e^2}{Q^2} \left[\Bar{u}(p^\prime)\gamma^\mu u(p)\right]\, 
\Big[\chi^\dag (p_p^\prime) v_\mu \chi (p_p)\Big]\,,
\label{eq:M0}
\end{eqnarray} 
where $\chi(p_p)$ and $\chi^\dagger(p^\prime_p)$ denote the initial- and final-state two-component Pauli-spinors for 
the non-relativistic proton, and $u(p)$ and $\bar{u}(p^\prime)$ are the corresponding Dirac spinors for the 
relativistic leptons. The OPE amplitudes due to insertions of NLO and NNLO interactions from 
$\mathcal{L}_{\pi N}^{(1,2)}$ are given by 
\begin{widetext}
\begin{eqnarray}
\mathcal{M}^{(1)}_\gamma \!&=&\! -\,\frac{e^2}{2 M Q^2}[{\bar u}(p^\prime)\gamma^\mu\,u(p)]\,\bigg[\chi^\dagger(p_p^\prime)
\Big\{(p_p + p_p^\prime)_\mu - v_\mu v \cdot (p_p + p_p^\prime) + (2+\kappa_s+\kappa_v)[S_\mu,S\cdot Q]\Big\}\, \chi(p_p)\bigg]\,,
\label{eq:M1}
\end{eqnarray}
and 
\begin{eqnarray} 
\mathcal{M}^{(2)}_{\gamma} \!&=&\! -\, \frac{e^2}{8M^2Q^2}[{\bar u}_l(p^\prime)\gamma^\mu u_l(p)]
\bigg[\chi^\dagger(p_p^\prime)\, \Big\{\left(2(v\cdot Q)^2-Q^2\right)v_\mu - (v\cdot Q)Q_\mu\Big\} \,\chi(p_p)\bigg]
\nonumber\\
&&\!\! -\, \frac{e^2}{ Q^2}[{\bar u}_l(p^\prime)\gamma^\mu u_l(p)]\,
\left[\chi^\dagger(p_p^\prime)\,{\cal V}^{(2)}_\mu \,\chi(p_p)\right]\,,
\label{eq:M2}
\end{eqnarray}
respectively. The proton's hadronic structure effectively renormalizes the proton-photon NNLO vertex {\it via} the
term~\cite{Talukdar:2020aui} 
\begin{eqnarray}
{\cal V}^{(2)}_\mu \!&=&\!   (F^p_1-1)v_\mu+\frac{1}{2M}\Bigg\{(F^p_1-1)\left(Q_\mu+\frac{Q^2}{2M}v_\mu \right)  
+2(F^p_1+F^p_2-1-\kappa_p)\left[S_\mu,S\cdot Q\right]\,\Bigg\}
\nonumber\\
&&\hspace{1.65cm} -\,\frac{Q^2}{8M^2}(F^p_1-2F^p_2-1)v_\mu+{\mathcal O}\left(\frac{1}{M^3}\right)\,. 
\label{eq:V2}
\end{eqnarray} 
\end{widetext}
This expression tacitly incorporates the contributions from UV-divergent pion-loops and counter-terms, as shown in 
Fig.~\ref{fig:pi_loop}, parameterizing the proton's hadronic structure {\it via} the Dirac and Pauli form factors 
$F^p_{1,2}$. At very low-$Q^2$ values, the form factors allow a Taylor expansion in terms of the mean-squared radii, 
$\langle r_{1,2}^2 \rangle \sim {\mathcal O}(1/\Lambda^2_\chi) \sim {\mathcal O}(1/M^2)$, and the anomalous magnetic
moment of the  proton, $\kappa_p=(\kappa_v+ \kappa_s)/2=1.795$~\cite{{Bernard1998}}:
\begin{figure}
\centering
\includegraphics[width=1\linewidth]{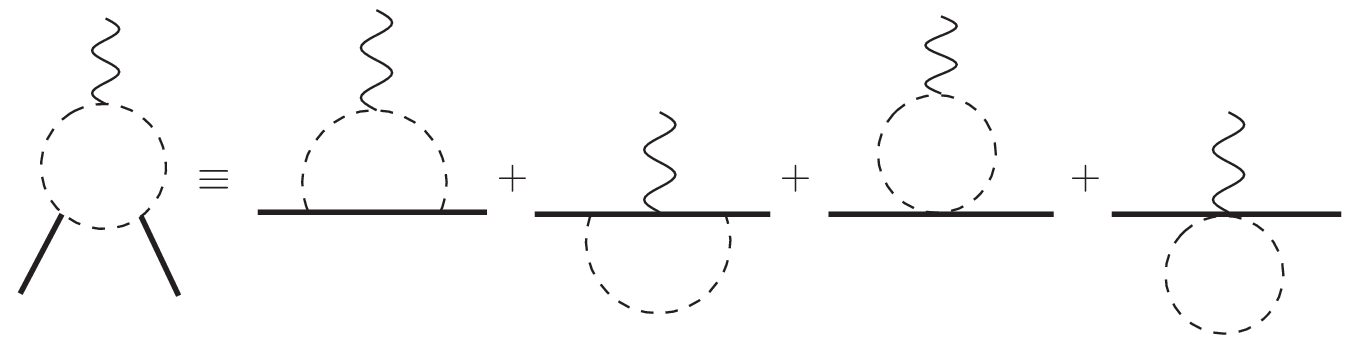}
    \caption{Pion-loops and counter-terms at NNLO in HB$\chi$PT with LO interactions at the proton's
             vertices contributing to its form factors. The solid, dashed, and wiggly lines denote 
             the proton, pion, and photon propagators, respectively.}
\label{fig:pi_loop}
\end{figure}
\begin{eqnarray}
F^p_{1,2}(Q^2) \!&=&\! 1 + \frac{Q^2}{6} \langle r_{1,2}^2\rangle + {\mathcal O}\left(\frac{1}{M^3}\right) \,.
\label{eq:Taylor}
\end{eqnarray}
However, to ${\mathcal O}(1/M^2)$ we only need the contribution from the mean-squared $\langle r_{1}^2\rangle$, 
which in turn could be related to the mean-squared charge or electric radius of the proton up to terms of 
${\mathcal O}(1/M^3)$, where the {\it root-mean squared} (rms) radius is given by
\begin{eqnarray}
r_p \equiv {\langle r_{E}^2\rangle}^{1/2} \approx \sqrt{\langle r_{1}^2\rangle + \frac{3\kappa_p}{2M^2}}\,.
\label{eq:r_p}
\end{eqnarray}
Thus, the full chirally corrected elastic Born differential cross section up-to-and-including NNLO can be 
expressed in the following form:
\begin{eqnarray} 
\left[\frac{{\rm d}\sigma_{el}(Q^2)}{{\rm d} \Omega^\prime_l}\right]_\gamma \!&=&\! 
\left[\frac{{\rm d}\sigma_{el}(Q^2)}{{\rm d} \Omega^\prime_l}\right]_0 \left\{1+\delta^{(2)}_\chi(Q^2, r_p) \right\}\,,\qquad
\end{eqnarray}
where the LO Born contribution is given by~\cite{Talukdar:2020aui}
\begin{eqnarray}
\left[\frac{{\rm d}\sigma_{el}(Q^2)}{{\rm d}\Omega^\prime_l}\right]_0 \!&=&\! 
\frac{\alpha^2 |{\bf p}^\prime|}{Q^2 |{\bf p}\,|}\left(1-\frac{Q^2}{4M^2}\right)\left[1+\frac{4EE^\prime}{Q^2}\right]\,. \quad\,
\label{eq:diff_LO}
\end{eqnarray}
Here the fractional contribution $\delta^{(2)}_{\chi}$ not only includes the {\it dynamical} hadronic 
contributions from the NNLO effective interaction vertex, Eq.~\eqref{eq:V2}, but also ${\mathcal O}(1/M^2)$ 
{\it kinematical recoil} corrections. The latter terms arise from the contributions of the amplitudes 
$\mathcal{M}^{(0,1)}_\gamma$, which do not receive contribution from the pion-loop effects, {\it vis-a-vis}, 
the proton's finite-size structure. Consequently, we have the following charge-even result for the hadronic
vertex corrections, as borrowed from the work of Talukdar {\it et al}.~\cite{Talukdar:2020aui}:
\begin{widetext}
\begin{eqnarray}
\delta^{(2)}_{\chi}(Q^2,r_p) = \frac{Q^2}{3}\left[r^2_p -\frac{3\kappa_p}{2M^2}\right] 
+ \frac{Q^2}{4M^2}\Bigg[1+2\kappa_p
+(1+\kappa_p)^2\left(\frac{Q^2+4m_l^2- 4E^2}{Q^2+4E^2}\right) \Bigg] + \mathcal{O}\left(\frac{1}{M^3}\right)\,.
\label{delta_chi}
\end{eqnarray}
For additional details regarding the calculation of the proton form factors $F^p_{1,2}$, we refer the reader 
to the work of Ref.~\cite{Bernard1998}.

\subsection{The soft-photon bremsstrahlung process} 
\label{Sec:II-b} 
The {\it lab.}-frame differential cross section for the single photon emission bremsstrahlung process is given by
\begin{eqnarray}
\left[{\rm d}\sigma_{br}\right]_{\gamma\gamma^*} =\frac{(2\pi)^4\delta^4\left(p+p_p-p^\prime-p^\prime_p-k\right)}{4M|{\bf p}|} 
\frac{{\rm d}^3{\bf p}^{\prime}}{(2\pi)^3 2E^\prime} \frac{{\rm d}^3{\bf p}^{\prime}_p}{(2\pi)^32E^\prime_p} 
\frac{{\rm d}^3{\bf k}^{\prime}}{(2\pi)^3 2E_{\gamma^*}}\, \frac{1}{4}\sum_{\rm spins}|\mathcal{M}_{\gamma\gamma^*}|^2\,,
\label{eq:dsigma_brem}
\end{eqnarray}   
\end{widetext} 
where $\mathcal{M}_{\gamma\gamma^*}$ denotes the sum of all possible chiral order amplitudes in the Lorentz
gauge, which can {\it potentially} contribute to the single photon emission cross section up-to-and-including
NLO  [i.e., ${\mathcal O}(\alpha^3/M)$] in HB$\chi$PT. These include amplitudes up to N${}^4$LO, whose 
general, frame-independent expressions -- presented without approximation -- are compiled in Appendix A. 
However, in this section, we will present only the subset of the non-vanishing amplitudes specific to the 
{\it lab.}-frame that pertains to a single soft-photon ($\gamma^*_{\rm soft}$) emission with energy within the
range, $0\leq E_{\gamma^*}\lesssim\Delta_{\gamma^*}$, which we will refer to as the {\it soft-photon limit} 
(SPL) of the bremsstrahlung amplitudes. It is important to emphasize that the SPL \underline{must not} be 
confused with the {\it soft-photon approximation} (SPA), which is commonly employed in TPE calculations -- see,
for example, our recent TPE analysis in Ref.~\cite{Goswami:2025yky}.  

\subsubsection{LO SPL bremsstrahlung amplitudes} 
\begin{figure*}[tbp]
\centering
\includegraphics[width=0.65\linewidth]{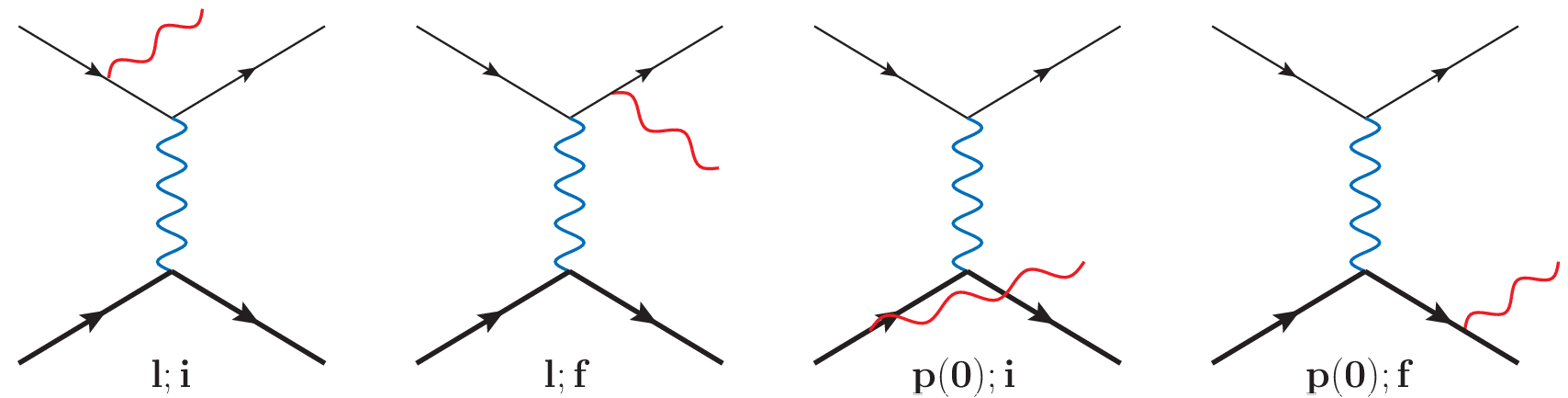}
    \caption{Photon bremsstrahlung diagrams at LO [i.e., ${\mathcal O}(e^3)$], contributing to the 
             radiative corrections of the $\ell$-p elastic scattering cross section 
             up-to-and-including NLO [i.e., ${\mathcal O}(\alpha^3/M)$] in HB$\chi$PT. The thin, 
             thick, and wiggly lines denote the propagators for the lepton, proton, and photon 
             (color online: red for the soft bremsstrahlung photon and blue for the exchanged 
             off-shell photon), respectively. The nomenclature of the individual diagrams is based
             on whether the radiated photon originates from the initial-state ``i” or final-state 
             ``f” lepton ``l” or proton ``p”, where the label ``(0)" indicates a LO 
             proton-radiating vertex. }
\label{fig:LO}
\end{figure*}
The four LO [i.e., ${\mathcal O}(e^3)$] amplitudes [cf. Eqs.~\eqref{eq:LO_1}-\eqref{eq:LO_4}] for the 
soft-photon bremsstrahlung processes, as displayed in Fig.~\ref{fig:LO}, are given by
\begin{eqnarray}
\label{LO_1}
\mathcal{M}^{[l;\,{\rm i}]}_{\gamma\gamma^*} \,&\stackrel{\gamma^*_{\rm soft}}{\leadsto}&\,
e\mathcal{M}^{(0)}_\gamma\left(\frac{p\cdot\varepsilon^*} {p\cdot k}\right)\,,
\\
\label{LO_2}
\mathcal{M}^{[l;\,{\rm f}]}_{\gamma\gamma^*} \,&\stackrel{\gamma^*_{\rm soft}}{\leadsto}&\,
 -\,e\mathcal{M}^{(0)}_\gamma\left(\frac{p^\prime\cdot\varepsilon^*} {p^\prime\cdot k}\right)\,,
\\
\label{LO_3}
\mathcal{M}^{[p(0);\,{\rm i}]}_{\gamma\gamma^*} \,&\stackrel{\gamma^*_{\rm soft}}{\leadsto}&\, 
-\,e\mathcal{M}^{(0)}_\gamma\left(\frac{v \cdot \varepsilon^*}{v\cdot k}\right)\,, \quad \text{and}
\\
\label{LO_4}
\mathcal{M}^{[p(0);\,{\rm f}]}_{\gamma\gamma^*} \,&\stackrel{\gamma^*_{\rm soft}}{\leadsto}&\,
e\mathcal{M}^{(0)}_\gamma\left(\frac{v \cdot \varepsilon^*}{v\cdot Q}\right)\,,
\end{eqnarray}
where the polarization vector of the emitted photon is $\varepsilon^{*}_\mu$. Here, we use the notation 
$\stackrel{\gamma^*_{\rm soft}}{\leadsto}$ to denote that these expressions are valid only in SPL. The 
contractions of the photon momentum $k^\mu$ with the external momenta $p^\mu$ and $p^{\prime \mu}$ in 
the numerators of the amplitudes are effectively set to zero, i.e., $k\cdot p\sim k\cdot p^\prime\sim 0$.
The resulting amplitudes are consistent with the well-known 
{\it Low’s soft-photon theorem}~\cite{Low:1958sn}, which asserts that in the soft limit, the 
bremsstrahlung amplitudes are proportional to the LO Born amplitude $\mathcal{M}^{(0)}_\gamma$, 
Eq.~\eqref{eq:M0}. The amplitudes, $\mathcal{M}^{[l;\,{\rm i/f}]}_{\gamma\gamma^*}$ and 
$\mathcal{M}^{[p(0);\,{\rm i/f}]}_{\gamma\gamma^*}$, respectively, represent the 
initial-state/final-state radiating lepton and proton, with LO proton-photon vertices. There are two key
points to be emphasized here:
\begin{enumerate}
\item 
In contrast to the Coulomb gauge used by Talukdar \textit{et al.}~\cite{Talukdar:2020aui}, the Lorentz 
gauge yields nonvanishing LO bremsstrahlung amplitudes proportional to $v\cdot \varepsilon^*$ -- 
specifically $\mathcal{M}^{[p(0);\,{\rm i}]}_{\gamma\gamma^*}$ and 
$\mathcal{M}^{[p(0);\,{\rm f}]}_{\gamma\gamma^*}$ -- in which the bremsstrahlung photon is emitted by 
the proton. These amplitudes, therefore, contribute to the NLO [i.e., ${\mathcal O}(\alpha^3/M)$] cross 
section.
\item 
Each of the two LO amplitudes, ${\mathcal M}^{[p(0);\,{\rm i}]}_{\gamma\gamma^*}$ and 
${\mathcal M}^{[p(0),\,{\rm f}]}_{\gamma\gamma^*}$, involves a ${\mathcal O}(M^0)$ proton propagator
($iS^{(0)}$) insertion characterized by the denominators, $-v\cdot k+i0$ and $v\cdot Q+i0$, respectively
[see Eq.~\eqref{eq:p_prop}]. As $v\cdot Q=-Q^2 /(2M)$, the latter amplitude corresponds to the 
\underline{final-state} proton radiation, and since the proton propagator scales as $\mathcal{O}(M)$, it
implies that the amplitude is chirally \underline{enhanced}. In contrast, the former amplitude with the 
initial-state radiating proton scales as expected, i.e., $\mathcal{O}(M^0)$. This modifies the naive 
hierarchy of ordering of Feynman diagrams as mandated by using chiral counting, enhancing the class of 
\underline{all} diagrams with the \underline{final-state} radiating proton by $\mathcal{O}(M)$, as 
compared to those with the initial-state radiating proton. A similar enhancement is also manifested in 
the subleading diagrams with an ${\mathcal O}(1/M)$ propagator ($iS^{(1)}$) insertion in the 
\underline{final-state} (however, not pertinent to $iS^{(\nu\geq 2)}$ insertions), as we subsequently 
demonstrate. 
\end{enumerate}

\subsubsection{NLO SPL bremsstrahlung amplitudes} 
\begin{figure*}[tbp]
\centering
\includegraphics[width=0.64\linewidth]{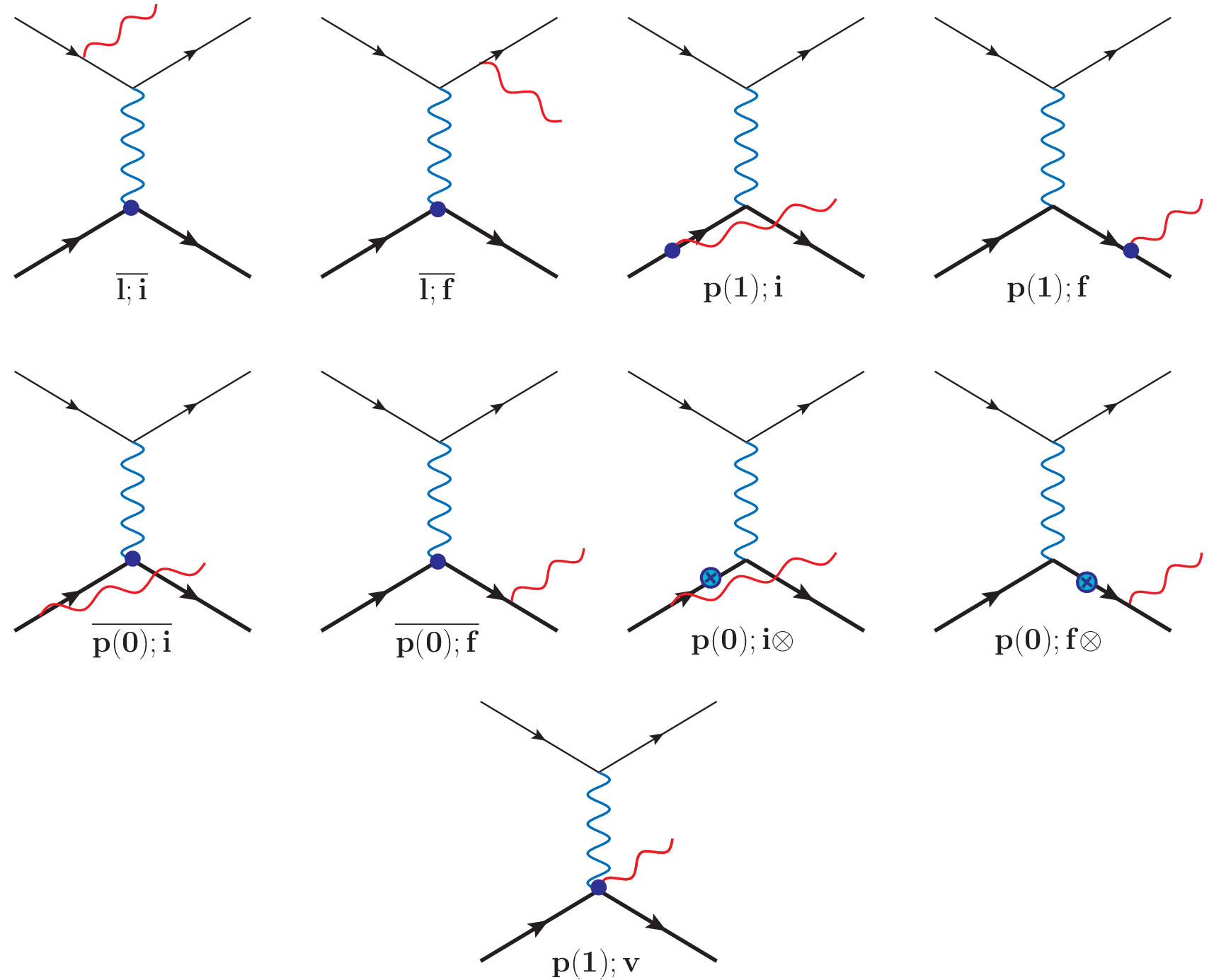}
    \caption{Photon bremsstrahlung diagrams at NLO [i.e., ${\mathcal O}(e^3/M)$] with insertions
             of NLO proton-photon interaction vertex (filled blob $\bullet$) or ${\mathcal O}(1/M)$ 
             proton propagator (crossed circle $\otimes$) insertions, contributing to the radiative 
             corrections for $\ell$-p elastic scattering cross section up-to-and-including NLO 
             [i.e., ${\mathcal O}(\alpha^3/M)$] in HB$\chi$PT. The thin, thick, and wiggly lines 
             denote the propagators for the lepton, proton, and photon (color online: red for the 
             soft bremsstrahlung photon and blue for the exchanged off-shell photon), 
             respectively. The nomenclature of the individual diagrams is based on whether the 
             radiated photon originates from the initial-state ``i” or final-state ``f” lepton 
             ``l” or proton ``p”. The labels ``(0)” and ``(1)” indicate LO and NLO proton-radiating
             vertices, respectively, while the overline signifies an NLO exchange-photon proton 
             vertex insertion.} 
\label{fig:NLO}
\end{figure*}

There are nine NLO [i.e., ${\mathcal O} (e^3/M)$] bremsstrahlung amplitudes [see Eqs.~\eqref{eq:NLO_5} - 
\eqref{eq:NLO_13}], as displayed in Fig.~\ref{fig:NLO}. In SPL, all but one decompose to yield either the
LO Born amplitude $\mathcal{M}^{(0)}_{\gamma}$, Eq.~\eqref{eq:M0}, or the amplitude,
\begin{eqnarray}
\mathcal{N}^{(1)}_\gamma = -\,\frac{e^2}{2 M Q^2}\left[\bar{u}(p^\prime) \gamma^\mu u(p) \right] 
\left[\chi(p_p^\prime)^\dagger Q_{\mu}\chi(p_p)\right]\,, 
\label{eq:NLO_N}
\end{eqnarray}
which in essence, forms a part of the full NLO Born amplitude, Eq.~\eqref{eq:M1}. The resulting 
amplitudes are given by
\begin{eqnarray}
\label{NLO_1}
\mathcal{M}^{\left[\,\overline{l;{\rm i}} \,\right]}_{\gamma\gamma^*} \,&\stackrel{\gamma^*_{\rm soft}}{\leadsto}&\! 
e\mathcal{N}^{(1)}_\gamma\left(\frac{p\cdot\varepsilon^*}{p\cdot k}\right) 
\nonumber\\
&& +\, \frac{e}{2M}\mathcal{M}^{(0)}_\gamma\left(\frac{p\cdot\varepsilon^*}{p\cdot k}\right)(v\cdot k - v\cdot Q)\,,\quad\, 
\\
\label{NLO_2}
\mathcal{M}^{\left[\, \overline{l;{\rm f}} \, \right]}_{\gamma\gamma^*} \,&\stackrel{\gamma^*_{\rm soft}}{\leadsto}&\! 
-\,e\mathcal{N}^{(1)}_\gamma\left(\frac{p^\prime\cdot\varepsilon^*}{p^\prime\cdot k}\right) 
\nonumber\\
&& -\, \frac{e}{2M}\mathcal{M}^{(0)}_\gamma
\left(\frac{p^\prime\cdot\varepsilon^*}{p^\prime\cdot k}\right)(v\cdot k - v\cdot Q)\,, \qquad\,
\\
\label{NLO_3}
\mathcal{M}^{\left[p(1);{\rm i}\right]}_{\gamma\gamma^*} \,&\stackrel{\gamma^*_{\rm soft}}{\leadsto}&\! 
\frac{-e}{2M}\mathcal{M}^{(0)}_\gamma\left[v\cdot \varepsilon^* - \frac{k\cdot\varepsilon^*}{v\cdot k}\right]\,,\quad\,
\\
\label{NLO_4}
\mathcal{M}^{\left[p(1);{\rm f}\right]}_{\gamma\gamma^*} \,&\stackrel{\gamma^*_{\rm soft}}{\leadsto}&\! 
\frac{e}{2M}\frac{\mathcal{M}^{(0)}_\gamma}{v\cdot Q} \Big[(2Q-k)\cdot\varepsilon^* 
\nonumber\\
&& \hspace{1.6cm} -\, [v\cdot(2Q-k)](v\cdot\varepsilon^*)\Big]\,,
\\
\label{NLO_5}
\mathcal{M}^{\left[\,\overline{p(0);{\rm i}}\,\right]}_{\gamma\gamma^*} \,&\stackrel{\gamma^*_{\rm soft}}{\leadsto}&\! 
-\,e\mathcal{N}^{(1)}_\gamma\left(\frac{v\cdot\varepsilon^*}{v\cdot k}\right) 
\nonumber\\
&& +\, \frac{e}{M}\mathcal{M}^{(0)}_\gamma\!\!
\left[\frac{(v\cdot Q)}{2(v\cdot k)}-1\right](v\cdot\varepsilon^*)\,,
\\
\label{NLO_6}
\mathcal{M}^{\left[\,\overline{p(0);{\rm f}}\,\right]}_{\gamma\gamma^*} \,&\stackrel{\gamma^*_{\rm soft}}{\leadsto}&\! 
e\mathcal{N}^{(1)}_\gamma\!\left(\frac{v\cdot\varepsilon^*}{v\cdot Q}\right) 
- \frac{e}{2M}\mathcal{M}^{(0)}_\gamma(v\cdot\varepsilon^*)\,,\quad\,\,
\end{eqnarray}
\begin{eqnarray}
\label{NLO_7}
\mathcal{M}^{\left[p(0);{\rm i}\otimes\right]}_{\gamma\gamma^*} \,&\stackrel{\gamma^*_{\rm soft}}{\leadsto}&\! 
\frac{e}{2M}\mathcal{M}^{(0)}_\gamma(v\cdot\varepsilon^*)\,, 
\\
\label{NLO_8}
\mathcal{M}^{\left[p(0);{\rm f}\otimes\right]}_{\gamma\gamma^*} \,&\stackrel{\gamma^*_{\rm soft}}{\leadsto}&\! 
\frac{e}{2M}\mathcal{M}^{(0)}_\gamma\left[1-\frac{Q^2}{(v\cdot Q)^2}\right](v\cdot\varepsilon^*)\,,\qquad\,
\end{eqnarray}
and
\begin{eqnarray}
\label{NLO_9}
\mathcal{M}^{\left[p(1);{\rm v}\right]}_{\gamma\gamma^*} \,&\stackrel{\gamma^*_{\rm soft}}{\leadsto}&\! 
\frac{e}{M}\mathcal{M}^{(0)}_\gamma(v\cdot\varepsilon^*) 
\\
&& +\, \frac{e^3}{M Q^2}\!\left[\Bar{u}(p^\prime)\gamma^\mu u(p)\right] 
\Big[\chi^\dagger(p^\prime_p) \varepsilon^*_\mu \chi(p_p)\Big]\,.
\nonumber
\end{eqnarray} 
A notable feature of the Born-like amplitude ${\mathcal N}^{(1)}_{\gamma}$ is that its interference with
the LO Born amplitude ${\mathcal M}^{(0)}_{\gamma}$ vanishes identically, i.e., 
\begin{eqnarray}
\sum\limits_{\rm spins}\left[{\mathcal M}^{(0)*}_{\gamma}{\mathcal N}^{(1)}_{\gamma}\right] = 0 \,,
\end{eqnarray}
leading to a considerable simplification in evaluating the NLO contributions to the cross sections. As 
shown in the Fig.~\ref{fig:NLO}, the amplitudes 
$\mathcal{M}^{\left[\,\overline{l;{\rm i/f}}\,\right]}_{\gamma\gamma^*}$ and 
$\mathcal{M}^{\left[\,\overline{p(0);{\rm i/f}}\,\right]}_{\gamma\gamma^*}$, respectively, represent the 
initial-state/final-state radiating lepton and proton processes, each having an insertion of NLO 
exchange-photon proton vertex. The amplitudes 
$\mathcal{M}^{\left[p(1);{\rm i/f} \right]}_{\gamma\gamma^*}$ and 
$\mathcal{M}^{\left[p(0);{\rm i/f}\otimes\right]}_{\gamma\gamma^*}$, respectively, represent the 
initial-state/final-state radiating proton processes with an insertion of an NLO proton-photon emission
vertex or an ${\mathcal O}(1/M)$ component of the proton's propagator ($iS^{(1)}$), as in 
Eq.~\ref{eq:p_prop}. The ninth amplitude $\mathcal{M}^{\left[p(1);{\rm v}\right]}_{\gamma\gamma^*}$ is 
the bremsstrahlung amplitude containing an insertion of the NLO seagull proton-photon vertex, which does
not factorize into one of the Born amplitudes~\cite{Talukdar:2020aui}. As previously noted, unlike the 
Coulomb gauge evaluation in Ref.~\cite{Talukdar:2020aui}, all terms in the amplitudes proportional to 
$v\cdot\varepsilon\neq 0$, remain non-vanishing in our Lorentz gauge analysis. Furthermore, as stated 
before, the presence of the term $v\cdot Q$ in the denominators of the LO and NLO components of the 
\underline{final-state} proton propagators ($iS^{(0,1)}$) in the three scattering amplitudes -- 
$\mathcal{M}^{\left[p(1);{\rm f}\right]}_{\gamma\gamma^*}$, 
$\mathcal{M}^{\left[\,\overline{p(0);{\rm f}}\,\right]}_{\gamma\gamma^*}$, and 
$\mathcal{M}^{\left[p(0);{\rm f}\otimes\right]}_{\gamma\gamma^*}$ -- modifies their naive ordering, 
kinematically enhancing them so that they effectively contribute as LO [i.e., ${\mathcal O}(M^0)$], or 
even higher [i.e., ${\mathcal O}(M)$]. It is precisely these \underline{final-state} propagator terms 
that motivate us to include contributions from diagrams with higher chiral order components -- formally 
appearing at NNLO and N${}^3$LO -- which could potentially contribute to the ${\mathcal O}(\alpha^3/M)$ 
NLO cross section. Such higher-order amplitudes are discussed next.

\subsubsection{NNLO SPL bremsstrahlung amplitudes} 
\begin{figure*}
\centering
\includegraphics[width=0.65\linewidth]{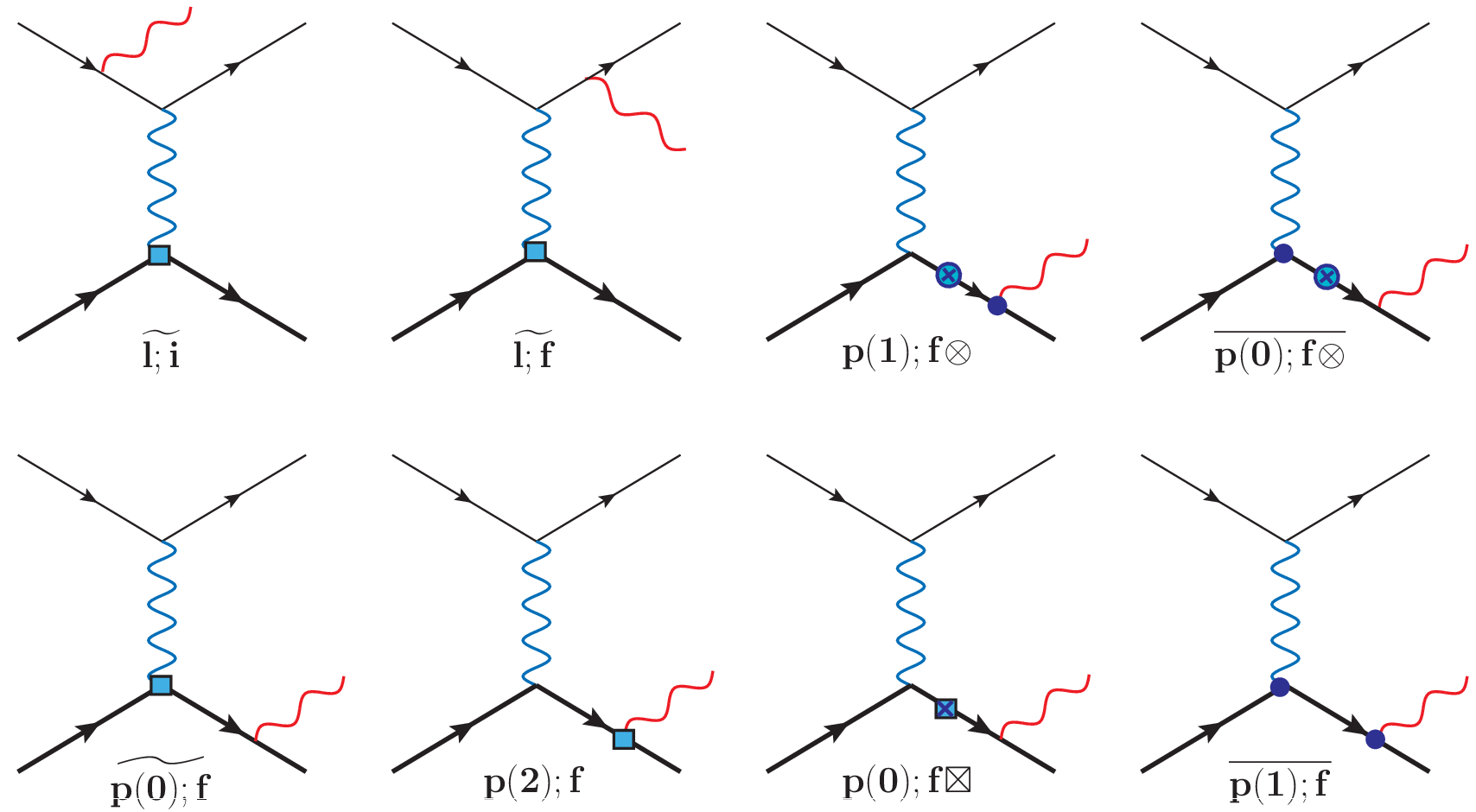}
    \caption{Chirally enhanced photon bremsstrahlung diagrams of NNLO [i.e., ${\mathcal O}(e^3/M^2)$]
             with insertions of NLO (filled blob $\bullet$) and NNLO (filled box $\square$) 
             proton-photon interaction vertices, and ${\mathcal O}(1/M)$ (crossed circle $\otimes$) 
             and ${\mathcal O}(1/M^2)$ (crossed box $\boxtimes$) proton propagator components, 
             contributing to the radiative corrections for the $\ell$-p elastic scattering cross 
             section up-to-and-including NLO [i.e., ${\mathcal O}(\alpha^3/M)$] in HB$\chi$PT. The 
             thin, thick, and wiggly lines denote the propagators for the lepton, proton, and photon 
             (color online: red for the soft bremsstrahlung photon and blue for the exchanged 
             off-shell photon), respectively. The nomenclature of the individual diagrams is based on
             whether the radiated photon originates from the initial-state ``i” or final-state ``f” 
             lepton ``(l)” or proton ``p”. The labels “(0)”, “(1)”, and “(2)” denote LO, NLO, and NNLO
             proton-radiating vertices, respectively, while the overline and tilde signify NLO and 
             NNLO exchange-photon proton vertex insertions, respectively. Topologies involving 
             initial-state proton radiation are not chirally enhanced, and hence, contribute beyond 
             NLO in the cross section; they are therefore deferred to Fig.~\ref{fig:NNLO_extra} in 
             Appendix A.}
\label{fig:NNLO}
\end{figure*}
At NNLO two types of dynamical contributions constitute the bremsstrahlung matrix elements: (i) the
${\mathcal O}(e^3/M^2)$ recoil-radiative corrections, and (ii) the ${\mathcal O}(e^3/\Lambda^2_\chi)$ 
hadronic corrections. The former contributions arise from insertions of subleading-order proton-photon
interaction vertices and proton propagator component insertions, while the latter arise from 
pion-loops and counterterms that determine the proton rms charge radius contributions (cf. 
Fig.~\ref{fig:pi_loop}). Below, we explore each of the two distinct types of contributions. 

\noindent $\bullet$ {\it Recoil-radiative corrections}: As detailed in Appendix A, these consist 
primarily of fifteen bremsstrahlung amplitudes that are formally expected to contribute at NNLO [see 
Eqs.~\eqref{eq:NNLO_14} - \eqref{eq:NNLO_28}]. However, as shown in Fig.~\ref{fig:NNLO}, only eight of 
these amplitudes, Eqs.~\eqref{eq:NNLO_14} - \eqref{eq:NNLO_21}, survive in SPL and contribute to the 
bremsstrahlung cross section at our intended NLO accuracy, namely, ${\mathcal O}(\alpha^3/M)$. The rest
of the amplitudes, Eqs.~\eqref{eq:NNLO_22} - \eqref{eq:NNLO_28}, either vanish or are neglected as they
generate higher-order terms. In SPL, these eight amplitudes either decompose to yield the LO Born 
amplitude $\mathcal{M}^{(0)}_{\gamma}$, Eq.~\eqref{eq:M0}, or the Born-like amplitude, 
$\mathcal{N}^{(1)}_\gamma$, Eq.~\eqref{eq:NLO_N}:
\begin{eqnarray}
\label{N2LO_1}
\mathcal{M}^{\left[\widetilde{\,l;{\rm i}\,}\right]}_{\gamma\gamma^*} \,&\stackrel{\gamma^*_{\rm soft}}{\leadsto}&\, 
\frac{-e}{4M}\mathcal{N}^{(1)}_\gamma\left(\frac{p\cdot\varepsilon^*}{p\cdot k}\right)[v\cdot(Q-k)] 
\nonumber\\
&& +\, \frac{e}{8M^2}\mathcal{M}^{(0)}_\gamma\left(\frac{p\cdot\varepsilon^*}{p\cdot k}\right) 
\nonumber\\
&& \times\, \Big[2[v\cdot(Q-k)]^2 - (Q-k)^2\Big]\,,
\\
\label{N2LO_2}
\mathcal{M}^{\left[\widetilde{\,l;{\rm f}}\right]\,}_{\gamma\gamma^*} \,&\stackrel{\gamma^*_{\rm soft}}{\leadsto}&\, 
\frac{e}{4M}\mathcal{N}^{(1)}_\gamma\left(\frac{p^\prime\cdot\varepsilon^*}{p^\prime\cdot k}\right)[v\cdot(Q-k)] 
\nonumber\\
&& -\, \frac{e}{8M^2}\mathcal{M}^{(0)}_\gamma\left(\frac{p^\prime\cdot\varepsilon^*}{p^\prime\cdot k}\right) 
\nonumber\\
&& \times\, \Big[2[v\cdot(Q-k)]^2 - (Q-k)^2\Big]\,,
\\
\label{N2LO_3}
\mathcal{M}^{\left[\,\overline{p(1);{\rm f}}\,\right]}_{\gamma\gamma^*} \,&\stackrel{\gamma^*_{\rm soft}}{\leadsto}&\, 
\frac{e}{2M} \left(\frac{\mathcal{N}^{(1)}_\gamma}{v\cdot Q} - \frac{\mathcal{M}^{(0)}_\gamma}{2M}\right) \Big[(2Q-k)\cdot\varepsilon^*
\nonumber\\
&& -\, (v\cdot\varepsilon^*)[v\cdot(2Q-k)]\Big]\,,  
\end{eqnarray}
\begin{eqnarray}
\label{N2LO_4}
\mathcal{M}^{\left[\,\widetilde{p(0);{\rm f}}\,\right]}_{\gamma\gamma^*} \,&\stackrel{\gamma^*_{\rm soft}}{\leadsto}&\, 
\frac{-e}{4M}\mathcal{N}^{(1)}_\gamma(v\cdot\varepsilon^*) 
\nonumber\\
&& -\, \frac{eQ^2}{8M^2}\mathcal{M}^{(0)}_\gamma \!\left(\frac{v\cdot\varepsilon^*}{v\cdot Q}\right) 
+  \mathcal{O}\left(\frac{1}{M^3}\right), \qquad\,
\\
\label{N2LO_5}
\mathcal{M}^{\left[p(2);{\rm f}\right]}_{\gamma\gamma^*} \,&\stackrel{\gamma^*_{\rm soft}}{\leadsto}&\, 
\frac{e}{8M^2}\frac{\mathcal{M}^{(0)}_\gamma}{v\cdot Q} 
\Big[2(v\cdot k)^2(v\cdot\varepsilon^*) 
\nonumber\\
&& \hspace{1.9cm} -\, (v\cdot k)(\varepsilon^*\cdot k)\Big]\,,
\\
\label{N2LO_6}
\mathcal{M}^{\left[p(0);{\rm f}\boxtimes\right]}_{\gamma\gamma^*} &\stackrel{\gamma^*_{\rm soft}}{\leadsto}& 
\frac{-eQ^2}{4M^2}\mathcal{M}^{(0)}_\gamma \!\left(\frac{v\cdot \varepsilon^*}{v\cdot Q}\right)
+ \mathcal{O}\left(\frac{1}{M^3}\right)\,,\,\, 
\\
\label{N2LO_7}
\mathcal{M}^{\left[p(1);{\rm f}\otimes\right]}_{\gamma\gamma^*} &\stackrel{\gamma^*_{\rm soft}}{\leadsto}& 
\frac{e}{4M^2}\mathcal{M}^{(0)}_\gamma\left(1 - \frac{Q^2}{(v\cdot Q)^2}\right) 
\nonumber\\
&& \!\! \times\, \Big[(2Q-k)\!\cdot\!\varepsilon^* - (v\!\cdot\!\varepsilon^*)[v\cdot(2Q-k)]\Big]\,,
\nonumber\\
\end{eqnarray}
and
\begin{eqnarray}
\label{N2LO_8}
\mathcal{M}^{\left[\,\overline{p(0);{\rm f}\otimes}\,\right]}_{\gamma\gamma^*} &\stackrel{\gamma^*_{\rm soft}}{\leadsto}& 
\frac{e}{2M}\left(\mathcal{N}^{(1)}_\gamma - \frac{v\cdot Q}{2M}\mathcal{M}^{(0)}_\gamma\right) 
\nonumber\\
&& \times\, \left(1 - \frac{Q^2}{(v\cdot Q)^2}\right)(v\cdot\varepsilon^*)\,. 
\end{eqnarray}
Above, we include all Feynman diagrams with proton-photon interaction vertices and propagator insertions 
up to $\mathcal{O}(1/M^2)$, capturing all possible topologies necessary to retain amplitude terms of the
same order, such that they will contribute to the NLO cross section of ${\mathcal O}(\alpha^3/M)$. As 
shown in Fig.~\ref{fig:NNLO}, the amplitude 
$\mathcal{M}^{\left[\widetilde{l;{\rm i/f}}\right]}_{\gamma\gamma^*}$ represents the 
initial-state/final-state radiating lepton process with the insertion of an NNLO proton-photon vertex. 
The amplitudes, $\mathcal{M}^{\left[\,\widetilde{p(0);{\rm f}}\, \right]}_{\gamma\gamma^*}$ and 
$\mathcal{M}^{\left[p(2);{\rm f}\right]}_{\gamma\gamma^*}$, represent diagrams with the LO 
\underline{final-state} proton propagator with NNLO proton-photon vertices, while the amplitude 
$\mathcal{M}^{\left[p(0); {\rm f}\boxtimes\right]}_{\gamma\gamma^*}$ has a ${\mathcal O}(1/M^2)$ 
\underline{final-state} proton propagator ($iS^{(2)}$) insertion. Finally, the amplitudes, 
$\mathcal{M}^{\left[p(1); {\rm f}\otimes\right]}_{\gamma\gamma^*}$, 
$\mathcal{M}^{\left[\,\overline{p(0);{\rm f}\otimes}\,\right]}_{\gamma\gamma^*}$, and 
$\mathcal{M}^{\left[ \,\overline{p(1);{\rm f}}\, \right]}_{\gamma\gamma^*}$, represent topologies with 
\underline{final-state} radiating proton, having up to two insertions of NLO components. It is again 
important to recognize that diagrams involving the \underline{final-state} proton propagators, although
classified as higher-order contributions [i.e., $\mathcal{O}(1/M^2)$], are chirally \underline{enhanced} 
by the factor $(v\cdot Q)^{-1}=-2M/Q^2$ appearing in the proton's propagator denominator. As indicated
earlier, this enhancement modifies their naive chiral ordering. Furthermore, it is notable that the 
genuine $\mathcal{O}(1/M^2)$ amplitudes, such as 
$\mathcal{M}^{\left[\widetilde{l;{\rm i/f}}\right]}_{\gamma\gamma^*}$, are also needed here so that their
interference with the chirally \underline{enhanced} LO amplitude 
$\mathcal{M}^{[p(0);\,{\rm f}]}_{\gamma\gamma^*}\sim \mathcal{O}(M)$, yields non-vanishing contribution 
to the ${\mathcal O}(\alpha^3/M)$ NLO cross section. In contrast, the seven other commensurate 
$\mathcal{O}(1/M^2)$ amplitudes, involving either the initial-state proton propagators [see 
Eqs.~\eqref{eq:NNLO_22} - \eqref{eq:NNLO_27}] or the NNLO seagull vertex, Eq.~\eqref{eq:NNLO_28} [cf. 
Fig.~\ref{fig:NNLO_extra} in appendix A), yield no contribution to the NLO cross section; they vanish 
upon integrating the soft-photon bremsstrahlung energies over the range 
$0\leq E_\gamma \leq \Delta_{\gamma^*}$. 

\begin{figure*}[tbp]
\centering
\includegraphics[width=0.65\linewidth]{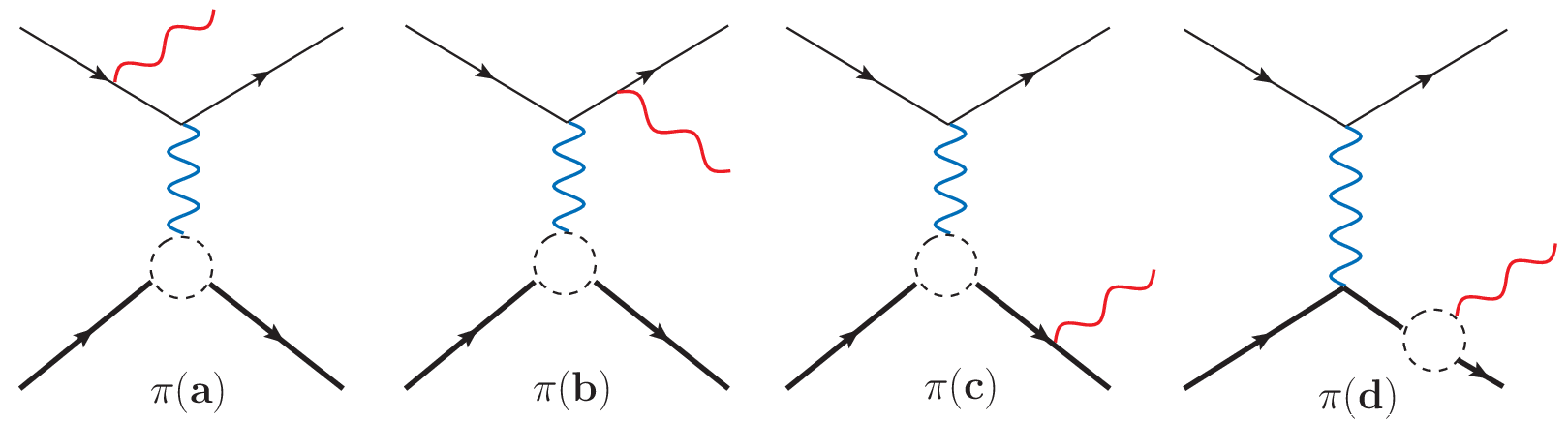}
    \caption{Photon bremsstrahlung diagrams at NNLO [i.e., ${\mathcal O}(e^3/\Lambda^2_\chi)$, with
             $\Lambda_\chi \sim M \approx 1$~GeV/$c$], contributing to the radiative corrections for 
             $\ell$-p elastic scattering cross section up-to-and-including NLO [i.e., 
             ${\mathcal O}(\alpha^3/M)$] in HB$\chi$PT. The thin, thick, and wiggly lines denote the
             propagators for the lepton, proton, and photon (color online: red for the real  
             soft bremsstrahlung photon and blue for the exchanged off-shell photon), respectively. 
             The dashed circular blob denotes the insertion of proton form factors (rms radius), 
             incorporated {\it via} renormalized pion-loops shown in Fig.~\ref{fig:pi_loop}. The 
             diagrams $\pi(c)$ and $\pi(d)$ are chirally enhanced compared to the first two diagrams.
             Topologies involving initial-state proton radiation are not enhanced, and hence, 
             contribute beyond NLO in the cross section; they are therefore deferred to 
             Fig.~\ref{fig:NNLO_extra} in Appendix A.}
\label{fig:NNLO_loop}
\end{figure*}
\noindent $\bullet$ {\it Hadronic corrections}: A total of six possible NNLO bremsstrahlung amplitudes, 
Eqs.~\eqref{eq:NNLO_29} - \eqref{eq:NNLO_34}, are given in Appendix A. They effectively renormalize the 
proton's NNLO recoil correction vertices {\it via} the hadronic interaction ${\cal V}^{(2)}_\mu$ [see 
Eqs.~\eqref{eq:M2} and \eqref{eq:V2}]. These terms entail UV-divergent pion-loops [cf. 
Fig.~\ref{fig:pi_loop}] as well as their counter-terms, and contribute to the proton's form factors 
$F^p_{1,2}$. In the works of Refs.~\cite{Bernard:1995dp} and \cite{Bernard1998}, the form factors were 
evaluated to NNLO accuracy in HB$\chi$PT, which in turn yield the proton's rms charge radius $r_p$ that
scales as ${\mathcal O}(1/M^2)$, Eq.~\eqref{eq:r_p}. In SPL, only the first four amplitudes 
${\mathcal M}^{\rm \left[\pi(a),...,\pi(d)\right]}_{\gamma\gamma^*}$, Eqs.~\eqref{eq:NNLO_29} - 
\eqref{eq:NNLO_32}, contribute to non-vanishing terms for the NLO bremsstrahlung cross section. The 
corresponding Feynman diagrams are displayed in Fig.~\ref{fig:NNLO_loop}. In the same vein as some of 
the genuine NNLO recoil-radiative amplitudes, the pion-loop amplitudes 
${\mathcal M}^{\rm \left[\pi(g),\pi(h)\right]}_{\gamma\gamma^*}$, Eqs.~\eqref{eq:NNLO_33} and 
\eqref{eq:NNLO_34} (cf. Fig.~\ref{fig:NNLO_extra}) do not yield any contributions to the NLO cross
section. The resulting SPL amplitudes read as follows:
\begin{eqnarray}
\label{eq:Mpi_a}
\mathcal{M}^{\left[\rm \pi (a)\right]}_{\gamma\gamma^*} &\stackrel{\gamma^*_{\rm soft}}{\leadsto}&  
e\mathcal{M}^{(0)}_\gamma \left(\frac{p\cdot\varepsilon^*}{p\cdot k}\right) {\mathcal G}^{p}_{12}(Q^2)
\nonumber\\
&& +\, 2e \mathcal{N}^{(1)}_\gamma \left(\frac{p\cdot\varepsilon^*}{p\cdot k}\right)\left[F^p_1(Q^2)-1\right] 
\nonumber\\
&& +\, \mathcal{O}\left(\frac{1}{M^3}\right),
\\
\mathcal{M}^{\left[\rm \pi (b)\right]}_{\gamma\gamma^*}  &\stackrel{\gamma^*_{\rm soft}}{\leadsto}&  
-\, e\mathcal{M}^{(0)}_\gamma \left(\frac{p^\prime\cdot\varepsilon^*}{p^\prime\cdot k}\right) {\mathcal G}^{p}_{12}(Q^2)
\nonumber\\
&& -\, 2e\mathcal{N}^{(1)}_\gamma \left(\frac{p^\prime\cdot\varepsilon^*}{p^\prime\cdot k}\right)\left[F^p_1(Q^2)-1\right] 
\nonumber\\
&& +\, \mathcal{O}\left(\frac{1}{M^3}\right),
\\
\mathcal{M}^{\left[\rm \pi (c)\right]}_{\gamma\gamma^*} &\stackrel{\gamma^*_{\rm soft}}{\leadsto}&  
-\, e{\mathcal M}^{(0)}_\gamma \left(\frac{v \cdot \varepsilon^*}{v\cdot Q}\right) \,{\mathcal G}^{p}_{12}(Q^2)
\nonumber\\
&& -\, 2e{\mathcal N}^{(1)}_\gamma \left(\frac{v\cdot\varepsilon^*}{v\cdot Q}\right)\left[F^p_1(Q^2)-1\right] 
\nonumber\\
&& +\, \mathcal{O}\left(\frac{1}{M^3}\right)\,,
\\
\nonumber
\end{eqnarray}
and
\begin{eqnarray}
\mathcal{M}^{\left[\rm \pi (d)\right]}_{\gamma\gamma^*} \!\!&\stackrel{\gamma^*_{\rm soft}}{\leadsto}&\!\!  
-\, e\mathcal{M}^{(0)}_\gamma\!\! 
\left(\frac{v\cdot\varepsilon^*}{v\cdot Q}\right) \,{\mathcal G}^{p}_{12}(Q^2)
\nonumber\\
&&\!\! -\, \frac{e}{M}\mathcal{M}^{(0)}_\gamma \!\! 
\left(\frac{Q\cdot\varepsilon^*}{v\cdot Q}\right)\!\!\left[F^p_1(Q^2)-1\right] \!+\! 
\mathcal{O}\!\left(\!\frac{1}{M^3}\!\right)\!,  
\nonumber\\ 
\label{eq:Mpi_d}
\end{eqnarray}
where 
\begin{eqnarray}
{\mathcal G}^{p}_{12}(Q^2) \!&=&\! F^p_1(Q^2)-1 + \frac{3Q^2}{8M^2}\left[F^p_1(Q^2)-1\right] 
\nonumber\\
&& \hspace{1.8cm} +\, \frac{Q^2}{4M^2}F^p_2(Q^2)\,,
\label{eq:G12}
\end{eqnarray}
is a function of the proton's form factors $F^p_{1,2}$. As evident from the Taylor expansion of the form
factors, Eq.~\eqref{eq:Taylor}, ${\mathcal G}^{p}_{12}\sim {\mathcal O}(1/M^2)$. Thus, we again find that
the amplitudes $\mathcal{M}^{\rm \left[\pi(c),\pi(d)\right]}_{\gamma\gamma^*}$ are \underline{enhanced} 
relative to $\mathcal{M}^{\rm \left[\pi(a),\pi(b)\right]}_{\gamma\gamma^*}$ by ${\mathcal O}(M)$, 
effectively rendering them NLO by the modified power-counting. 

\subsubsection{N${\,}^3$LO SPL bremsstrahlung amplitudes} 
\begin{figure*}[tbh]
\centering
\includegraphics[width=0.65\linewidth]{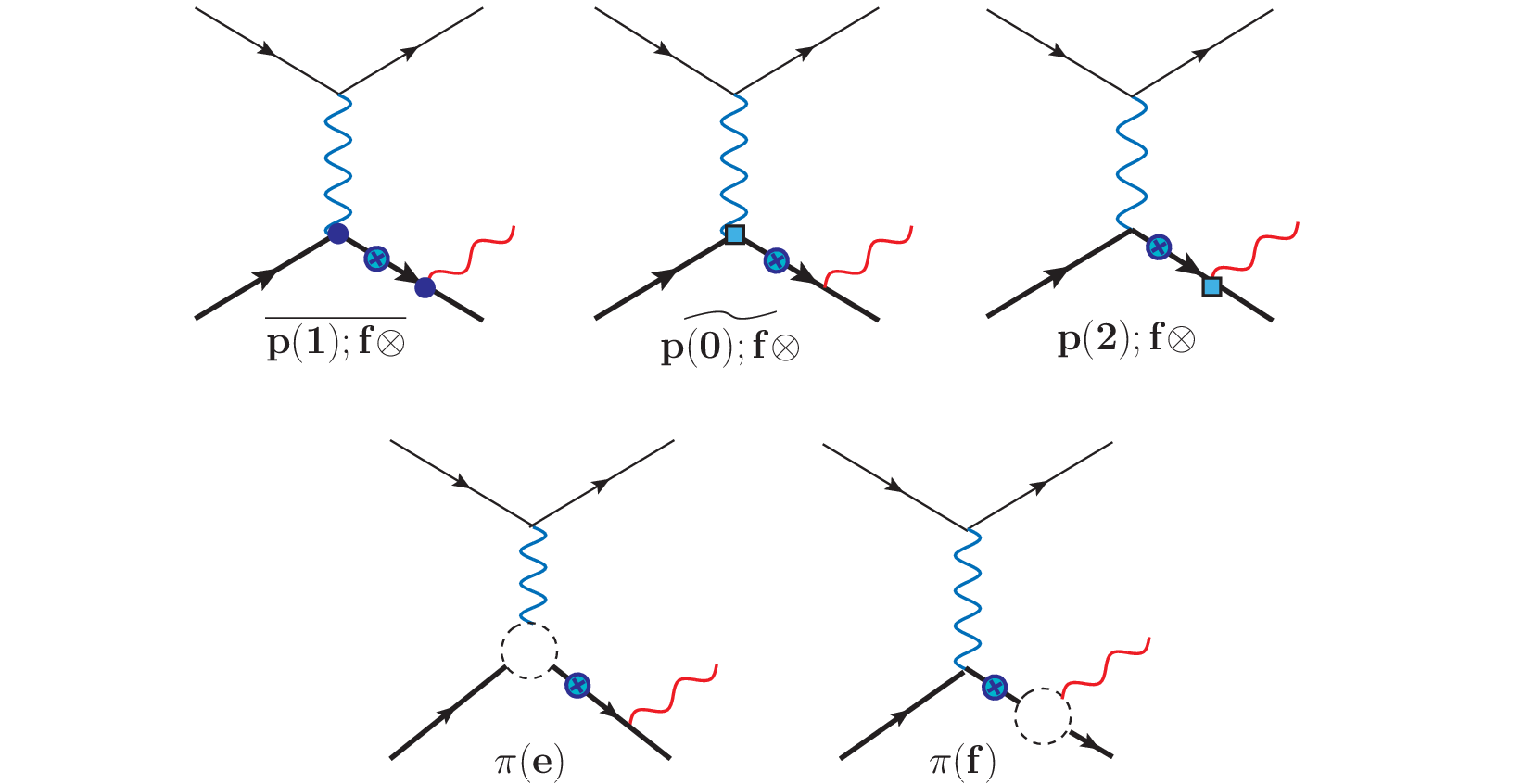}
    \caption{Chirally enhanced photon bremsstrahlung diagrams at N${}^3$LO [ie., 
             ${\mathcal O}(e^3/M^3)\sim {\mathcal O}\left(e^3/(M\Lambda^2_\chi)\right)$]. The insertions
             are: NLO (filled blob $\bullet$) and NNLO (filled box $\square$) proton-photon interaction 
             vertices, proton's form factors (dashed circular blob), and ${\mathcal O}(1/M)$ proton 
             propagator component (crossed circle $\otimes$), contributing to the radiative corrections 
             for $\ell$-p elastic scattering cross section up-to-and-including NLO [i.e., 
             ${\mathcal O}(\alpha^3/M)$] in HB$\chi$PT. The thin, thick, and wiggly lines denote 
             propagators for the lepton, proton, and photon (color online: red for the soft 
             bremsstrahlung photon and blue for the exchanged off-shell photon), respectively. The 
             nomenclature of the individual diagrams is based on whether the radiated photon originates 
             from the initial-state ``(i)” or final-state ``(f)” proton ``(p)”. Diagrams with form 
             factors have the symbol ``$\pi$", namely, ``$\pi(\rm e)$" and ``$\pi(\rm f)$”. Topologies
             involving the insertions of final-state proton propagator components and $\nu=3$ 
             chiral-order vertex, which do not contribute to the NLO cross section, are deferred to 
             Fig.~\ref{fig:N3LO_extra} in Appendix A. Whereas, Topologies involving initial-state proton
             radiation lack enhancement, contributing only at higher-orders, and are thus omitted. }
\label{fig:N3LO}
\end{figure*}
At N$^{3}$LO there are nine possible recoil-radiative amplitudes of ${\mathcal O}(e^3/M^3)$, 
Eqs.~\eqref{eq:N3LO_35} - \eqref{eq:N3LO_43}, involving NLO, NNLO and N${}^{3}$LO proton-photon 
interaction vertices, and  ${\mathcal O}(1/M)$ and ${\mathcal O}(1/M^2)$ proton propagator 
($iS^{(1,2)}$) insertions. In addition, we have five pion-loop amplitudes of 
${\mathcal O}(e^3/M\Lambda^2_\chi)$, Eqs.~\eqref{eq:N3LO_44} - \eqref{eq:N3LO_48}, involving the proton's
form factors. However, in SPL, only three of the former and two of the latter amplitudes survive to 
contribute to the ${\mathcal O}(\alpha^3/M)$ NLO cross section (cf. Fig.~\ref{fig:N3LO}). They read as 
follows:
\begin{eqnarray}
\mathcal{M}^{\left[\,\overline{p(1);{\rm f}\otimes}\,\right]}_{\gamma\gamma^*} 
&\stackrel{\gamma^*_{\rm soft}}{\leadsto}& 
\frac{e}{4M^2}\bigg(\mathcal{N}^{(1)}_\gamma - \frac{v\cdot Q}{2M}\mathcal{M}^{(0)}_\gamma\bigg)
\nonumber\\
&& \times\, \left(1 - \frac{Q^2}{(v\cdot Q)^2}\right) \Big[\varepsilon^*\cdot(2Q-k) 
\nonumber\\
&& -\, (v\cdot \varepsilon^*)[v\cdot(2Q-k)]\Big]\, ,
\label{N3LO_1}
\end{eqnarray}
\begin{eqnarray}
\mathcal{M}^{\left[\,\widetilde{p(0);{\rm f}\otimes}\,\right]}_{\gamma\gamma^*} 
&\stackrel{\gamma^*_{\rm soft}}{\leadsto}&
\frac{-e}{8M^2}\bigg(\mathcal{N}^{(1)}_\gamma(v\cdot Q) 
- \frac{1}{2M}\mathcal{M}^{(0)}_\gamma 
\nonumber\\
&& \hspace{1.2cm} \times\, \left[2(v\cdot Q)^2 - Q^2 \right]\bigg) 
\nonumber\\
&& \times\, \left(1-\frac{Q^2}{(v\cdot Q)^2}\right)(v\cdot\varepsilon^*)\,,
\label{N3LO_2}
\\
\nonumber\\
\mathcal{M}^{\left[p(2);{\rm f}\otimes\right]}_{\gamma\gamma^*} 
\,&\stackrel{\gamma^*_{\rm soft}}{\leadsto}&\, 
\frac{-e}{16M^3}\mathcal{M}^{(0)}_\gamma \frac{Q^2}{\left(v\cdot Q\right)^2} 
\Big[2(v\cdot k)^2(v\cdot\varepsilon^*) 
\nonumber\\
&& -\, (v\cdot k)(k\cdot\varepsilon^*)\Big] + \mathcal{O}\left(\frac{1}{M^3}\right),
\label{N3LO_3}
\\
\nonumber\\
\mathcal{M}^{\left[\rm \pi(e)\right]}_{\gamma\gamma^*} 
\,&\stackrel{\gamma^*_{\rm soft}}{\leadsto}&\, 
\frac{-e}{M}\bigg(\mathcal{N}^{(1)}_\gamma \left[F^p_1(Q^2)-1\right] 
\nonumber\\
&& \hspace{0.8cm} +\, \frac{1}{2}\mathcal{M}^{(0)}_\gamma {\mathcal G}^p_{12}(Q^2) \bigg) 
(v\cdot\varepsilon^*)
\nonumber\\
&& \times\, \left(1-\frac{Q^2}{(v\cdot Q)^2}\right) + \mathcal{O}\left(\frac{1}{M^3}\right),
\label{N3LO_4}
\\
\nonumber\\
\mathcal{M}^{\left[\rm \pi(f)\right]}_{\gamma\gamma^*} \,&\stackrel{\gamma^*_{\rm soft}}{\leadsto}&\, 
\frac{-e}{2M}\mathcal{M}^{(0)}_\gamma \bigg[\left[F^p_1(Q^2)-1\right](Q\cdot\varepsilon^*) 
\nonumber\\
&& \hspace{1.6cm} +\, {\mathcal G}^p_{12}(Q^2) (v\cdot\varepsilon^*)\bigg] 
\nonumber\\
&& \times\, \left(1-\frac{Q^2}{(v\cdot Q)^2}\right) + \mathcal{O}\left(\frac{1}{M^3}\right)\,,
\label{N3LO_5}
\end{eqnarray}
where the expression for ${\mathcal G}^p_{12}$ incorporating the proton's form factors was already given in 
Eq.~\eqref{eq:G12}. Here, the amplitude 
$\mathcal{M}^{\left[\,\overline{p(1);{\rm f}\otimes}\,\right]}$ has two NLO proton-photon vertices and one
${\mathcal O}(1/M)$ proton propagator ($iS^{(1)}$) insertion, while the amplitudes 
$\mathcal{M}^{\left[\,\widetilde{p(0);{\rm f}\otimes}\,\right]}_{\gamma\gamma^*}$ and 
$\mathcal{M}^{\left[p(2):{\rm f}\otimes\right]}_{\gamma\gamma^*}$ have one NNLO proton-photon vertex and 
${\mathcal O}(1/M)$ propagator ($iS^{(1)}$) insertions each. All the remaining chirally \underline{enhanced}
N$^3$LO amplitudes, except for $\mathcal{M}^{\left[\widehat{p(0);f}\right]}$, as shown in 
Fig~\ref{fig:N3LO_extra} of Appendix A vanish upon integrating over the soft-photon bremsstrahlung energies
in our NLO cross section calculation. As for the amplitude 
$\mathcal{M}^{\left[\widehat{p(0);f}\right]}_{\gamma\gamma^*}$, the $\mathcal{O}(M)$ enhancement by the 
\underline{final-state} proton propagator is equally compensated by the $v\cdot Q\sim {\mathcal O}(1/M)$ 
term stemming from the N${}^3$LO vertex, thereby restoring it to the N${}^3$LO chiral counting scaling, i.e.,
$\mathcal{O}(1/M^3)$. Thus, the amplitude contributes to the cross section only at $\mathcal{O}(1/M^2)$, i.e., 
beyond NLO. Moreover, it is worth noting that all commensurate amplitudes with insertions of 
${\mathcal O(1/M^{\nu})}$ ($\nu\geq2$) final-state proton propagator components, effectively scale as 
$iS^{(\nu)}(v\cdot Q)\sim {\mathcal O(1/M^{\nu-1})}$, and therefore, do not lead to any type of enhancements.

\subsubsection{N${}^{\,4}$LO SPL bremsstrahlung amplitudes} 
At N${}^4$LO there are two chirally \underline{enhanced} amplitudes, 
$\mathcal{M}^{[p(3);{\rm f}\otimes]}_{\gamma\gamma^*}$ and 
$\mathcal{M}^{\left[\widehat{p(0)\,;{\rm f}\otimes}\right]}_{\gamma\gamma^*}$, presented in 
Eqs.~\eqref{eq:N4LO_49} and \eqref{eq:N4LO_50}, respectively (cf. Fig.~\ref{fig:N4LO_extra} in  Appendix A). 
These amplitudes appear to scale as $\mathcal{O}(1/M^2)$, suggesting that they can potentially contribute to 
the ${\mathcal O}(\alpha^3/M)$ NLO cross section. However, akin to the N${}^3$LO diagrams shown in 
Fig~\ref{fig:N3LO_extra}, the amplitude $\mathcal{M}^{[p(3);{\rm f}\otimes]}_{\gamma\gamma^*}$ vanishes when
computing the cross section. As for the amplitude 
$\mathcal{M}^{\left[\widehat{p(0)\,;{\rm f}\otimes}\right]}_{\gamma\gamma^*}$, the expected $\mathcal{O}(1/M^4)$
scaling re-emerges, rendering it irrelevant at our intended ${\mathcal O}(\alpha^3/M)$ accuracy of NLO cross 
section.

\subsubsection{Sum of SPL bremsstrahlung amplitudes} 
Before evaluating the soft-photon bremsstrahlung cross section, we summarize the complete set of relevant 
amplitudes required to achieve accuracy up-to-and-including NLO [i.e., ${\mathcal O}(\alpha^3/M)$]. To this 
end, if $\widetilde{\mathcal M}_{\gamma\gamma^*}\subset {\mathcal M}_{\gamma\gamma^*}$, denotes the 
corresponding sum of the relevant subset of bremsstrahlung amplitudes up to ${\mathcal O}(e^3/M^3)$, needed at 
our working accuracy to evaluate the NLO cross section [cf. Eq.~\eqref{eq:dsigma_brem}], then the contributing
amplitudes LO through N${}^3$LO are consolidated as
\begin{eqnarray}
\widetilde{\mathcal M}_{\gamma\gamma^*} \!&=&\! \mathcal{M}_{\rm LO} + \mathcal{M}_{\rm NLO} 
+ \widetilde{\mathcal M}_{\rm NNLO} + \widetilde{\mathcal M}_{\rm N{}^3LO}\,,\quad\,
\end{eqnarray}
where the tildes indicate only those parts considered in this section that contribute non-vanishingly to our 
{\it lab.}-frame ${\mathcal O}(\alpha^3/M)$ cross section, as summarized below:
\begin{eqnarray}
\label{eq:M_LO}
\mathcal{M}_{\rm LO} \!&=&\! \mathcal{M}^{\left[l;{\rm i}\right]}_{\gamma\gamma^*} 
+\mathcal{M}^{\left[l;{\rm f}\right]}_{\gamma\gamma^*} 
+ \mathcal{M}^{\left[p(0);{\rm i}\right]}_{\gamma\gamma^*} 
+ \mathcal{M}^{\left[p(0);{\rm f}\right]}_{\gamma\gamma^*}\,, 
\\
\label{eq:M_NLO}
\mathcal{M}_{\rm NLO} \!&=&\! \mathcal{M}^{\left[\,\overline{l;{\rm i}}\,\right]}_{\gamma\gamma^*} 
+ \mathcal{M}^{\left[\,\overline{l;{\rm f}}\,\right]}_{\gamma\gamma^*} 
+ \mathcal{M}^{\left[p(1);{\rm i}\right]}_{\gamma\gamma^*} 
+ \mathcal{M}^{\left[p(1);{\rm f}\right]}_{\gamma\gamma^*} 
\nonumber\\
&& +\, \mathcal{M}^{\left[\,\overline{p(0);{\rm i}}\,\right]}_{\gamma\gamma^*} 
+ \mathcal{M}^{\left[\,\overline{p(0);{\rm f}}\,\right]}_{\gamma\gamma^*} 
+ \mathcal{M}^{\left[p(0);{\rm f}\otimes\right]}_{\gamma\gamma^*} 
\nonumber\\
&& +\, \mathcal{M}^{\left[p(0);{\rm i}\otimes\right]}_{\gamma\gamma^*} 
+ \mathcal{M}^{\left[p(1);{\rm v}\right]}_{\gamma\gamma^*} \,,
\\
\label{eq:M_NNLO}
\widetilde{\mathcal{M}}_{\rm NNLO} \!&=&\! \mathcal{M}^{\left[\widetilde{\,l;{\rm i}\,}\right]}_{\gamma\gamma^*} 
+ \mathcal{M}^{\left[\widetilde{\,l;{\rm f}\,}\right]}_{\gamma\gamma^*} 
+ \mathcal{M}^{\left[\,\overline{p(1);f}\,\right]}_{\gamma\gamma^*} 
+ \mathcal{M}^{\left[\,\widetilde{p(0);f}\,\right]}_{\gamma\gamma^*} 
\nonumber\\
&& +\, \mathcal{M}^{\left[p(2);f\right]}_{\gamma\gamma^*} 
+ \mathcal{M}^{\left[p(0);f\boxtimes\right]}_{\gamma\gamma^*} 
+\mathcal{M}^{\left[p(1);f \otimes\right]}_{\gamma\gamma^*} 
\nonumber\\
&& +\, \mathcal{M}^{\left[\,\overline{p(0);f \otimes}\,\right]}_{\gamma\gamma^*} 
+ \mathcal{M}^{\left[\rm \pi (a)\right]}_{\gamma\gamma^*} + \mathcal{M}^{\left[\rm \pi (b)\right]}_{\gamma\gamma^*} 
+ \mathcal{M}^{\left[\rm \pi (c)\right]}_{\gamma\gamma^*}
\nonumber\\
&& +\, \mathcal{M}^{\left[\rm \pi (d)\right]}_{\gamma\gamma^*} \,, \quad \text {and}
\\
\label{eq:M_N3LO}
\widetilde{\mathcal{M}}_{\rm N^3LO} \!&=&\! \mathcal{M}^{\left[\,\overline{p(1);{\rm f}\otimes}\,\right]}_{\gamma\gamma^*} 
+ \mathcal{M}^{\left[\,\widetilde{p(0);{\rm f}\otimes}\,\right]}_{\gamma\gamma^*} 
+ \mathcal{M}^{\left[p(2);{\rm f}\otimes\right]}_{\gamma\gamma^*}
\nonumber\\
&& +\, \mathcal{M}^{\left[\rm \pi (e)\right]}_{\gamma\gamma^*} + \mathcal{M}^{\left[\rm \pi (f)\right]}_{\gamma\gamma^*} \,.
\end{eqnarray}
Finally, we note that the total squared matrix element for the bremsstrahlung process (${\mathcal S}_{br}$) 
could be decomposed into distinct charge-even (${\mathcal S}^{(\nu)}_{\rm even}$) and charge-odd 
(${\mathcal S}^{(\nu)}_{\rm odd}$) chiral order ($\nu$) components based on lepton charge symmetry of the 
products of bremsstrahlung amplitudes, i.e.,
\begin{eqnarray}
{\mathcal S}_{br} \equiv \sum_{\rm spins}|\mathcal{M}_{\gamma\gamma^*}|^2 
= \sum^{3}_{\nu=0}\mathcal{S}^{(\nu)}_{\rm even} + \sum^{3}_{\nu=0}\mathcal{S}^{(\nu)}_{\rm odd} \,.
\end{eqnarray} 
As we subsequently demonstrate, at the desired level of accuracy, the matrix elements of chiral order 
components $S^{(\nu\geq 3)}_{\rm even}$ and $S^{(\nu\geq 4)}_{\rm odd}$, arising from the interference of the 
aforementioned amplitudes, do not contribute to the NLO cross section. Next, we separately determine these 
charge-even and charge-odd components.

\subsection{The charge-even bremsstrahlung component} 
\label{Sec:C} 
The component of the cross section for inelastic charge-even (anti-)lepton-proton bremsstrahlung process is 
obtained from the {\it direct-interference} (see introduction) product of Feynman diagrams in which the 
bremsstrahlung photon is emitted from either the (anti-)lepton or the proton legs of both graphs. As a 
result, the cross section remains symmetric under conjugation of the lepton’s charge. In particular, we 
evaluate the charge-even differential cross section, Eq.~\eqref{eq:dsigma_brem}, corresponding to the single
soft-photon emission process contributing to the lowest order QED corrections to the elastic $\ell^\pm$-p 
scattering up-to-and-including NLO accuracy [i.e., ${\mathcal O}(\alpha^3/M)$] in HB$\chi$PT. The only 
non-vanishing contributions to the charge-even results stem from the LO and NLO diagrams (cf. 
Figs.~\ref{fig:LO} and ~\ref{fig:NLO}). Apart from a few chirally \underline{enhanced} diagrams beyond NLO,
which contribute higher-order terms [i.e., ${\mathcal O}(\alpha^3/M^2)$] to the charge-even cross section, 
most yield vanishing contributions when integrated over the soft-photon emission energies within the range, 
$0\leq E_\gamma\leq \Delta_{\gamma^*}$. Their non-contribution relies on the fact that in dimensional 
regularization (DR) {\it scaleless} loop-integrals vanish identically, i.e., 
\begin{eqnarray}
\int\frac{{\rm d}^{d}{k}}{(2\pi)^d} \frac{(k^2)^m}{(v \cdot k+i0)^n} \xrightarrow{\mathrm{DR}} 0\,.
\label{dr_condition}
\end{eqnarray}
 
\subsubsection{$\nu=0$ charge-even contribution}
The charge-even cross section component for soft-photon bremsstrahlung at LO [i.e., 
${\mathcal O}(\alpha^3)$] is obtained from the following charge-even components of the squared matrix 
elements, stemming from the LO bremsstrahlung amplitudes (cf. Fig.~\ref{fig:LO}):
\begin{eqnarray}
{\mathcal S}^{(\nu=0)}_{\rm even}  \!&=&\!
\sum_{\rm spins}\left | \mathcal{M}_{\rm LO}\right |^2_{\rm even}
\nonumber\\
&=&\! \sum_{\rm spins} \bigg[ \left |\mathcal{M}^{\left[\,l;{\rm i}\,\right]}_{\gamma\gamma^*} \right |^2 
+ \left |\mathcal{M}^{\left[\,l;{\rm f}\,\right]}_{\gamma\gamma^*} \right |^2 
+ \left |\mathcal{M}^{\left[\,p(0);{\rm i}\,\right]}_{\gamma\gamma^*}\right |^2
\nonumber\\
&& \hspace{0.65cm} +\,\left |\mathcal{M}^{\left[\,p(0);{\rm f}\,\right]}_{\gamma\gamma^*} \right |^2 
+ 2\mathcal{R}e\Big( {\mathcal{M}^{\left[\,l;{\rm i}\,\right]}_{\gamma\gamma^*}}^\dagger 
\mathcal{M}^{\left[\,l;{\rm f}\,\right]}_{\gamma\gamma^*} 
\nonumber\\
&& \hspace{0.65cm} +\, {\mathcal{M}^{\left[\,p(0);{\rm i}\,\right]}_{\gamma\gamma^*}}^\dagger 
\mathcal{M}^{\left[\,p(0);{\rm f}\,\right]}_{\gamma\gamma^*} \Big )\bigg]\,, 
\label{eq:Se_LO}
\end{eqnarray}

After integrating over the phase-space, the LO differential cross section for charge-even bremsstrahlung 
corrections in SPL can be expressed as~\cite{Talukdar:2020aui},
\begin{eqnarray}
\left[\frac{{\rm d}\sigma^{\rm (LO;even)}_{br}(Q^2)}{{\rm d}\Omega^\prime_l} 
\right]^{(E_{\gamma^*}\leq \Delta_{\gamma^*})}_{\gamma\gamma^*} \hspace{-0.5cm} \!&=&\!\!
-\,\frac{\alpha}{2 \pi^2} \left[\frac{{\rm d}\sigma_{el}(Q^2)}{{\rm d}\Omega^\prime_l}\right]_{0} 
\nonumber\\
&& \times\, \Big[L_{\rm ii} + L_{\rm ff} - L_{\rm if} \Big]\,, \quad\,\,
\end{eqnarray} 
where $\left[\frac{{\rm d}\sigma_{el}(Q^2)}{\rm d \Omega^\prime_l}\right]_{0}$ denotes the LO Born/OPE 
differential cross section for the elastic lepton-proton scattering process, Eq.~\eqref{eq:diff_LO}. 
$L_{\rm ii}$, $L_{\rm ff}$, and $L_{\rm if}$ are IR-divergent integrals which results from the
integration over the soft-photon energy $0\leq E_{\gamma^*}\leq \Delta_{\gamma^*}$. Following the 
methodology adopted in 
Refs.~\cite{Mo:1968cg,Maximon:1969nw,Maximon:2000hm,Vanderhaeghen:2000ws,Bucoveanu:2018soy,Tsai:1961zz}, 
we boost to the so-called {\it S-frame} (the center-of-mass frame of the recoiling proton and emitted 
soft-photon), where such phase-space integrations are conveniently evaluated. This approach facilitates 
the extraction of the corresponding IR-finite contributions, $\tilde{L}_{\rm ii}$, $\tilde{L}_{\rm ff}$, 
and $\tilde{L}_{\rm if}$ using DR. However, here we quote the integration results borrowed
from Ref.~\cite{Talukdar:2020aui}, as expressed in {\it lab.}-frame coordinates (see Appendix B for 
details of the evaluations in the S-frame):
\begin{widetext}
\begin{eqnarray}
L_{\rm ii} \!&=&\! \pi\left[\frac{1}{|\varepsilon_{\rm IR}|}+\gamma_E - 
\ln\left(\frac{4\pi\mu^2}{m_l^2}\right)\right] +2\pi\widetilde{L}_{\rm ii}\,;
\qquad
\widetilde{L}_{\rm ii} = \frac{1}{2}\ln\left(\frac{4\Delta^{2}_{\gamma^*}E^2}{m_l^2 E^{\prime 2}}\right)
-\frac{1}{\beta}\ln\sqrt{\frac{1+\beta}{1-\beta}}\,,
\label{eq:Lii}
\\
L_{\rm ff}  \!&=&\! \pi\left[\frac{1}{|\varepsilon_{\rm IR}|}+\gamma_E 
- \ln\left(\frac{4\pi\mu^2}{m_l^2}\right)\right]+2\pi\widetilde{L}_{\rm ff}\,;
\qquad
\widetilde{L}_{\rm ff} = \frac{1}{2}\ln\left(\frac{4\Delta^{2}_{\gamma^*}E^2}{m_l^2E^{\prime 2}}\right)
-\frac{1}{\beta^\prime}\ln\sqrt{\frac{1+\beta^\prime}{1-\beta^\prime}}\,, \quad \text{and}
\label{eq:Lff}
\end{eqnarray}
\begin{eqnarray}
\label{eq:Lif}
L_{\rm if} \!&=&\! \pi\left[\frac{1}{|\varepsilon_{\rm IR}|}+\gamma_E 
- \ln\left(\frac{4\pi\mu^2}{m_l^2}\right)\right] \left(\frac{\nu^2_l+1}{\nu_l}\right) 
\ln\left[\frac{\nu_l+1}{\nu_l-1}\right]+2\pi\widetilde{L}_{\rm if}\,;
\nonumber\\
\widetilde{L}_{\rm if} \!&=&\! 
\frac{\nu^2_l+1}{2\nu_l}\Bigg[\ln\left(\frac{4\Delta^{2}_{\gamma^*}E^2}{m_l^2E^{\prime 2}}\right)
\ln\left[\frac{\nu_l+1}{\nu_l-1}\right]+\ln^2\sqrt{\frac{1+\beta^\prime}{1-\beta^\prime}} 
- \ln^2\sqrt{\frac{1+\beta}{1-\beta}} 
+ \text{Li}_2\left(1-\frac{\lambda_\nu E^\prime-E}{(1-\beta^\prime)E^\prime\xi_\nu}\right)
\nonumber\\
&& \hspace{1.15cm} +\,\text{Li}_2\left(1-\frac{\lambda_\nu E^\prime-E}{(1+\beta^\prime)E^\prime\xi_\nu}\right)
-\text{Li}_2\left(1-\frac{\lambda_\nu E^\prime-E}{(1-\beta)E \lambda_\nu\xi_\nu}\right)
-\text{Li}_2\left(1-\frac{\lambda_\nu E^\prime-E}{(1+\beta)E\lambda_\nu\xi_\nu}\right)\Bigg]\,, 
\end{eqnarray}
\begin{eqnarray*}
\text{where} \quad \nu_l(Q^2) = \sqrt{1 - \frac{4 m_l^2}{Q^2}}\,, \quad \xi_\nu(Q^2) = \frac{2 \nu_l}{(\nu_l + 1)(\nu_l -1)}\,, \quad 
\text{and} \quad  \lambda_\nu(Q^2)= \frac{\nu_l-1}{\nu_l+1}\,.
\end{eqnarray*}
\end{widetext} 
$\beta=|{\bf p}\,|/E$ and $\beta^\prime=|{{\bf p}^\prime}\,|/E^\prime$ denote the initial and final 
lepton velocities, $\gamma_E=0.577216...$ is the Euler-Mascheroni constant, and 
$\varepsilon_{IR}=(4-D)/2 < 0$. Furthermore, the well-known di-logarithm or Spence function that 
appears in Eq.~\eqref{eq:Lif} is defined by 
\begin{eqnarray}
{\rm Li}_2(z) = - \int_0^z {\rm d}t\, \frac{\ln(1-t)}{t}\,,\quad  \forall z\in {\mathbb C}\,.
\label{eq:Li2}
\end{eqnarray}
We note that in Eq.~\eqref{eq:Se_LO}, none of the terms in the squared matrix elements, 
$\left |\mathcal{M}^{\left[\,p(0);{\rm i}\,\right]}_{\gamma\gamma^*}
+\mathcal{M}^{\left[\,p(0);{\rm f}\,\right]}_{\gamma\gamma^*}\right |^2$, contribute to the cross 
section as they lead to scaleless integrals vanishing according to the condition given in 
Eq.~\eqref{dr_condition}. Thus, at LO, there is no contribution from amplitudes where the proton 
radiates a single soft-photon to the charge-even bremsstrahlung cross section. The corresponding 
fractional contribution to the charge-even differential cross section at LO is expressed 
as follows\footnote{Regarding our expression in Eq.~\eqref{eq:deltaLO_brem_even}, a few 
typographical errors were identified in the corresponding expression, Eq.~(67) in 
Ref.~\cite{Talukdar:2020aui}. These errors have been rectified in the recent SPA-based TPE work 
of Goswami {\it et al}.~\cite{Goswami:2025yky}.}:
\begin{widetext}
\begin{eqnarray}
\label{eq:deltaLO_brem_even}
\delta^{\rm (LO;even)}_{\gamma\gamma^*}(Q^2,\Delta_{\gamma^*}) \!&=&\! 
\left[\frac{{\rm d}\sigma^{\rm (LO;even)}_{br}(Q^2)}{{\rm d}\Omega^\prime_l} 
\right]^{(E_{\gamma^*}\leq \Delta_{\gamma^*})}_{\gamma\gamma^*}\!\!\!\!\bigg{/}
\left[\frac{{\rm d}\sigma_{el}(Q^2)}{{\rm d}\Omega^\prime_l} \right]_{0}
= \text{\bf IR}^{(0)}_{\gamma\gamma^*}(Q^2) - \frac{\alpha}{\pi} \left(\widetilde{L}_{\rm ii} 
+ \widetilde{L}_{\rm ff}-\widetilde{L}_{\rm if}\right)
\nonumber\\
&=&\! \text{\bf IR}^{(0)}_{\gamma\gamma^*}(Q^2) 
+ \frac{\alpha}{\pi}\Bigg[\left(\frac{\nu^2_l+1}{2\nu_l}\ln\left[\frac{\nu_l+1}{\nu_l-1}\right]-1\right) 
\ln{\left(\frac{4 \Delta^2_{\gamma^*}}{m^2_l}\right)}+\frac{2}{\beta}\ln\sqrt{\frac{1+\beta}{1-\beta}}
+ \left(\frac{\nu^2_l+1}{2\nu_l}\right)
\nonumber\\
&& \hspace{2.5cm} \times\, \Bigg\{\text{Li}_2\left(1-\frac{\lambda_\nu-1}{(1-\beta)\xi_\nu}\right) 
+ \text{Li}_2\left(1-\frac{\lambda_\nu-1}{(1+\beta)\xi_\nu}\right)
- \text{Li}_2\left(1-\frac{\lambda_\nu -1}{(1-\beta)\lambda_\nu\xi_\nu}\right)
\nonumber\\
&& \hspace{3.1cm} -\, \text{Li}_2\left(1-\frac{\lambda_\nu-1}{(1+\beta)\lambda_\nu\xi_\nu}\right)\Bigg\}\Bigg] 
+ \delta^{({\rm LO-}1/M;{\rm even})}_{\gamma\gamma^*}(Q^2,\Delta_{\gamma^*})\,,
\end{eqnarray} 
where the ${\mathcal O}(1/M)$ component of the charge-even LO result is given by
\begin{eqnarray}
\label{eq:deltaLO_NLO_brem_even}
\delta^{({\rm LO-}1/M;{\rm even})}_{\gamma\gamma^*}(Q^2,\Delta_{\gamma^*}) \!&=&\! 
\frac{\alpha Q^2}{2\pi ME\beta^2}\Bigg[1-2 \beta^ 2\left(\frac{\nu^2_l+1}{2\nu_l}\ln\left[\frac{\nu_l+1}{\nu_l-1}\right]-1\right)  
- \left\{\frac{1}{\beta}-\beta \left(\frac{\nu^2_l+\nu_l+1}{\nu_l}\right)\right\}\ln{\sqrt{\frac{1+\beta}{1-\beta}}} 
\nonumber\\
&& \hspace{1.6cm} +\, \beta\left(\frac{\nu^2_l+1}{2\nu_l}\right)\Bigg\{\frac{\lambda_\nu(1+\beta)-1}{\xi_\nu(1-\beta)-\lambda_\nu+1}\!
\ln{\left(\frac{\lambda_\nu-1}{(1-\beta)\xi_\nu}\right)} - \frac{\lambda_\nu(1-\beta)-1}{\xi_\nu(1+\beta)-\lambda_\nu+1}\! 
\nonumber\\
&& \hspace{4cm} \times\, \ln{\left(\frac{\lambda_\nu-1}{(1+\beta)\xi_\nu}\right)} 
- \frac{\lambda_\nu \beta}{(1+\beta)\lambda_\nu \xi_\nu-\lambda_\nu+1}
\ln{\left(\frac{\lambda_\nu-1}{(1+\beta)\xi_\nu \lambda_\nu}\right)} 
\nonumber\\
&&  \hspace{4cm} -\, \frac{\lambda_\nu \beta}{(1-\beta)\lambda_\nu \xi_\nu-\lambda_\nu+1}
\ln{\left(\frac{\lambda_\nu-1}{(1-\beta)\xi_\nu \lambda_\nu}\right)}\Bigg\}\Bigg]  
+ {\mathcal O}\left(\frac{1}{M^2}\right)\,,
\nonumber\\
\end{eqnarray} 
and the LO IR-divergent part of the charge-even soft-photon bremsstrahlung component is given as 
\begin{eqnarray}
\text{\bf IR}^{(0)}_{\gamma\gamma^*}(Q^2)
= \frac{\alpha}{\pi}\left[\frac{1}{|\varepsilon_{\rm IR}|}+\gamma_E-\ln\left(\frac{4\pi\mu^2}{m_l^2}\right)\right]
\left(\frac{\nu^2_l+1}{2\nu_l}\ln\left[\frac{\nu_l+1}{\nu_l-1}\right]-1\right)\,.
\label{eq:IR0_brem_even}
\end{eqnarray}
\end{widetext}
The same charge-even result also applies to anti-lepton scattering off the proton. Note that the above
expression is not genuinely LO in the strict sense of chiral power-counting, as they include
the ${\mathcal O}\left(1/M\right)$ kinematically suppressed terms from quantities such as $E^\prime$ 
and $\beta^\prime$, where $E^\prime=E-Q^2/(2M)$ and 
\begin{equation*}
\beta^\prime = \beta - \frac{Q^2}{2ME}\left(\beta-\frac{1}{\beta}\right) + {\mathcal O}\left(\frac{1}{M^2}\right)\,.
\end{equation*}
Such higher-order terms must be systematically accounted for to be consistent with the NLO 
contributions, which we subsequently examine. We further note that the above LO Lorentz gauge 
charge-even bremsstrahlung result is identical to the one obtained previously by Talukdar 
\textit{et al.}~\cite{Talukdar:2020aui} adopting the Coulomb gauge. The IR divergence of 
Eq.~\eqref{eq:IR0_brem_even} is specifically canceled when added with the virtual photon-loop 
corrections (attributed to the one-loop self-energy, vertex, and vacuum polarization corrections) to 
the elastic cross section in HB$\chi$PT. These virtual QED radiative corrections 
$\delta^{\rm (LO;even)}_{\gamma\gamma}$ were analytically computed in several earlier works
(see e.g., Refs.~\cite{Bucoveanu:2018soy,Talukdar:2020aui,Vanderhaeghen:2000ws}). Thus, we have the
following corrections to the charge-even elastic cross section arising from all the LO amplitudes 
involving the virtual photon-loops:
\begin{widetext}
\begin{eqnarray}
\delta^{\rm (LO;even)}_{\gamma\gamma}(Q^2) \!&=&\! 
\left[\frac{{\rm d}\sigma^{\rm (LO;even)}_{virt}(Q^2)}{{\rm d}\Omega^\prime_l}\right]_{\gamma\gamma}
\bigg{/} \left[\frac{{\rm d}\sigma_{el}(Q^2)}{{\rm d}\Omega^\prime_l} \right]_{0}
\nonumber\\
&=&\! 
\text{\bf IR}^{(0)}_{\gamma\gamma}(Q^2)+\frac{\alpha}{\pi}\Bigg[\frac{{\nu_l}^2+1}{4\nu_l}
\ln\left[\frac{{\nu_l}+1}{{\nu_l}-1}\right]\ln\left[\frac{{\nu_l}^2-1}{4{\nu_l}^2}\right]
+\frac{2{\nu_l}^2+1}{2{\nu_l}}\ln\left[\frac{{\nu_l}+1}{{\nu_l}-1}\right]-2 
\nonumber\\
&& \hspace{2.35cm} +\, \frac{{\nu_l}^2+1}{2{\nu_l}} \left\{\text{Li}_2\left(\frac{{\nu_l}+1}{2{\nu_l}}\right)
-\text{Li}_2\left(\frac{{\nu_l}-1}{2{\nu_l}}\right)\right\} 
+ \frac{1}{{\nu_l}}\left(\frac{2m^2_l}{Q^2+4 E^2}\right) \ln\left[\frac{{\nu_l}+1}{{\nu_l}-1}\right]
\nonumber\\
&& \hspace{2.35cm} +\!\!\!\! \sum_{f=e,\mu,\tau}\!\!\left\{\frac{2}{3}\!\left(\nu^2_f-\frac{8}{3}\right)
+\nu_f\!\left(\frac{3-\nu^2_f}{3}\right)\ln\left[\frac{\nu_f+1}{\nu_f-1}\right]\right\} 
- \frac{2}{3}\!\left(\nu^2_\pi+\frac{1}{3}\right) + \frac{\nu^3_\pi}{3}\ln\!\left[\frac{\nu_\pi+1}{\nu_\pi-1}\right]\Bigg] 
\nonumber\\
&& \hspace{1.6cm} -\, \frac{\alpha Q^2E}{\pi M}\left(\frac{2 m_l}{Q^2+4E^2}\right)^2 \frac{1}{\nu_l}
\ln\left[\frac{{\nu_l}+1}{{\nu_l}-1}\right] + {\mathcal O}\left(\frac{1}{M^2}\right)\,,
\label{eq:deltaLO_virt_even}
\end{eqnarray}
where, $\nu_{f,\pi}=\sqrt{1-4m_{f,\pi}^2/Q^2}$. We find that the IR-divergent component is exactly equal
in magnitude and opposite in sign to that arising from the charge-even bremsstrahlung correction: 
\begin{eqnarray}
\text{\bf IR}^{(0)}_{\gamma\gamma}(Q^2) = - \text{\bf IR}^{(0)}_{\gamma\gamma^*}(Q^2)
= -\,\frac{\alpha}{\pi}\left[\frac{1}{|\varepsilon_{\rm IR}|} + \gamma_E-\ln\left(\frac{4\pi\mu^2}{m^2_l}\right)\right] 
\left(\frac{\nu^2+1}{2\nu}\ln\left[\frac{\nu+1}{\nu-1}\right]-1\right) \,.
\label{eq:IR0_virt_even}
\end{eqnarray}
Following the consistent subtraction of infrared singularities, we obtain the total one-loop IR-finite
fractional radiative correction to the charge-even (anti-)lepton-proton elastic differential cross 
section stemming from all LO amplitudes~\cite{Talukdar:2020aui}: 
{\small
\begin{eqnarray}
\label{eq:deltaLO_rad_even}
\delta^{\rm (LO;even)}_{2\gamma} \hspace{-0.3cm}&&\hspace{-0.3cm} (Q^2,\Delta_{\gamma^*}) =
\left[\frac{{\rm d}\sigma^{\rm (LO;even)}_{el}(Q^2)}{{\rm d}\Omega^\prime_l}\right]_{2\gamma}
\bigg{/}\left[\frac{{\rm d}\sigma_{el}(Q^2)}{{\rm d}\Omega^\prime_l} \right]_{0}
\\
&=&\! \delta^{\rm (LO;even)}_{\gamma\gamma}(Q^2) + \delta^{\rm (LO;even)}_{\gamma\gamma^*}(Q^2,\Delta_{\gamma^*})
\nonumber\\
&=&\! \frac{\alpha}{\pi}\Bigg[\frac{\nu_l^2+1}{4\nu_l}
\ln{\left[\frac{\nu_l+1}{\nu_l-1}\right]}\ln{\left[\frac{\nu_l^2-1}{4\nu_l^2}\right]} 
- 2 + \frac{2\nu_l^2+1}{2\nu_l}\ln{\left[\frac{\nu_l+1}{\nu_l-1}\right]} 
- \frac{\nu_l^2+1}{2\nu_l}\bigg\{{\rm Li}_2\left(\frac{\nu_l-1}{2\nu_l}\right)
- {\rm Li}_2\left(\frac{\nu_l+1}{2\nu_l}\right)\bigg\}
\nonumber\\
&& \hspace{0.3cm} +\,\frac{1}{\nu_l}\left(\frac{2m_l^2}{Q^2+4E^2 }\right)\ln{\left[\frac{\nu_l+1}{\nu_l-1}\right]} 
+ \sum_{f=e,\mu,\tau} \left\{\frac{2}{3}\left(\nu_f^2-\frac{8}{3}\right) + \nu_f\left(\frac{3-\nu_f^2}{3}\right)
\ln{\left[\frac{\nu_f+1}{\nu_f-1}\right]}\right\} - \frac{2}{3}\left(\nu_{\pi}^2+\frac{1}{3}\right) 
\nonumber\\
&& \hspace{0.3cm} +\, \frac{\nu_{\pi}^2}{3}\ln{\left[\frac{\nu_{\pi}+1}{\nu_{\pi}-1}\right]}
+ \bigg\{\left(\frac{\nu_l^2+1}{2\nu_l}\right)\ln{\left[\frac{\nu_l+1}{\nu_l-1}\right]}-1\bigg\} 
\ln{\left(\frac{4\Delta^2_{\gamma^*}}{m_l^2}\right)} + \frac{2}{\beta}\ln{\sqrt{\frac{1+\beta}{1-\beta}}} 
+ \left(\frac{\nu_l^2+1}{2\nu_l}\right)
\nonumber\\
&& \hspace{0.3cm}  \times\, \left\{{\rm Li}_2\left(1+\frac{1-\lambda_\nu }{(1-\beta)\xi_\nu}\right) 
+ {\rm Li}_2\left(1+\frac{1-\lambda_\nu }{(1+\beta)\xi_\nu}\right)
- {\rm Li}_2\left(1+\frac{1-\lambda_\nu }{ (1-\beta)\lambda_\nu\xi_\nu}\right)
- {\rm Li}_2\left(1+\frac{1- \lambda_\nu }{(1+\beta)\lambda_\nu\xi_\nu}\right)\right\}\Bigg]
\nonumber\\
&&\!\!\! -\, \frac{\alpha Q^2}{2\pi ME\beta^2}\Bigg[-1 -2\beta^2 + \frac{\beta^2}{\nu_l}\left\{1+\nu^2_l 
+ 2\left(\frac{2m_l E\beta}{Q^2+4E^2}\right)^2\right\}\ln\left[\frac{\nu_l+1}{\nu_l-1}\right]
+ \left\{\frac{1}{\beta}-\beta \left(\frac{\nu^2_l+\nu_l+1}{\nu_l}\right)\right\}\ln{\sqrt{\frac{1+\beta}{1-\beta}}}
\nonumber\\
&& \hspace{1.7cm} -\, \beta\left(\frac{\nu^2_l+1}{2\nu_l}\right)\Bigg\{\frac{\lambda_\nu(1+\beta)-1}{\xi_\nu(1-\beta)-\lambda_\nu+1}
\ln{\left(\frac{\lambda_\nu-1}{(1-\beta)\xi_\nu}\right)} - \frac{\lambda_\nu(1-\beta)-1}{\xi_\nu(1+\beta)-\lambda_\nu+1} 
\ln{\left(\frac{\lambda_\nu-1}{(1+\beta)\xi_\nu}\right)} 
\nonumber\\
&& \hspace{1.7cm} -\, \frac{\lambda_\nu \beta}{(1+\beta)\lambda_\nu \xi_\nu-\lambda_\nu+1}
\ln{\left(\frac{\lambda_\nu-1}{(1+\beta)\xi_\nu \lambda_\nu}\right)}
- \frac{\lambda_\nu \beta}{(1-\beta)\lambda_\nu \xi_\nu-\lambda_\nu+1} 
\ln{\left(\frac{\lambda_\nu-1}{(1-\beta)\xi_\nu \lambda_\nu}\right)}\Bigg\}\Bigg] + {\mathcal O}\left(\frac{1}{M^2}\right)\,, 
\nonumber
\end{eqnarray}
}where $\delta^{\rm (LO;even)}_{\gamma\gamma^*}$ is obtained from Eqs.~\eqref{eq:deltaLO_brem_even} and  
\eqref{eq:deltaLO_NLO_brem_even}.
 
\subsubsection{$\nu=1$ charge-even contribution} 
Here we compute the {\it genuinely dynamical} NLO charge-even contributions by involving the NLO 
soft-photon bremsstrahlung amplitudes. This requires the following charge-even component of the 
interference of the LO and NLO matrix elements to be determined (cf. 
Figs.~\ref{fig:LO} and~\ref{fig:NLO}):
\begin{eqnarray}
{\mathcal S}^{(\nu=1)}_{\rm even}  \!&=&\!
\sum_{\rm spins}\left[\mathcal{M}^\dagger_{\rm LO}\, \mathcal{M}_{\rm NLO} + {\rm h.c}\right]_{\rm even}
\nonumber\\
&=&\! 2 \mathcal{R}e \sum_{\rm spins}\Bigg [\left(\mathcal{M}^{\left[\,l;{\rm i}\,\right]}_{\gamma\gamma^*} 
+\mathcal{M}^{\left[\,l;{\rm f}\,\right]}_{\gamma\gamma^*}\right)^\dagger 
\left(\mathcal{M}^{\left[\,\overline{l;{\rm i}} \,\right]}_{\gamma\gamma^*} 
+ \mathcal{M}^{\left[\,\overline{l;{\rm f}} \,\right]}_{\gamma\gamma^*} \right) 
+ \left( \mathcal{M}^{\left[\,p(0);{\rm i}\,\right]}_{\gamma\gamma^*} 
+ \mathcal{M}^{\left[\,p(0);{\rm f}\,\right]}_{\gamma\gamma^*}\right)^\dagger
\bigg(\mathcal{M}^{\left[\,p(1);{\rm i}\,\right]}_{\gamma\gamma^*}
+\mathcal{M}^{\left[\,p(1);{\rm f}\,\right]}_{\gamma\gamma^*} 
\nonumber\\
&& \hspace{4.3cm} +\,\mathcal{M}^{\left[\,\overline{p(0);{\rm i}}\,\right]}_{\gamma\gamma^*} 
+ \mathcal{M}^{\left[\,\overline{p(0);{\rm f}}\,\right]}_{\gamma\gamma^*} 
+ \mathcal{M}^{\left[\,p(0);{\rm i}\otimes\,\right]}_{\gamma\gamma^*} 
+\mathcal{M}^{\left[\,p(0);{\rm f}\otimes \,\right]}_{\gamma\gamma^*} 
+ \mathcal{M}^{\left[\,p(1);{\rm v}\,\right]}_{\gamma\gamma^*}\bigg)\Bigg]\,.
\label{eq:Se_NLO}
\end{eqnarray}
\end{widetext}

After integration over the bremsstrahlung phase-space, we obtain the following expression for the NLO 
charge-even differential cross section in SPL:
\begin{eqnarray}
\left[\frac{{\rm d}\sigma^{\rm (NLO;even)}_{br}(Q^2)}{{\rm d}\Omega^\prime_l} 
\right]^{(E_{\gamma^*}\leq \Delta_{\gamma^*})}_{\gamma\gamma^*} \hspace{-0.5cm} 
&=&\! -\,\frac{\alpha}{2\pi^2 M}\left[\frac{{\rm d}\sigma_{el}(Q^2)}{{\rm d} \Omega^\prime_l}\right]_{0}
\nonumber\\
&& \times\, \Big[LL_{\rm ii} + LL_{\rm ff} - LL_{\rm if} \Big]\,,
\nonumber\\
\end{eqnarray}
where $LL_{\rm ii}$, $LL_{\rm ff}$ and $LL_{\rm if}$ are IR-finite integrals, conveniently evaluated
in the S-frame (see Appendix B for details). However, here we require their expressions in the 
{\it lab.}-frame, which are given as follows: 
\begin{eqnarray}
\label{eq:LLii}
LL_{\rm ii} \!&=&\! LL_{\rm ff} = \frac{2\pi\Delta_{\gamma^*} E}{E^{\prime}}\,, \quad \text{and}
\\
LL_{\rm if} \!&=&\!  \frac{4 \pi \Delta_{\gamma^*}E}{E^\prime} \left(\frac{1+\nu_l^2}{2 \nu_l}\right)\ln{\left[\frac{\nu_l+1}{\nu_l-1}\right]} \,.
\label{eq:LLif}
\end{eqnarray}
As in the LO case, we find that all NLO matrix elements corresponding to the proton emitting a single
soft-photon in Eq.~\eqref{eq:Se_NLO}, lead to scaleless integrals that vanish in DR. The corresponding NLO 
correction to the charge-even differential cross section is finite, and given by  
{\small
\begin{eqnarray}
\delta^{\rm (NLO;even)}_{\gamma\gamma^*} (Q^2,\Delta_{\gamma^*}) \!&=&\! 
\frac{\left[\frac{{\rm d}\sigma^{\rm (NLO;even)}_{br}(Q^2)}{{\rm d}\Omega^\prime_l}
\right]^{(E_{\gamma^*}\leq \Delta_{\gamma^*})}_{\gamma\gamma^*}}{\left[
\frac{{\rm d}\sigma_{el}(Q^2)}{{\rm d}\Omega^\prime_l} \right]_{0}}
\nonumber\\
&=&\! \frac{2 \alpha \Delta_{\gamma^*}}{\pi M} 
\left[\left(\frac{1+\nu_l^2}{2 \nu_l}\right) \ln\left[{\frac{\nu_l+1}{\nu_l-1}}\right]-1\right]
\nonumber\\
&& +\, \mathcal{O}\left(\frac{1}{M^2}\right)\,. 
\label{eq:deltaNLO_brem_even}
\end{eqnarray} 
}Notably, unlike at LO, our charge-even result involving the NLO amplitudes differs from the calculation
of Talukdar {\it at al.}~\cite{Talukdar:2020aui} where by adopting the Coulomb gauge condition, 
$v\cdot k=0$, it yielded higher-order terms in the cross section, namely, 
$\delta^{\rm (NLO;even)}_{\gamma\gamma^*}\sim \mathcal O(1/M^2)$. Beyond these NLO matrix elements, no 
further charge-even bremsstrahlung contributions to the NLO cross sections arise. All possible 
interference matrix elements involving the products of chirally \underline{enhanced} NNLO and N${}^3$LO 
bremsstrahlung amplitudes, namely, those contributing to ${\mathcal S}^{(\nu=2)}_{\rm even}$ and 
${\mathcal S}^{(\nu=3)}_{\rm even}$, either yield cross section terms of ${\mathcal O}(1/M^2)$, or vanish
upon integrating over the bremsstrahlung phase-space. Thus, we obtain the total corrections to the 
charge-even (anti-)lepton-proton differential cross section due to the soft-photon bremsstrahlung 
calculated up-to-and-including NLO [i.e., ${\mathcal O}(\alpha^3/M)$], as below:
\begin{widetext}
\begin{eqnarray}
\label{eq:delta_brem_even}
\delta^{\rm (even)}_{\gamma\gamma^*} (Q^2,\Delta_{\gamma^*}) \!&=&\! 
\left[\frac{{\rm d}\sigma^{\rm (even)}_{br}(Q^2)}{{\rm d}\Omega^\prime_l} 
\right]^{(E_{\gamma^*}\leq \Delta_{\gamma^*})}_{\gamma\gamma^*}\!\!\!\!\bigg{/}
\left[\frac{{\rm d}\sigma_{el}(Q^2)}{{\rm d}\Omega^\prime_l} \right]_{0} 
\nonumber\\
&=&\! \delta^{\rm (LO;even)}_{\gamma\gamma^*}(Q^2,\Delta_{\gamma^*}) + \delta^{\rm (NLO;even)}_{\gamma\gamma^*}(Q^2,\Delta_{\gamma^*})
\nonumber\\
&\equiv&\!  \text{\bf IR}^{(0)}_{\gamma\gamma^*}(Q^2) + \delta^{\rm (0;\,even)}_{\gamma\gamma^*}(Q^2,\Delta_{\gamma^*}) 
+  \delta^{\rm (1;\,even)}_{\gamma\gamma^*}(Q^2,\Delta_{\gamma^*}) + \mathcal{O}\left(\frac{1}{M^2}\right)\,.
\end{eqnarray}
where $\delta^{\rm (LO;even)}_{\gamma\gamma^*}$ was already given by Eq.~\eqref{eq:deltaLO_brem_even}. For 
convenience of displaying our numerical results (cf. Fig.~\ref{fig:even_brem}), we have equivalently 
expressed the above contributions by isolating the genuine LO ($\nu=0$) and NLO ($\nu=1$) chiral components
of the finite soft-photon bremsstrahlung correction, as given by 
$\delta^{\rm (0;\,even)}_{\gamma\gamma^*}\equiv \delta^{\rm (LO;even)}_{\gamma\gamma^*}-\delta^{({\rm LO-}1/M;{\rm even})}_{\gamma\gamma^*}$ 
and
$\delta^{\rm (1;\,even)}_{\gamma\gamma^*}\equiv \delta^{({\rm LO-}1/M;{\rm even})}_{\gamma\gamma^*}+\delta^{\rm (NLO;even)}_{\gamma\gamma^*}$, 
respectively, namely,  
\begin{eqnarray}
\label{eq:delta_brem_even_0}
\delta^{\rm (0;\,even)}_{\gamma\gamma^*}(Q^2,\Delta_{\gamma^*}) \!&=&\! \frac{\alpha}{\pi}\Bigg[\left(\frac{\nu^2_l+1}{2\nu_l}\ln\left[\frac{\nu_l+1}{\nu_l-1}\right]-1\right) 
\ln{\left(\frac{4 \Delta^2_{\gamma^*}}{m^2_l}\right)}+\frac{2}{\beta}\ln\sqrt{\frac{1+\beta}{1-\beta}}
+ \left(\frac{\nu^2_l+1}{2\nu_l}\right)
\nonumber\\
&& \hspace{0.4cm} \times\, \Bigg\{\text{Li}_2\left(1-\frac{\lambda_\nu-1}{(1-\beta)\xi_\nu}\right) 
+ \text{Li}_2\left(1-\frac{\lambda_\nu-1}{(1+\beta)\xi_\nu}\right)
- \text{Li}_2\left(1-\frac{\lambda_\nu -1}{(1-\beta)\lambda_\nu\xi_\nu}\right)
\nonumber\\
&& \hspace{2.1cm} -\, \text{Li}_2\left(1-\frac{\lambda_\nu-1}{(1+\beta)\lambda_\nu\xi_\nu}\right)\Bigg\}\Bigg]\,, \quad \text{and}
\\
\nonumber\\
\label{eq:delta_brem_even_1}
\delta^{\rm (1;\,even)}_{\gamma\gamma^*}(Q^2,\Delta_{\gamma^*}) \!&=&\! \frac{\alpha Q^2}{2\pi ME\beta^2}\Bigg[1-2 \beta^ 2\left(\frac{\nu^2_l+1}{2\nu_l}\ln\left[\frac{\nu_l+1}{\nu_l-1}\right]-1\right)  
- \left\{\frac{1}{\beta}-\beta \left(\frac{\nu^2_l+\nu_l+1}{\nu_l}\right)\right\}\ln{\sqrt{\frac{1+\beta}{1-\beta}}} 
\nonumber\\
&& \hspace{1.6cm} +\, \beta\left(\frac{\nu^2_l+1}{2\nu_l}\right)\Bigg\{\frac{\lambda_\nu(1+\beta)-1}{\xi_\nu(1-\beta)-\lambda_\nu+1}\!
\ln{\left(\frac{\lambda_\nu-1}{(1-\beta)\xi_\nu}\right)} - \frac{\lambda_\nu(1-\beta)-1}{\xi_\nu(1+\beta)-\lambda_\nu+1}\! 
\nonumber\\
&& \hspace{4cm} \times\, \ln{\left(\frac{\lambda_\nu-1}{(1+\beta)\xi_\nu}\right)} 
- \frac{\lambda_\nu \beta}{(1+\beta)\lambda_\nu \xi_\nu-\lambda_\nu+1}
\ln{\left(\frac{\lambda_\nu-1}{(1+\beta)\xi_\nu \lambda_\nu}\right)} 
\nonumber\\
&&  \hspace{4cm} -\, \frac{\lambda_\nu \beta}{(1-\beta)\lambda_\nu \xi_\nu-\lambda_\nu+1}
\ln{\left(\frac{\lambda_\nu-1}{(1-\beta)\xi_\nu \lambda_\nu}\right)}\Bigg\}
\nonumber\\
&& \hspace{4cm} +\, \frac{4 \Delta_{\gamma^*}E \beta^2}{Q^2} 
\left[\left(\frac{1+\nu_l^2}{2 \nu_l}\right) \ln\left[{\frac{\nu_l+1}{\nu_l-1}}\right]-1\right]\Bigg]\,. 
\end{eqnarray}
Similarly, it should be interesting to note that there are no genuine NLO amplitudes involving the virtual 
photon-loops that contribute to charge-even correction to the NLO cross section, i.e., 
$\delta^{\rm (NLO;even)}_{\gamma\gamma}=0$, as previously reported in Ref.~\cite{Talukdar:2020aui}. Hence,
the total one-loop IR-finite radiative correction to the charge-even (anti-)lepton-proton elastic 
differential cross section calculated up-to-and-including NLO [i.e, ${\mathcal O}(\alpha^3/M)$] is given as 
%
\begin{figure*}[tbp]
\begin{center}
\includegraphics[width=0.48\linewidth]{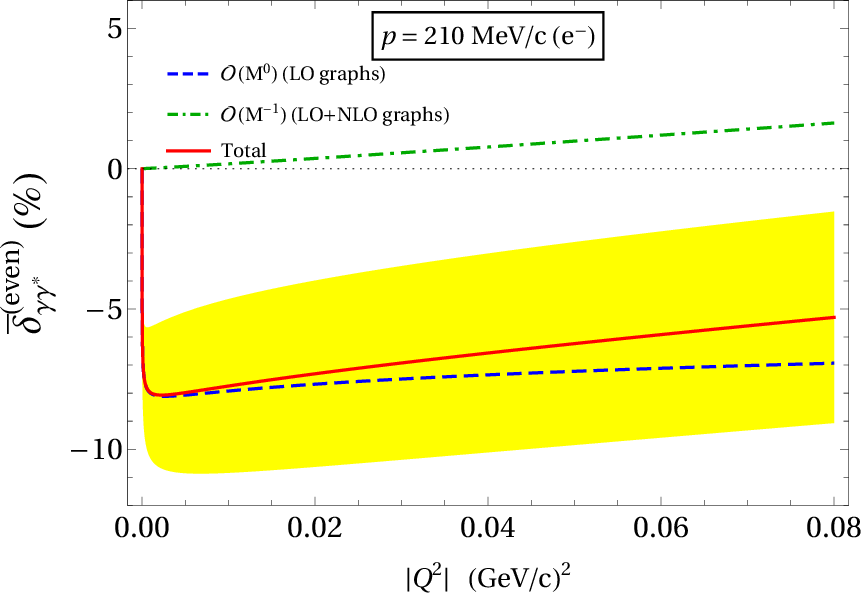}~\quad~\includegraphics[width=0.48\linewidth]{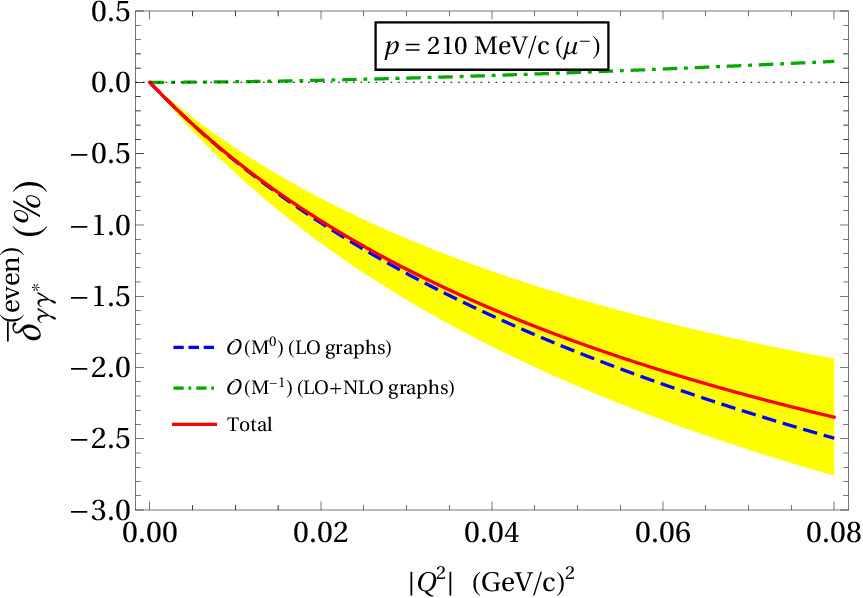}
    
\vspace{0.6cm}
    
\includegraphics[width=0.48\linewidth]{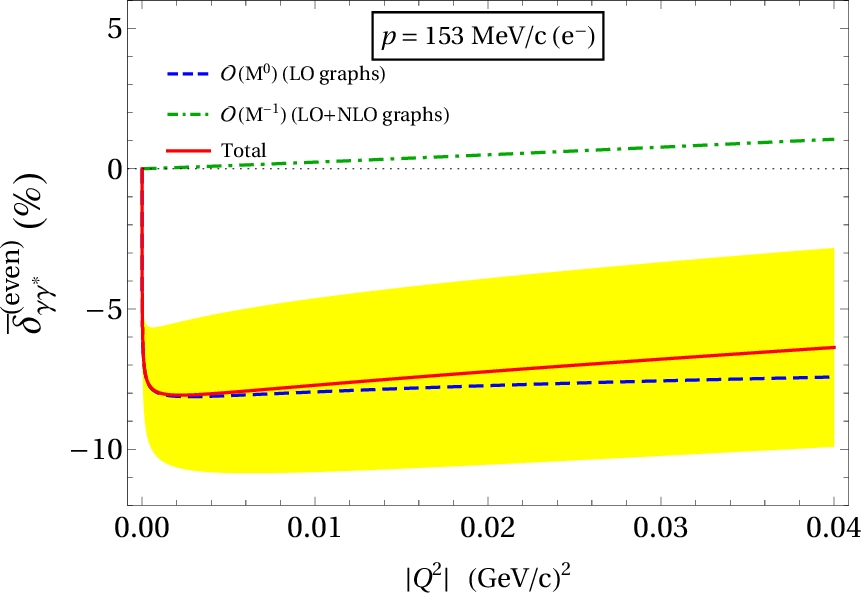}~\quad~\includegraphics[width=0.48\linewidth]{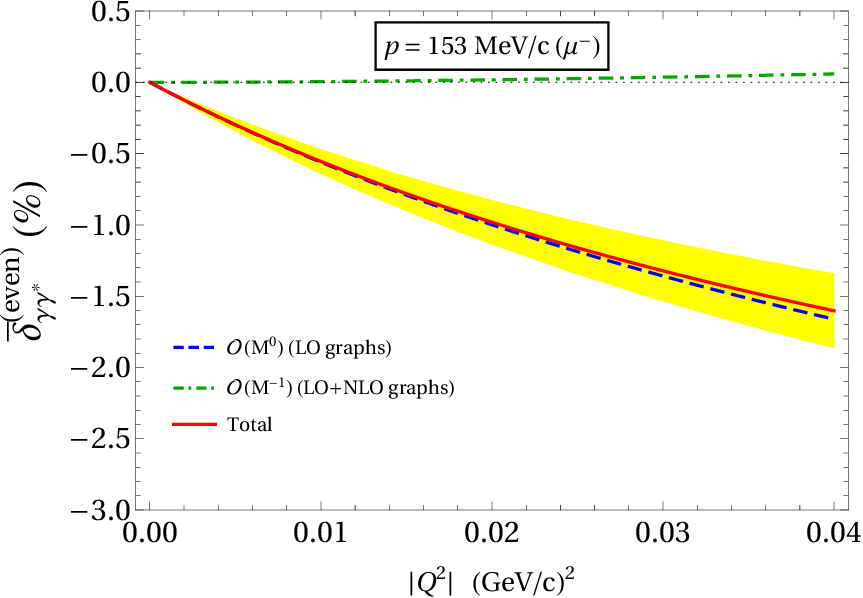}

\vspace{0.6cm}
    
\includegraphics[width=0.48\linewidth]{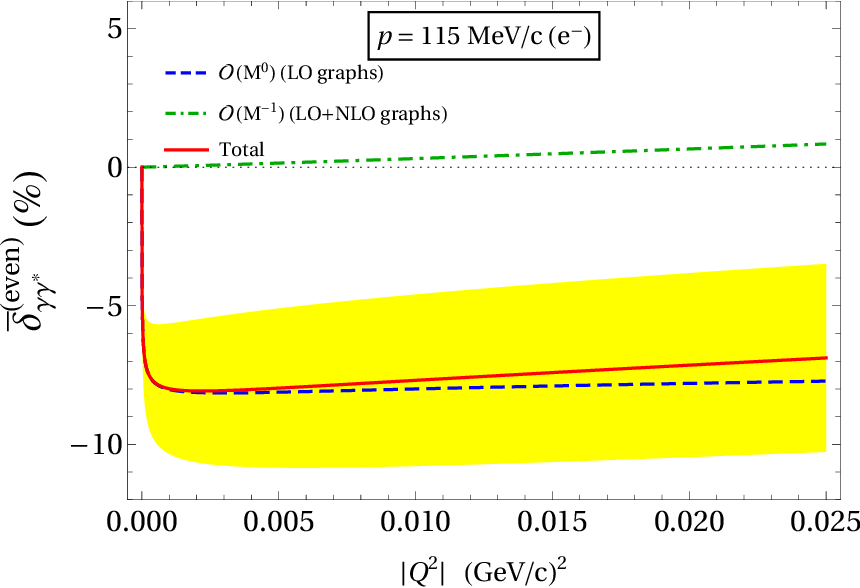}~\quad~\includegraphics[width=0.48\linewidth]{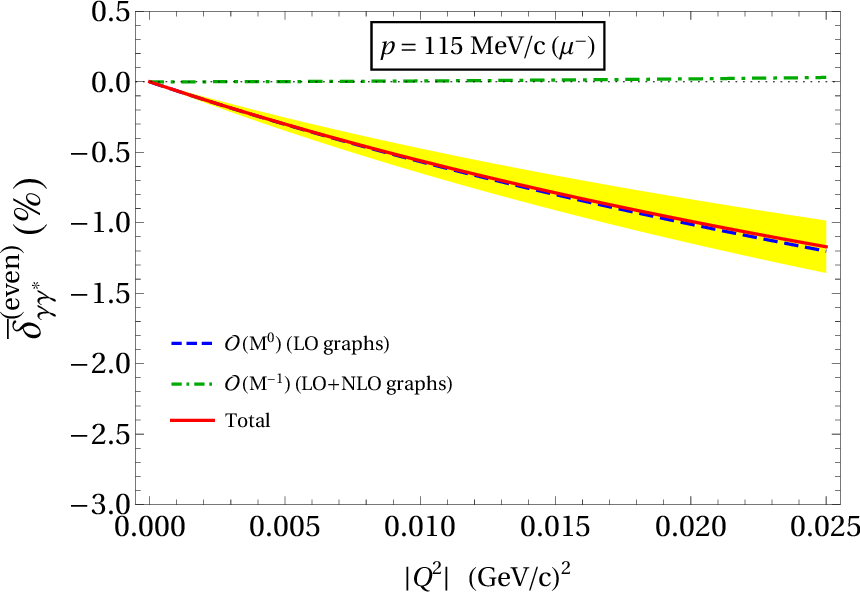}
    \caption{The finite fractional \underline{charge-even} soft-photon bremsstrahlung corrections, 
             $\overline{\delta}^{\rm (even)}_{\gamma\gamma^*}=\delta^{\rm (even)}_{\gamma\gamma^*}
             -{\bf IR}^{(0)}_{\gamma\gamma^*}$ [expressed as percentage relative to LO Born/OPE 
             differential cross section, Eq.~\eqref{eq:diff_LO}], to the $e^\pm$-p (left panel) and 
             $\mu^\pm$-p (right panel) unpolarized elastic scatterings is presented as a function of 
             $\left|Q^2\right|$ in HB$\chi$PT.The results correspond to three fixed values of the 
             incident lepton beam momenta proposed by the MUSE experiment: 210~MeV/$c$, 153~MeV/$c$, and 
             115~MeV/$c$. Color online: The blue [green] dashed [dashed-dot] curve represents the sum of 
             all genuine chiral LO [NLO], i.e., $\mathcal{O}(M^0)$ [$\mathcal{O}(M^{-1})$] terms 
             $\delta^{\rm (0;\,even)}_{\gamma\gamma^*}$ [$\delta^{\rm (1;\,even)}_{\gamma\gamma^*}$], 
             given in Eq.~\eqref{eq:delta_brem_even_0} [Eq.~\eqref{eq:delta_brem_even_1}], and the red 
             solid curve is their sum (LO+NLO) 
             $\delta^{\rm (0;\,even)}_{\gamma\gamma^*}+\delta^{\rm (1;\,even)}_{\gamma\gamma^*}$ (labeled
             ``Total"). All curves are evaluated at a nominal detector cutoff of $\Delta_{\gamma^*}=1\%$ 
             of the incident lepton beam energy $E$. The yellow shaded band indicates the sensitivity of 
             the full result to the acceptance in the range, $0.5\% E<\Delta_{\gamma^*}<2\% E$.}
\label{fig:even_brem} 
\end{center}
\end{figure*} 
%
\begin{eqnarray}
\delta^{\rm (even)}_{2\gamma} (Q^2,\Delta_{\gamma^*}) \!&=&\!
\left[\frac{{\rm d}\sigma^{\rm (even)}_{el}(Q^2)}{{\rm d}\Omega^\prime_l}\right]_{2\gamma}\!\!\!\!\bigg{/}
\left[\frac{{\rm d}\sigma_{el}(Q^2)}{{\rm d}\Omega^\prime_l} \right]_{0}  
\nonumber\\
&=&\! \delta^{\rm (LO;even)}_{2\gamma}(Q^2,\Delta_{\gamma^*}) 
+ \delta^{\rm (NLO;even)}_{\gamma\gamma^*}(Q^2,\Delta_{\gamma^*})\,,
\label{eq:delta_rad_even} 
\end{eqnarray}
\end{widetext}
where $\delta^{\rm (LO;even)}_{2\gamma}$ was already given by Eq.~\eqref{eq:deltaLO_rad_even}. 
Figure.~\ref{fig:even_rad} displays the numerical estimates of $\delta^{\rm (even)}_{2\gamma}$ for 
$e^\pm$-p and $\mu^\pm$-p scatterings.

\begin{figure*}[tbp]
\begin{center}
\includegraphics[width=0.48\linewidth]{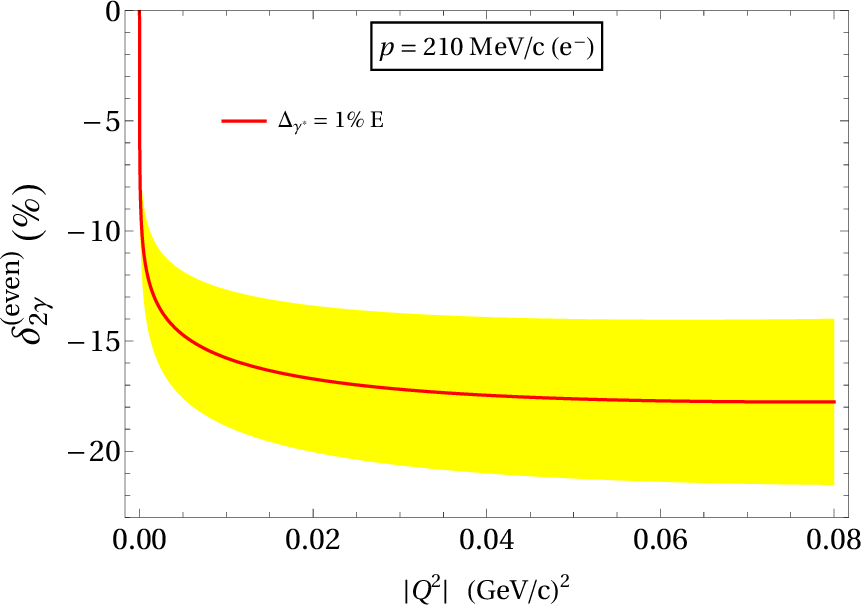}~\quad~\includegraphics[width=0.48\linewidth]{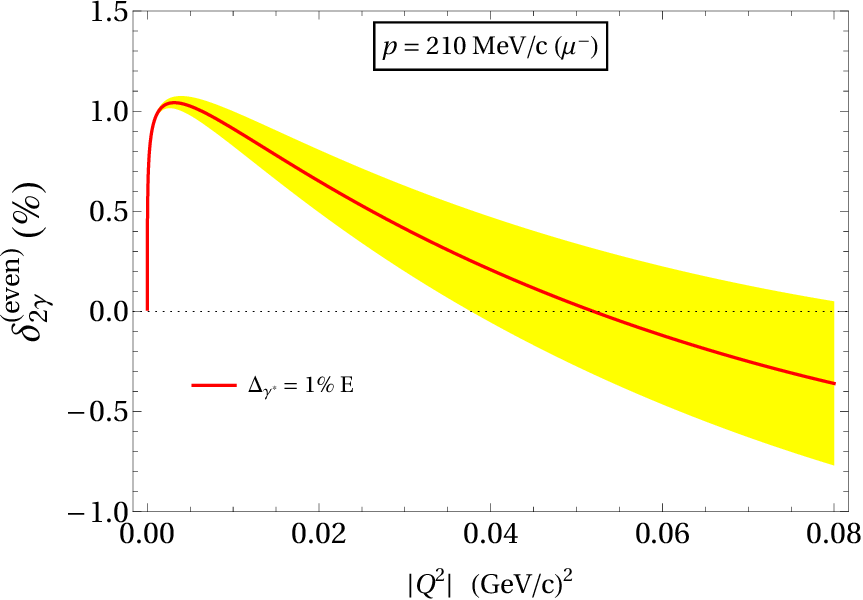}

\vspace{0.6cm}
    
\includegraphics[width=0.48\linewidth]{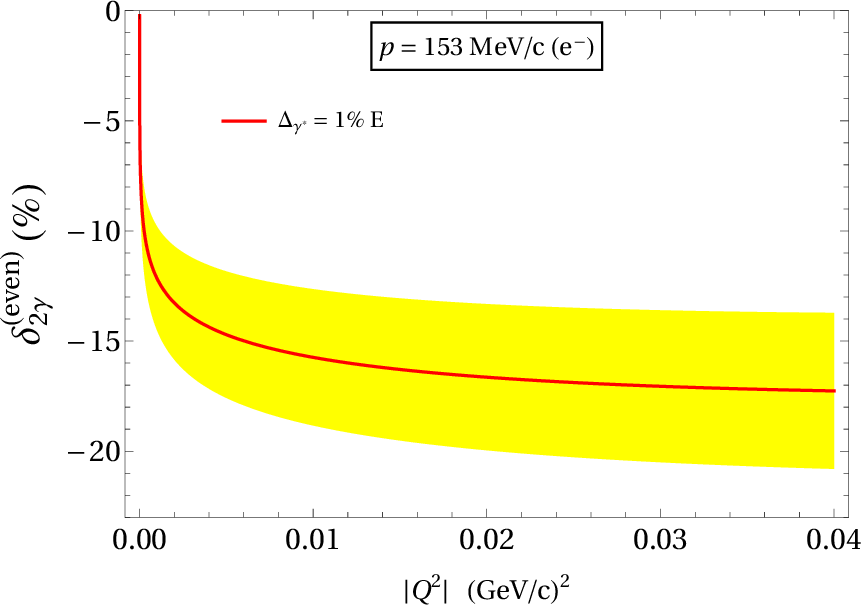}~\quad~\includegraphics[width=0.48\linewidth]{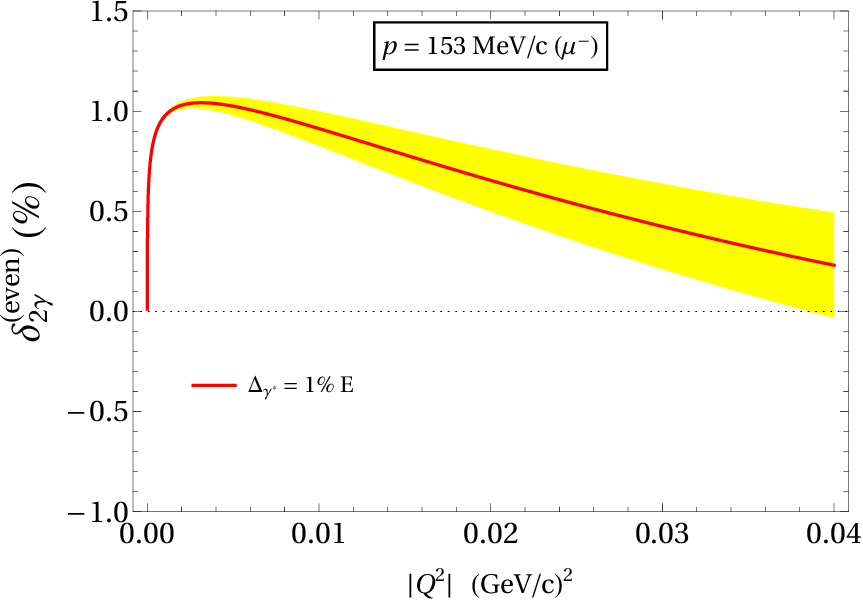}

\vspace{0.6cm}
    
\includegraphics[width=0.48\linewidth]{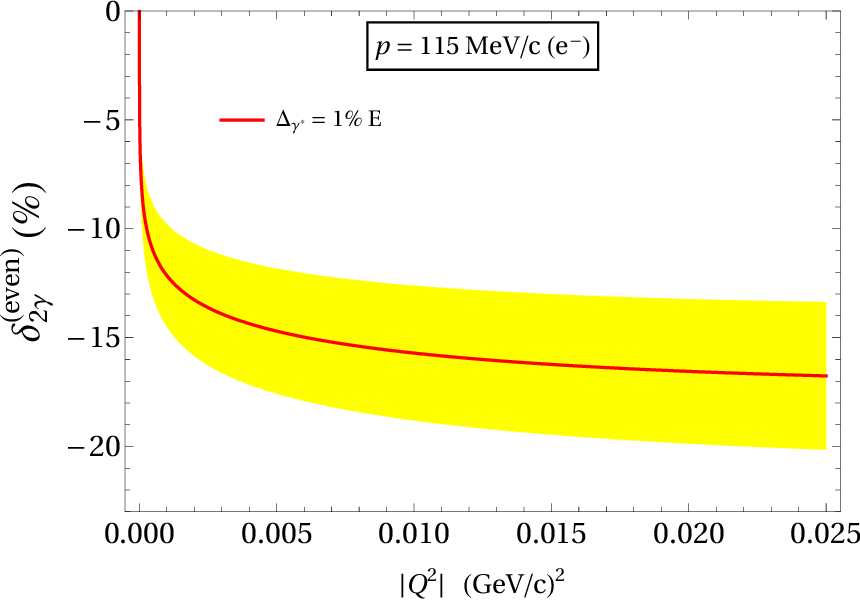}~\quad~\includegraphics[width=0.48\linewidth]{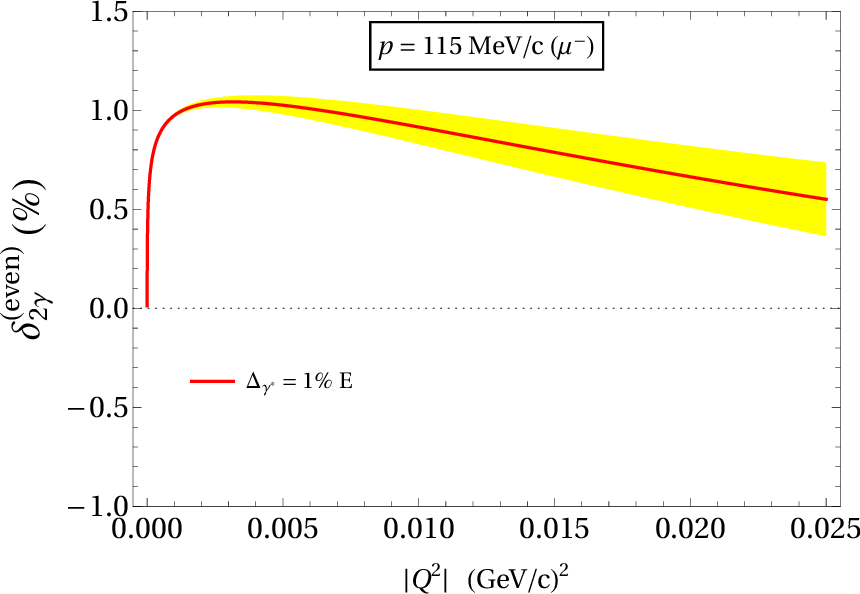}
    \caption{The finite fractional \underline{charge-even} radiative corrections [expressed as 
             percentage relative to LO Born/OPE differential cross section, Eq.~\eqref{eq:diff_LO}] 
             to the $e^\pm$-p (left panel) and $\mu^\pm$-p (right panel) unpolarized elastic scatterings 
             is presented as a function of $\left|Q^2\right|$ in HB$\chi$PT. The results correspond to 
             three fixed values of the incident lepton beam momenta proposed by the MUSE experiment: 
             210~MeV/$c$, 153~MeV/$c$, and 115~MeV/$c$. Color online: The red solid curve represents 
             the total charge-even radiative corrections, $\delta_{2\gamma}^{\rm (even)}$, 
             Eq.~\eqref{eq:delta_rad_even}, evaluated at the nominal detector cutoff 
             $\Delta_{\gamma^*}=1\%$ of the incident lepton beam energy $E$. The yellow shaded band 
             indicates the sensitivity of our result to the detector acceptance in the range,
             $0.5\%E < \Delta_{\gamma^*} < 2\% E$.}
\label{fig:even_rad}
\end{center}
\end{figure*} 

%

\subsection{The charge-odd bremsstrahlung component} 
\label{Sec:II-d} 
The cross section for the charge-odd (anti-)lepton-proton bremsstrahlung process is obtained from the 
{\it exchange-interference} (see introduction) products of amplitudes, where the photon is emitted by the
(anti-)lepton in one and the proton in the other. Consequently, the cross section changes sign when the 
lepton's charge polarity is reversed. Below, we evaluate the charge-odd differential cross section in 
Eq.~\eqref{eq:dsigma_brem}, corresponding to the single soft-photon emission process, which contributes to
the lowest order QED corrections of the $\ell^\pm$-p cross section, up-to-and-including NLO accuracy 
[i.e., ${\mathcal O}(\alpha^3/M)$] in HB$\chi$PT. In contrast to the charge-even bremsstrahlung cross 
section, we will now obtain non-vanishing contributions from chirally \underline{enhanced} NNLO and N${}^3$LO 
amplitudes with a single soft-photon radiated from the \underline{final-state} proton.

\subsubsection{$\nu=0$ charge-odd contribution}
To compute the charge-odd cross section for soft-photon bremsstrahlung at LO, it is necessary to determine 
the following charge-odd component of the squared LO matrix elements (cf. Fig.~\ref{fig:LO}):
\begin{widetext}
\begin{eqnarray}
\mathcal{S}^{(\nu=0)}_{\rm odd} = \sum_{\rm spins} \left |{\mathcal M}_{\rm LO} \right|^2_{\rm odd} 
= 2\mathcal{R}e \sum_{\rm spins} \bigg[ \left(\mathcal{M}^{\left[l;{\rm i}\right]}_{\gamma\gamma^*} 
+\mathcal{M}^{\left[l;{\rm f}\right]}_{\gamma\gamma^*}\right)^\dagger 
\left(\mathcal{M}^{\left[p(0);{\rm i}\right]}_{\gamma\gamma^*} 
+ \mathcal{M}^{\left[p(0);{\rm f}\right]}_{\gamma\gamma^*}\right) \bigg] \,.
\label{eq:So_LO}
\end{eqnarray}

After phase-space integration, we obtain the LO charge-odd lepton-proton differential cross section in SPL as below:
\begin{eqnarray}
\left[\frac{{\rm d}\sigma^{\rm (LO;odd)}_{\gamma\gamma^*}(Q^2)}{{\rm d}\Omega^\prime_l} 
\right]^{(E_{\gamma^*}\leq \Delta_{\gamma^*})}_{\gamma\gamma^*}   \hspace{-0.5cm} = 
\frac{\alpha M}{\pi^2 Q^2} \left[\frac{{\rm d}\sigma_{el}(Q^2)}{{\rm d}\Omega^\prime_l} \right]_{0}
\left[E I_1 - E^\prime I_2 + \frac{Q^2}{2M}\left(E I_3 - E^\prime I_4\right)\right]\,,\quad\,
\end{eqnarray}
where $I_{1,2}$ are IR-finite and $I_{3,4}$ are IR-divergent integrals. The integrals are most conveniently evaluated
in the S-frame [cf. Appendix B]. However, here we need their corresponding expressions in the {\it lab.}-frame, namley,  
\begin{eqnarray}
\label{eq:I1}
&& \hspace{3cm} I_1 = \frac{4\pi \Delta_{\gamma^*}}{\beta E^\prime} \ln\sqrt{\frac{1+\beta}{1-\beta}} \,,
\\
\label{eq:I2}
&& \hspace{3cm} I_2 = \frac{4\pi \Delta_{\gamma^*}E}{\beta^\prime E^{\prime 2}} \ln\sqrt{\frac{1+\beta^\prime}{1-\beta^\prime}}\,,
\\
\label{eq:I3}
I_3 \!&=&\! \frac{2 \pi}{\beta E} \left[ \frac{1}{|\varepsilon_{\rm IR}|} + \gamma_E 
- \ln\left(\frac{4\pi\mu^2}{m_l^2}\right) \right] \ln\sqrt{\frac{1+\beta}{1-\beta}} 
+ 2\pi\widetilde{I_3}\,;
\nonumber\\
\widetilde{I_3} \!&=&\! \frac{1}{\beta E} 
\Bigg[\ln\left(\frac{4 \Delta_{\gamma^*}^2 E^2}{m_l^2 E^{\prime 2}} \right) \ln\sqrt{\frac{1+\beta}{1-\beta}} 
+ \frac{1}{2}{\rm Li}_2\left( \frac{2\beta}{\beta - 1} \right) 
- \frac{1}{2}{\rm Li}_2\left( \frac{2\beta}{\beta+1}\right)\Bigg] \,, \quad \text{and}
\\
\label{eq:I4}
I_4 \!&=&\! \frac{2 \pi}{\beta^\prime E^\prime} \left[ \frac{1}{|\varepsilon_{\rm IR}|} + \gamma_E 
- \ln\left(\frac{4\pi\mu^2}{m_l^2}\right) \right] \ln\sqrt{\frac{1+\beta^\prime}{1-\beta^\prime}} 
+ 2\pi\widetilde{I_4}\,;
\nonumber
\\
 \widetilde{I_4} \!&=&\! \frac{1}{\beta^\prime E^\prime} 
\Bigg[\ln\left(\frac{4 \Delta_{\gamma^*}^2 E^2}{m_l^2 E^{\prime 2}} \right) \ln\sqrt{\frac{1+\beta^\prime}{1-\beta^\prime}} 
+ \frac{1}{2} \operatorname{Li}_2\left(\frac{2\beta^\prime}{\beta^\prime-1}\right) 
- \frac{1}{2}{\rm Li}_2\left( \frac{2\beta^\prime}{\beta^\prime+1}\right)\Bigg]\,.
\end{eqnarray}
The corresponding fractional contribution to the charge-odd lepton-proton differential cross section arising solely 
from the LO soft-photon bremsstrahlung amplitudes is given as 
\begin{eqnarray}
\delta^{\rm (LO;odd)}_{\gamma\gamma^*}(Q^2,\Delta_{\gamma^*}) \!&=&\! 
\left[\frac{{\rm d}\sigma^{\rm (LO;odd)}_{br}(Q^2)}{{\rm d}\Omega^\prime_l} 
\right]^{(E_{\gamma^*}\leq \Delta_{\gamma^*})}_{\gamma\gamma^*}\!\!\!\!\bigg{/}
\left[\frac{{\rm d}\sigma_{el}(Q^2)}{{\rm d}\Omega^\prime_l} \right]_{0}
\nonumber\\
&=&\! \text{\bf IR}^{(1)}_{\gamma\gamma^*}(Q^2) 
+ \frac{\alpha M}{\pi^2 Q^2} \left[E I_1-E^\prime I_2 + \frac{\pi Q^2}{M}\left(E\widetilde{I_3} 
- E^\prime\widetilde{I_4}\right)\right] + 
\nonumber\\
&=&\! \text{\bf IR}^{(1)}_{\gamma\gamma^*}(Q^2) 
- \frac{2\alpha\Delta_{\gamma^*}}{\pi E\beta^2} 
\Bigg[1+\left(\beta-\frac{1}{\beta}\right)\ln\sqrt{\frac{1+\beta}{1-\beta}}\,\Bigg] 
+ \delta^{({\rm LO}-1/M;{\rm odd})}_{\gamma\gamma^*}(Q^2,\Delta_{\gamma^*})\,, 
\label{eq:deltaLO_brem_odd}
\end{eqnarray}
where  
\begin{eqnarray}
\delta^{({\rm LO}-1/M;{\rm odd})}_{\gamma\gamma^*}(Q^2,\Delta_{\gamma^*}) \!&=&\! 
\frac{-\alpha Q^2}{2\pi ME\beta^2}\Bigg[\!\ln\left(\frac{4\Delta_{\gamma^*}^2}{m^2_l}\right)\! 
\left\{1+\left(\beta-\frac{1}{\beta}\right)\ln{\sqrt{\frac{1+\beta}{1-\beta}}}\,\right\} 
- \frac{3\Delta_{\gamma^*}}{E\beta^2} + \frac{1}{2}\left(\beta-\frac{1}{\beta}\right) 
\left\{{\rm Li}_2\left(\frac{2\beta}{\beta-1}\right) \right.
\nonumber \\
&& \left. \hspace{1.6cm} -\, {\rm Li}_2\!\left(\frac{2\beta}{1+\beta}\right) \right\} 
- \frac{2}{\beta}\!\left\{\!1+\frac{\Delta_{\gamma^*}\beta}{E}\!\left(1+\frac{3}{2\beta^2}\right)\!\!
\left(\beta-\frac{1}{\beta}\right)\right\}\!\ln\sqrt{\frac{1+\beta}{1-\beta}}\,\Bigg] 
\nonumber\\
&& +\, \mathcal{O}\left(\frac{1}{M^2}\right)\,,
\label{eq:deltaLO-1/M_brem_odd}
\end{eqnarray}
with the IR-divergent part given by
\begin{eqnarray}
\text{\bf IR}^{(1)}_{\gamma\gamma^*}(Q^2) \!&=&\!\frac{\alpha}{\pi}\left[\frac{1}{|\varepsilon_{\rm IR}|}+\gamma_E 
- \ln\left(\frac{4\pi\mu^2}{m_l^2}\right)\right]\Bigg[\frac{1}{\beta}\ln{\sqrt{\frac{1+\beta}{1-\beta}}} 
- \frac{1}{\beta^\prime}\ln{\sqrt{\frac{1+\beta^\prime}{1-\beta^\prime}}}\,\Bigg]
\nonumber\\
&=&\! -\, \frac{\alpha Q^2}{2\pi ME\beta^2}
\left[\frac{1}{|\varepsilon_{\rm IR}|}+\gamma_E - \ln\left(\frac{4\pi\mu^2}{m_l^2}\right)\right] 
\Bigg[ 1 + \left(\beta-\frac{1}{\beta}\right)\ln{ \sqrt{\frac{1+\beta}{1- \beta}}}\,\Bigg] 
+ \mathcal{O}\left(\frac{1}{M^2}\right)\,.
\label{eq:IR1_brem_odd}
\end{eqnarray}
\end{widetext}
The following two features of the charge-odd soft-photon bremsstrahlung result merit attention: 
\begin{itemize}
\item First, akin to the charge-even contributions, we again observe that the above expression is 
\underline{not} genuinely LO, since it also includes ${\mathcal O}\left(1/M\right)$ terms which are 
commensurate with the NLO corrections. Moreover, the LO contribution receives additional modification from
chirally \underline{enhanced} graphs of ${\mathcal O}(M)$ and ${\mathcal O}(M^2)$, which, under standard
power-counting, would be classified as NLO and NNLO terms.
\item Secondly, in our charge-odd bremsstrahlung cross section up-to-and-including NLO, the IR divergence
arises solely from the LO amplitudes. However, again at the genuine LO, i.e., ${\mathcal O}(M^0)$ 
correction to the cross section exhibit complete cancellation of the IR-divergent terms, leaving a residual
${\mathcal O}(1/M)$ IR divergence that effectively contributes to the NLO cross section.\footnote{This 
feature is reminiscent of the analogous singularity structure of the LO TPE cross section, as discussed in 
the work of Choudhary {\it et al}.~\cite{Choudhary:2023rsz}, constituting the charge-odd virtual correction
counterpart of the {\it exchange-interference} (see introduction) bremsstrahlung cross section at LO.} As 
revealed in our subsequent analysis, no additional charge-odd IR divergences arise from the genuine NLO 
corrections to the soft-photon bremsstrahlung cross section. 
\end{itemize} 

To cancel the residual ${\mathcal O}\left(1/M\right)$ IR-singularity, it is necessary to include the TPE 
counterpart, such that the total charge-odd radiative correction to the elastic cross section is obtained 
as a finite physical quantity. Recently, the low-energy TPE corrections to the elastic cross section were 
analytically evaluated exactly up-to-and-including NLO in HB$\chi$PT by Choudhury 
{\it et al.}~\cite {Choudhary:2023rsz}. It turns out that in the corrections to the TPE cross section up 
to ${\mathcal O}\left(1/M\right)$, the IR divergences arise solely from the LO TPE box-amplitudes with no 
additional singularities arising from the genuine NLO TPE amplitudes. The two LO IR-divergent parts of the
TPE exactly cancel, leaving only a residual IR divergence of $\mathcal{O}\left(1/M\right)$ that 
contributes to the NLO TPE cross section~\cite {Choudhary:2023rsz}. From Eq.~(33) of 
Ref.~\cite {Choudhary:2023rsz}, we find the following analytical expression for the IR-divergent part of 
the TPE cross section up-to-and-including NLO in HB$\chi$PT:
\begin{widetext}
\begin{eqnarray}
\text{\bf IR}^{(1)}_{\rm TPE}(Q^2) = -\text{\bf IR}^{(1)}_{\gamma\gamma^*}(Q^2) 
&=&\! -\,\frac{\alpha}{\pi}\left[\frac{1}{|\varepsilon_{\rm IR}|}+\gamma_E 
- \ln\left(\frac{4\pi\mu^2}{m_l^2}\right)\right]\Bigg[\frac{1}{\beta}\ln{\sqrt{\frac{1+\beta}{1-\beta}}} 
- \frac{1}{\beta^\prime}\ln{\sqrt{\frac{1+\beta^\prime}{1-\beta^\prime}}}\,\Bigg]
\nonumber\\
&=&\! \frac{\alpha Q^2}{2\pi ME\beta^2} \left[\frac{1}{|\varepsilon_{\rm IR}|}+\gamma_E 
- \ln\left(\frac{4\pi\mu^2}{m_l^2}\right)\right] 
\Bigg[1+\left(\beta-\frac{1}{\beta}\right)\ln{ \sqrt{\frac{1+\beta}{1- \beta}}}\,\Bigg] 
+ \mathcal{O}\left(\frac{1}{M^2}\right).\qquad\,  
\label{eq:TPE_IR}
\end{eqnarray}
\end{widetext}
With equal and opposite charge-odd IR-divergent terms arising from the TPE and {\it exchange-interference} (see 
introduction) soft-photon bremsstrahlung counterparts, the net charge-odd radiative correction to the elastic 
cross section up-to-and-including NLO is therefore finite. The work of Choudhary 
{\it et al.}~\cite{Choudhary:2023rsz} used analytical techniques, such as integration-by-parts and partial 
fractions, to evaluate the TPE box diagrams, which from the LO amplitudes yielded the following form of the 
result:
\begin{eqnarray}
\delta^{\rm (LO;odd)}_{\gamma\gamma}(Q^2) \!&=&\! 
\frac{\left[\frac{{\rm d}\sigma^{\rm (LO;odd)}_{virt}(Q^2)}{{\rm d}\Omega^\prime_l} 
\right]_{\gamma\gamma}}{\left[\frac{{\rm d}\sigma_{el}(Q^2)}{{\rm d}\Omega^\prime_l} \right]_{0}} 
\nonumber\\
&=&\! \text{\bf IR}^{(1)}_{\rm TPE}(Q^2) + \pi\alpha\frac{\sqrt{{-Q^2}}}{2E}\left[\frac{1}{1+\frac{Q^2}{4 E^2}}\right] 
\nonumber\\
&& +\, \delta^{(ab;1/M)}_{\gamma\gamma}(Q^2)\,.
\label{eq:deltaLO_TPE_odd}
\end{eqnarray}
Here, the second term is our HB$\chi$PT version of the Feshbach-McKinley term~\cite{McKinley:1948zz}, and the 
IR-finite term $\delta^{(ab;1/M)}_{\gamma\gamma}\sim \mathcal{O}\left(1/M\right)$ is taken from Choudhary 
{\it et al.}~\cite{Choudhary:2023rsz}, wherein one can find the detailed expression. In particular, the 
latter ${\mathcal O}(1/M)$ correction effectively modifies the contributions from the seven other genuine NLO
TPE diagrams~\cite{Choudhary:2023rsz}. Hence, the total one-loop finite fractional radiative correction to the 
charge-odd elastic differential cross section is
\begin{eqnarray}
\delta^{\rm (LO;odd)}_{2\gamma}(Q^2,\Delta_{\gamma^*}) \!&=&\! 
\frac{\left[\frac{{\rm d}\sigma^{\rm (LO;odd)}_{el}(Q^2)}{{\rm d}\Omega^\prime_l} 
\right]_{2\gamma}}{\left[\frac{{\rm d}\sigma_{el}(Q^2)}{{\rm d}\Omega^\prime_l} \right]_{0}}
\nonumber\\
&=&\! \delta^{\rm (LO;odd)}_{\gamma\gamma^*}(Q^2,\Delta_{\gamma^*}) 
\nonumber\\
&& +\, \delta^{\rm (LO;odd)}_{\gamma\gamma}(Q^2)\,,
\end{eqnarray}
where $\delta^{\rm (LO;odd)}_{\gamma\gamma^*}$ is given by Eqs.~\eqref{eq:deltaLO_brem_odd} and 
\eqref{eq:deltaLO-1/M_brem_odd}, and $\delta^{\rm (LO;odd)}_{\gamma\gamma}$ is given by 
Eq.~\eqref{eq:deltaLO_TPE_odd}. 

 \subsubsection{$\nu=1$ charge-odd contribution} 
Next, we compute the {\it genuinely dynamical} NLO charge-odd contributions to the cross section by involving 
the NLO soft-photon bremsstrahlung amplitudes. The following charge-odd component of the interference of the LO
and NLO matrix elements must be determined in this case (cf. Figs.~\ref{fig:LO} and~\ref{fig:NLO}):
\begin{widetext}
\begin{eqnarray}
\label{eq:So_NLO}
\mathcal{S}^{(\nu=1)}_{\rm odd} \!&=&\! 
\sum_{\rm spins} \left[{\mathcal{M}^\dagger_{\rm LO}} \mathcal{M}_{\rm NLO} + {\rm h.c}\right]_{\rm odd} 
\nonumber\\
&=&\! 2\mathcal{R}e \sum_{\rm spins}\Bigg[\left(\mathcal{M}^{\left[l;{\rm i}\right]}_{\gamma\gamma^*} 
+\mathcal{M}^{\left[l;{\rm f}\right]}_{\gamma\gamma^*}\right)^\dagger 
\bigg(\mathcal{M}^{\left[p(1);{\rm i}\right]}_{\gamma\gamma^*} 
+ \mathcal{M}^{\left[p(1);{\rm f}\right]}_{\gamma\gamma^*}  
+ \mathcal{M}^{\left[\,\overline{p(0);{\rm i}}\,\right]}_{\gamma\gamma^*}
+ \mathcal{M}^{\left[\,\overline{p(0);{\rm f}}\,\right]}_{\gamma\gamma^*} 
+ \mathcal{M}^{\left[p(0);{\rm i}\otimes\right]}_{\gamma\gamma^*} 
\nonumber\\
&& \hspace{4.7cm} +\, \mathcal{M}^{\left[p(0);{\rm f}\otimes\right]}_{\gamma\gamma^*} 
+ \mathcal{M}^{\left[p(1);{\rm v}\right]}_{\gamma\gamma^*}\bigg) 
+\left(\mathcal{M}^{\left[p(0);{\rm i}\right]}_{\gamma\gamma^*} 
+ \mathcal{M}^{\left[p(0);{\rm f}\right]}_{\gamma\gamma^*}\right)^\dagger
\left(\mathcal{M}^{\left[\,\overline{l;{\rm i}}\,\right]}_{\gamma\gamma^*} 
+ \mathcal{M}^{\left[\,\overline{l;{\rm f}}\,\right]}_{\gamma\gamma^*}\right)\Bigg] \,.
\nonumber\\
\end{eqnarray}

After the integration over the bremsstrahlung phase-space, the corresponding charge-odd lepton-proton 
differential cross section in SPL can be expressed as
\begin{eqnarray}
\left[\frac{{\rm d}\sigma^{\rm (NLO;odd)}_{br}(Q^2)}{{\rm d}\Omega^\prime_l} 
\right]^{(E_{\gamma^*}\leq \Delta_{\gamma^*})}_{\gamma\gamma^*} && \hspace{-0.5cm} = 
\frac{\alpha M}{\pi^2 Q^2} \left[\frac{{\rm d}\sigma_{el}(Q^2)}{{\rm d}\Omega^\prime_l} \right]_{0} 
\Bigg[EI_1-E^\prime I_2 + \frac{Q^2}{2M}\left\{I_1 + I_2 + \frac{2}{Q^2}\left(EI_5 -E^\prime I_6\right)\right\} 
\nonumber\\
&& \hspace{3cm} +\, \frac{Q^2}{M^2}\left\{EI_1 - E^\prime I_2 + \frac{4m^2_l(E+E^\prime)}{Q^2+4 E E^\prime} 
\left(I_1 - I_2\right)\right\}\Bigg] + \mathcal{O}\left(\frac{1}{M^2}\right),
\nonumber\\
\end{eqnarray}
where $I_{1,2,5,6}$ are all IR-finite integrals. These integrals are most conveniently evaluated in the 
S-frame whose expressions are presented in Appendix B. Translating those results to the {\it lab.}-frame, 
the corresponding expressions for $I_{1,2}$ have already been presented in Eqs.~\eqref{eq:I1} and 
\eqref{eq:I2}, while those for $I_{5}$ and $I_{6}$ are respectively given as
\begin{eqnarray} 
\label{eq:I5}
I_5 \!&=&\! \frac{2\pi\Delta^2_{\gamma^*} E^2}{\beta EE^{\prime 2}} \ln\sqrt{\frac{1+\beta}{1-\beta}} \,, \quad \text{and}
\\
\label{eq:I6}
I_6 \!&=&\! \frac{2\pi\Delta_{\gamma^*}^2 E^2}{\beta^\prime E^{\prime 3}} \ln\sqrt{\frac{1+\beta^\prime}{1-\beta^\prime}} \,.
\end{eqnarray}
We then obtain the fractional contribution to the charge-odd lepton-proton differential cross section at 
NLO [i.e., ${\mathcal O}(\alpha^3/M)$] arising from the LO and NLO bremsstrahlung amplitudes. As mentioned
earlier, the NLO correction is IR-finite and given by the following expression:
\begin{eqnarray}
\label{eq:deltaNLO_brem_odd}
\delta^{\rm (NLO;odd)}_{\gamma\gamma^*}(Q^2,\Delta_{\gamma^*}) \!&=&\! 
\left[\frac{{\rm d}\sigma^{\rm (NLO;odd)}_{br}(Q^2)}{{\rm d}\Omega^\prime_l} 
\right]^{(E_{\gamma^*}\leq \Delta_{\gamma^*})}_{\gamma\gamma^*}\!\!\!\!\bigg{/}
\left[\frac{{\rm d}\sigma_{el}(Q^2)}{{\rm d}\Omega^\prime_l} \right]_{0}
\nonumber\\
&=&\! -\,\frac{2\alpha\Delta_{\gamma^*}}{\pi E\beta^2} 
\Bigg[1-\left(\beta+\frac{1}{\beta}\right) \ln\sqrt{\frac{1+\beta}{1-\beta}}\,\Bigg] 
+ \delta^{({\rm NLO-}1/M;{\rm odd})}_{\gamma\gamma^*}(Q^2,\Delta_{\gamma^*})\, \qquad \text{where} 
\nonumber\\
 \delta^{({\rm NLO-}1/M;{\rm odd})}_{\gamma\gamma^*}(Q^2,\Delta_{\gamma^*}) \!&=&\!  
 -\frac{\alpha Q^2}{2\pi ME\beta^2} \Bigg[\frac{2\Delta^2_{\gamma^*}}{Q^2} 
- \frac{3\Delta_{\gamma^*}}{E\beta^2}\left(1+\frac{2\beta^2}{3}\right) 
+\left\{\frac{2\Delta^2_{\gamma^*}}{Q^2}\left(\beta-\frac{1}{\beta}\right) \right.
\nonumber\\
&& \left. \hspace{1.85cm} +\,\frac{\Delta_{\gamma^*}}{E}\left(\frac{3}{\beta^3} 
+\frac{1}{\beta}+2\beta\right)\right\}\ln\sqrt{\frac{1+\beta}{1-\beta}} \,\Bigg]
+ \mathcal{O}\left(\frac{1}{M^2}\right) \,.
\end{eqnarray} 
\end{widetext} 
We find that the above expression is not genuinely NLO [i.e., ${\mathcal O}(1/M)$]. Interestingly, the 
first term contains a LO [i.e., ${\mathcal O}(M^0)$] term due to the chiral enhancement. This contribution 
originates from the NLO amplitudes involving the \underline{final-state} proton radiation, each containing
a proton propagator that scales as ${\mathcal O}(M)$, namely, 
$\mathcal{M}^{\left[p(1);{\rm f}\right]}_{\gamma\gamma^*} \,,\, 
\mathcal{M}^{\left[\,\overline{p(0);{\rm f}}\,\right]}_{\gamma\gamma^*}\,,\,$ and
$\mathcal{M}^{\left[p(0);{\rm f}\otimes\right]}_{\gamma\gamma^*}$, which yield the above LO component. It 
is important to recognize that the ${\mathcal O}(1/M)$ terms from the previously considered LO charge-odd 
bremsstrahlung correction, namely, $\delta^{({\rm LO}-1/M;{\rm odd})}_{\gamma\gamma^*}$ in 
Eq.~\eqref{eq:deltaLO-1/M_brem_odd}, must be retained alongside $\delta^{\rm (NLO;odd)}_{\gamma\gamma^*}$ 
to properly account for the complete NLO contribution. In addition, one must also include the potential 
contributions to the NLO cross section arising from the ${\mathcal O}(M)$ chirally \underline{enhanced} 
NNLO and N${}^3$LO bremsstrahlung amplitudes, which we examine in the following analysis.

\subsubsection{$\nu=2$ charge-odd contribution}
The computation of the charge-odd component of the cross section for soft-photon bremsstrahlung 
up-to-and-including NLO, entails the inclusion of additional exchange-interference matrix elements 
involving products of LO and NLO amplitudes with chirally \underline{enhanced} N$^2$LO amplitudes (cf. 
Figs.~\ref{fig:LO}, \ref{fig:NLO} and \ref{fig:NNLO}):
\begin{widetext}
\begin{eqnarray}
\mathcal{S}^{(\nu=2)}_{\rm odd} \!&=&\! \sum_{\rm spins} \left|{\mathcal M}_{\rm NLO}\right|^2_{\rm odd} 
+ \sum_{\rm spins} \left[{\mathcal M}^\dagger_{\rm LO} \widetilde{\mathcal M}_{\rm NNLO} + {\rm h.c}\right]_{\rm odd} 
\nonumber\\
&=&\! 2 \mathcal{R}e \sum_{\rm spins} 
\Bigg[\left(\mathcal{M}^{\left[\,\overline{l;{\rm i}}\,\right]}_{\gamma\gamma^*} 
+ \mathcal{M}^{\left[\,\overline{l;{\rm f}}\,\right]}_{\gamma\gamma^*}\right)^\dagger 
\left({\mathcal{M}^{\left[p(1);{\rm f}\right]}}_{\gamma\gamma^*} 
+ \mathcal{M}^{\left[\,\overline{p(0);{\rm f}}\,\right]}_{\gamma\gamma^*} 
+ \mathcal{M}^{\left[p(0);{\rm f}\otimes\right]}_{\gamma\gamma^*} \right)
+ \left(\mathcal{M}^{\left[l;{\rm i}\right]}_{\gamma\gamma^*} 
+ \mathcal{M}^{\left[l;{\rm f}\right]}_{\gamma\gamma^*}\right)^\dagger 
\nonumber\\
&& \hspace{0.75cm} \times\, \left(\mathcal{M}^{\left[\,\overline{p(1);f}\,\right]}_{\gamma\gamma^*} 
+ \mathcal{M}^{\left[\,\widetilde{p(0);f}\,\right]}_{\gamma\gamma^*} 
+ \mathcal{M}^{\left[p(2);f\right]}_{\gamma\gamma^*} 
+ \mathcal{M}^{\left[p(0);f\boxtimes\right]}_{\gamma\gamma^*} 
+ \mathcal{M}^{\left[p(1);f \otimes\right]}_{\gamma\gamma^*}  
+ \mathcal{M}^{\left[\,\overline{p(0);f \otimes}\,\right]}_{\gamma\gamma^*} 
+ \mathcal{M}^{\left[\rm \pi (c)\right]}_{\gamma\gamma^*}+\mathcal{M}^{\left[\rm \pi (d)\right]}_{\gamma\gamma^*} \right) 
\nonumber\\
&& \hspace{0.75cm} +\, {\mathcal M}^{\left[p(0);{\rm f}\right]\dagger}_{\gamma\gamma^*}
\left(\mathcal{M}^{\left[\widetilde{\,l;{\rm i}\,}\right]}_{\gamma\gamma^*} 
+ \mathcal{M}^{\left[\widetilde{\,l;{\rm f}\,}\right]}_{\gamma\gamma^*} 
+ \mathcal{M}^{\left[\rm \pi (a)\right]}_{\gamma\gamma^*}
+\mathcal{M}^{\left[\rm \pi(b)\right]}_{\gamma\gamma^*}\right)\Bigg] 
+ \mathcal{O}\left(\frac{1}{M^2}\right)\,.
\label{eq:So_NNLO}
\end{eqnarray}
\end{widetext}
Here, the ${\mathcal O}(1/M^2)$ terms correspond to neglected products of amplitudes lacking chiral 
enhancement, resulting in higher-order contributions to the charge-odd bremsstrahlung cross section.
We have relegated all such ${\mathcal O}(1/M^2)$ term $\in S^{(\nu=2)}_{\rm odd}$ to Appendix A [see 
Eq.~\eqref{eq:So_NNLO_dropped}]. 

After the phase-space integration, the terms displayed in Eq.~\eqref{eq:So_NNLO} yield the following 
additional correction terms to the charge-odd lepton-proton differential cross section
up-to-and-including NLO in SPL:
\begin{widetext}
\begin{eqnarray}
\left[\frac{{\rm d}\sigma^{\rm (NNLO;odd)}_{br}(Q^2)}{{\rm d}\Omega^\prime_l}
\right]^{(E_{\gamma^*}\leq \Delta_{\gamma^*})}_{\gamma\gamma^*} \hspace{-0.3cm} \!&=&\! 
\frac{\alpha}{2\pi^2 Q^2} \left[\frac{{\rm d}\sigma_{el}(Q^2)}{{\rm d}\Omega^\prime_l} \right]_{0} 
\Bigg[ Q^2(I_1 + I_2) + 2(EI_5-E^\prime I_6)  + \frac{Q^2}{2M} \Bigg\{2\left(\frac{r^2_p M^2}{3} +1\right) 
\nonumber\\
&& \hspace{3.3cm} \times\, \left(EI_1 -E^\prime I_2\right) + (I_5 +I_6) + \frac{3}{Q^2}\left(EI_7 -E^\prime I_8\right) 
\nonumber\\
&& \hspace{3.3cm} +\, \frac{1}{Q^2}\left(EI_9 -E^\prime I_{10}\right) \Bigg\} \Bigg] 
+ \mathcal{O}\left(\frac{1}{M^2}\right)
\nonumber\\
\!&=&\! \frac{\alpha}{2\pi^2 Q^2} \left[\frac{{\rm d}\sigma_{el}(Q^2)}{{\rm d}\Omega^\prime_l} \right]_{0} 
\Bigg[ Q^2(I_1 + I_2) + 2(EI_5-E^\prime I_6)  + \frac{Q^2}{2M}\left(I_5 +I_6\right) \Bigg] 
+ \mathcal{O}\left(\frac{1}{M^2}\right)\,,
\nonumber\\
\label{eq:dsigmaNNLO_brem_odd}
\end{eqnarray}
\end{widetext}
where the finite phase-space integrals $I_{1,2,5,6}$, have already been presented earlier. The integrals 
$I_{7,8,9,10}$ were evaluated by boosting to the S-frame [cf. Appendix B], and subsequently 
translated back into the {\it lab.}-frame after their evaluation. They are given below:
\begin{eqnarray}
\label{eq:I7}
I_7 \!&=&\! \frac{4\pi\Delta_{\gamma^*}^3 E^2}{3\beta E^{\prime 3}} \ln\sqrt{\frac{1+\beta}{1-\beta}} \,,
\\
\label{eq:I8}
I_8 \!&=&\!  \frac{4\pi\Delta_{\gamma^*}^3 E^3}{3\beta^\prime E^{\prime 4}} 
\ln\sqrt{\frac{1+\beta^\prime}{1-\beta^\prime}} \,,
\\
\label{eq:I9}
I_9 \!&=&\! \frac{4\pi \Delta_{\gamma^*}^2 E}{2\beta E^\prime}  
\left[\beta^\prime-\left(1-\frac{\beta^\prime}{\beta}\right) \ln\sqrt{\frac{1+\beta}{1-\beta}}\,\right]\,,\quad\,
\\
&& \hspace{-1cm} \text{and} \nonumber\\
\label{eq:I10}
I_{10} \!&=&\!  -\, \frac{4\pi\Delta_{\gamma^*}^2 E^3}{2\beta^\prime E^{\prime 3}}  
\left[\beta-\left(1-\frac{\beta}{\beta^\prime} \right) 
\ln\sqrt{\frac{1+\beta^\prime}{1-\beta^\prime}}\,\right]\,.\qquad\, 
\end{eqnarray} 
Here we reiterate that in our analysis, the proton’s rms charge radius $r_p$ is determined as a 
phenomenological fixed LEC. Since it implicitly accounts for all possible pion-loop contributions
responsible for the proton's hadronic structure, it must therefore scale as 
$r^2_p\sim{\mathcal O}(1/M^2)$. Thus, all terms enclosed within the curly brackets 
``$\left\{ \cdots \right\}$” in Eq.~\eqref{eq:dsigmaNNLO_brem_odd}, appear to scale as 
${\mathcal O}(1)$. However, the subtlety here is that all the following combinations of integrals
scale as, $\left(EI_1 -E^\prime I_2\right)\sim \left(EI_7-E^\prime I_8\right)\sim 
\left(EI_9 -E^\prime I_{10}\right)\sim {\mathcal O}(1/M)$, namely, they are all kinematically 
suppressed with ${\mathcal O}(1)$ terms canceling out. Multiplied by the factor $Q^2/(2M)$, all 
these terms within the curly brackets, except for $I_{5}\sim I_{6}\sim {\mathcal O}(1)$, scale as
${\mathcal O}(1/M^2)$, and hence, ultimately discarded. Accordingly, the only additional finite 
fractional contribution to the charge-odd lepton-proton differential cross section 
up-to-and-including NLO, involving the chirally \underline{enhanced} NNLO amplitudes, is given as   
\begin{widetext}
\begin{eqnarray}
\label{eq:deltaNNLO_brem_odd}
\delta^{\rm (NNLO;odd)}_{\gamma\gamma^*}  (Q^2,\Delta_{\gamma^*}) \!&=&\!
\left[\frac{{\rm d}\sigma^{\rm (NNLO;odd)}_{br}(Q^2)}{{\rm d}\Omega^\prime_l} 
\right]^{(E_{\gamma^*}\leq \Delta_{\gamma^*})}_{\gamma\gamma^*}\!\!\!\!\bigg{/}
\left[\frac{{\rm d}\sigma_{el}(Q^2)}{{\rm d}\Omega^\prime_l} \right]_{0}
\nonumber\\
&=&\! \frac{4\alpha\Delta_{\gamma^*}}{\pi E\beta}\ln{\sqrt{\frac{1+\beta}{1-\beta}}} 
+ \delta^{({\rm NNLO-}1/M;{\rm odd})}_{\gamma\gamma^*} (Q^2,\Delta_{\gamma^*})\, \qquad \text{where}
\nonumber\\
\delta^{({\rm NNLO-}1/M;{\rm odd})}_{\gamma\gamma^*}  (Q^2,\Delta_{\gamma^*}) \!&=&\! 
- \frac{\alpha Q^2}{2\pi ME\beta^2}\Bigg[\frac{2\Delta^2_{\gamma^*}}{Q^2}\left(1-\frac{\beta^2 E^2}{Q^2}\right)
-\frac{2\Delta_{\gamma^*}}{E} 
\nonumber\\
&& \hspace{1.9cm} -\,\left\{\frac{2\Delta^2_{\gamma^*}}{Q^2\beta} 
- \frac{2\Delta_{\gamma^*}}{E}\left(2\beta+\frac{1}{\beta}\right)\right\}
\ln{\sqrt{\frac{1+\beta}{1-\beta}}}\,\Bigg] + \mathcal{O}\left(\frac{1}{M^2}\right)\,.
\end{eqnarray} 
\end{widetext}
It is indeed interesting that even a LO [i.e., ${\mathcal O}(M^0)$] contribution to the charge-odd 
bremsstrahlung cross section transcends into the otherwise naively counted NNLO correction as 
obtained above. The origin of such a term is attributed to the NNLO \underline{final-state} 
radiating proton amplitude $\mathcal{M}^{\left[p(1);{\rm f}\otimes\right]}_{\gamma\gamma^*}$ [cf. 
Eq.~\eqref{N2LO_2}] interfering with the two LO radiating lepton amplitudes 
$\mathcal{M}^{\left[l;{\rm i/f}\right]}_{\gamma\gamma^*}$, as seen in  Eq.~\eqref{eq:So_NNLO}. 
Moreover, we find that the hadronic structure-dependent terms -- those proportional to the proton 
charge radius $r_p$ -- arising from the NNLO bremsstrahlung amplitudes with renormalized pion-loops 
ultimately cancel out and do not contribute to the ${\mathcal O}(\alpha^3/M)$ NLO cross section.  

\subsubsection{$\nu=3$ charge-odd contribution}
Finally, we account for the contribution from the chirally \underline{enhanced} N${}^3$LO amplitudes 
(cf. Fig.~\ref{fig:N3LO}) in the evaluation of the charge-odd NLO cross section for soft-photon 
bremsstrahlung. This involves computing the following interference among the LO, NLO, and the chirally 
\underline{enhanced} NNLO and N$^3$LO amplitudes:
\begin{widetext}
\begin{eqnarray}
\label{eq:So_N3LO}
\mathcal{S}^{(\nu=3)}_{\rm odd} \!&=&\! 
\sum_{\rm spins} \left[{\mathcal M}^\dagger_{\rm NLO} \widetilde{\mathcal M}_{\rm NNLO} 
+ {\mathcal M}^\dagger_{\rm LO} \widetilde{\mathcal M}_{\rm N^3LO} + {\rm h.c}\right]_{\rm odd} 
\\
&=&\! 2\mathcal{R}e \Bigg[ \left(\mathcal{M}^{\left[\,\overline{l;{\rm i}}\,\right]}_{\gamma\gamma^*} 
+ \mathcal{M}^{\left[\,\overline{l;{\rm f}}\,\right]}_{\gamma\gamma^*}\right)^\dagger \!\!
\bigg(\mathcal{M}^{[p(1);f \otimes]}_{\gamma\gamma^*}  
+ \mathcal{M}^{[\,\overline{p(0);f \otimes}\,]}_{\gamma\gamma^*} \bigg) 
+ \mathcal{M}^{[p(0);f\otimes]\dagger}_{\gamma\gamma^*} 
\bigg(\mathcal{M}^{\left[\widetilde{\,l;{\rm i}\,}\right]}_{\gamma\gamma^*}  
+ \mathcal{M}^{\left[\widetilde{\,l;{\rm f}\,} \,\right]}_{\gamma\gamma^*} 
+ \mathcal{M}^{\left[\rm \pi (a)\right]}_{\gamma\gamma^*}+\mathcal{M}^{\left[\rm \pi (b)\right]}_{\gamma\gamma^*}\bigg) 
\nonumber\\
&& \hspace{0.75cm} +\, \left(\mathcal{M}^{\left[l;{\rm i}\right]}_{\gamma\gamma^*} 
+\mathcal{M}^{\left[l;{\rm f}\right]}_{\gamma\gamma^*}\right)^\dagger\!\!
\left(\mathcal{M}^{\left[\,\overline{p(1);{\rm f}\otimes}\,\right]}_{\gamma\gamma^*} 
+ \mathcal{M}^{\left[\,\widetilde{p(0);{\rm f}\otimes}\,\right]}_{\gamma\gamma^*} 
+ \mathcal{M}^{\left[p(2);{\rm f}\otimes\right]}_{\gamma\gamma^*} 
+ \mathcal{M}^{\left[\rm \pi (e)\right]}_{\gamma\gamma^*}+\mathcal{M}^{\left[\rm \pi(f)\right]}_{\gamma\gamma^*} \right)\Bigg] 
+ \mathcal{O}\left(\frac{1}{M^2}\right)\,,
\nonumber
\end{eqnarray}
\end{widetext}
where ${\mathcal O}(1/M^2)$ terms again denote neglected higher-order contributions to the cross 
section, arising from the interference between the NLO amplitudes $\in {\mathcal M}_{\rm NLO}$ and 
the renormalized pion-loop NNLO amplitudes $\in \widetilde{\mathcal M}_{\rm NNLO}$. Such interference 
terms $\in S^{(\nu=3)}_{\rm odd}$ are not chirally \underline{enhanced}, and hence, excluded in the 
above expression, relegating them to Appendix A [see Eq.~\eqref{eq:So_NNNLO_dropped}].

Upon integrating over the  bremsstrahlung phase-space, all contributions to the cross section from 
the interference terms
$\sum_{\rm spins} \left[{\mathcal M}^\dagger_{\rm LO} \widetilde{\mathcal M}_{\rm N^3LO}+{\rm h.c}\right]$, 
either correspond to higher-order terms of ${\mathcal O}(1/M^2)$, or lead to scaleless integrals that
vanish in dimensional regularization, as per the condition given in Eq.~\eqref{dr_condition}. 
{\it Consequently, none of the chirally \underline{enhanced} N${\,}^3$LO amplitudes ultimately 
contribute to the ${\mathcal O}(\alpha^3/M)$ NLO cross section} (cf. Fig.~\ref{fig:N3LO}). The 
remaining interference terms between ${\mathcal M}_{\rm NLO}$ and the \underline{enhanced} 
$\widetilde{\mathcal M}_{\rm NNLO}$ amplitudes grouped in ${\mathcal S}^{(\nu=3)}_{\rm odd}$, solely 
account for the following correction to the NNLO counterpart, i.e., 
Eq.~\eqref{eq:dsigmaNNLO_brem_odd}, contributing to the charge-odd soft-photon bremsstrahlung 
differential cross section in the {\it lab.}-frame:
\begin{widetext}
\begin{eqnarray}
\label{eq:dsigmaN3LO_brem_odd}
\Delta\left[\frac{{\rm d}\sigma^{\rm (NNLO;odd)}_{br}(Q^2)}{{\rm d}\Omega^\prime_l} \right]_{\gamma\gamma^*} \!&=&\! 
\frac{\alpha}{4\pi^2M} \left[\frac{{\rm d}\sigma_{el}(Q^2)}{{\rm d}\Omega^\prime_l} \right]_{0} 
\Bigg[\left(\frac{2r_p^2M^2}{3}-1\right)\left(EI_1 -E^\prime I_2 \right) + (I_5 + I_6)  
\nonumber\\
&& \hspace{3cm} +\, \frac{3}{Q^2}(EI_7 - E^\prime I_8) + \frac{1}{Q^2}(EI_9 -E^\prime I_{10})\Bigg] + \mathcal{O}\left(\frac{1}{M^2}\right)
\nonumber\\
&=&\! \frac{\alpha}{4\pi^2M} \left[\frac{{\rm d}\sigma_{el}(Q^2)}{{\rm d}\Omega^\prime_l} \right]_{0} \left(I_5 + I_6\right) 
+ \mathcal{O}\left(\frac{1}{M^2}\right)\,.
\end{eqnarray}
Here, we again find that the proton's hadronic structure-dependent terms, arising from the NNLO 
bremsstrahlung amplitudes with renormalized pion-loops, ultimately cancel out and do not contribute 
to the ${\mathcal O}(\alpha^3/M)$ NLO cross section. Accordingly, the additional finite fractional 
correction to Eq.~\eqref{eq:deltaNNLO_brem_odd}, arising from the interference between the NLO and
chirally \underline{enhanced} NNLO amplitudes and contribute to the charge-odd lepton-proton 
differential cross section at NLO [i.e., ${\mathcal O}(\alpha^3/M)$], is given as
\begin{eqnarray}
\label{eq:deltaN3LO_brem_odd} 
\widetilde{\,\delta\,}^{\rm (NNLO;odd)}_{\gamma\gamma^*}(Q^2,\Delta_{\gamma^*}) \!&=&\!
\Delta\left[\frac{{\rm d}\sigma^{\rm (NNLO;odd)}_{br}(Q^2)}{{\rm d}\Omega^\prime_l} 
\right]^{(E_{\gamma^*}\leq \Delta_{\gamma^*})}_{\gamma\gamma^*}\!\!\!\!\bigg{/}
\left[\frac{{\rm d}\sigma_{el}(Q^2)}{{\rm d}\Omega^\prime_l} \right]_{0}
\nonumber\\
&=&\! -\,\frac{\alpha Q^2}{2\pi ME\beta^2} \left[-\frac{4\Delta_{\gamma^*}^2 E^2\beta^2}{Q^4} 
- \frac{2\Delta_{\gamma^*}^2\beta}{Q^2}\ln{\sqrt{\frac{1+\beta}{1-\beta}}}\,\right] 
+ \mathcal{O}\left(\frac{1}{M^2}\right)\,, 
\end{eqnarray}
\end{widetext}
where, as stated above, 
$\widetilde{\,\delta\,}^{\rm (NNLO;odd)}_{\gamma\gamma^*}\sim {\mathcal O}(\alpha/M)$ originates from 
the two first terms in Eq.~\eqref{eq:dsigmaN3LO_brem_odd} which are generated by 
\underline{enhanced} NNLO terms, while the genuine N$^3$LO terms of ${\mathcal O}(1/M^2)$ have been
dropped. Hence, the total fractional corrections to the charge-odd (anti-)lepton-proton differential
cross section due to the soft-photon bremsstrahlung, calculated up-to-and-including NLO in HB$\chi$PT,
can be consolidated by the sum of contributions that we obtained in Eqs.~\eqref{eq:deltaLO_brem_odd},
~\eqref{eq:deltaLO-1/M_brem_odd},~\eqref{eq:deltaNLO_brem_odd},~\eqref{eq:deltaNNLO_brem_odd}, and 
\eqref{eq:deltaN3LO_brem_odd} (cf. Fig.~\ref{fig:odd_brem} for numerical estimates for $e^-$-p and 
$\mu^-$-p scatterings):

\begin{figure*}[tbp]
\begin{center}
\includegraphics[width=0.48\linewidth]{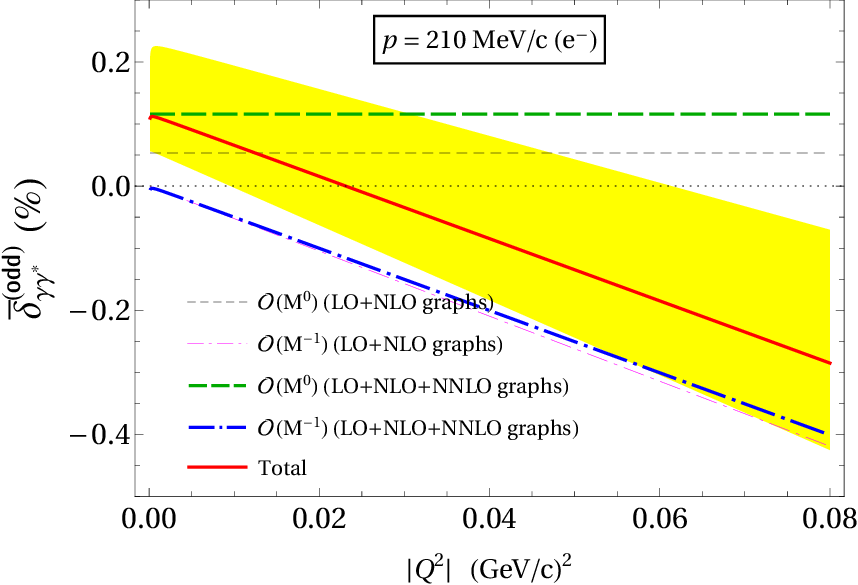}~\quad~\includegraphics[width=0.48\linewidth]{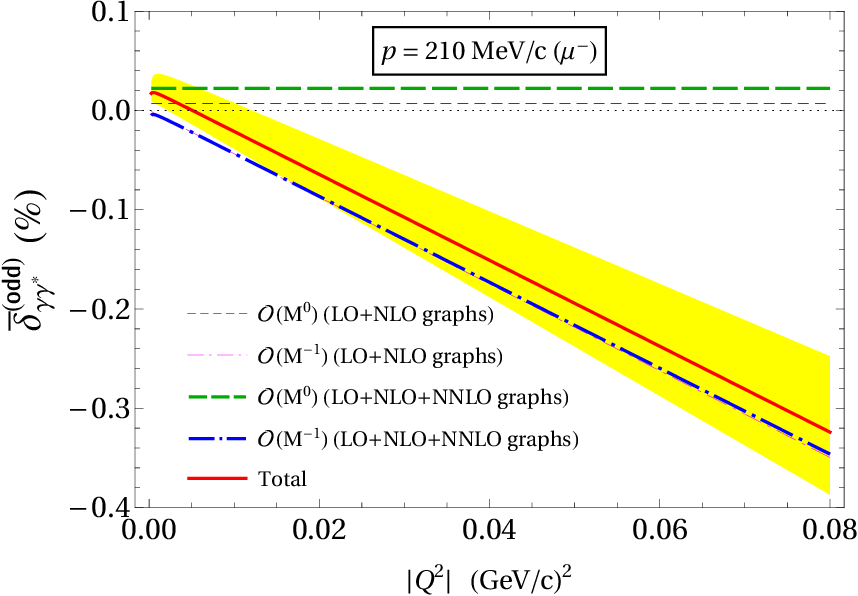}

\vspace{0.6cm}
    
\includegraphics[width=0.48\linewidth]{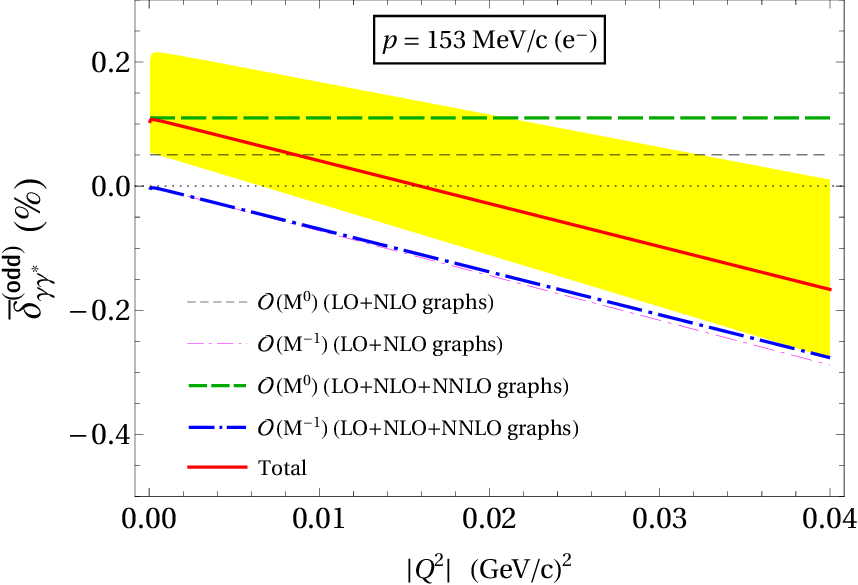}~\quad~\includegraphics[width=0.48\linewidth]{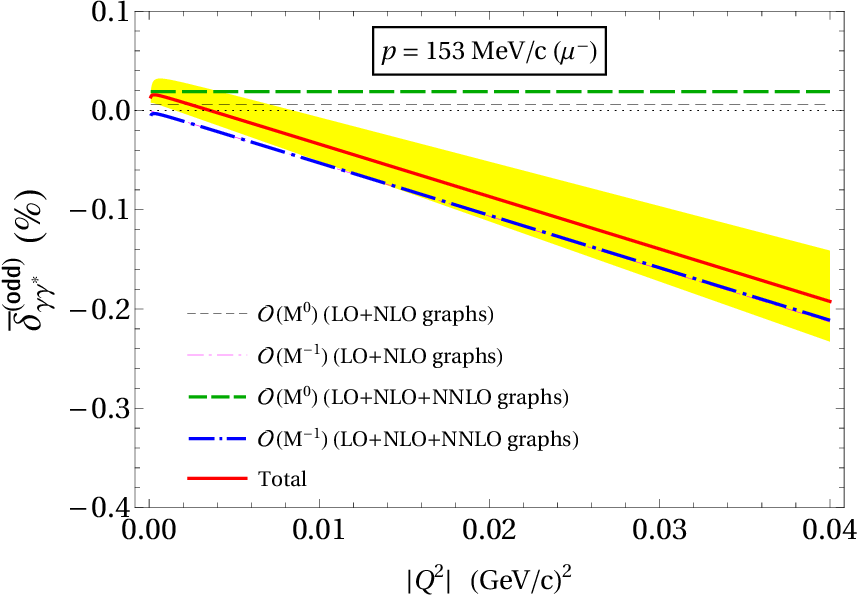}

\vspace{0.6cm}
    
\includegraphics[width=0.48\linewidth]{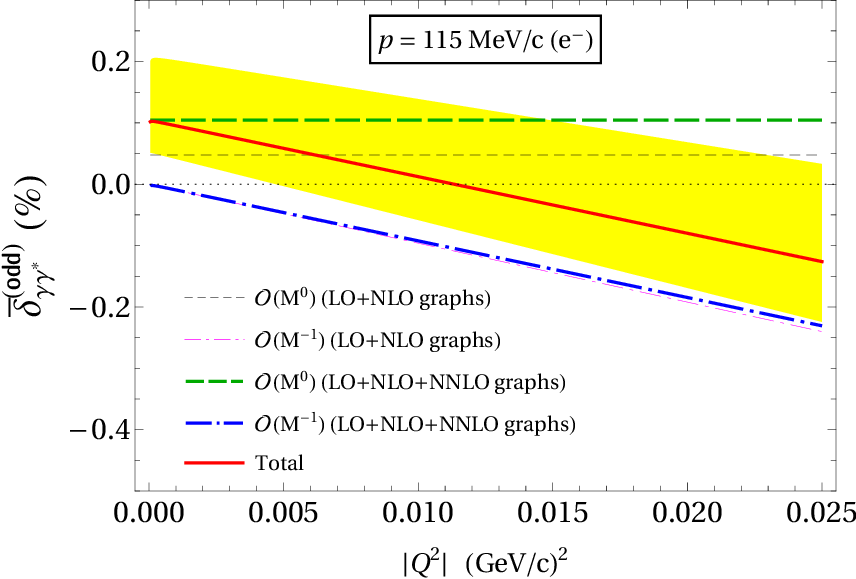}~\quad~\includegraphics[width=0.48\linewidth]{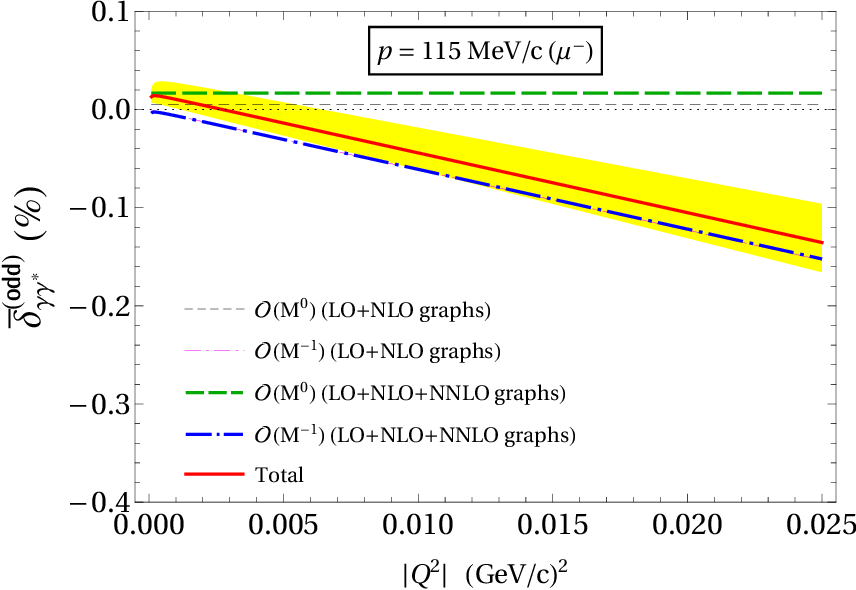}
    \caption{The finite fractional \underline{charge-odd} soft-photon bremsstrahlung corrections 
             $\overline{\delta}^{\rm (odd)}_{\gamma\gamma^*}=\delta^{\rm (odd)}_{\gamma\gamma^*}
             -{\bf IR}^{(1)}_{\gamma\gamma^*}$[expressed as percentage relative to LO Born/OPE 
             differential cross section, Eq.~\eqref{eq:diff_LO}] to the $e^{-}$-p (left panel) and $\mu^{-}$-p
             (right panel) unpolarized elastic scatterings is presented as a function of $\left|Q^2\right|$ in
             HB$\chi$PT. The results correspond to three fixed values of the incident lepton beam momenta 
             proposed by the MUSE experiment: 210~MeV/$c$, 153~MeV/$c$, and 115~MeV/$c$. Color online: The 
             green [blue] long dashed [long dashed-dot] curve represents the sum of all genuine chiral LO [NLO],
             i.e., $\mathcal{O}(M^0)$ [$\mathcal{O}(M^{-1})$] terms $\delta^{\rm (0;\,odd)}_{\gamma\gamma^*}$ 
             [$\delta^{\rm (1;\,odd)}_{\gamma\gamma^*}$], given in Eq.~\eqref{eq:delta_brem_odd_0} 
             [Eq.~\eqref{eq:delta_brem_odd_1}], which \underline{includes} the chirally enhanced contributions
             from NNLO bremsstrahlung graphs with final-state proton propagator components. The black [magenta] 
             short dashed [short dashed-dot] curve represents the corresponding results \underline{without} the
             enhanced contributions from the NNLO graphs. The red solid curve finally represents the complete 
             chirally enhanced sum (LO+NLO), namely, 
             $\delta^{\rm (0;\,odd)}_{\gamma\gamma^*}+\delta^{\rm (1;\,odd)}_{\gamma\gamma^*}$ (labeled as 
             ``Total"). All curves are evaluated at a nominal detector cutoff $\Delta_{\gamma^*} = 1\%$ of the
             incident lepton beam energy $E$. The yellow shaded band indicates the sensitivity of our full 
             result to the detector acceptance in the range, $0.5\% E < \Delta_{\gamma^*} < 2\% E$.}  
\label{fig:odd_brem}
\end{center}
\end{figure*} 
%
\begin{widetext}
\begin{eqnarray}
\delta^{\rm (odd)}_{\gamma\gamma^*}(Q^2,\Delta_{\gamma^*}) \!&=&\! 
\left[\frac{{\rm d}\sigma^{\rm (odd)}_{br}(Q^2)}{{\rm d}\Omega^\prime_l} 
\right]^{(E_{\gamma^*}\leq \Delta_{\gamma^*})}_{\gamma\gamma^*}\!\!\!\!\bigg{/}
\left[\frac{{\rm d}\sigma_{el}(Q^2)}{{\rm d}\Omega^\prime_l} \right]_{0}
\nonumber\\
&=&\!  \delta^{\rm (LO;odd)}_{\gamma\gamma^*}(Q^2,\Delta_{\gamma^*}) 
+  \delta^{\rm (NLO;odd)}_{\gamma\gamma^*}(Q^2,\Delta_{\gamma^*})  
+  \delta^{\rm (NNLO;odd)}_{\gamma\gamma^*}(Q^2,\Delta_{\gamma^*}) 
+  \widetilde{\,\delta\,}^{\rm (NNLO;odd)}_{\gamma\gamma^*}(Q^2,\Delta_{\gamma^*}) 
\nonumber\\
&\equiv&\!  \text{\bf IR}^{(1)}_{\gamma\gamma^*}(Q^2) + \delta^{\rm (0;\,odd)}_{\gamma\gamma^*}(Q^2,\Delta_{\gamma^*}) 
+  \delta^{\rm (1;\,odd)}_{\gamma\gamma^*}(Q^2,\Delta_{\gamma^*}) + \mathcal{O}\left(\frac{1}{M^2}\right)\,,
\label{eq:delta_brem_odd}
\end{eqnarray}  
where, the genuine chiral LO ($\nu=0$) and NLO ($\nu=1$) IR-finite soft-photon bremsstrahlung corrections 
are given as
\begin{eqnarray}
\label{eq:true_LO_odd}
\delta^{\rm (0;\,odd)}_{\gamma\gamma^*} (\Delta_{\gamma^*}) \!&=&\! 
\left[\delta^{\rm (LO;odd)}_{\gamma\gamma^*}(Q^2,\Delta_{\gamma^*}) 
- \delta^{({\rm LO-}1/M;{\rm odd})}_{\gamma\gamma^*}(Q^2,\Delta_{\gamma^*})\right] 
+ \left[\delta^{\rm (NLO;odd)}_{\gamma\gamma^*}(Q^2,\Delta_{\gamma^*}) 
- \delta^{({\rm NLO-}1/M;{\rm odd})}_{\gamma\gamma^*}(Q^2,\Delta_{\gamma^*}) \right]
\nonumber\\
&& + \left[\delta^{\rm (NNLO;odd)}_{\gamma\gamma^*}(Q^2,\Delta_{\gamma^*}) 
- \delta^{({\rm NNLO-}1/M;{\rm odd})}_{\gamma\gamma^*}(Q^2,\Delta_{\gamma^*}) \right] 
\nonumber\\
&=&\! -\,\frac{4 \alpha \Delta_{\gamma^*}}{\pi E \beta^2} 
\Bigg[1-\left(\beta+\frac{1}{\beta}\right) \ln{\sqrt{\frac{1+\beta}{1-\beta}}}\,\Bigg]\,, 
\label{eq:delta_brem_odd_0}
\end{eqnarray}
which is interestingly a $Q^2$-independent result, and  
{\small
\begin{eqnarray}
\label{eq:true_NLO_odd}
\delta^{\rm (1;\,odd)}_{\gamma\gamma^*}(Q^2,\Delta_{\gamma^*}) \!&=&\! 
\delta^{({\rm LO-}1/M;{\rm odd})}_{\gamma\gamma^*}(Q^2,\Delta_{\gamma^*}) 
+ \delta^{({\rm NLO-}1/M;{\rm odd})}_{\gamma\gamma^*}(Q^2,\Delta_{\gamma^*}) 
+ \delta^{({\rm NNLO-}1/M;{\rm odd})}_{\gamma\gamma^*}(Q^2,\Delta_{\gamma^*}) 
+  \widetilde{\,\delta\,}^{\rm (NNLO;odd)}_{\gamma\gamma^*}(Q^2,\Delta_{\gamma^*}) 
\nonumber\\
&=&\! -\,\frac{\alpha Q^2}{2\pi ME\beta^2} \Bigg[\ln\left(\frac{4 \Delta_{\gamma^*}^2}{m^2_l}\right) \!\! 
\left\{1+\left(\beta-\frac{1}{\beta}\right)\ln{\sqrt{\frac{1+\beta}{1-\beta}}}\,\right\} 
+ \frac{1}{2}\left(\beta-\frac{1}{\beta}\right)\!\! \left\{{\rm Li}_2\left(\frac{2\beta}{\beta-1}\right) 
- {\rm Li}_2\left(\frac{2\beta}{1 + \beta}\right) \right\} 
\nonumber\\
&& \hspace{1.8cm} +\, \frac{4\Delta^2_{\gamma^*}}{Q^2}\left(1-\frac{3\beta^2 E^2}{2Q^2}\right) 
- \frac{4\Delta_{\gamma^*}}{E}\left(1+\frac{3}{2\beta^2}\right) - \left\{\frac{4\Delta^2_{\gamma^*}}{Q^2\beta} 
- \frac{2\Delta_{\gamma^*}}{E} \left(2\beta+\frac{1}{\beta}+\frac{3}{\beta^3}\right) + \frac{2}{\beta}\right\} 
\nonumber\\
&&  \hspace{1.8cm} \times\,\ln{\sqrt{\frac{1+\beta}{1-\beta}}}\,\Bigg]\,,
\label{eq:delta_brem_odd_1}
\end{eqnarray}
}
\end{widetext}
respectively. The above result is consistent with the standard expectation in HB$\chi$PT analysis that
up to corrections of $\mathcal O(1/M)$ (i.e., chiral order $\nu\!=\!1$) in the radiative corrections, 
the proton's hadronic finite-size structure effectively does not play any role. As previously noted,
the residual ${\mathcal O}(1/M)$ IR-divergence $\text{\bf IR}^{(1)}_{\gamma\gamma^*}$, 
Eq.~\eqref{eq:IR1_brem_odd}, is exactly canceled by the charge-odd counterpart stemming from the 
residual ${\mathcal O}(1/M)$ IR divergences of the LO TPE box (a) and crossed-box (b) amplitudes (see 
Refs.~\cite{Talukdar:2019dko,Talukdar:2020aui,Choudhary:2023rsz} for detailed discussions of the TPE 
diagrams in HB$\chi$PT), i.e., $\text{\bf IR}^{(1)}_{\rm TPE}=-\text{\bf IR}^{(1)}_{\gamma\gamma^*}$. 
The net contribution to the charge-odd radiative correction to the elastic cross section is therefore 
a finite physical quantity. In the recent work by Choudhary {\it et al.}~\cite{Choudhary:2023rsz}, the
contributions to the elastic cross section from the sum of seven genuine NLO TPE amplitudes -- 
specifically, the box diagrams (c), (e), and (g); the crossed-box diagrams (d), (f), and (h); and the
seagull diagram (i), which are not shown in this paper -- were evaluated exactly using analytical 
methods. In this work, we will denote these contributions collectively as 
\begin{widetext}
\begin{eqnarray}
\delta^{\rm (NLO;odd)}_{\gamma\gamma} (Q^2) \equiv \delta^{(c)}_{\rm box} (Q^2) + \delta^{(d)}_{\rm xbox} (Q^2) 
+ \delta^{(e)}_{\rm box}(Q^2) + \delta^{(f)}_{\rm xbox}(Q^2) +\delta^{(g)}_{\rm box} (Q^2) + \delta^{(h)}_{\rm xbox}(Q^2)
+ \delta^{(i)}_{\rm seagull}(Q^2)\,.
\end{eqnarray}
\end{widetext}
Details of the analytical expressions for the individual TPE amplitudes are provided in 
Ref.~\cite{Choudhary:2023rsz}. The complete analytical expression of ${\mathcal O}(1/M)$ is too lengthy 
to be reproduced here. However, for the sake of brevity, we only provide a symbolic expression for the 
total fractional contribution to the charge-odd lepton–proton differential cross section arising from 
virtual corrections up-to-and-including NLO in HB$\chi$PT, which effectively encapsulates the complete 
impact of the virtual TPE corrections:
\begin{eqnarray}
\delta^{\rm (odd)}_{\gamma\gamma}(Q^2) \!&=&\! 
\frac{\left[\frac{{\rm d}\sigma^{\rm (odd)}_{virt}(Q^2)}{{\rm d}\Omega^\prime_l}
\right]_{\gamma\gamma}}{\left[\frac{{\rm d}\sigma_{el}(Q^2)}{{\rm d}\Omega^\prime_l} \right]_{0}}
\nonumber\\
&=&\! \delta^{\rm (LO;odd)}_{\gamma\gamma}(Q^2) + \delta^{\rm (NLO;odd)}_{\gamma\gamma}(Q^2)
\nonumber\\
&=&\! \text{\bf IR}^{(1)}_{\rm TPE}(Q^2) + \delta^{\rm (0;odd)}_{\gamma\gamma}(Q^2) 
+ \delta^{\rm (1;odd)}_{\gamma\gamma}(Q^2)  
\nonumber\\
&& +\, \mathcal{O}\left(\frac{1}{M^2}\right) \,,
\label{eq:delta_TPE}
\end{eqnarray} 
where, the genuine chiral LO ($\nu=0$) and NLO ($\nu=1$) IR-finite TPE corrections are given, respectively, 
as~\cite{Choudhary:2023rsz} 
\begin{eqnarray}
\delta^{\rm (0;odd)}_{\gamma\gamma}(Q^2) 
= \pi\alpha\frac{\sqrt{{-Q^2}}}{2E}\left[\frac{1}{1+\frac{Q^2}{4 E^2}}\right]\,,
\end{eqnarray}
and 
\begin{eqnarray}
\delta^{\rm (1;odd)}_{\gamma\gamma}(Q^2) = \delta^{(ab;1/M)}_{\gamma\gamma}(Q^2) 
+ \delta^{\rm (NLO;odd)}_{\gamma\gamma} (Q^2)\,.
\end{eqnarray} 
Hence, it follows that the total one-loop IR-finite fractional radiative correction to the charge-odd 
lepton-proton elastic differential cross section calculated up-to-and-including NLO [i.e, 
${\mathcal O}(\alpha^3/M)$] in HB$\chi$PT is given as (cf. Fig.~\ref{fig:odd_rad} for numerical 
estimates for $e^-$-p and $\mu^-$-p scatterings):
\begin{eqnarray}
\delta^{\rm (odd)}_{2\gamma}(Q^2,\Delta_{\gamma^*}) \!&=&\! 
\frac{\left[\frac{{\rm d}\sigma^{\rm (LO;odd)}_{el}(Q^2)}{{\rm d}\Omega^\prime_l} 
\right]_{2\gamma}}{\left[\frac{{\rm d}\sigma_{el}(Q^2)}{{\rm d}\Omega^\prime_l} \right]_{0}}
\nonumber\\
&=&\! \delta^{\rm (odd)}_{\gamma\gamma^*}(Q^2,\Delta_{\gamma^*}) + \delta^{\rm (odd)}_{\gamma\gamma}(Q^2)\,,\qquad\,\,
\label{eq:delta_rad_odd2}
\end{eqnarray}
where $\delta^{\rm (odd)}_{\gamma\gamma^*}$ is total charge-odd bremsstrahlung contribution 
up-to-and-including NLO, Eq.~\eqref{eq:delta_brem_odd}, and $\delta^{\rm (odd)}_{\gamma\gamma}$ is the
corresponding total TPE contribution, Eq.~\eqref{eq:delta_TPE}.

In the interest of evaluating the total differential cross section for elastic (anti-)lepton-proton 
scattering, we incorporate all QED and recoil-radiative corrections of ${\mathcal O}(\alpha/M)$ 
accurately up-to-and-including NLO, as well as the hadronic corrections of 
${\mathcal O}(1/\Lambda^2_{\chi})$ arising at NNLO in HB$\chi$PT, by summing the charge-odd and 
charge-even fractional contributions with respect to the LO Born contribution, i.e.,
\begin{eqnarray} 
\label{eq:Charge-difference}  
\left[ \frac{{\rm d} \sigma_{el}(Q^2)}{{\rm d}\Omega^\prime_{\ell}} \right]^{(\mp)} \hspace{-0.3cm} \!&=&\!  
\left[ \frac{{\rm d} \sigma_{el}(Q^2)}{{\rm d}\Omega^\prime_\ell} \right]_{0} 
\nonumber\\
&& \!\!\! \times \Big[1 + \delta^{\rm (even)}_{el}(Q^2,r_p) \pm \delta^{\rm (odd)}_{2\gamma}(Q^2,\Delta_{\gamma^*})\Big], 
\nonumber\\
\end{eqnarray}
where the charge-odd contributions, $+\delta^{\rm (odd)}_{\gamma\gamma^*}$ and 
$-\delta^{\rm (odd)}_{\gamma\gamma^*}$, correspond to the case of lepton-proton ($\ell^-$-p) and 
anti-lepton-proton ($\ell^+$-p) scattering, respectively. In particular, the total charge-even 
contribution to the elastic cross section $\delta^{\rm (even)}_{el}$, includes not only the 
charge-even real and virtual radiative corrections components, but also the hadronic corrections to 
the LO OPE arising at NNLO in chiral expansion, 
$\delta^{(2)}_{\chi} \sim \mathcal{O}(1/\Lambda^2_{\chi})$, as given in Eq.~\eqref{delta_chi}. Thus,
the total charge-even fractional corrections (with respect to the LO Born cross section) can be 
quantified by the following sum:
\begin{equation}
\delta^{(\rm even)}_{el}(Q^2,\Delta_{\gamma^*},r_p) = \delta^{(2)}_{\chi}(Q^2,r_p) 
+ \delta^{(\rm even)}_{2\gamma}(Q^2,\Delta_{\gamma^*}) \,,
\end{equation}
where $\delta^{(\rm even)}_{2\gamma}$ represents out total charge-even radiative correction result,
as obtained in Eq.~\eqref{eq:delta_rad_even} [cf. Fig.~\ref{fig:even_rad}]. In 
Fig.~\ref{fig:Total_radiative}, we summarize our numerical estimates for the total radiative 
corrections calculated up-to-and-including NLO [i.e., ${\mathcal O}(\alpha^3/M)$] in HB$\chi$PT for 
lepton proton ($\ell^-$-p) and anti-lepton proton ($\ell^+$-p) scatterings, namely,
\begin{eqnarray}
\delta^{(\ell^\mp{\rm p})}_{2\gamma}(Q^2,\Delta_{\gamma^*}) \!&=&\! \delta^{\text{(even)}}_{2\gamma}(Q^2,\Delta_{\gamma^*}) 
\nonumber\\
&& \pm\, \delta^{\text{(odd)}}_{2\gamma}(Q^2,\Delta_{\gamma^*})\,;\quad \ell\equiv e,\,\mu\,,
\nonumber\\
\label{eq:leptonProton}
\end{eqnarray}
where $\delta^{\text{(odd)}}_{2\gamma}$ was obtained in Eq.~\eqref{eq:delta_rad_odd2}.

\begin{figure*}[tbp]
\begin{center}
\includegraphics[width=0.48\linewidth]{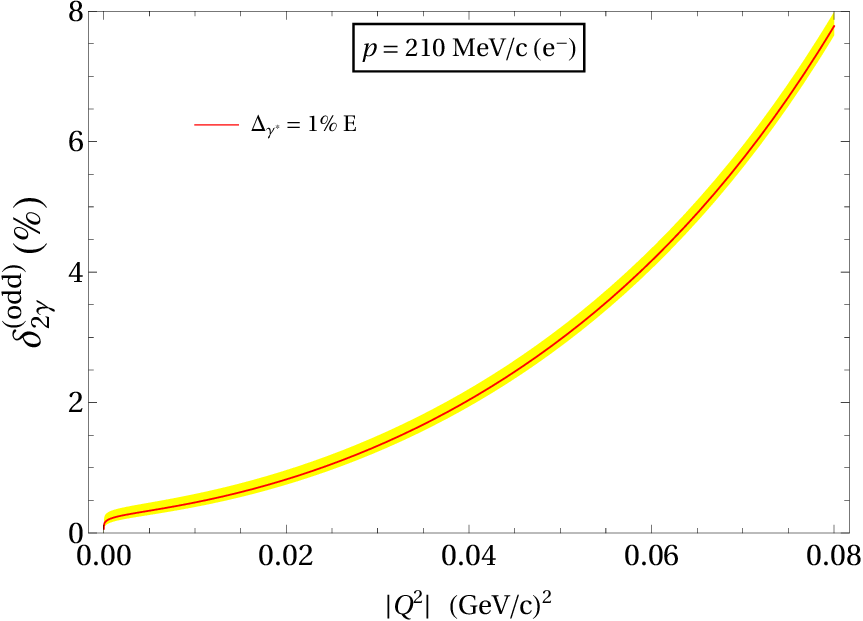}~\quad~\includegraphics[width=0.48\linewidth]{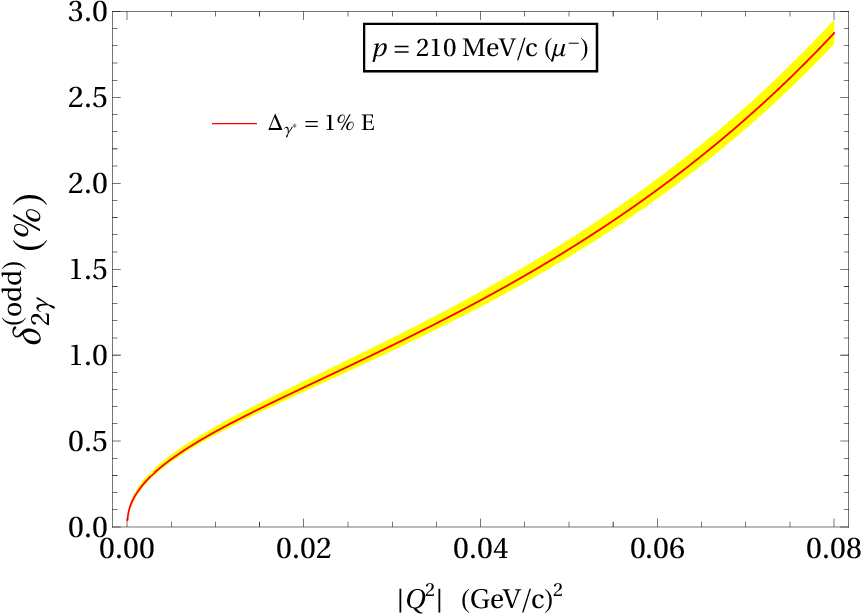}

\vspace{0.6cm}
    
\includegraphics[width=0.48\linewidth]{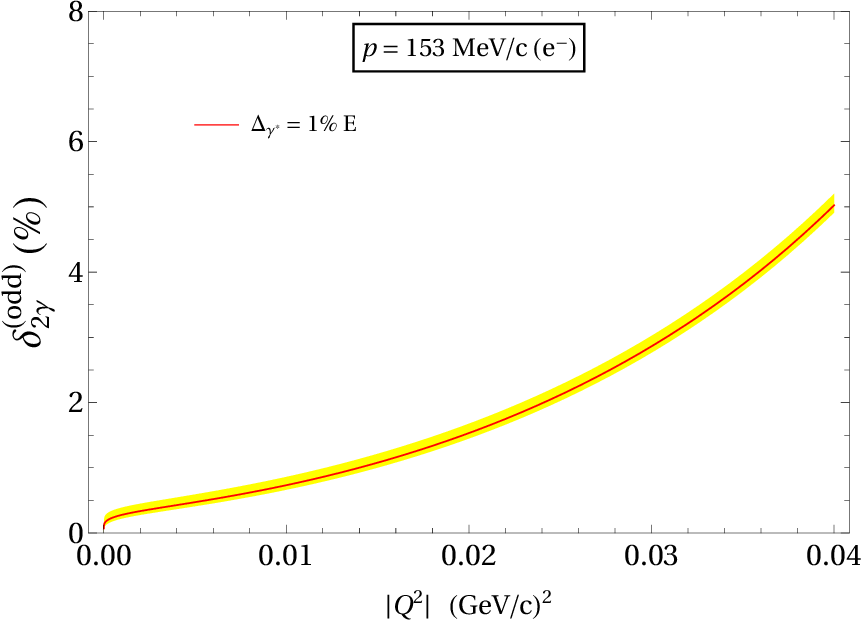}~\quad~\includegraphics[width=0.48\linewidth]{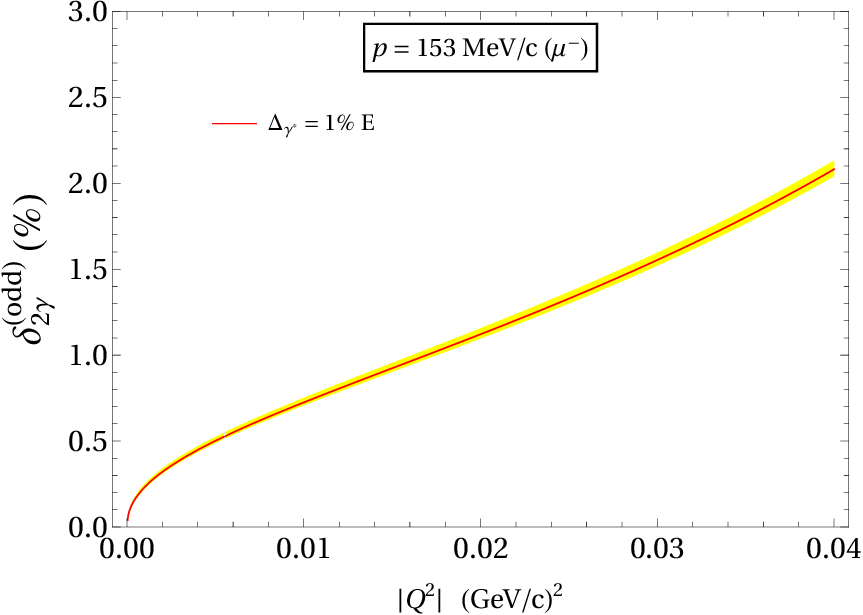}

\vspace{0.6cm}
    
\includegraphics[width=0.48\linewidth]{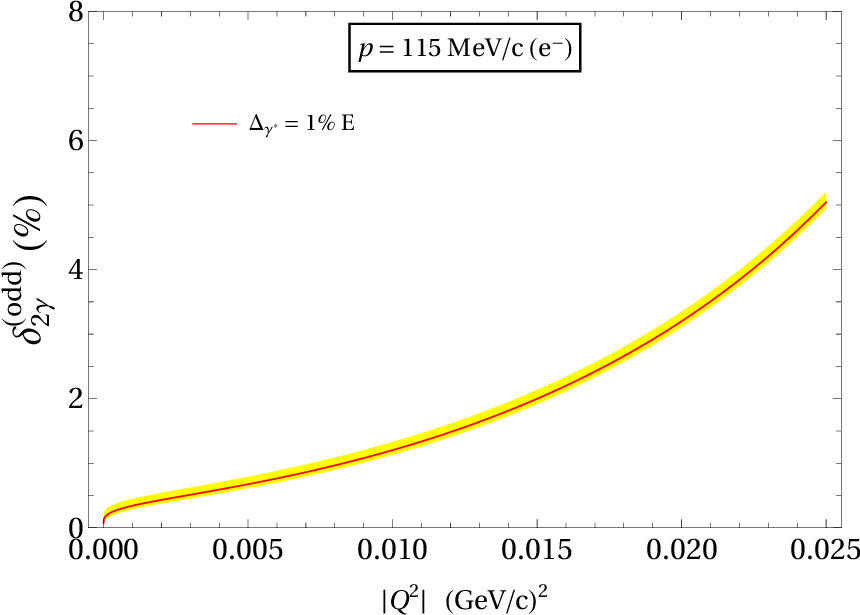}~\quad~\includegraphics[width=0.48\linewidth]{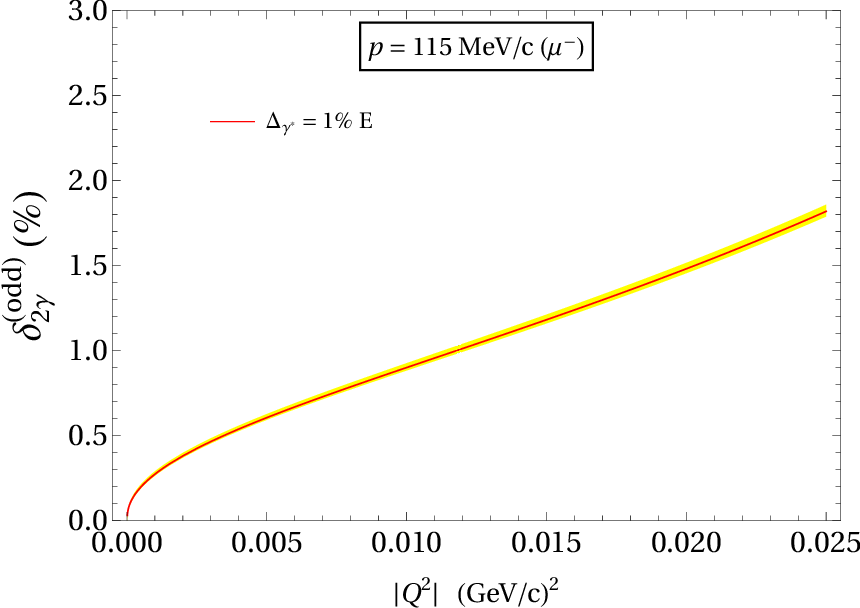}
    \caption{The finite fractional \underline{charge-odd} radiative corrections [expressed in 
             percentage relative to LO Born/OPE differential cross section, Eq.~\eqref{eq:diff_LO}]
             to the $e^{-}$-p (left panel) and $\mu^{-}$-p (right panel) unpolarized elastic 
             scatterings is presented as a function of $\left|Q^2\right|$ in HB$\chi$PT. The results 
             correspond to three fixed values of the incident lepton beam momenta proposed by the MUSE
             experiment: 210~MeV/$c$, 153~MeV/$c$, and 115~MeV/$c$. Color online: The red solid curve 
             represents the total charge-odd radiative corrections, $\delta_{2\gamma}^{\rm (odd)}$ 
             given in Eq.~\eqref{eq:delta_rad_odd2}, evaluated at a nominal detector cutoff 
             $\Delta_{\gamma^*}=1\%$ of the incident lepton beam energy $E$. The shaded yellow band 
             indicates the sensitivity of our result to the detector acceptance in the range, 
             $0.5\%E <\Delta_{\gamma^*}< 2\% E$.}
\label{fig:odd_rad}
\end{center}
\end{figure*} 
%
\begin{figure*}[tbp]
\begin{center}
\includegraphics[width=0.48\linewidth]{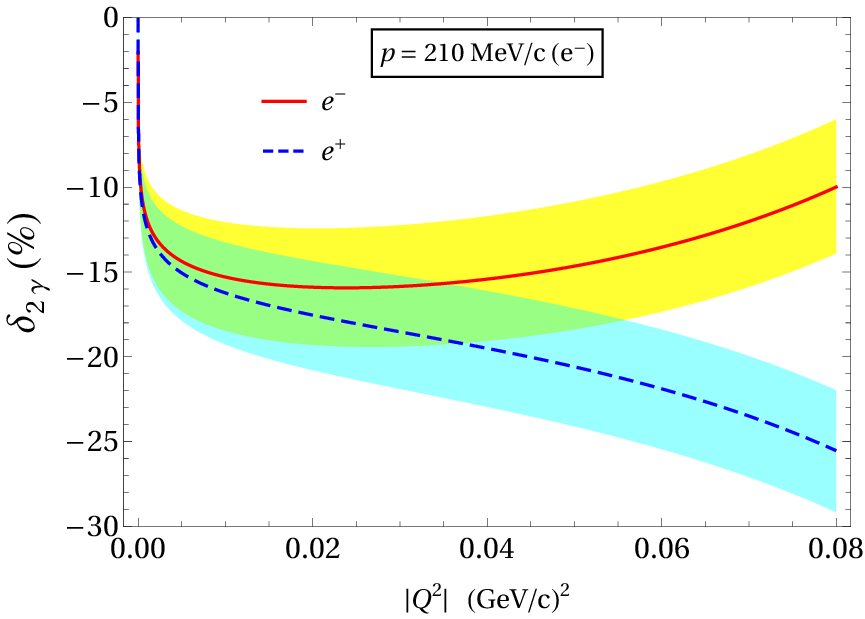}~\quad~\includegraphics[width=0.48\linewidth]{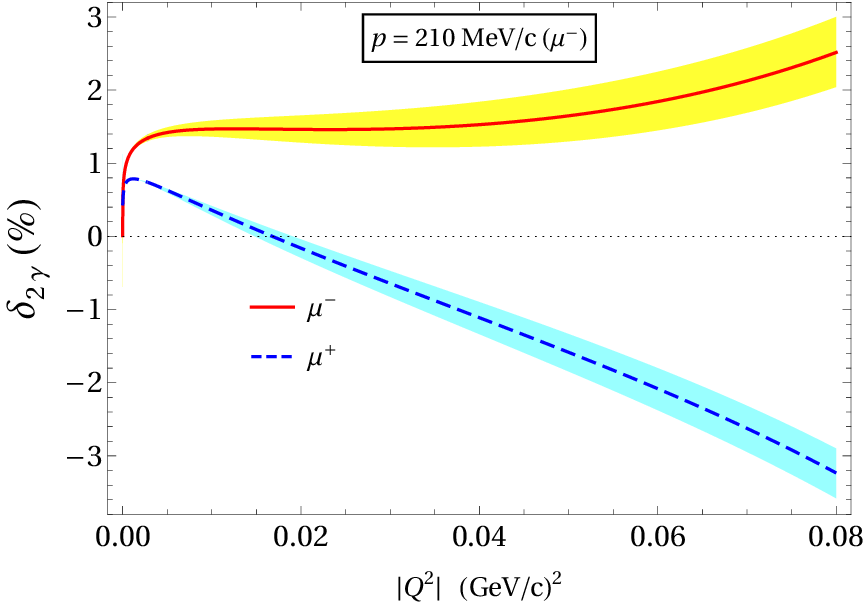}

\vspace{0.6cm}
    
\includegraphics[width=0.48\linewidth]{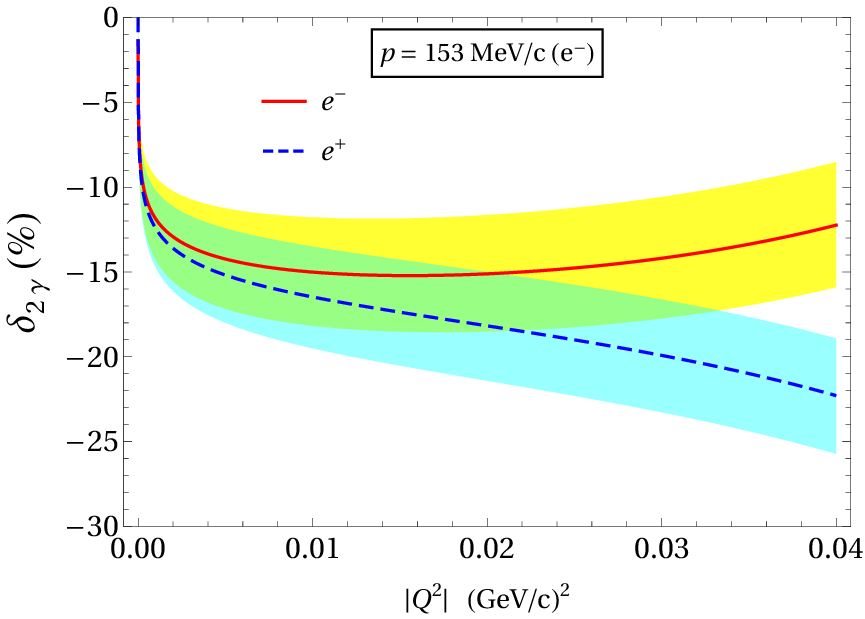}~\quad~\includegraphics[width=0.48\linewidth]{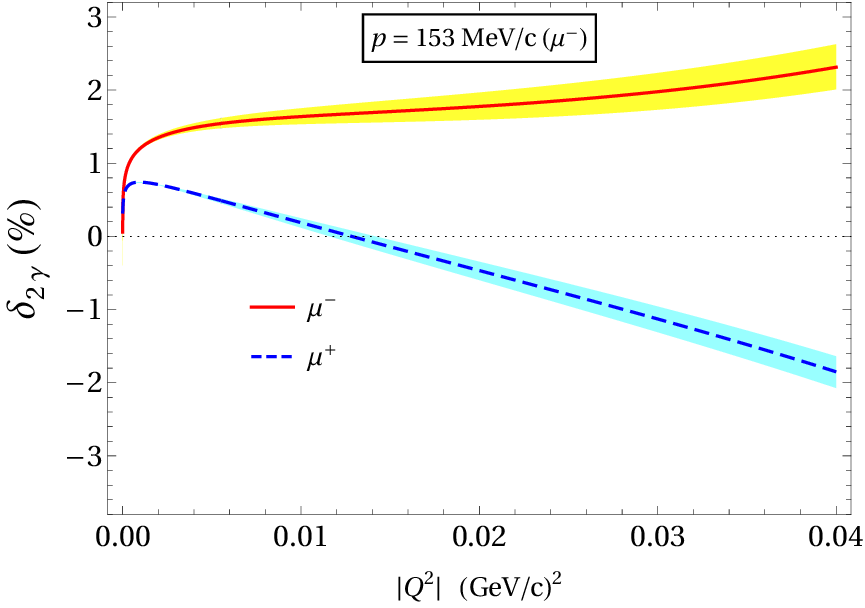}

\vspace{0.6cm}
    
\includegraphics[width=0.48\linewidth]{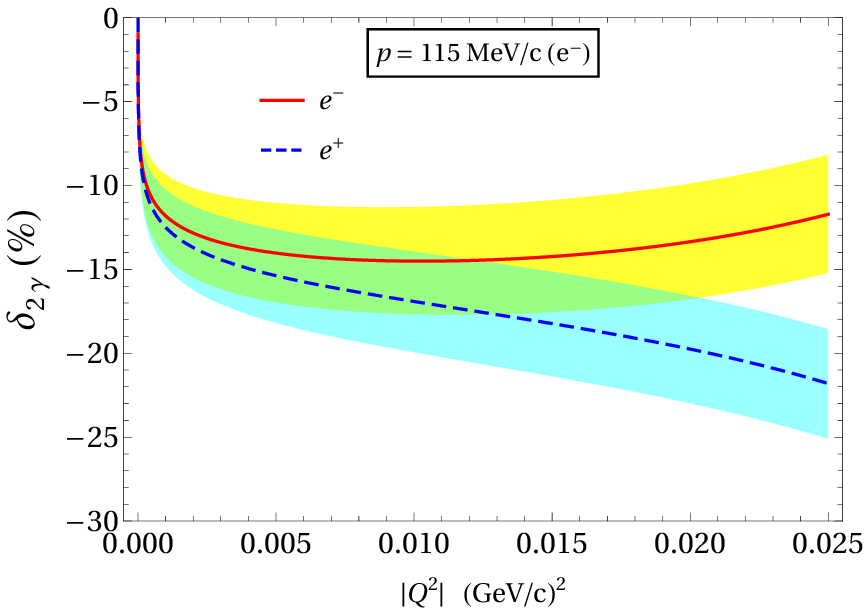}~\quad~\includegraphics[width=0.48\linewidth]{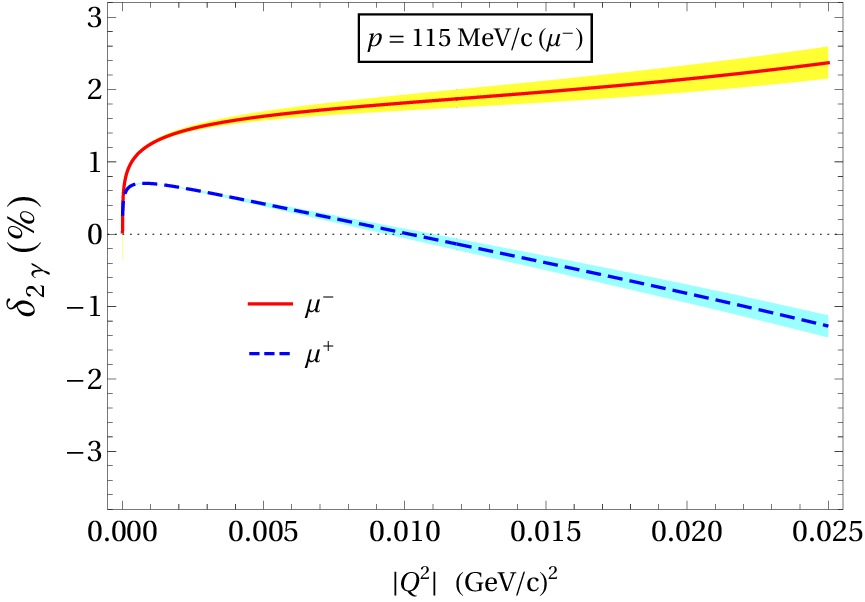}
    \caption{Using Eq.~\eqref{eq:leptonProton}, the total finite fractional radiative corrections for
             the $\ell^-$-p (solid red line) and $\ell^+$-p (blue dashed line) unpolarized elastic 
             scatterings [expressed in percentage relative to LO OPE or Born differential cross section, 
             Eq.~\eqref{eq:diff_LO}] up-to-and-including NLO [i.e., ${\mathcal O}(\alpha^3/M)$] is 
             presented as a function of $|Q^2|$ in HB$\chi$PT. The $e^\pm$-p results are on the left 
             panel, and $\mu^\pm$-p is on the right panel. The results correspond to the three fixed 
             values of the incident lepton beam momenta proposed by the MUSE experiment: 210~MeV/$c$, 
             153~MeV/$c$, and 115~MeV/$c$. All curves are evaluated at a nominal detector cutoff 
             $\Delta_{\gamma^*} = 1\%$ of the incident (anti-)lepton beam energy $E$. Color online: 
             The yellow [cyan] shaded band indicates the sensitivity of the lepton [anti-lepton] 
             scattering results with a detector acceptance in the range, 
             $0.5\% E < \Delta_{\gamma^*} < 2\% E$.}
\label{fig:Total_radiative}
\end{center}
\end{figure*} 
%

\subsection{The lepton-anti-lepton charge asymmetry}  
\begin{figure*}[tbp]
\begin{center}
\includegraphics[width=0.48\linewidth]{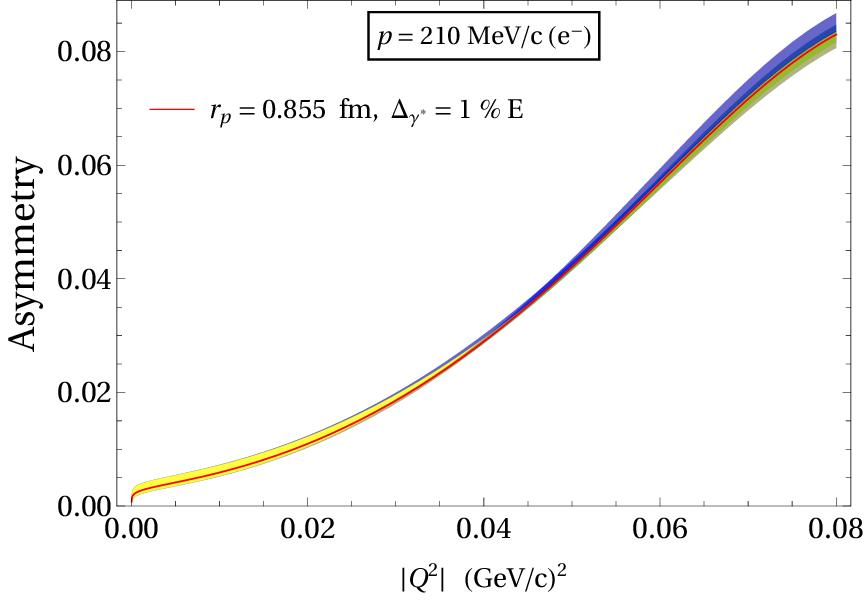}~\quad~\includegraphics[width=0.48\linewidth]{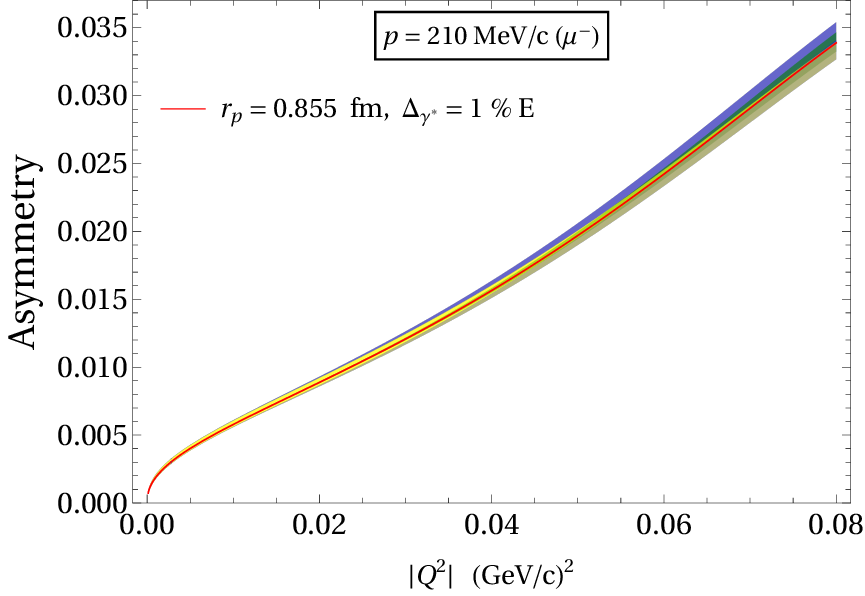}

\vspace{0.6cm}
    
\includegraphics[width=0.48\linewidth]{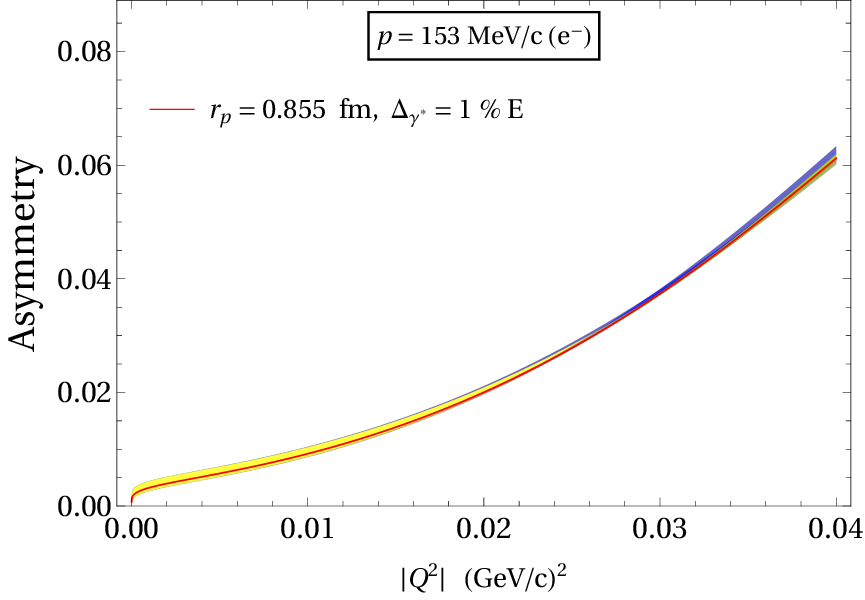}~\quad~\includegraphics[width=0.48\linewidth]{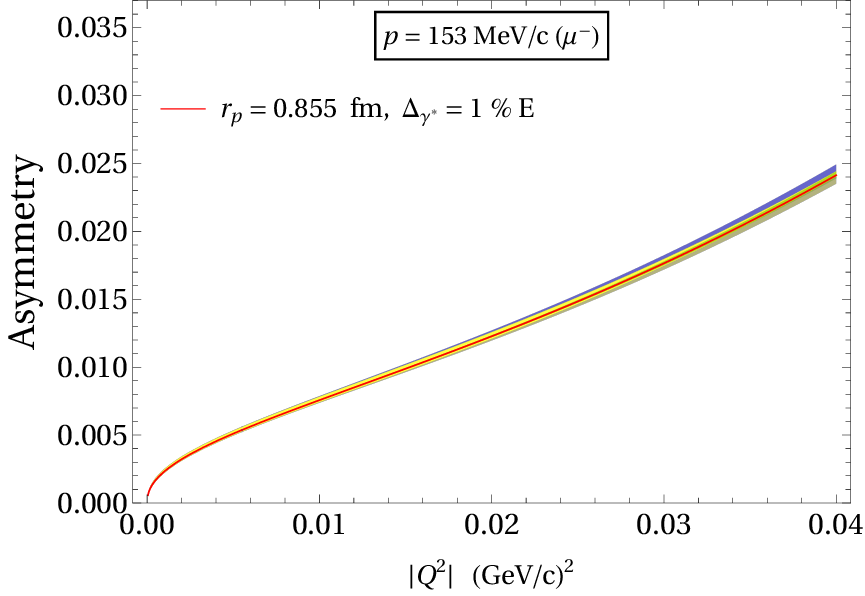}

\vspace{0.6cm}

\includegraphics[width=0.48\linewidth]{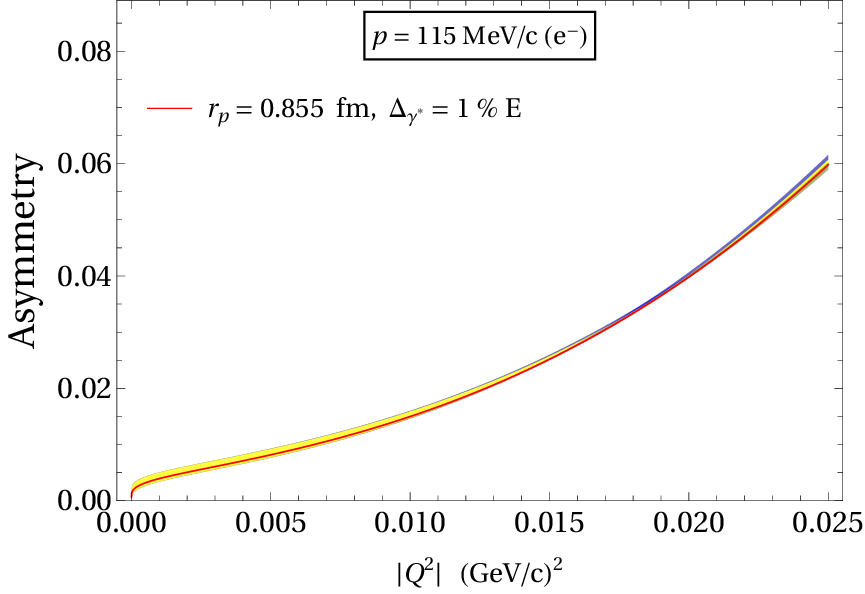}~\quad~\includegraphics[width=0.48\linewidth]{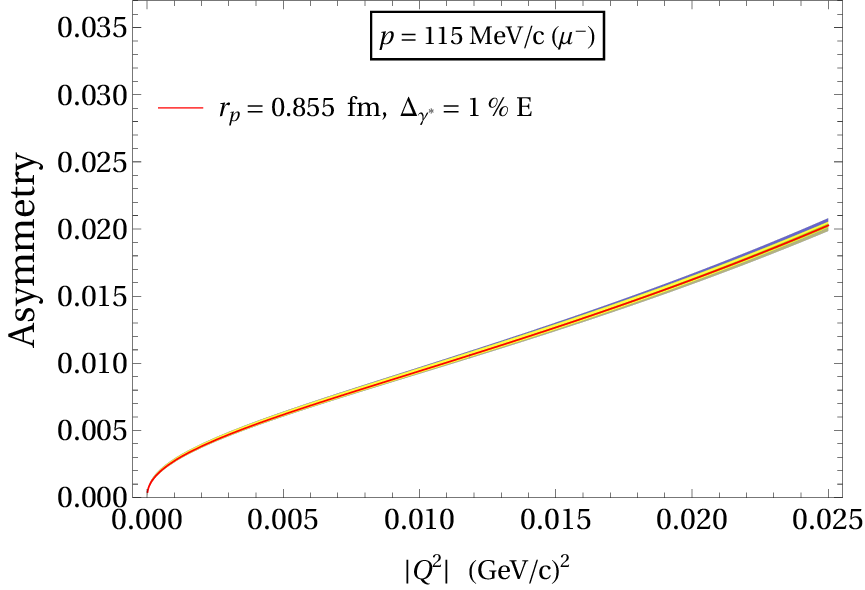}
   \caption{The charge asymmetry ($\mathcal{A}_{\ell^\pm}$), Eq.~\eqref{eq:Asy_result}, for $e^\pm$-p 
            (left panels) and $\mu^\pm$-p (right panels) unpolarized elastic scattering cross sections
            up-to-and-including NLO in HB$\chi$PT is presented as a function of $|Q^2|$. The results 
            correspond to three fixed values of the incident lepton beam momenta proposed by the MUSE
            experiment: 210~MeV/$c$, 153~MeV/$c$, and 115~MeV/$c$. Color online: The yellow shaded 
            band indicates the sensitivity of our results with a detector acceptance in the range, 
            $0.5\% < \Delta_{\gamma^*} < 2\%$ of the incident (anti-)lepton beam energy $E$, 
            while the blue shaded band reflects the uncertainty for a wide range of experimentally 
            extracted values of the proton's rms charge radius ($r_p$) between the 
            CREMA~\cite{Antognini:2013txn,Pohl:2013yb} [$r_p=0.84087(39)$~fm] and MAMI 
            ISR~\cite{Mihovilovic:2019jiz} 
            [$r_p=0.87\pm (0.014){\rm stat.}\pm (0.024){\rm syst.}\pm (0.003)_{\rm mod.}$~fm] 
            measurements. The solid red curve corresponds to our prototype result, 
            evaluated at a nominal detector cutoff $\Delta_{\gamma^*}=1\%E$ and an rms radius 
            $r_p = 0.855$~fm, averaged over the given extracted range.}
\label{fig:asymmetry}
\end{center}
\end{figure*}

As inferred in the introduction, one major goal of our paper is to predict the $Q^2$-dependence of the charge 
asymmetry $\mathcal{A}_{\ell^\pm}$. It yields a measure of the difference between higher-order quantum corrections 
affecting the respective cross sections of lepton and anti-lepton elastic scatterings with the proton. Such an 
asymmetry arises primarily due to the interference between the OPE and TPE amplitudes and is expected to be small
in magnitude, yet non-negligible. Its determination provides one of the most effective means of isolating TPE 
effects from the corresponding scattering cross sections. We expect that the MUSE collaboration will publish their
results in this context in the near future, with an expected systematic uncertainty of 0.1\%. To achieve comparable
accuracy, theoretical predictions within low-energy EFT must account for contributions at least up to NLO [i.e., 
${\mathcal O}(\alpha^3/M)$], if not beyond. To that end, as per the definition, Eq.~\eqref{eq:Asy}, based on our 
aforementioned HB$\chi$PT analysis, we provide the following NLO determination of the asymmetry observable: 
\begin{eqnarray}
\mathcal{A}_{\ell^\pm}(Q^2,\Delta_{\gamma^*},r_p) \!&=&\!   
\frac{{\rm d}\sigma^{(\rm odd)}_{\rm tot}}{{\rm d}\sigma^{(\rm even)}_{\rm tot}}
= \frac{{\rm d}\sigma^{(-)}_{el}-{\rm d}\sigma^{(+)}_{el}}{{\rm d}\sigma^{(-)}_{el}+{\rm d}\sigma^{(+)}_{el}} \,
\nonumber \\
\!&=&\! \frac{\delta^{\rm (odd)}_{2\gamma}(Q^2,\Delta_{\gamma^*})}{1+ \delta^{\rm (even)}_{el}(Q^2,\Delta_{\gamma^*},r_p)} \,.
\label{eq:Asy_result}
\end{eqnarray} 
In Figure~\ref {fig:asymmetry}, we display our evaluated $\mathcal{A}_{\ell^\pm}$ results as a function of $|Q^2|$, 
especially highlighting the asymmetry's sensitivity to the incident lepton beam momenta pertinent to the proposed 
MUSE kinematics. 
%

\section{Results and Discussion}
\label{sec:III}

In this section, we present our numerical results pertaining to the elastic (anti-)lepton–proton scattering cross 
section computed within the framework of HB$\chi$PT, up-to-and-including NLO [i.e., ${\mathcal O}(\alpha^3/M)$] 
correction. Our numerical results for the soft-photon bremsstrahlung calculations are obtained using SPL, where 
the emitted real soft photon four-momentum $k\to 0$. This corresponds to photon energies below a certain detector
threshold ($\Delta_{\gamma^*}$). In all our subsequent numerical estimates, $\Delta_{\gamma^*}$ will be treated
as a free parameter to assess the sensitivity of our EFT results to external factors, such as variations arising
from specific experimental design choices.

Before outlining our numerical results in the different figures, we briefly discuss some key aspects of our 
bremsstrahlung amplitudes that underpin our cross section calculations. They are in order below: 
\begin{itemize}
\item 
As presented in Section~\ref{sec:II}, the emergence of chiral \underline{enhancement} of certain amplitudes which
involve the insertion of \underline{final-state} proton propagator components ($iS^{\nu}$) containing the factor 
$(v\cdot Q)^{-1} =-2M/ Q^2$ in their denominators. This highlights the significance of contributions that go 
beyond the naive diagrammatical hierarchy implied by chiral power-counting expectations. Specifically, radiative 
processes involving the \underline{final-state} photon emission necessitate the inclusion of chirally 
\underline{enhanced} amplitudes that could formally arise from contributions up to N$^4$LO ($\nu=4$ chiral order)
since they could {\it per se} yield a non-negligible contribution to the intended ${\mathcal O}(\alpha^3/M)$ 
accuracy of the NLO differential cross section. Consequently, to maintain consistency in evaluating the cross 
section up-to-and-including NLO accuracy, it is essential to account for all relevant amplitudes -- starting from
the lowest-order contribution at ${\mathcal O}(e^3M)$, stemming from the LO chirally \underline{enhanced} 
amplitude, ${\mathcal M}^{[p(0);f]}_{\gamma\gamma^*}$ [see Eq.~\eqref{LO_4}], and extending up to the highest 
contributing order at $\mathcal{O}(e^3/M^2)$. The latter can originate from the chirally \underline{enhanced} 
amplitudes, including potentially those as high as N$^4$LO, which effectively contribute at 
${\mathcal O}(1/M^2)$ -- specifically, $\mathcal{M}^{[p(3);{\rm f}\otimes]}_{\gamma\gamma^*}$ and 
$\mathcal{M}^{\left[\widehat{p(0)\,;{\rm f}\otimes}\right]}_{\gamma\gamma^*}$ [see Eqs.~\eqref{eq:N4LO_49} and 
\eqref{eq:N4LO_50} in Appendix A]. In this process, we only retain the relevant amplitudes that: (i) do not 
vanish in the {\it lab.}-frame using SPL, (ii) yield non-vanishing contributions either to the charge-even or 
charge-odd bremsstrahlung cross section upon phase-space integration using DR, i.e., 
subject to the condition, Eq.~\eqref{dr_condition}, and (iii) could potentially yield cross section terms of 
$\mathcal{O}(\alpha^3/M)$ accuracy in the SPL. Figures.~\ref{fig:LO} - \ref{fig:N3LO}, display all such graphs
that satisfy these three criteria. For, completeness, Appendix A provides the general frame-independent 
expressions for all contributing amplitudes, including those numerous other ${\mathcal O}(e^3/M^2)$ amplitudes
that do not enter our NLO soft-bremsstrahlung cross section in the {\it lab.}-frame.
\item 
It is only upon performing the bremsstrahlung phase-space integrations that two additional subtle features 
underlying our analytical result for the $\mathcal{O}(\alpha^3/M)$ NLO bremsstrahlung differential cross become
apparent -- features that are not readily apparent before integration. Firstly, all N${}^3$LO amplitudes shown
in Fig.~\ref{fig:N3LO} drop out for the charge-even case due to the DR condition, Eq.~\eqref{dr_condition}, while 
for the charge-odd case, some contributions similarly vanish in DR, while others result in 
$\mathcal{O}(\alpha^3/M^2)$ expressions [cf. Eqs.~\eqref{eq:dsigmaN3LO_brem_odd} and 
\eqref{eq:deltaN3LO_brem_odd}]. Secondly, all NNLO hadronic amplitudes (those with renormalized pion-loops), 
shown in Fig.~\ref{fig:NNLO_loop} drop out as they lead to $\mathcal{O}(1/M^2)$ cross section [see 
Eqs.~\eqref{eq:dsigmaNNLO_brem_odd} and \eqref{eq:deltaNNLO_brem_odd}]. {\it Consequently, the bremsstrahlung 
cross section up-to-and-including NLO receives no contribution from the proton's hadronic finite-size structure
effects. In other words, at this level of accuracy, the proton effectively behaves as a point-like particle.}
\end{itemize}

We now turn to the numerical results shown in Figs.~\ref{fig:even_brem}–\ref{fig:asymmetry}, which correspond 
to the proposed low-energy kinematics relevant to the ongoing MUSE experiment~\cite{MUSE:2017dod,Cline:2021ehf,Strauch:2024imt}.
First, we consider our results for the finite part of the fractional charge-even bremsstrahlung correction to 
the unpolarized elastic (anti-)lepton-proton scattering cross sections, defined by
\begin{equation}
\overline{\delta}^{\rm (even)}_{\gamma\gamma^*}(Q^2,\Delta_{\gamma^*})
=\delta^{\rm (even)}_{\gamma\gamma^*}(Q^2,\Delta_{\gamma^*}) - {\bf IR}^{(0)}_{\gamma\gamma^*}(Q^2)\,.
\end{equation}
As displayed in Figure~\ref{fig:even_brem}, the red (color online) solid line represents the full result 
up-to-and-including NLO [i.e., $\mathcal{O}(\alpha/M)$] in HB$\chi$PT. The following salient characteristics
emerge from the plots:
\begin{enumerate} 
    \item
    The corrections for both electron and muon scatterings are predominantly insensitive to the variation of 
    the incident lepton beam momentum in the MUSE experimental kinematic range, $p=|{\bf p}|\in[115,\, 210]$~MeV/$c$.
    \item 
    The genuine LO [i.e., $\mathcal{O}(M^0)$] correction $\delta^{(0,\rm even)}_{\gamma\gamma^*}$,
    Eq.~\eqref{eq:delta_brem_even_0},  without subleading terms, is almost a negative constant and relatively
    large $\sim 7\%$, for the entire range of squared four-momentum transfers $-Q^2 > 0$ for electron 
    scattering. The $\mathcal{O}(1/M)$ corrections $\delta^{(1,\rm even)}_{\gamma\gamma^*}$,
    Eq.~\eqref{eq:delta_brem_even_1}, on the other hand, are quite small but positive, monotonically 
    increasing very gradually in the same range.  
    \item 
    For the muon scattering, the corresponding genuine LO correction has, in contrast, a sharp negative slope 
    with a maximum of about $2.5\%$, while for the $\mathcal{O}(1/M)$ corrections, the behavior is analogous, 
    albeit a factor of 10 smaller than the electron case. 
    \item 
    The full charge-even bremsstrahlung correction for electron scattering is much more sensitive to the
    variation with the detector threshold, $\Delta_{\gamma^*}\in[0.5\%,\,2\%]$, than in muon scattering, as 
    evident from the contrasting widths of the yellow shaded band. This indicates the suppression of 
    charge-even bremsstrahlung effects for heavier leptons.
    \item 
    It is important to emphasize that our calculated charge-even bremsstrahlung correction is smaller than 
    what was reported in the work of Talukdar \textit{et al.}~\cite{Talukdar:2020aui}. This is primarily due
    to the adoption of a different subtraction scheme for eliminating the IR divergences. In our analysis,
    the IR-divergent contribution is a constant. This contrasts the $|Q^2|$-dependent subtraction adopted by
    Talukdar \textit{et al.}
\end{enumerate} 

Second, we consider our results displayed in Fig.~\ref{fig:even_rad}, where we plot the total fractional 
charge-even radiative corrections $\delta^{\rm (even)}_{2\gamma}$, Eq.~\eqref{eq:delta_rad_even}, 
calculated up-to-and-including NLO [i.e., $\mathcal{O}(\alpha/M)$] to the unpolarized elastic 
(anti-)lepton-proton scattering cross sections. The results include the analytically evaluated charge-even
virtual photon-loop corrections (i.e., self-energy, vertex, and vacuum polarization one-loop effects) 
counterpart taken from Talukdar {\it et al}.~\cite{Talukdar:2020aui}. The key features observed in the plots 
are summarized below:  
\begin{enumerate}
    \item 
    There is little sensitivity of both electron and muon scattering results on the variation of the incident
    lepton beam momentum ($p=|{\bf p}|$) in the MUSE kinematic range. 
    \item 
    For electron scattering, the total correction remains substantially negative throughout the range of 
    $|Q^2|$ values considered at MUSE, exhibiting a very gradual increase in magnitude with rising $|Q^2|$. 
    \item 
    In contrast to the electron scattering, the result for muon scattering exhibits a change in sign at the 
    low-momentum transfer region, $-Q^2\sim 0.05$~GeV${}^2\!/c^2$. Consequently, at the lowest MUSE beam 
    momentum of $p = 115~\mathrm{MeV}/c$, the correction stays positive across the entire kinematic range
    below $|Q^2|\simeq 0.025$~GeV${}^2\!/c^2$. While for the largest beam momentum of $p=210~\mathrm{MeV}/c$,
    the correction is positive only within the range $|Q^2|\lesssim 0.05$~GeV${}^2\!/c^2$. Beyond that, it
    turns negative, with magnitude increasing monotonically with a negative slope. However, the overall
    magnitude remains strongly suppressed, staying within the $1\%$ level throughout the kinematic interval. 
    \item For the same reason discussed in point 4 of Fig.~\ref{fig:even_brem}, the magnitude of the 
    correction in our analysis is smaller than that reported by Talukdar 
    \textit{et al.}~\cite{Talukdar:2020aui}.
\end{enumerate}

Third, we consider our results for the finite part of the fractional charge-odd bremsstrahlung correction 
to the unpolarized elastic electron-proton and muon-proton scattering cross sections, defined by
\begin{equation}
\overline{\delta}^{\rm (odd)}_{\gamma\gamma^*} (Q^2,\Delta_{\gamma^*})
=\delta^{\rm (odd)}_{\gamma\gamma^*}(Q^2,\Delta_{\gamma^*}) - {\bf IR}^{(1)}_{\gamma\gamma^*}(Q^2)\,.
\end{equation}
Given the charge-odd character, the corresponding correction for anti-lepton scattering is obtained by 
inverting the overall sign. As displayed in Fig.~\ref{fig:odd_brem}, the red (color online) solid line 
represents our full lepton scattering result up-to-and-including NLO [i.e., $\mathcal{O}(\alpha/M)$], 
obtained from the sum of Eqs.~\eqref{eq:true_LO_odd} and \eqref{eq:true_NLO_odd}. The salient features of 
the plots are summarized in order:
\begin{enumerate}
    \item 
    A comparison with the charge-even bremsstrahlung correction displayed in Fig.~\ref{fig:even_brem} 
    reveals a striking contrast -- the charge-odd bremsstrahlung correction is smaller than more than an
    order of magnitude.
    \item
    Similar to the charge-even case, there is little sensitivity of both electron and muon scattering 
    results on the variation of the incident lepton beam momentum. 
    \item 
    Unlike the charge-even bremsstrahlung correction, the magnitudes of the full (including enhancement) 
    charge-odd bremsstrahlung correction for the electron and muon scatterings are similar (they are 
    slightly larger for the electron case). They become more negative with increasing $|Q^2|$, and have 
    comparable slopes. 
    \item 
    The genuine LO [i.e., $\mathcal{O}(M^0)$] correction $\delta^{(0,\rm odd)}_{\gamma\gamma^*}$, 
    Eq.~\eqref{eq:true_LO_odd}, is a positive constant, independent of $|Q^2|$ for both electron and muon 
    scatterings, whereas the ${\mathcal O}(1/M)$ correction $\delta^{(1,\rm odd)}_{\gamma\gamma^*}$, 
    Eq.~\eqref{eq:true_NLO_odd}, yields a negative slope in each case.
    \item 
    The observed difference between the ``enhanced" (color online: green long dashed and blue long 
    dashed-dot curves) and ``unenhanced" (color online: black short dashed and magenta short dashed-dot
    curves) results, arises from the inclusion of additional ${\mathcal O}(M^0)$ and ${\mathcal O}(1/M)$ 
    correction terms in the NLO cross section -- specifically, the contributions
    $\delta^{\rm (NNLO;odd)}_{\gamma\gamma^*}$, Eq.~\eqref{eq:deltaNNLO_brem_odd} and 
    $\widetilde{\delta}^{\,\rm (NNLO;odd)}_{\gamma\gamma^*}$, Eq.~\eqref{eq:deltaN3LO_brem_odd}, which
    stem from $\nu=2,3$ chiral order matrix elements $\mathcal S^{\nu=2,3}_{\rm odd}$ involving chirally 
    \underline{enhanced} NNLO amplitudes (the contribution from the N${}^3$LO amplitudes drops out, as we
    discussed). {\it We find that the enhanced contribution substantially modifies (almost doubles) the 
    ${\mathcal O}(M^0)$ results, while the ${\mathcal O}(1/M)$ results show very nominal change that is 
    hardly discernible.}
    \item
    A comparison of the yellow shaded band widths reveals that the full result exhibits greater 
    sensitivity to $\Delta_{\gamma^*}$ variation in electron scattering than in muon scattering.
\end{enumerate} 

Fourth, we consider our results in Fig.~\ref{fig:odd_rad} for the total fractional charge-odd radiative 
corrections $\delta^{\rm (odd)}_{2\gamma}$, Eq.~\eqref{eq:delta_rad_odd2}, to the unpolarized elastic 
electron-proton and muon-proton scattering cross sections, up-to-and-including NLO in HB$\chi$PT. The 
results include the analytically evaluated exact TPE virtual corrections counterpart 
$\delta^{\rm (odd)}_{\gamma\gamma}$, as given in Eq.~\eqref{eq:delta_TPE}, taken from Fig.~5 of 
Choudhary {\it et al}.~\cite{Choudhary:2023rsz}.\footnote{In this work, we consider an updated numerical
estimate of the exact TPE result, incorporating a correction for an error identified in our 
Mathematica-based code used to generate Fig.~5 of Ref.~\cite{Choudhary:2023rsz}. The revised plots yield
a substantial increase (up to nearly $3\%$ for electron scattering and $0.5\%$ for muon scattering) in 
the magnitude of the ${\mathcal O}(1/M)$ corrections to the elastic cross section arising from the LO TPE 
diagrams (a) and (b).} The following are the key features observed in the plots:
\begin{enumerate}
    \item
    As demonstrated in the TPE analysis by Choudhary {\it et al}., unlike other types of radiative 
    corrections, the TPE contributions for both electron and muon scatterings exhibit a pronounced 
    sensitivity to the variation in the incident lepton beam momentum in the MUSE kinematic range, 
    $p=|{\bf p}|\in[115,\,210]$~MeV/$c$. Consequently, the dominance of the finite part of the TPE 
    contribution, i.e.,  
    $\overline{\delta}^{\text{(odd)}}_{\gamma\gamma}=\delta^{\text{(odd)}}_{\gamma\gamma}-{\bf IR}^{(1)}_{\rm TPE}$, 
    is likewise manifested in the results for the charge-odd radiative correction.
    \item 
    For both electron and muon scatterings, the corrections stay positive and exhibit a monotonic 
    increase with increasing $|Q^2|$, indicating a net enhancement to the cross section. Here, the TPE 
    corrections $\delta^{\text{(odd)}}_{\gamma \gamma}$ dominates over the charge-odd bremsstrahlung 
    corrections given in Eq.~\eqref{eq:delta_brem_odd}. {\it Indeed, the negligible impact of the 
    bremsstrahlung correction suggests that the charge-odd radiative correction shown in 
    Fig.~\ref{fig:odd_rad} could serve as a reliable proxy for the ``physical TPE" contribution, 
    making the latter amenable to extraction from asymmetry measurements.}
    \item 
    The magnitude of $\delta^{\text{(odd)}}_{2\gamma}$ is more than a factor of two larger for electron
    scattering than for muon scattering at the same beam momenta. This discrepancy primarily arises from
    the greater sensitivity of electrons to TPE effects, which are significantly suppressed for the 
    heavier muons.
    \item 
    The rather narrow yellow shaded bands -- particularly in the case of the muon scattering -- 
    indicate that the TPE corrections are largely insensitive to variations in the detector threshold
    $\Delta_{\gamma^*}$.
\end{enumerate}

Fifth, we consider Fig~\ref{fig:Total_radiative} where we display the total fractional radiative 
correction $\delta^{(\ell^\mp{\rm p})}_{2\gamma}$, Eq.~\eqref{eq:leptonProton}, to the unpolarized elastic 
(anti-)lepton-proton cross section of $e^\pm$–p and $\mu^\pm$–p scattering processes, up-to-and-including
NLO in HB$\chi$PT. An analysis of the plots reveals the following key features:
\begin{enumerate}
    \item 
    The corrections are sizable and negative across the entire range of low-energy MUSE kinematics for 
    both electron and positron scatterings, with the latter exhibiting a larger magnitude. This difference
    becomes increasingly pronounced with rising $|Q^2|$, driven by the \underline{enhanced} charge-odd 
    contributions.
    \item 
    In contrast to electron and positron scatterings, the radiative corrections for muon and anti-muon 
    scatterings are about a magnitude smaller. For muon scattering, the correction is positive and 
    increases gradually with rising $|Q^2|$. Interestingly, however, for anti-muon scattering, the 
    correction is positive only for a limited range of momentum transfers, below 
    $|Q^2| \lesssim 0.02$~GeV${}^2\!/c^2$, beyond which it turns negative for all $|Q^2|$ values pertinent
    to MUSE.
    \item 
    In general, the dependence on the incident lepton beam momentum is weak. 
    \item 
    The variation with $\Delta_{\gamma^*}$, as indicated by the yellow and cyan shaded bands for lepton 
    and anti-lepton scattering, points toward a notably higher sensitivity in the electron/positron 
    scattering process relative to the muon/anti-muon counterpart. 
\end{enumerate}

Finally, we consider our result for the charge asymmetry observable $\mathcal{A}_{\ell^\pm}$, 
Eq.~\eqref{eq:Asy_result}, which incorporates the total charge-odd and charge-even radiative corrections 
up-to-and-including NLO in HB$\chi$PT. These are as illustrated in Fig.~\ref{fig:asymmetry} (see also 
Tables.~\ref{tab1} and \ref{tab2}). The former correction $\delta^{\rm (odd)}_{2\gamma}$ arises from the 
{\it exchange-interference} bremsstrahlung (see introduction) and TPE effects, while the latter 
$\delta^{\rm (even)}_{el}$ arises from the {\it direct-interference} bremsstrahlung, QED one-loop virtual
corrections (other than the charge-odd TPE), and the leading hadronic corrections 
$\delta^{(2)}_\chi$ to the OPE process of $\mathcal{O}(1/\Lambda^2_\chi)\sim \mathcal{O}(1/M^2)$. 
Consequently, our asymmetry results rely on two free parameters, namely the detector cutoff 
$\Delta_\gamma^*$, which enters {\it via} the bremsstrahlung corrections, and the proton's rms charge 
radius $r_p$, which enters into the hadronic OPE correction given in Eq.~\eqref{delta_chi}. To quantify 
the sensitivity of our prediction to these two inputs, we examine the variation of our results across a 
reasonable range of parameter values, namely, $\Delta_{\gamma^*}\in [0.5\%,\,2\%]$ of the incident lepton
beam energy $E$, and $r_p=0.87\pm (0.014){\rm stat.}\pm (0.024){\rm syst.}\pm (0.003)_{\rm mod.}$~fm to 
$0.84087(39)$~fm. The former $r_p$ value corresponds to the recent precision extractions from the 
electron-proton scattering ISR measurements by the A1 Collaboration at MAMI~\cite{Mihovilovic:2019jiz}, 
and the latter corresponds to the well-acclaimed high-precision muonic hydrogen spectroscopy measurements
by the CREMA Collaboration at PSI~\cite{Antognini:2013txn,Pohl:2013yb} about a decade ago. These choices
facilitate an assessment of the extent to which the current phenomenological discrepancy in the proton 
radius contributes to the theoretical error in our asymmetry predictions. The following key features
are evident from the plots in Fig.~\ref{fig:asymmetry} and data presented in Tables.~\ref{tab1} and 
\ref{tab2}: 
\begin{enumerate}
\item 
The asymmetry remains positive and exhibits a monotonic increase with $|Q^2|$ across all incident lepton 
beam momenta and species. For electron (muon) scattering, the asymmetry increases beyond $\gtrsim 0.08$ 
($\lesssim 0.035$) for the highest momentum transfers at MUSE, $|Q^2|\approx 0.08$~GeV${}^2\!/c^2$.   
\item 
A clear, increasing trend with a narrow width is observed with increasing lepton beam momentum for both
electrons and muons. 
\item
The asymmetry exhibits a rather small sensitivity to either parameter variation of $\Delta_{\gamma^*}$ 
and $r_p$, given their rapid increase with $|Q^2|$ for both electron and muon scatterings. 
\item 
At low values of $|Q^2|$, the results show greater sensitivity to the variations in the detector threshold
$\Delta_{\gamma^*}$, but this sensitivity diminish as $|Q^2|$ increase. 
\item 
In contrast, the sensitivity to changes in the proton's rms radius [where $r_p\sim{\mathcal O}(1/M^2)$] and
to some extent $\Delta_{\gamma^*}$ becomes more significant at higher $|Q^2|$. 
\item
The dependence of the electron-positron asymmetry on the detector threshold $\Delta_{\gamma^*}$ is, 
however, quite subtle as evidenced by both Fig~\ref{fig:asymmetry} and also Table.~\ref{tab1}. The
asymmetry first decreases and then increases again with increasing $\Delta_{\gamma^*}$. Notably, at 
small $|Q^2|$ values, the yellow band exhibits a visible width, which constricts at intermediate $|Q^2|$,
and then becomes prominent again at higher $|Q^2|$ values. For muon-anti-muon asymmetry, the behavior is
more or less straightforward with the width appearing to increase monotonically with increasing 
$\Delta_{\gamma^*}$ across all MUSE kinematical settings.
\item
Table.~\ref{tab2} reveals that, for fixed $|Q^2|$ and incident lepton beam momenta, the asymmetry increases
monotonically with the proton's rms charge radius across all MUSE kinematical settings and for both lepton
species.
\end{enumerate}
%
\begin{table*}[tbp]
\centering
\renewcommand{\arraystretch}{1.2}
\begin{tabular}{|c|c|c||c|c|}
\hline
$|Q^2|$ (GeV${}^2\!/c^2$) & $p$ (MeV/$c$) & $\Delta_{\gamma^*}$ (in $\%$ of $E$) 
&  $\mathcal{A}_{e^\pm}$ (error)  & $\mathcal{A}_{\mu^\pm}$ (error)\\
\hline\hline
\multirow{9}{*}{0.025} 
  & \multirow{3}{*}{115}  & 0.5 & 0.061(1) & 0.020(1) \\
  &                       & 1   & 0.059(1) & 0.020(1) \\
  &                       & 2   & 0.059(1) & 0.021(1) \\
\cline{2-5}
  & \multirow{3}{*}{153}  & 0.5 & 0.028(1) & 0.014(1) \\
  &                       & 1   & 0.028(1) & 0.015(1) \\
  &                       & 2   & 0.028(1) & 0.015(1) \\
\cline{2-5}
  & \multirow{3}{*}{210}  & 0.5 & 0.014(1) & 0.010(1) \\
  &                       & 1   & 0.014(1) & 0.010(1) \\
  &                       & 2   & 0.015(1) & 0.011(1) \\
\hline
\multirow{6}{*}{0.04}
  & \multirow{3}{*}{153}  & 0.5 & 0.062(1) & 0.024(1) \\
  &                       & 1   & 0.061(1) & 0.024(1) \\
  &                       & 2   & 0.060(1) & 0.024(1) \\
\cline{2-5}
  & \multirow{3}{*}{210}  & 0.5 & 0.029(1) & 0.015(1) \\
  &                       & 1   & 0.029(1) & 0.016(1) \\
  &                       & 2   & 0.030(1) & 0.016(1) \\
\hline
\multirow{3}{*}{0.08}
  & \multirow{3}{*}{210}  & 0.5 & 0.084(1) & 0.033(1) \\
  &                       & 1   & 0.082(1) & 0.034(1) \\
  &                       & 2   & 0.081(1) & 0.034(1) \\
\hline\hline
\end{tabular}
\caption{A study of the charge asymmetry ($\mathcal{A}_{\ell^\pm}$) in unpolarized elastic 
$e^\pm$-p and $\mu^\pm$-p scattering cross sections, calculated up-to-and-including NLO in 
HB$\chi$PT. The analysis is performed for a median value of the proton's rms charge radius,
$r_p \approx 0.855$~fm, between the CREMA~\cite{Antognini:2013txn,Pohl:2013yb} 
[$r_p=0.84087(39)$~fm] and MAMI ISR~\cite{Mihovilovic:2019jiz} 
[$r_p=0.87\pm (0.014){\rm stat.}\pm (0.024){\rm syst.}\pm (0.003)_{\rm mod.}$~fm] 
measurement results, while varying the detector cutoff $\Delta_{\gamma^*}$ within the 
range $[0.5\%, 2\%]$ of the incident lepton beam energy $E$. Numerical results are 
presented at the maximum allowed four-momentum transfer, $|Q^2| = |Q^2_{\max}|$, 
corresponding to each of the three MUSE kinematical domains, defined by the incident lepton 
beam momenta, $p = 115, 153$, and $210$~MeV/$c$. }
\label{tab1}
\end{table*}
\begin{table*}
\centering
\renewcommand{\arraystretch}{1.2}
\begin{tabular}{|c|c|c||c|c|}
\hline
$|Q^2|$ (GeV${}^2\!/c^2$) & $p$ (MeV/$c$) & $r_p$ (fm) {\cal O}(1/M$^2$) 
& $\mathcal{A}_{e^\pm}$ (error) & $\mathcal{A}_{\mu^\pm}$ (error) \\
\hline\hline
\multirow{6}{*}{0.025} 
  & \multirow{2}{*}{115} & 0.84087 (CREMA) & 0.059(1) & 0.020(1) \\
  &                      & 0.87 (MAMI ISR) & 0.060(1) & 0.020(1)\\
\cline{2-5}
  & \multirow{2}{*}{153} & 0.84087 (CREMA) & 0.027(1) & 0.015(1) \\
  &                      & 0.87 (MAMI ISR) & 0.028(1) & 0.015(1)\\
\cline{2-5}
  & \multirow{2}{*}{210} & 0.84087 (CREMA) & 0.014(1) & 0.010(1) \\
  &                      & 0.87 (MAMI ISR) & 0.014(1) & 0.010(1) \\
\hline
\multirow{4}{*}{0.04} 
  & \multirow{2}{*}{153} & 0.84087 (CREMA) & 0.060(1) & 0.024(1) \\
  &                      & 0.87 (MAMI ISR) & 0.061(1) & 0.024(1) \\
\cline{2-5}
  & \multirow{2}{*}{210} & 0.84087 (CREMA) & 0.028(1) & 0.015(1) \\
  &                      & 0.87 (MAMI ISR) & 0.029(1) & 0.016(1) \\
\hline
\multirow{2}{*}{0.08} 
  & \multirow{2}{*}{210} & 0.84087 (CREMA) & 0.081(1) & 0.033(1) \\
  &                      & 0.87 (MAMI ISR) & 0.084(1) & 0.034(1) \\
\hline\hline
\end{tabular}
\caption{A study of the charge asymmetry ($\mathcal{A}_{\ell^\pm}$) in unpolarized elastic 
$e^\pm$-p and $\mu^\pm$-p scattering cross sections, calculated up-to-and-including NLO in
HB$\chi$PT. The analysis is performed for a fixed benchmark value of the detector cutoff, 
$\Delta_{\gamma^*}=1\%$ of the incoming lepton beam energy $E$, while varying the proton's
rms charge radius ($r_p$) [which scales as ${\mathcal O}(1/M^2)$]  between the 
CREMA~\cite{Antognini:2013txn,Pohl:2013yb} [$r_p=0.84087(39)$~fm] and 
MAMI ISR~\cite{Mihovilovic:2019jiz} 
[$r_p=0.87\pm (0.014){\rm stat.}\pm (0.024){\rm syst.}\pm (0.003)_{\rm mod.}$~fm] 
measurement values. Numerical results are presented at the maximum allowed four-momentum 
transfer, $|Q^2| = |Q^2_{\max}|$, corresponding to each of the three MUSE kinematical 
domains, defined by incident lepton beam momenta, $p = 115, 153$, and $210$~MeV/$c$.}
\label{tab2}
\end{table*}
\begin{table*}[tbp]
\centering
\small
\renewcommand{\arraystretch}{1.3}
\begin{tabular}{|c||c|c||c|c|c||c|c|c||c|}
\hline
$\ell^\pm$ & $\Delta\bar{\delta}^{\rm(even)}_{\gamma\gamma^*}$ ($\%$) & $\Delta\bar{\delta}^{\rm (odd)}_{\gamma\gamma^*}$ ($\%$) &
$\Delta\bar{\delta}^{\rm (even)}_{\gamma\gamma}$ ($\%$) & $\Delta\bar{\delta}^{\rm (odd)}_{\gamma\gamma}$($\%$) & $\Delta{\delta}^{(2)}_{\chi}$ ($\%$) &
$\Delta{\delta}^{\rm (even)}_{2\gamma}$ ($\%$) & $\Delta{\delta}^{\rm (even)}_{el}$ ($\%$) & $\Delta{\delta}^{\rm (odd)}_{2\gamma}$ ($\%$) &  
$\Delta{\mathcal A}_{\ell^\pm}$ ($\%$)\\
\hline\hline
$e^\pm$   & 0.2  & 0.02 & 0.00023 & 0.4 & 2.3 & 0.2 & 2.31 & 0.4 & 0.5 \\
\hline
$\mu^\pm$ & 0.01 & 0.01 &  0.88   & 0.2 & 2.3 & 0.88 & 2.46 & 0.2 & 0.3\\
\hline\hline
\end{tabular}
\caption{A summary of our EFT error estimates (in percentage, relative to the Born 
cross section at MUSE kinematical settings) in the finite part of the NLO charge-even 
($\Delta\bar{\delta}^{\rm(even)}_{\gamma\gamma^*}$) and charge-odd 
($\Delta\bar{\delta}^{\rm (odd)}_{\gamma\gamma^*}$) bremsstrahlung, NLO TPE 
($\Delta\delta^{\rm (odd)}_{\gamma\gamma}$) and charge-even virtual/photon-loop 
($\Delta\delta^{\rm (even)}_{\gamma\gamma}$), and the NNLO hadronic OPE
($\Delta\delta^{(2)}_{\chi}$) corrections for (anti-)lepton-proton scatterings. 
Hence, applying the quadrature rule for combining uncorrelated uncertainties, the 
corresponding errors in the charge-even radiative 
($\Delta{\delta}^{\rm (even)}_{2\gamma}$), charge-even total 
($\Delta{\delta}^{\rm (even)}_{el}$), charge-odd radiative 
($\Delta{\delta}^{\rm (odd)}_{2\gamma}$) corrections, and the asymmetry observable 
($\Delta{\mathcal A}_{\ell^\pm}$) are determined. }
\label{tab_error}
\end{table*}

Prior to concluding this section, we address the theoretical uncertainties associated with our analysis. 
Besides the parametric sensitivities to $\Delta_{\gamma^*}$ and $r_p$, our results are subject to another
principal source of systematic theoretical uncertainty associated with higher-order EFT contributions. 
Specifically, the charge-odd bremsstrahlung correction $\delta^{\rm (odd)}_{\gamma\gamma^*}$ in electron
scattering, and the total charge-even radiative correction $\delta^{\rm (even)}_{2\gamma}$ in muon 
scattering, both exhibit a zero crossing in the small $|Q^2|$ domain. This trend underscores a kinematic
regime where the approximation within the EFT framework breaks down, rendering NNLO contributions 
essential for a more robust estimation of the corrections. While a complete NNLO calculation of all 
radiative corrections lies beyond the scope of this work, a naive estimate of the associated uncertainty
can still be made. This is done by retaining all kinematically suppressed terms in the ${\mathcal O}(1/M)$ 
cross section that depend on the energy $E^\prime$ and velocity $\beta^\prime$ of the outgoing lepton -- 
i.e., without making the approximations $E^\prime\to E$ and $\beta^\prime\to \beta$. In this way, we may 
obtain an estimate of the percentage error associated with the different charge-odd and charge-even 
radiative corrections, including that of the TPE, as displayed in Table.~\ref{tab_error}. In particular,
the table suggests that the NNLO [i.e., ${\mathcal O}(1/M^2)$] hadronic and higher-order radiative effects
induce deviations (relative to the LO Born cross sections at MUSE kinematical settings) of up to 
\begin{eqnarray}
\Delta{\delta}^{\rm (even)}_{el} \!&=&\! 
\sqrt{\left(\Delta\bar{\delta}^{\rm(even)}_{\gamma\gamma^*}\right)^2 
+ \left(\Delta\bar{\delta}^{\rm (even)}_{\gamma\gamma}\right)^2 + \left(\Delta\delta^{(2)}_{\chi} \right)^2}
\nonumber\\
&=&\! \begin{cases}
& \!\!\!\!\!2.31 \% \quad {\rm for}\,\,\,\ell=e \,, \\
& \!\!\!\!\!2.46 \% \quad {\rm for}\,\,\,\ell=\mu\,, \\
\end{cases}
\\
\Delta{\delta}^{\rm (even)}_{2\gamma} \!&=&\! 
\sqrt{\left(\Delta\bar{\delta}^{\rm(even)}_{\gamma\gamma^*}\right)^2 
+ \left(\Delta\bar{\delta}^{\rm (even)}_{\gamma\gamma}\right)^2}
\nonumber\\
&=&\! \begin{cases}
& \!\!\!\!\!0.2 \% \quad {\rm for}\,\,\,\ell=e \,, \\
& \!\!\!\!\!0.88 \% \quad {\rm for}\,\,\,\ell=\mu\,, \qquad \text{and}\\
\end{cases}  
\\
\Delta{\delta}^{\rm (odd)}_{2\gamma} \!&=&\! 
\sqrt{\left(\Delta\bar{\delta}^{\rm(odd)}_{\gamma\gamma^*}\right)^2 
+ \left(\Delta\bar{\delta}^{\rm (odd)}_{\gamma\gamma}\right)^2}
\nonumber\\
&=&\! \begin{cases}
& \!\!\!\!\!0.4 \% \quad {\rm for}\,\,\,\ell=e\,, \\
& \!\!\!\!\!0.2 \% \quad {\rm for}\,\,\,\ell=\mu\,, \\
\end{cases}
\end{eqnarray}
where the uncertainty in our NNLO hadronic correction result, stemming from the N${}^3$LO corrections, is 
based on a naive dimensional assessment, namely, $\Delta\delta^{(2)}_\chi\sim (\sqrt{-Q^2}/M)^3$. 
Consequently, our charge asymmetry predictions are subject to deviations of up to 
\begin{eqnarray}
\Delta{\mathcal A}_{\ell^\pm}  \!&=&\! 
{\mathcal A}_{\ell^\pm} \sqrt{\left(\frac{\Delta{\delta}^{\rm (even)}_{el}}{1+{\delta}^{\rm (even)}_{el}}\right)^2 
+ \left(\frac{\Delta{\delta}^{\rm (odd)}_{2\gamma}}{{\delta}^{\rm (odd)}_{2\gamma}}\right)^2}
\nonumber\\
&=&\! \begin{cases}
& \!\!\!\!\!0.5 \% \quad {\rm for}\,\,\,\ell=e\,, \\
& \!\!\!\!\!0.3 \% \quad {\rm for}\,\,\,\ell=\mu\,. \\
\end{cases}
\end{eqnarray}
It is also worth noting that while the proton's rms charge radius does not affect the $\mathcal{O}(1/M)$
cross section, its influence nevertheless reappears in the higher-order uncertainty estimates, which now
have been incorporated.

\section{Summary and Conclusion}
\label{sec:IV}
In this work, we have carried out a comprehensive analysis of the soft-photon bremsstrahlung radiative 
corrections to the unpolarized elastic lepton-proton scattering within the framework of HB$\chi$PT. The 
study includes all contributions to the cross section up-to-and-including NLO  [i.e., 
${\mathcal O}(\alpha^3/M)$], systematically incorporating both charge-even and charge-odd bremsstrahlung
corrections. Subsequently, the charge asymmetry observable $A_{\ell^\pm}$ is predicted by incorporating
the NLO virtual photon radiative corrections -- both charge-even (self-energy, vacuum polarization, and 
vertex) and the charge-odd TPE -- alongside the NNLO hadronic corrections to the LO Born/OPE process. Our 
result thereby represents a model-independent prediction of the charge asymmetry in (anti-)lepton-proton
scattering, specifically in scenarios where finite lepton mass effects cannot be neglected.

In our evaluation of the bremsstrahlung contributions from the undetected soft-photons -- i.e., photons
with energies below the detector threshold $\Delta_{\gamma^*}$ -- we employed the so-called SPL 
methodology, wherein the emitted photon four-momentum is taken to vanishingly small ($k \to 0$). This 
approximation, a well-established technique in QED, enables a systematic and transparent evaluation of the
IR 
divergences~\cite{Yennie:1961ad,Mo:1968cg,Vanderhaeghen:2000ws, Maximon:1969nw,Maximon:2000hm,Tsai:1961zz}. 
Specifically, working in the Lorentz gauge, we identified a class of chirally \underline{enhanced} 
amplitudes involving the final-state radiating proton, notably those containing ${\mathcal O}(M^0)$ and
${\mathcal O}(1/M)$ components of the proton propagator propagator. As a result, certain kinematically
\underline{enhanced} NLO $(\nu = 1)$ and NNLO $(\nu = 2)$ amplitudes were needed to be promoted to lower 
effective orders, thereby contributing to the  ${\mathcal O}(\alpha^3/M)$ NLO cross section. This outcome
departs from the standard chiral power-counting typically used to classify Feynman diagrams in HB$\chi$PT.
While such \underline{enhanced} NNLO amplitudes do not affect the charge-even bremsstrahlung corrections 
at our working precision, the charge-odd results get substantially modified with the genuine LO [i.e, 
${\mathcal O}(\alpha^3 M^0)$ corrections almost getting doubled (cf. Fig.~\ref{fig:odd_brem}) within the 
MUSE kinematic regime. The NLO [i.e.,${\mathcal O}(\alpha^3/M)$] result, in contrast, exhibits only a very
nominal change.

Our general finding for both lepton scattering processes is that the finite part of the charge-even 
bremsstrahlung correction $\overline{\delta}^{(\text{even})}_{\gamma\gamma^*}$ exceeds the charge-odd 
counterpart $\overline{\delta}^{(\text{odd})}_{\gamma\gamma^*}$ by more than an order of magnitude 
(compare Figs.~\ref{fig:even_brem} and \ref{fig:odd_brem}). The finite charge-odd bremsstrahlung 
correction $\overline{\delta}^{(\text{odd})}_{\gamma\gamma^*}$ in the MUSE regime was found to reach 
values only about $0.3\%$ for both $e^{-}$–p and $\mu^{-}$–p scattering, assuming a benchmark value of the
detector cutoff parameter, $\Delta_{\gamma^*} = 1\%$ of the incident lepton beam energy. For electron 
scattering, in particular, the observed zero crossing of the correction within the relevant energy range 
suggests that the NNLO [i.e., ${\mathcal O}(\alpha^3/M^2)$] terms may become important, pointing to the
potential influence of hadronic structure effects not captured at NLO (recalling that all dependence on the
proton's rms charge radius $r_p$ drops out of the NLO charge-odd cross section). Therefore, a more 
comprehensive evaluation, systematically incorporating all NNLO contributions to the cross section, is 
left for future work.

As for the charge-even bremsstrahlung correction $\overline{\delta}^{(\text{even})}_{\gamma\gamma^*}$, our
results with the benchmark detector cutoff, $\Delta_{\gamma^*} = 1\%$, indicate that the total correction
can be as high as $8\%$ for $e^{\pm}$–p scattering, while remaining below $2.5\%$ for the $\mu^{\pm}$–p 
case under identical detector conditions (cf. Fig.~\ref{fig:even_brem}). These values are smaller than 
those reported in a prior radiative corrections work of Talukdar {\it et al}.~\cite{Talukdar:2020aui}, 
owing to the adoption of the specific $|Q^2|$-dependent IR subtraction scheme in that analysis, as opposed
to the momentum-independent subtraction scheme adopted in this work, as elucidated in Sec.~\ref{sec:III}.

We next combined our bremsstrahlung results, $\overline{\delta}^{(\text{even})}_{\gamma\gamma^*}$ and 
$\overline{\delta}^{(\text{odd})}_{\gamma\gamma^*}$, with the NLO charge-even virtual correction 
$\delta^{(\text{even})}_{\gamma\gamma}$ from Talukdar {\it et al}.~\cite{Talukdar:2020aui}, and the 
exact NLO charge-odd TPE correction $\delta^{\rm (odd)}_{\gamma\gamma}$ from Choudhary 
{\it et al}.~\cite{Choudhary:2023rsz} to determine the full charge-even 
$\delta^{(\text{even})}_{2\gamma}$ and charge-odd $\delta^{(\text{odd})}_{2\gamma}$ radiative correction
components of the elastic cross section (cf. Figs.~\ref{fig:even_rad} and \ref{fig:odd_rad}). Because our
charge-odd bremsstrahlung results are relatively small, the latter correction is overwhelmingly dominated 
by the much larger TPE contribution -- exceeding it by nearly two orders of magnitude). This ensures that
a precise extraction of the TPE remains feasible in the near future, as it should ostensibly 
dominate asymmetry measurements.

Furthermore, these corrections enabled us to determine the corresponding total radiative corrections, 
$\delta^{(\ell^-{\rm p})}_{2\gamma}$ for the leptons and $\delta^{(\ell^+{\rm p})}_{2\gamma}$ for the 
anti-leptons, using Eq.~\eqref{eq:leptonProton}. The results reveal that within the MUSE regime, the total
radiative correction up-to-and-including NLO could reach up to about $10\%$ for $e^{-}-$p scattering and 
$25\%$ for $e^{+}-$p scattering, whereas for $\mu^{-}-$p scattering, they go up to $2.5\%$ and for 
$\mu^{+}-$p scattering up to $3\%$ (cf. Fig.~\ref{fig:Total_radiative}). Compared to the earlier results 
of Talukdar \textit{et al.} (cf. the unresummed results represented in Fig.~13 of 
Ref.~\cite{Talukdar:2020aui}), our findings show notable qualitative and quantitative differences -- 
reaching up to $15\%$ for $e^-$-p and $4\%$ for $\mu^-$-p scattering.

The final part of our analysis dealt with the prediction of the charge asymmetry between lepton and 
anti-lepton scatterings with the proton. As indicated in Eq.~\eqref{eq:Asy_result}, the asymmetry result 
is obtained from combining our NLO (even and odd) radiative corrections with the NNLO hadronic correction
$\delta^{(2)}_{\chi}$ to the elastic Born/OPE cross section. The latter result, which contributes to the
total charge-even component $\delta^{(\text{even})}_{el}$, has also been adopted from Talukdar
{\it et al}.~\cite{Talukdar:2020aui}. The asymmetry result, displayed in Fig.~\ref{fig:asymmetry}, reveals 
that it increases with $|Q^2|$ for both electron and muon scatterings. Our prediction indicates that the
asymmetry could reach levels of approximately 8\% for $e^\pm$-p and 3\%  for $\mu^\pm$-p scattering 
processes -- magnitudes that should be accessible to experimental verification by MUSE in the near term. 
Looking forward, we plan to improve upon Tsai’s soft-photon bremsstrahlung framework~\cite{Tsai:1961zz}
by dispensing with approximations such as the SPL and the use of the artificial cutoff $\Delta_{\gamma^*}$. 
Our goal is to develop a more robust method capable of extracting soft-photon contributions directly from 
the elastic end-point of the radiative-tail differential spectrum, allowing a more precise theoretical 
prediction of the charge asymmetry. From the experimental standpoint, the forthcoming MUSE results are 
highly anticipated, as they will offer a crucial test of our predicted asymmetry.

\section*{Acknowledgment}
We are thankful to Steffen Strauch for useful discussions pertaining to the ongoing developments with the 
MUSE experiment. RG acknowledges the organizers of the HADRON2025 International Conference at Osaka 
University for their financial support and hospitality. UR acknowledges financial support from the Science 
and Engineering Research Board, Republic of India, grant  CRG/2022/000027. He is also grateful to Daniel 
Phillips and the Institute of Nuclear and Particle Physics, Ohio University, Athens, for their local 
hospitality and travel support during the initial stages of the work.

\appendix
\section{Bremsstrahlung in HB$\chi$PT} 
\label{sec:amplitudes}
In this work, our primary goal is to evaluate the single soft-photon bremsstrahlung cross section for 
the (anti-)lepton-proton inelastic scattering process in the {\it lab.}-frame, up-to-and-including NLO 
[i.e., ${\mathcal O}(\alpha^3/M)$] in HB$\chi$PT. The analysis entails evaluating a wide range of 
amplitudes up to ${\mathcal O}(e^3/M^2)$, which may originate from as high as N${}^4$LO, i.e., 
$\nu=4$ chirally \underline{enhanced} amplitudes. To that end, we collect in this Appendix all such 
relevant amplitudes without approximation, in a manifestly covariant (frame-independent) manner, 
adopting the Lorentz gauge. These amplitudes are organized according to their standard chiral ordering,
beginning with the LO, which naively scale as ${\mathcal O}(e^3M^0)$, and extending up to N$^4$LO, 
with a naive scaling of ${\mathcal O}(e^3/M^4)$.\\

\subsection{LO [i.e., ${\mathcal O}(e^3M^0)$] amplitudes}
\begin{figure*}[tbp]
\centering
\includegraphics[width=0.68\linewidth]{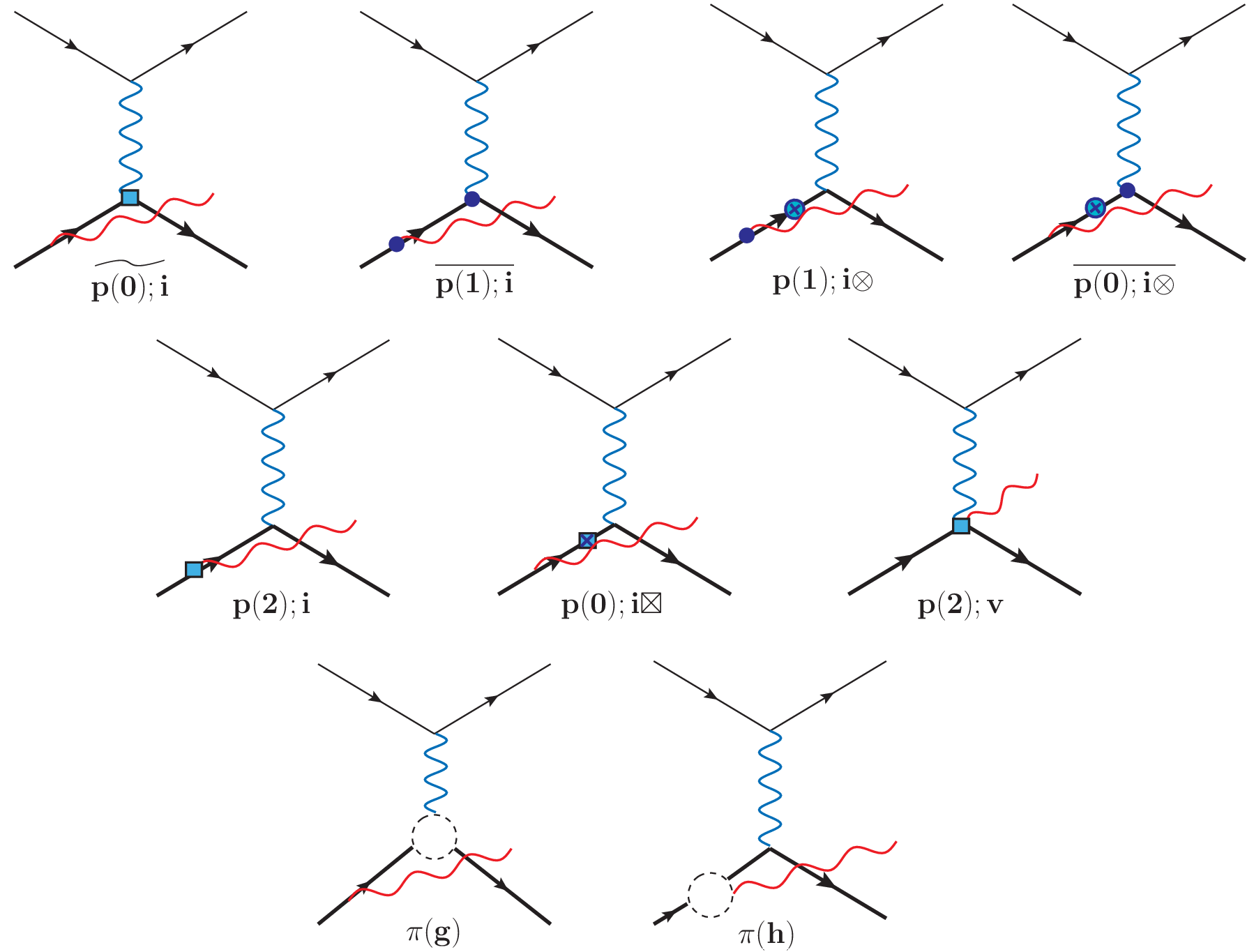}
    \caption{Other photon bremsstrahlung diagrams of NNLO [i.e., 
             ${\mathcal O}(e^3/M^2)\sim {\mathcal O}(e^3/\Lambda^2_\chi)$] with insertions of NLO 
             (filled blob $\bullet$) and NNLO (filled box $\square$) proton-photon interaction vertices, 
             proton's form factors (dashed circular blob), and ${\mathcal O}(1/M)$ (crossed circle 
             $\otimes$) and ${\mathcal O}(1/M^2)$ (crossed box $\boxtimes$) proton propagator components,
             \underline{not} contributing to our $\ell$-p elastic scattering cross section 
             up-to-and-including NLO [i.e., ${\mathcal O}(\alpha^3/M)$] in HB$\chi$PT. The thin, thick, 
             and wiggly lines denote the propagators for the lepton, proton, and photon (color online: 
             red for the soft bremsstrahlung photon and blue for the exchanged off-shell photon), 
             respectively. The nomenclature of the individual diagrams is based on whether the radiated
             photon originates from the initial-state ``i” or final-state proton ``p”. The labels “(0)”,
             “(1)”, and “(2)” denote LO, NLO, and NNLO proton-radiating vertices, respectively, while 
             the overline and tilde signify NLO and NNLO exchange-photon 
             proton vertex insertions, respectively. }
    \label{fig:NNLO_extra}
\end{figure*}

At this order, four relevant amplitudes contribute to the bremsstrahlung cross section 
up-to-and-including NLO [i.e., ${\mathcal O}(\alpha^3/M)$] in HB$\chi$PT. They are given as follows (cf.
Fig.~\ref{fig:LO}):
\begin{widetext}
\begin{eqnarray}
\label{eq:LO_1}
i \mathcal{M}^{\left[l;\rm{i}\right]}_{\gamma\gamma^*} 
\!&=&\!  \frac{i e^3}{(Q-k)^2}\left[\Bar{u}(p^\prime) \gamma^\mu  
\frac{ (\slashed p - \slashed k + m_l)}{(p-k)^2 - m_l ^2} \slashed\varepsilon^* u(p) \right] 
\Big[\chi^\dagger(p_p^\prime)  v_\mu \chi(p_p)\Big]\,,
\\
\label{eq:LO_2}   
i \mathcal{M}^{\left[l;\rm{f}\right]}_{\gamma\gamma^*}
\!&=&\! \frac{ie^3}{(Q-k)^2}\left[\Bar{u}(p^\prime)\slashed\varepsilon^* 
\frac{(\slashed p^\prime + \slashed k + m_l)}{(p^\prime+ k)^2 - m_l ^2} \gamma^\mu u(p) \right] 
\Big[\chi^\dagger(p_p^\prime)  v_\mu \chi(p_p)\Big]\,,
\\
\label{eq:LO_3}   
i \mathcal{M}^{\left[p(0);\rm{i}\right]}_{\gamma\gamma^*}
\!&=&\! \frac{-i e^3}{Q^2\left[v \cdot (p_p -k) + i0\right]}  \left[\Bar{u}(p^\prime)\gamma^\mu u(p) \right]
\Big[\chi^\dagger(p_p^\prime)  v_\mu (v\cdot \varepsilon^*) \chi(p_p)\Big]\,,  \qquad \text{and} 
\\
\label{eq:LO_4}   
i \mathcal{M}^{\left[p(0);\rm{f}\right]}_{\gamma\gamma^*}
\!&=&\! \frac{-i e^3}{Q^2\left[v \cdot (p_p^\prime + k) + i0\right]}  \left[\Bar{u}(p^\prime)\gamma^\mu u(p) \right]
\Big[\chi^\dagger(p_p^\prime) (v\cdot\varepsilon^*) v_\mu  \chi(p_p)\Big]\,.
\end{eqnarray}
Especially, in the {\it lab.}-frame, the amplitude 
$\mathcal{M}^{\left[p(0);\rm{f}\right]}_{\gamma\gamma^*}$ involving the LO [i.e., ${\mathcal O}(M^0)$] 
final-state proton propagator component, is chirally \underline{enhanced} under the {\it soft-photon limit} (SPL). 

\subsection{NLO [i.e., ${\mathcal O}(e^3/M)$] amplitudes}
At this order, nine relevant amplitudes can contribute to the bremsstrahlung cross section 
up-to-and-including NLO [i.e., ${\mathcal O}(\alpha^3/M)$] in HB$\chi$PT. They are given as follows (cf.
Fig.~\ref{fig:NLO}):
\begin{eqnarray}
\label{eq:NLO_5}  
i \mathcal{M}^{\left[\,\overline{l;\rm{i}}\,\right]}_{\gamma\gamma^*}
\!&=&\!  \frac{i e^3}{2 M (Q-k)^2}\left[\Bar{u}(p^\prime) \gamma^\mu  
\frac{ (\slashed p - \slashed k + m_l)}{(p-k)^2 - m_l ^2} \slashed\varepsilon^*  u(p) \right] 
\bigg[\chi^\dagger(p_p^\prime)  \Big\{(p_p+ p_p^\prime)_\mu 
- v_\mu [v\cdot (p_p +p_p^\prime)]\Big\} \chi(p_p)\bigg] \,,
\end{eqnarray}
\begin{eqnarray} 
\label{eq:NLO_6}  
i \mathcal{M}^{\left[\,\overline{l;\rm{f}}\,\right]}_{\gamma\gamma^*}
\!&=&\!  \frac{i e^3}{2M(Q-k)^2 }\left[\Bar{u}(p^\prime) \slashed\varepsilon^*
\frac{(\slashed p^\prime + \slashed k + m_l)}{(p^\prime+k)^2 - m_l ^2} \gamma ^\mu  u(p) \right]
\bigg[\chi^\dagger(p_p^\prime)  \Big\{(p_p+p_p^\prime)_\mu 
- v_\mu [v\cdot (p_p +p_p^\prime)]\Big\}\chi(p_p)\bigg]\,,\qquad\,\,
\\  
\label{eq:NLO_7}  
i \mathcal{M}^{\left[p(1);\rm{i}\right]}_{\gamma\gamma^*} 
\!&=&\! \frac{-i e^3}{2 M Q^2\left[v \cdot (p_p -k) + i0\right]}  
\left[\Bar{u}(p^\prime)\gamma^\mu u(p) \right]
\bigg[\chi^\dagger(p_p^\prime) v_\mu \Big\{(2 p_p - k)\cdot \varepsilon^* 
- (v\cdot \varepsilon^*) [v\cdot (2p_p-k)]\Big\}\chi(p_p)\bigg] \,,\qquad\,\,
\\
\label{eq:NLO_8}  
i \mathcal{M}^{\left[p(1);\rm{f}\right]}_{\gamma\gamma^*} 
\!&=&\! \frac{-i e^3}{2 M Q^2\left[v \cdot (p_p^\prime+k)+i0\right]}  
\left[\Bar{u}(p^\prime)\gamma^\mu u(p) \right]
\bigg[\chi^\dagger(p_p^\prime)\Big\{(2 p_p^\prime+ k)\cdot \varepsilon^*
- (v\cdot \varepsilon^*) [v\cdot (2p_p^\prime + k)]\Big\} v_\mu\chi(p_p)\bigg] \,,\qquad\,\,
\\
\label{eq:NLO_9}  
i \mathcal{M}^{\left[\,\overline{p(0);\rm{i}}\,\right]}_{\gamma\gamma^*} 
\!&=&\! \frac{-i e^3}{2 M Q^2\left[v \cdot (p_p -k) + i0\right]}  
\left[\Bar{u}(p^\prime)\gamma^\mu u(p) \right] 
\bigg[\chi^\dagger(p_p^\prime) \Big\{(p_p + p_p^\prime - k)_\mu 
\nonumber\\
&& \hspace{7.1cm} -\, v_\mu [v\cdot (p_p + p_p^\prime-k)]\Big\} (v\cdot\varepsilon^*)\chi(p_p)\bigg]\,,\qquad\,\,
\\
\label{eq:NLO_10}  
i \mathcal{M}^{\left[\,\overline{p(0);\rm{f}}\,\right]}_{\gamma\gamma^*} 
\!&=&\! \frac{-ie^3}{2 M Q^2\left[v \cdot (p_p^\prime + k) + i0\right]} 
\left[\Bar{u}(p^\prime)\gamma^\mu u(p) \right] 
\bigg[\chi^\dagger(p_p^\prime)  (v\cdot \varepsilon^*) \Big\{(p_p+p_p^\prime +k)_\mu 
\nonumber\\
&& \hspace{8.3cm} -\, v_\mu [v\cdot (p_p+p_p^\prime +k)]\Big\} \chi(p_p)\bigg]\,,\qquad\,\,
\\
\label{eq:NLO_11}  
i \mathcal{M}^{\left[p(1);\rm{v}\right]}_{\gamma\gamma^*} 
\!&=&\! \frac{-i e^3}{ M Q^2} \left[\Bar{u}(p^\prime)\gamma^\mu u(p) \right] 
\bigg[\chi^\dagger(p_p^\prime)\Big\{ \varepsilon^*_\mu 
- v_\mu (v\cdot\varepsilon^*)\Big\}\chi(p_p)\bigg] \,,\qquad\,\,
\\
\label{eq:NLO_12}
i \mathcal{M}^{\left[p(0);\rm{i}\otimes\right]}_{\gamma\gamma^*}
\!&=&\! \frac{-ie^3}{ 2 M Q^2} \left(1 - \frac{(p_p-k)^2}{\left[v \cdot (p_p-k)+i0\right]^2}\right) 
\left[\Bar{u}(p^\prime)\gamma^\mu u(p) \right]
\Big[\chi^\dagger(p_p^\prime) v_\mu (v\cdot\varepsilon^*)\chi(p_p)\Big] \,, \qquad \text{and}
\\
\label{eq:NLO_13}
i \mathcal{M}^{\left[p(0);\rm{f}\otimes\right]}_{\gamma\gamma^*}
\!&=&\! \frac{-i e^3}{2MQ^2}\left(1-\frac{(p_p^\prime+k)^2}{\left[v\cdot(p_p^\prime +k)+i0\right]^2}\right) 
\left[\Bar{u}(p^\prime)\gamma^\mu u(p) \right] 
\Big[\chi^\dagger(p_p^\prime)(v\cdot\varepsilon^*)v_\mu\chi(p_p)\Big]\,.
\end{eqnarray}
Especially, in the {\it lab.}-frame, the amplitudes $\mathcal{M}^{\left[p(1);\rm{f}\right]}_{\gamma\gamma^*}$
and $\mathcal{M}^{\left[\,\overline{p(0);\rm{f}}\,\right]}_{\gamma\gamma^*}$, involving the LO [i.e., 
${\mathcal O}(M^0)$], and $\mathcal{M}^{\left[p(0);\rm{f}\otimes\right]}_{\gamma\gamma^*}$, involving the NLO
[i.e., ${\mathcal O}(1/M)$] components of the final-state proton propagator, are all chirally 
\underline{enhanced} in SPL.

\subsection{NNLO [i.e., ${\mathcal O}(e^3/M^2)$] amplitudes }
At this order, fifteen relevant amplitudes can contribute to the bremsstrahlung cross section 
up-to-and-including NLO [i.e., ${\mathcal O}(\alpha^3/M)$] in HB$\chi$PT. They are given as follows (cf. 
Fig.~\ref{fig:NNLO} and the first two rows in Fig.~\ref{fig:NNLO_extra}):
\begin{eqnarray}
\label{eq:NNLO_14}
i \mathcal{M}^{\left[\widetilde{\,l;\rm{i}\,}\right]}_{\gamma\gamma^*} 
\!&=&\! \frac{i e^3}{8 M^2 (Q-k)^2}\left[\Bar{u}(p^\prime) \gamma^\mu  
\frac{ (\slashed p - \slashed k + m_l)}{(p-k)^2 - m_l ^2} \slashed\varepsilon^*  u(p) \right] 
\nonumber \\
&& \hspace{1.8cm} \times\, \bigg[\chi^\dagger(p_p^\prime) \Big\{2 [v\cdot (Q-k)]^2 v_\mu 
- [v\cdot (Q-k)](Q-k)_\mu -(Q-k)^2 v_\mu \Big\} \chi(p_p)\bigg]\,,\qquad\,\,
\\
\label{eq:NNLO_15}
i \mathcal{M}^{\left[\widetilde{\,l;\rm{f}\,}\right]}_{\gamma\gamma^*} 
\!&=&\!  \frac{i e^3}{8 M^2 (Q-k)^2 }\left[\Bar{u}(p^\prime) \slashed\varepsilon^*
\frac{(\slashed p^\prime + \slashed k + m_l)}{(p^\prime+k)^2 - m_l ^2} \gamma ^\mu  u(p) \right]
\nonumber \\
&& \hspace{1.8cm} \times\, \bigg[\chi^\dagger(p_p^\prime) \Big\{2 [v\cdot (Q-k)]^2 v_\mu 
- [v\cdot (Q-k)](Q-k)_\mu -(Q-k)^2 v_\mu \Big\} \chi(p_p)\bigg]\,, \qquad\,\,
\\
\label{eq:NNLO_16}
i \mathcal{M}^{\left[p(1);\rm{f} \otimes\right]}_{\gamma\gamma^*}
\!&=&\! \frac{-ie^3}{4 M^2 Q^2} \left(1-\frac{ (p_p^\prime+k)^2}{\left[v \cdot( p_p^\prime+k)+i0\right]^2} \right)\!\! 
\left[\Bar{u}(p^\prime) \gamma^\mu u(p) \right]\!\! 
\bigg[\chi^\dagger(p_p^\prime) \Big\{(2p_p^\prime + k)\cdot \varepsilon^* 
\nonumber\\
&& \hspace{8.5cm} -\, (v\cdot \varepsilon^*) [v\cdot (2p_p^\prime + k)]\Big\} v_\mu \chi(p_p) \Big]\,,\qquad\,\,
\\
\label{eq:NNLO_17}
i \mathcal{M}^{\left[\,\overline{p(0);\rm{f} \otimes}\,\right]}_{\gamma\gamma^*} 
\!&=&\! 
\frac{-ie^3}{4 M^2 Q^2} \!\left(1-\frac{(p_p^\prime+k)^2}{\left[v\cdot(p_p^\prime+k)+i0\right]^2} \right) \!
\left[\Bar{u}(p^\prime) \gamma^\mu u(p) \right] \!
\bigg[\chi^\dagger(p_p^\prime) (v\cdot\varepsilon^*) \Big\{(p_p+p_p^\prime+k)_\mu 
\nonumber\\
&& \hspace{9.7cm} -\, v_\mu [v\cdot (p_p+p_p^\prime+k)] \Big\} \chi(p_p) \bigg],\qquad\,\,
\\
\label{eq:NNLO_18}
i \mathcal{M}^{\left[\,\widetilde{p(0);\rm{f}}\,\right]}_{\gamma\gamma^*}
\!&=&\! \frac{-i e^3}{8 M^2 Q^2\left[v \cdot (p_p^\prime + k) + i0\right]}\! 
\left[\Bar{u}(p^\prime) \gamma^\mu u(p) \right] \!
\bigg[\chi^\dagger(p_p^\prime) (v\cdot\varepsilon^*) \Big\{2 (v\cdot Q)^2 v_\mu 
- (v\cdot Q) Q_\mu -Q^2 v_\mu \Big\}\chi(p_p)\bigg],\qquad\,\,
\end{eqnarray}
\begin{eqnarray} 
\label{eq:NNLO_19}
i \mathcal{M}^{\left[p(2);\rm{f}\right]}_{\gamma\gamma^*}
\!&=&\! \frac{-i e^3}{8 M^2 Q^2\left[v \cdot (p_p^\prime + k) + i0\right]}  
\left[\Bar{u}(p^\prime) \gamma^\mu u(p) \right]
\bigg[\chi^\dagger(p_p^\prime) \Big\{2 (v\cdot k)^2 (v\cdot \varepsilon^*) 
\nonumber\\
&& \hspace{7.4cm} -\, (v\cdot k) (k\cdot\varepsilon^*) 
- k^2 (v\cdot\varepsilon^*) \Big\}v_\mu \chi(p_p)\bigg]\,,\qquad\,\,
\\
\label{eq:NNLO_20}
i \mathcal{M}^{\left[p(0);\rm{f}\boxtimes\right]}_{\gamma\gamma^*}
\!&=&\! \frac{-ie^3}{4 M^2 Q^2}\left(\frac{\left[v\cdot (p_p^\prime+k)\right]^3 
- [v \cdot (p_p^\prime+k)] (p_p^\prime+k)^2}{\left[v \cdot (p_p^\prime+k)+i0\right]^2} \right)
\left[\Bar{u}(p^\prime) \gamma^\mu u(p) \right] 
\Big[\chi^\dagger(p_p^\prime) (v\cdot \varepsilon^*)v_\mu \chi(p_p)\Big]\,,
\\
\label{eq:NNLO_21}
i \mathcal{M}^{\left[\,\overline{p(1);\rm{f}}\,\right]}_{\gamma\gamma^*} 
\!&=&\! \frac{-i e^3}{4 M^2 Q^2\left[v \cdot (p_p^\prime + k) + i0\right]} 
\left[\Bar{u}(p^\prime) \gamma^\mu u(p) \right]
\bigg[\chi^\dagger(p_p^\prime) \Big\{(2p_p^\prime + k)\cdot\varepsilon^* 
- (v\cdot\varepsilon^*) [v\cdot(2p_p^\prime + k)]\Big\}
\nonumber\\
&& \hspace{6.9cm} \times\, \Big\{(p_p+p_p^\prime+k)_\mu 
- v_\mu [v\cdot(p_p+p_p^\prime+k)] \Big\} \chi(p_p) \bigg]\,,
\\
\label{eq:NNLO_22}
i \mathcal{M}^{\left[\,\widetilde{p(0);\rm{i}}\,\right]}_{\gamma\gamma^*} 
\!&=&\! \frac{i e^3}{8 M^2 Q^2 \left[v \cdot (p_p-k) + i0\right]} 
\left[\Bar{u}(p^\prime) \gamma^\mu u(p) \right]
\bigg[\chi^\dagger(p_p^\prime) \Big\{2 (v\cdot Q)^2 v_\mu 
- (v\cdot Q) Q_\mu -Q^2 v_\mu \Big\}(v\cdot\varepsilon^*)\chi(p_p)\bigg]\,,
\\
\label{eq:NNLO_23}
i \mathcal{M}^{\left[\,\overline{p(1);\rm{i}}\,\right]}_{\gamma\gamma^*} 
\!&=&\! \frac{i e^3}{4 M^2 Q^2 \left[v \cdot (p_p -k) + i0\right]} 
\left[\Bar{u}(p^\prime) \gamma^\mu u(p) \right] 
\bigg[\chi^\dagger(p_p^\prime)  \Big\{(p_p+p_p^\prime-k)_\mu 
- v_\mu [v\cdot (p_p+p_p^\prime-k)]\Big\} 
\nonumber\\
&& \hspace{6.8cm} \times\, \Big\{(2p_p-k)\cdot\varepsilon^* 
- (v\cdot\varepsilon^*) [v\cdot (2p_p-k)]\Big\} \chi(p_p)\bigg]\,,
\\
\label{eq:NNLO_24}
i \mathcal{M}^{\left[p(1);\rm{i} \otimes\right]}_{\gamma\gamma^*} 
\!&=&\!  \frac{-ie^3}{4 M^2 Q^2}\left(1-\frac{(p_p-k)^2}{\left[v\cdot(p_p-k)+i0\right]^2}\right) 
\left[\Bar{u}(p^\prime) \gamma^\mu u(p) \right] 
\bigg[\chi^\dagger(p_p^\prime) v_\mu \Big\{(2p_p-k)\cdot \varepsilon^* 
\nonumber\\
&& \hspace{9.1cm} -\, (v\cdot \varepsilon^*) [v\cdot(2p_p- k)]\Big\}\chi(p_p) \bigg]\,,\qquad\,\,
\\
\label{eq:NNLO_25}
i \mathcal{M}^{\left[\,\overline{p(0);\rm{i} \otimes}\,\right]}_{\gamma\gamma^*} 
\!&=&\! \frac{-ie^3}{4 M^2 Q^2}\left(1-\frac{(p_p-k)^2}{\left[v\cdot(p_p-k)+i0\right]^2}\right)
\left[\Bar{u}(p^\prime) \gamma^\mu u(p) \right] 
\bigg[\chi^\dagger(p_p^\prime) \Big\{(p_p+p_p^\prime - k)_\mu 
\nonumber\\
&& \hspace{8.7cm} -\, v_\mu [v\cdot(p_p+p_p^\prime - k)] \Big\} (v\cdot\varepsilon^*)\chi(p_p)\bigg]\,,\qquad\,\,
\\
\label{eq:NNLO_26}
i \mathcal{M}^{\left[p(2);\rm{i}\right]}_{\gamma\gamma^*}
\!&=&\! \frac{i e^3}{8 M^2 Q^2 \left[v \cdot (p_p -k) + i0\right]} 
\left[\Bar{u}(p^\prime) \gamma^\mu u(p) \right]
\bigg[\chi^\dagger(p_p^\prime) v_\mu  \Big\{2 (v\cdot k)^2 (v\cdot \varepsilon^*) 
\nonumber\\
&& \hspace{7.7cm} -\, (v\cdot k) (k\cdot \varepsilon^*)-k^2 (v\cdot\varepsilon^*)\Big\} \chi(p_p)\bigg]\,,
\\
\label{eq:NNLO_27}
i \mathcal{M}^{\left[\,p(0);\rm{i}\boxtimes\,\right]}_{\gamma\gamma^*} 
\!&=&\! \frac{-ie^3}{4 M^2 Q^2} \left(\frac{\left[v\cdot(p_p-k)\right]^3 
- [v\cdot(p_p-k)](p_p-k)^2}{\left[v\cdot(p_p-k)+i0\right]^2}\right) 
\left[\Bar{u}(p^\prime) \gamma^\mu u(p) \right] 
\Big[\chi^\dagger(p_p^\prime) v_\mu (v\cdot\varepsilon^*) \chi(p_p)\Big]\,,
\end{eqnarray}
and
\begin{eqnarray} 
\label{eq:NNLO_28}
i \mathcal{M}^{\left[p(2);{\rm v}\right]}_{\gamma\gamma^*} 
\!&=&\! \frac{-ie^3}{4 M^2 Q^2} \left[\Bar{u}(p^\prime) \gamma^\mu u(p) \right]
\bigg[\chi^\dagger(p_p^\prime) \Big\{ 3[v\cdot(Q -k)] v_\mu (v\cdot\varepsilon^*) 
- (Q-k)_\mu (v\cdot\varepsilon^*)  - v_\mu (Q-k)\cdot \varepsilon^* 
\nonumber \\
&& \hspace{4.45cm} -\, [v \cdot (Q-k)]\varepsilon^*_\mu \Big\}\chi(p_p)\bigg]\,.
\end{eqnarray}
\end{widetext}
%
%
Of the fifteen recoil-radiative amplitudes listed above, only the first eight amplitudes (cf. 
Fig.~\ref{fig:NNLO}), namely, $\mathcal{M}^{\left[\widetilde{\,l;{\rm i}\,}\right]}_{\gamma\gamma^*}$, 
$\mathcal{M}^{\left[\widetilde{\,l;{\rm f}\,}\right]}_{\gamma\gamma^*}$,
$\mathcal{M}^{\left[p(1);f \otimes\right]}_{\gamma\gamma^*}$, 
$\mathcal{M}^{\left[\,\overline{p(0);f \otimes}\,\right]}_{\gamma\gamma^*}$
$\mathcal{M}^{\left[\,\widetilde{p(0);f}\,\right]}_{\gamma\gamma^*}$, 
$\mathcal{M}^{\left[p(2);f\right]}_{\gamma\gamma^*}$,
$\mathcal{M}^{\left[p(0);f\boxtimes\right]}_{\gamma\gamma^*}$, and
$\mathcal{M}^{\left[\,\overline{p(1);f}\,\right]}_{\gamma\gamma^*}$, Eqs.~\eqref{eq:NNLO_14} – 
\eqref{eq:NNLO_21}, contribute to the soft-photon bremsstrahlung cross sections in the {\it lab.}-frame at 
our intended $\mathcal{O}(1/M)$ accuracy. Apart from 
$\mathcal{M}^{\left[\widetilde{\,l;{\rm i}\,}\right]}_{\gamma\gamma^*}$ and 
$\mathcal{M}^{\left[\widetilde{\,l;{\rm f}\,}\right]}_{\gamma\gamma^*}$, the rest of these contributing 
amplitudes in the SPL, with either LO or NLO components of the final-state proton propagator, are chirally 
\underline{enhanced} in the {\it lab.}-frame. 

In addition to the recoil-radiative amplitudes, six pion-loop amplitudes (cf. Fig.~\ref{fig:NNLO_loop} and
the last row in Fig.~\ref{fig:NNLO_extra}) arise from insertions of the proton's Dirac ($F_{1}$) and Pauli
($F_{2}$) form factors, which can also contribute to the  ${\mathcal O}(\alpha^3/M)$ NLO cross section. 
The following are their expressions: 
\begin{widetext}
\begin{eqnarray} 
\label{eq:NNLO_29} 
i \mathcal{M}^{\left[\rm \pi (a)\right]}_{\gamma\gamma^*}  \!&=&\!  
\frac{i e^3}{(Q-k)^2}\left[\Bar{u}(p^\prime) \gamma^\mu  
\frac{(\slashed p - \slashed k + m_l)}{(p-k)^2 - m_l ^2} \slashed\varepsilon^* u(p) \right] 
\Big[\chi^\dagger(p_p^\prime)  \mathcal{V}^{(2)}_\mu \chi(p_p)\Big]\,,
\\
\label{eq:NNLO_30}
i \mathcal{M}^{\left[\rm \pi (b)\right]}_{\gamma\gamma^*}  \!&=&\! 
\frac{i e^3}{(Q-k)^2}\left[\Bar{u}(p^\prime)\gamma^\mu 
\frac{ (\slashed p - \slashed k + m_l)}{(p-k)^2 - m_l ^2} \slashed\varepsilon^* u(p) \right] 
\Big[\chi^\dagger(p_p^\prime) \mathcal{V}^{(2)}_\mu \chi(p_p)\Big]\,,
\\
\label{eq:NNLO_31}
i \mathcal{M}^{\left[\rm \pi (c)\right]}_{\gamma\gamma^*}  \!&=&\! 
\frac{-i e^3}{Q^2\left[v \cdot (p_p^\prime + k) + i0\right]}
\left[\Bar{u}(p^\prime) \gamma^\mu u(p) \right] 
\Big[\chi^\dagger(p_p^\prime) \mathcal{V}^{(2)}_\mu (v \cdot \varepsilon^*)\chi(p_p)\Big]\,,
\\
\label{eq:NNLO_32}
i \mathcal{M}^{\left[\rm \pi (d)\right]}_{\gamma\gamma^*}  \!&=&\! 
\frac{-ie^3}{Q^2\left[v \cdot (p_p^\prime + k) + i0\right]} 
\left[\Bar{u}(p^\prime) \gamma^\mu u(p) \right] 
\bigg[\chi^\dagger (p_p^\prime) \left(\mathcal{V}^{(2)}\cdot\varepsilon^*\right)v_\mu\chi(p_p)\bigg]\,,
\\
\label{eq:NNLO_33}
i \mathcal{M}^{\left[\rm \pi (g)\right]}_{\gamma\gamma^*}  \!&=&\! 
\frac{-i e^3}{Q^2\left[v \cdot (p_p - k) + i0\right]} 
\left[\Bar{u}(p^\prime) \gamma^\mu u(p) \right]
\Big[\chi^\dagger(p_p^\prime) \mathcal{V}^{(2)}_\mu \left(v\cdot\varepsilon^*\right)\chi(p_p)\Big]\,, \qquad\text{and}
\\
\label{eq:NNLO_34} 
i \mathcal{M}^{\left[\rm \pi (h)\right]}_{\gamma\gamma^*}  \!&=&\! 
\frac{-ie^3}{Q^2\left[v \cdot (p_p - k) + i0\right]} 
\left[\Bar{u}(p^\prime) \gamma^\mu u(p) \right]
\bigg[\chi^\dagger(p_p^\prime) v_\mu \left(\mathcal{V}^{(2)}_\mu\cdot \varepsilon^* \right) \chi(p_p)\bigg]\,, 
\end{eqnarray} 
\end{widetext}
where $\mathcal{V}^{(2)}_\mu$ is the effective hadronic corrections vertex, given in Eq.~\eqref{eq:V2} in 
the main text, which renormalizes the LO proton-photon vertex. Of the six pion-loop amplitudes listed 
above, only the first four amplitudes, $\mathcal{M}^{\left[\rm \pi(a),...,\pi(d)\right]}_{\gamma\gamma^*}$
(cf. Fig.~\ref{fig:NNLO_loop}), Eqs.~\eqref{eq:NNLO_29} – \eqref{eq:NNLO_32}, contribute to the 
soft-photon bremsstrahlung cross sections in the {\it lab.}-frame at our intended $\mathcal{O}(1/M)$ 
accuracy. In particular, the amplitudes, $\mathcal{M}^{\left[\rm \pi(c),\pi(d)\right]}_{\gamma\gamma^*}$,
Eqs.~\eqref{eq:NNLO_29} and \eqref{eq:NNLO_30}, involving the LO component of the final-state proton 
propagator, are chirally \underline{enhanced} in the {\it lab.}-frame in the SPL. 

The remaining nine amplitudes, Eqs.~\eqref{eq:NNLO_22} – \eqref{eq:NNLO_28}, \eqref{eq:NNLO_33} and 
\eqref{eq:NNLO_34}, shown in Fig.~\ref{fig:NNLO_extra}), are discarded in our main text analysis, as they 
do not contribute to the soft-photon bremsstrahlung cross section in the {\it lab.}-frame. A detailed 
investigation reveals that for the computation of the cross sections, they all either lead to scaleless 
phase-space integrals vanishing under dimensional regularization (DR) (charge-even case), or yield terms 
of $\mathcal{O}(1/M^2)$, which are beyond our working precision (charge-odd case).

\begin{figure*}
\centering
\includegraphics[width=0.64\linewidth]{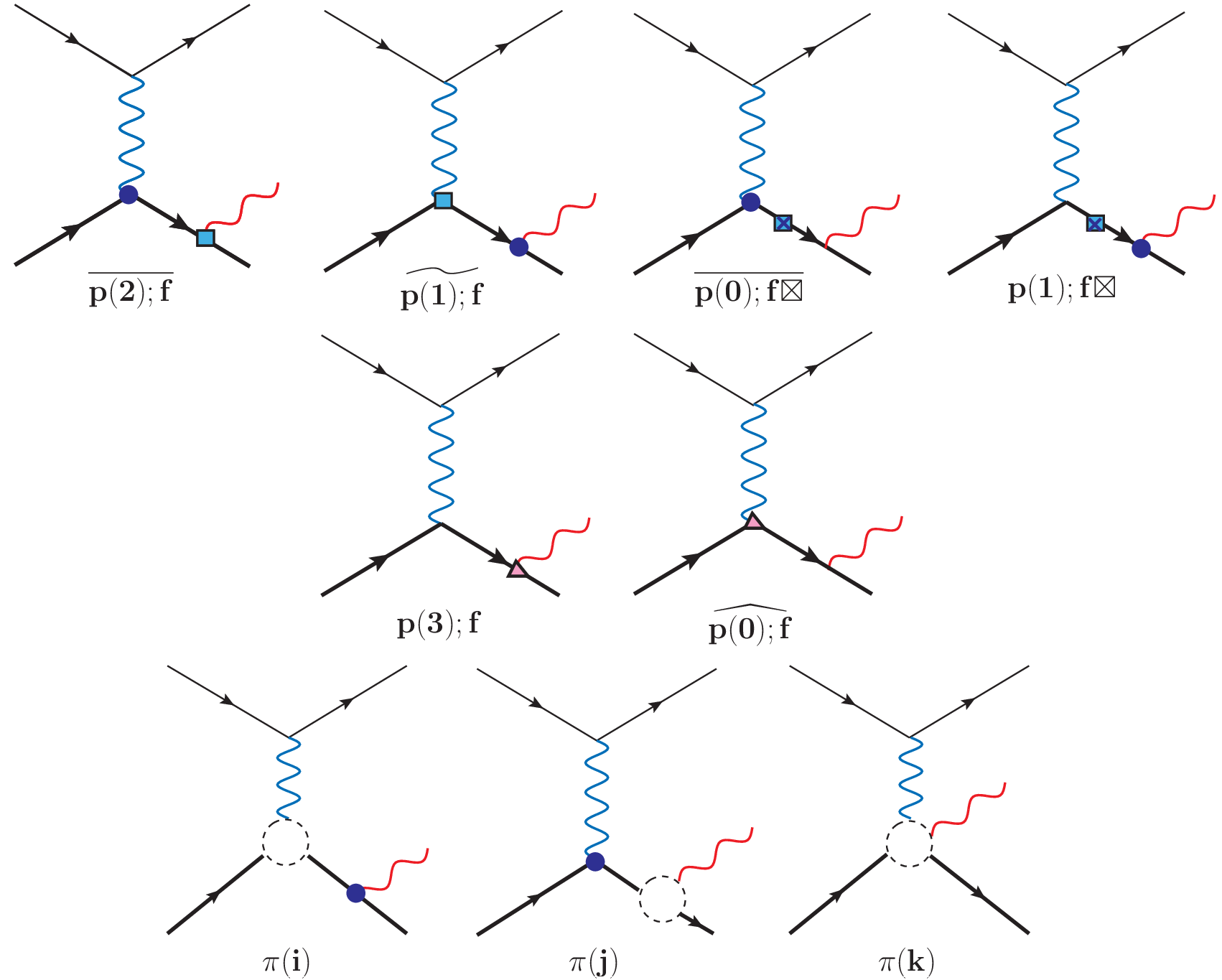}
    \caption{Photon bremsstrahlung diagrams of N${}^3$LO [i.e., 
             ${\mathcal O}(e^3/M^3)\sim {\mathcal O}\left(e^3/(M \Lambda^2_\chi)\right)]$ with 
             insertions of NLO (filled blob $\bullet$), NNLO (filled box $\square$), and $N^3$LO 
             (triangle $\triangle$) proton-photon interaction vertices, proton's form factors (dashed
             circular blob) and ${\mathcal O}(1/M^2)$ (crossed box $\boxtimes$) proton propagator 
             components, \underline{not} contributing to our $\ell$-p elastic scattering cross section
             up-to-and-including NLO [i.e., ${\mathcal O}(\alpha^3/M)$] in HB$\chi$PT. The thin, 
             thick, and wiggly lines denote the propagators for the lepton, proton, and photon (color
             online: red for the soft bremsstrahlung photon and blue for the exchanged off-shell 
             photon), respectively. The nomenclature of the individual diagrams is based on the 
             radiated photon originating from the final-state ``f” proton ``p”. The labels “(0)”, 
             “(1)”, “(2)”, and “(3)” denote LO, NLO, NNLO, and N${}^3$LO proton-radiating vertices, 
             respectively, while the overline, tilde, and wedge signify NLO, NNLO, and N${}^3$LO 
             exchange-photon proton vertex insertions, respectively. All of the above diagrams --
             except for the last one involving the two-photon pion-loop -- correspond to chirally 
             enhanced amplitudes. }
\label{fig:N3LO_extra}
\end{figure*}

Consequently, in the analysis of the charge-even soft-photon bremsstrahlung cross section at NLO, all
relevant contributions arising from the $\nu\geq2$ chiral order matrix elements scale as
$\mathcal{S}^{(\nu\geq 2)}_{\rm even}\sim {\mathcal O}(1/M^2)$, and hence, discarded. In contrast, the
analysis presented in Sec.~\ref{sec:II}, addresses the non-trivial contribution of chirally 
\underline{enhanced} NNLO amplitudes to the charge-odd soft-photon bremsstrahlung cross section at NLO. 
This necessitates the evaluation of the interference terms in ${\mathcal S}^{(\nu=2)}_{\rm odd}$, 
Eq.~\eqref{eq:So_NNLO}, namely,
\begin{eqnarray}
\mathcal{S}^{(\nu=2)}_{\rm odd} \!&=&\! \sum_{\rm spins} \left|{\mathcal M}_{\rm NLO}\right|^2_{\rm odd} 
\nonumber\\
&& +\, \sum_{\rm spins} \left[{\mathcal M}^\dagger_{\rm LO} \widetilde{\mathcal M}_{\rm NNLO} 
+ {\rm h.c}\right]_{\rm odd}\,, \quad\,
\end{eqnarray}
where ${\mathcal M}_{\rm LO}$, ${\mathcal M}_{\rm NLO}$ and $\widetilde{\mathcal M}_{\rm NNLO}$ are the 
sum of the soft-photon bremsstrahlung amplitudes in the {\it lab.}-frame, as given in Eqs.~\eqref{eq:M_LO}
- \eqref{eq:M_NNLO}, in the main text. However, the general structure of some of the amplitudes 
contributing to charge-odd $\nu=2$ chiral order matrix elements, namely, 
\begin{widetext}
\begin{eqnarray}
S^{(\nu=2)}_{\rm odd} \supset 
2\mathcal{R}e \sum_{\rm spins} \Bigg[\left(\mathcal{M}^{\left[\,\overline{l;{\rm i}}\,\right]}_{\gamma\gamma^*} 
+ \mathcal{M}^{\left[\,\overline{l;{\rm f}}\,\right]}_{\gamma\gamma^*}\right)^\dagger \hspace{-0.4cm}&&\!
\left(\mathcal{M}^{\left[p(1);{\rm i}\right]}_{\gamma\gamma^*} 
+ \mathcal{M}^{\left[\,\overline{p(0);{\rm i}}\,\right]}_{\gamma\gamma^*} 
+ \mathcal{M}^{\left[p(0);{\rm i}\otimes\right]}_{\gamma\gamma^*}+ \mathcal{M}^{\left[\,p(1);{\rm v}\,\right]}_{\gamma\gamma^*} \right)
\nonumber\\
+\, {\mathcal M}^{\left[p(0);{\rm i}\right]\dagger}_{\gamma\gamma^*} \hspace{-0.3cm}&&\!
\left(\mathcal{M}^{\left[\widetilde{\,l;{\rm i}\,}\right]}_{\gamma\gamma^*} 
+ \mathcal{M}^{\left[\widetilde{\,l;{\rm f}\,}\right]}_{\gamma\gamma^*} 
+ \mathcal{M}^{\left[\rm \pi (a)\right]}_{\gamma\gamma^*}+\mathcal{M}^{\left[\rm \pi (b)\right]}_{\gamma\gamma^*}\right)\Bigg] 
\sim \mathcal{O}\left(\frac{1}{M^2}\right)\,, 
\label{eq:So_NNLO_dropped} 
\end{eqnarray} 
makes it clear that they yield ${\mathcal O}(1/M^2)$ terms in the cross section which lie beyond the 
accuracy of our present analysis. Accordingly, the above combination of amplitudes is excluded from the 
subsequent treatment of the charge-odd component.

\subsection{N$\mathbf{{}^{3}}$LO [i.e., ${\mathcal O}(e^3/M^3)$] amplitudes }
At this order, it is necessary to include Feynman amplitudes having the contributions from the N$^3$LO
proton-photon interactions stemming from the chiral Lagrangian of $\nu=3$ order. The relevant 
$\mathcal{O}(1/M^{3})$ interaction terms are
\begin{eqnarray}
{\mathcal{L}}_{\pi N}^{(3)} \!&=&\! \bar{N}_v \left[ \frac{1}{8M^3}
\Big\{( v\cdot \mathcal{D})\mathcal{D}_{\mu}\mathcal{D}^{\mu}(v \cdot \mathcal{D}) 
- (v\cdot \mathcal{D})^4\Big\}+\cdots\right] N_v\,, \qquad \text{with}
\nonumber\\
\mathcal{D}_\mu \!&=&\! \partial_\mu +\frac{1}{2}\bigg[u^\dag \left({\mathbb I}\partial_\mu+i\tau^3\frac{e}{2}A_\mu\right)u 
+ u\left({\mathbb I}\partial_\mu+i\tau^3\frac{e}{2}A_\mu\right) u^\dag\bigg] + i {\mathbb I}\frac{e}{2} A_\mu\,.
\label{eq:LpiN-3}
\end{eqnarray}
The above ellipsis indicates additional operators in the chiral Lagrangian not relevant for the present
analysis. In this case, nine relevant amplitudes can contribute to the bremsstrahlung cross section 
up-to-and-including NLO [i.e., ${\mathcal O}(\alpha^3/M)$] in HB$\chi$PT. Especially, in the 
{\it lab.}-frame, all such amplitudes in SPL are chirally \underline{enhanced} with LO and NNLO 
components of the final-state proton propagator. They are given as follows (cf. the first row in 
Fig.~\ref{fig:N3LO} and the first two rows in Fig.~\ref{fig:N3LO_extra}):
\begin{eqnarray}
\label{eq:N3LO_35}
i \mathcal{M}^{\left[\,\overline{p(1);\rm{f} \otimes}\,\right]}_{\gamma\gamma^*} 
\!&=&\! \frac{-ie^3}{8 M^3 Q^2}\!\left(1-\frac{(p_p^\prime+k)^2}{\left[v\cdot(p_p^\prime+k)+i0\right]^2}\right) \! 
\left[\Bar{u}(p^\prime) \gamma^\mu u(p) \right] \!
\bigg[\chi^\dagger(p_p^\prime) \Big\{(2 p_p^\prime+k)\cdot\varepsilon^* 
- (v\cdot\varepsilon^*)[v\cdot(2 p_p^\prime +k)]\Big\}
\nonumber\\
&& \hspace{7.5cm} \times\, \Big\{(p_p+p_p^\prime+k)_\mu 
- v_\mu [v\cdot(p_p+p_p^\prime+k)]\Big\}\chi(p_p)\bigg] \,,\quad\,   
\nonumber\\
\\
\label{eq:N3LO_36}
i \mathcal{M}^{\left[\,\widetilde {p(0);\rm{f} \otimes}\,\right]}_{\gamma\gamma^*}
\!&=&\! \frac{-ie^3}{16 M^3 Q^2} \left(1-\frac{(p_p^\prime+k)^2}{\left[v \cdot( p_p^\prime+k)+i0\right]^2} \right) 
\left[\Bar{u}(p^\prime) \gamma^\mu u(p) \right] 
\bigg[\chi^\dagger(p_p^\prime) (v\cdot \varepsilon^*) \Big\{2(v\cdot Q)^2 v_\mu 
\nonumber\\
&&  \hspace{10.1cm} -\, (v\cdot Q) Q_\mu -Q^2 v_\mu\Big\}\chi(p_p)\bigg] \,,
\end{eqnarray}
\begin{eqnarray} 
\label{eq:N3LO_37}
i \mathcal{M}^{\left[p(2);\rm{f} \otimes\right]}_{\gamma\gamma^*}
\!&=&\! 
\frac{-ie^3}{16 M^3 Q^2} \left(1-\frac{ (p_p^\prime+k)^2}{\left[v \cdot( p_p^\prime+k)+i0\right]^2} \right) 
\left[\Bar{u}(p^\prime) \gamma^\mu u(p) \right]
\bigg[\chi^\dagger(p_p^\prime) \Big\{2 (v\cdot k)^2 (v\cdot\varepsilon^*) 
\\
&& \hspace{9cm} -\, (v\cdot k) (k\cdot \varepsilon^*) - k^2 (v\cdot \varepsilon^*)\Big\}v_\mu\chi(p_p)\bigg],\qquad\,
\nonumber\\
\label{eq:N3LO_38}
i \mathcal{M}^{\left[\,\overline {p(2);\rm{f}}\,\right]}_{\gamma\gamma^*}
\!&=&\!
\frac{-i e^3}{16 M^3 Q^2\left[v \cdot (p_p^\prime + k) + i0\right]} 
\left[\Bar{u}(p^\prime) \gamma^\mu u(p) \right] 
\bigg[\chi^\dagger(p_p^\prime) \Big\{2 (v\cdot k)^2 (v\cdot\varepsilon^*) 
- (v\cdot k) (k\cdot \varepsilon^*)-k^2 (v\cdot \varepsilon^*)\Big\} 
\nonumber\\
&& \hspace{7.1cm} \times\, \Big\{(p_p+p_p^\prime+k)_\mu 
- v_\mu [v\cdot (p_p+p_p^\prime+k)]\Big\}\chi(p_p)\bigg] \,,\qquad\,
\\
\label{eq:N3LO_39}
i \mathcal{M}^{\left[\,\widetilde {p(1);\rm{f}}\,\right]}_{\gamma\gamma^*}
\!&=&\!
\frac{-ie^3}{16 M^3 Q^2\left[v\cdot (p_p^\prime + k) + i0\right]} 
\left[\Bar{u}(p^\prime) \gamma^\mu u(p) \right] 
\bigg[\chi^\dagger(p_p^\prime) \Big\{(2p_p^\prime +k)\cdot\varepsilon^* 
- (v\cdot\varepsilon^*)[v\cdot(2p_p^\prime +k)]\Big\} 
\nonumber\\
&& \hspace{7.1cm} \times\, \Big\{2(v\cdot Q)^2 v_\mu - (v\cdot Q)Q_\mu -Q^2 v_\mu\Big\}\chi(p_p)\bigg] \,,\qquad\,
\\
\label{eq:N3LO_40}
i \mathcal{M}^{\left[\,\overline {p(0);\rm{f}\boxtimes}\,\right]}_{\gamma\gamma^*} 
\!&=&\! 
\frac{-ie^3}{8 M^3 Q^2}\left(\frac{\left[v \cdot (p_p^\prime+k)\right]^3 
- [v \cdot (p_p^\prime+k)] (p_p^\prime+k)^2}{\left[v \cdot (p_p^\prime+k)+i0\right]^2}\right) 
\left[\Bar{u}(p^\prime) \gamma^\mu u(p) \right]
\nonumber\\
&& \hspace{4cm} \times\, \bigg[\chi^\dagger(p_p^\prime)(v \cdot \varepsilon^*) \Big\{(p_p+p_p^\prime+k)_\mu 
- v_\mu [v\cdot(p_p+p_p^\prime+k)] \Big\}\chi(p_p)\bigg] \,,\qquad\,
\\
\label{eq:N3LO_41}
i \mathcal{M}^{\left[p(1);\rm{f} \boxtimes\right]}_{\gamma\gamma^*}
\!&=&\!  
\frac{-ie^3}{8 M^3 Q^2} \left( \frac{\left[v \cdot (p_p^\prime+k)\right]^3 
- [v \cdot (p_p^\prime+k)](p_p^\prime+k)^2}{\left[v\cdot(p_p^\prime+k)+i0\right]^2} \right)  
\left[\Bar{u}(p^\prime) \gamma^\mu u(p) \right] 
\nonumber\\
&& \hspace{4cm} \times\, \bigg[\chi^\dagger(p_p^\prime)\Big\{(2p_p^\prime +k)\cdot\varepsilon^* 
- (v\cdot \varepsilon^*)[v\cdot(2p_p^\prime +k)]\Big\}v_\mu\chi(p_p) \bigg] \,, \qquad\,\,
\\
\label{eq:N3LO_42}
i \mathcal{M}^{[p(3);{\rm f}]}_{\gamma\gamma^*} 
\!&=&\!
\frac{i e^3}{8 M^3 Q^2\left[v \cdot (p_p^\prime + k) + i0\right]}
\left[\Bar{u}(p^\prime) \gamma^\mu u(p) \right] \bigg[\chi^\dagger(p_p^\prime) 
\Big\{(v\cdot k)k^2-(v\cdot k)^3\Big\}  (v\cdot \varepsilon^*) v_\mu  \chi(p_p) \Big] \,, \qquad \text{and}
\\
\label{eq:N3LO_43}
i \mathcal{M}^{\left[\widehat{p(0);f}\right]}_{\gamma\gamma^*}
\!&=&\!\frac{ie^3}{8 M^3 Q^2\left[v\cdot (p_p^\prime + k) +i0\right]} 
\left[\Bar{u}(p^\prime) \gamma^\mu u(p) \right] \bigg[ \chi^\dagger(p^\prime_p)
(v\cdot\varepsilon^*) \Big\{(v\cdot Q)Q^2 - (v\cdot Q )^3 \Big\} v_\mu \chi(p_p) \bigg] \,. \qquad\,\,
\end{eqnarray} 
\end{widetext}
In particular, the amplitudes incorporating the N$^3$LO vertex structure are presented in 
Eqs.~\eqref{eq:N3LO_42} and \eqref{eq:N3LO_43}. Of the nine amplitudes listed above, only the 
first three \underline{enhanced} amplitudes (cf. first row in Fig.~\ref{fig:N3LO}), namely, 
$ \mathcal{M}^{\left[\,\overline{p(1);\rm{f} \otimes}\,\right]}_{\gamma\gamma^*}$, 
$\mathcal{M}^{\left[\,\widetilde {p(0);\rm{f} \otimes}\,\right]}_{\gamma\gamma^*}$, and 
$\mathcal{M}^{\left[p(2);\rm{f} \otimes\right]}_{\gamma\gamma^*}$, Eqs.~\eqref{eq:N3LO_35} - 
\eqref{eq:N3LO_37}, involving the NLO component of the final-state proton propagator, contribute
to the soft-photon bremsstrahlung cross sections in the {\it lab.}-frame at our intended 
$\mathcal{O}(1/M)$ accuracy.
%

In addition to the recoil-radiative amplitudes, there are five pion-loop amplitudes (cf. the last rows in 
Figs.~\ref{fig:N3LO} and \ref{fig:N3LO_extra}) arising from the insertions of the proton's form factors 
($F^p_{1,2}$), which can potentially contribute to the ${\mathcal O}(\alpha^3/M)$ NLO bremsstrahlung 
cross section. The following are their expressions: 
\begin{widetext}
\begin{eqnarray}
\label{eq:N3LO_44}
i \mathcal{M}^{\left[\rm \pi (e)\right]}_{\gamma\gamma^*} \!&=&\! 
\frac{-ie^3}{2 M Q^2} \left(1-\frac{(p_p^\prime+k)^2}{\left[v\cdot(p_p^\prime+k+i0)\right]^2} \right)
\left[\Bar{u}(p^\prime) \gamma^\mu u(p) \right] 
\bigg[\chi^\dagger(p_p^\prime) \left(\mathcal{V}^{(2)}\cdot\varepsilon^*\right) v_\mu \chi(p_p)\bigg]\,, \qquad\,\,
\\
\label{eq:N3LO_45}
i \mathcal{M}^{\left[\rm \pi (f)\right]}_{\gamma\gamma^*} \!&=&\! 
\frac{-ie^3}{2 M Q^2} \left(1-\frac{ (p_p^\prime+k)^2}{\left[v\cdot( p_p^\prime+k+i0)\right]^2}\right)
\left[\Bar{u}(p^\prime) \gamma^\mu u(p) \right] 
\Big[\chi^\dagger(p_p^\prime) \left(v\cdot\varepsilon^*\right)\mathcal{V}^{(2)}_\mu\chi(p_p) \Big]\,, \qquad\,\,
\\
\label{eq:N3LO_46}
i \mathcal{M}^{\left[\rm \pi (i)\right]}_{\gamma\gamma^*} \!&=&\!  
\frac{-ie^3}{2 M Q^2\left[v \cdot (p_p^\prime + k) + i0\right]}\! 
\left[\Bar{u}(p^\prime) \gamma^\mu u(p) \right] \!
\bigg[\chi^\dagger(p_p^\prime) \left(\mathcal{V}^{(2)}\cdot\varepsilon^*\right) 
\Big\{(p_p+p_p^\prime+k)_\mu  
\nonumber\\
&& \hspace{8.9cm} -\, v_\mu [v\cdot(p_p+p_p^\prime+k)]\Big\} \chi(p_p)\bigg]\,, \qquad\,\,
\\
\label{eq:N3LO_47}
i \mathcal{M}^{\left[\rm \pi (j)\right]}_{\gamma\gamma^*} \!&=&\! 
\frac{-ie^3}{2 M Q^2\left[v \cdot (p^\prime_p + k) + i0\right]} \left[\bar{u}(p^\prime) \gamma^\mu u(p) \right] 
\bigg[\chi^\dagger(p^\prime_p) \Big\{(2p_p^\prime+k)\cdot\varepsilon^* 
- (v\cdot\varepsilon^*)[v\cdot (2p_p^\prime+k)]\Big\} \mathcal{V}^{(2)}_\mu \chi(p_p)\bigg]\,,  \qquad\,\,
\end{eqnarray}
and
\begin{eqnarray} 
\label{eq:N3LO_48}
i \mathcal{M}^{\left[\rm \pi (k)\right]}_{\gamma\gamma^*} 
\!&=&\! \frac{-i e^3}{ M Q^2} \left[\Bar{u}(p^\prime)\gamma^\mu u(p) \right] 
\bigg[\chi^\dagger(p_p^\prime)\Big\{ \Gamma_1(Q^2) \varepsilon^*_\mu 
- \Gamma_2(Q^2) v_\mu (v\cdot\varepsilon^*)\Big\}\chi(p_p)\bigg] \,.
\end{eqnarray} 
\end{widetext}
Here, $\Gamma_1\sim {\mathcal O}(1/\Lambda^2_{\chi})$ and $\Gamma_2\sim {\mathcal O}(1/\Lambda^2_{\chi})$,
are two other generic $Q^2$-dependent form factor functions that could be generated from evaluating the 
pion-loop diagrams that renormalize the NLO two-photon seagull vertex (cf. the last graph in 
Fig.~\ref{fig:NLO}). Except for the pion-loop seagull amplitude 
$\mathcal{M}^{\left[\rm \pi (k)\right]}_{\gamma\gamma^*}$, Eq.~\eqref{eq:N3LO_48}, all four other listed 
amplitudes involve LO and NLO components of the final-state proton propagator. Hence, in SPL, these 
amplitudes are expected to be chirally \underline{enhanced} in the {\it lab.}-frame. However, only the 
first two amplitudes, $\mathcal{M}^{\left[\rm \pi(e),\pi(f)\right]}_{\gamma\gamma^*}$ (cf. 
Fig.~\ref{fig:N3LO}), Eqs.~\eqref{eq:N3LO_44} and \eqref{eq:N3LO_45}, contribute to the soft-photon 
bremsstrahlung cross sections in the {\it lab.}-frame at our intended $\mathcal{O}(1/M)$ accuracy. 

The remaining nine amplitudes, Eqs.~\eqref{eq:N3LO_38} – \eqref{eq:N3LO_43} and Eqs.~\eqref{eq:N3LO_46} - 
\eqref{eq:N3LO_48}, shown in Fig.~\ref{fig:N3LO_extra}, are discarded in the main text analysis, since 
they do not contribute to the soft-photon bremsstrahlung cross section in the {\it lab.}-frame. They
either lead to scaleless phase-space integrals vanishing under DR, subject to the condition, 
Eq.~\eqref{dr_condition} (charge-even case), or yield cross section terms of $\mathcal{O}(1/M^2)$ which 
are beyond our working precision (charge-odd case).

\begin{figure}[tbp]
\centering
\includegraphics[scale=0.37]{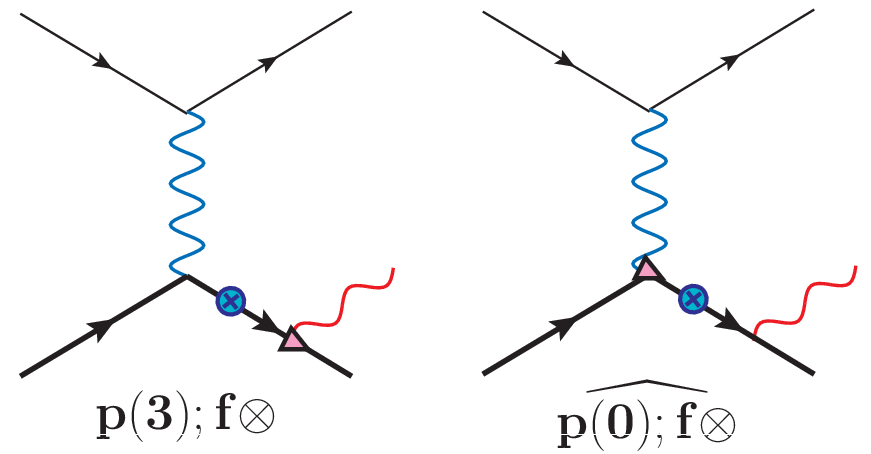}
    \caption{Chirally enhanced photon bremsstrahlung diagrams of N${}^4$LO [i.e., 
             ${\mathcal O}(e^3/M^4)]$ with insertions of $N^3$LO proton-photon interaction vertex 
             (triangle $\triangle$) and ${\mathcal O}(1/M)$ proton propagator component (crossed box
             $\otimes$), \underline{not} contributing to our $\ell$-p elastic scattering cross 
             section up-to-and-including NLO [i.e., ${\mathcal O}(\alpha^3/M)$] in HB$\chi$PT. The 
             thin, thick, and wiggly lines denote the propagators for the lepton, proton, and photon
             (color online: red for the soft bremsstrahlung photon and blue for the exchanged 
             off-shell photon), respectively. The nomenclature of the individual diagrams is based 
             on the radiated photon originating from the final-state ``f” proton ``p”. The labels 
             “(0)” and “(3)” denote LO and N${}^3$LO proton-radiating vertices, respectively, while 
             the wedge signifies an N${}^3$LO exchange-photon proton vertex insertion. }
\label{fig:N4LO_extra}
\end{figure}

In our analysis presented in Sec.~\ref{sec:II}, we examine the contribution of additional chirally 
\underline{enhanced} N${}^3$LO amplitudes to the charge-odd soft-photon bremsstrahlung cross 
section at NLO. This requires us to evaluate the interference matrix elements of $\nu=3$ chiral order
in ${\mathcal S}^{(\nu=3)}_{\rm odd}$, Eq.~\eqref{eq:So_N3LO}, namely,
\begin{eqnarray}
\mathcal{S}^{(\nu=3)}_{\rm odd} \!&=&\! 
\sum_{\rm spins} \left[{\mathcal M}^\dagger_{\rm NLO} \widetilde{\mathcal M}_{\rm NNLO} 
+ {\mathcal M}^\dagger_{\rm LO} \widetilde{\mathcal M}_{\rm N^3LO} + {\rm h.c}\right]_{\rm odd},
\nonumber\\
\end{eqnarray}
where ${\mathcal M}_{\rm LO}$, ${\mathcal M}_{\rm NLO}$, $\widetilde{\mathcal M}_{\rm NNLO}$ and 
$\widetilde{\mathcal M}_{\rm N{}^3LO}$ are the sum of the soft-photon bremsstrahlung amplitudes in the 
{\it lab.}-frame, as given in Eqs.~\eqref{eq:M_LO} - \eqref{eq:M_N3LO}, in the main text. However, the 
general structure of some of the amplitudes contributing to the first interference term of the 
amplitudes (along with their hermitian conjugates), namely, 
\begin{widetext}
\begin{eqnarray} 
\sum_{\rm spins} \left[{\mathcal M}^\dagger_{\rm NLO} \widetilde{\mathcal M}_{\rm NNLO} 
+ {\rm h.c}\right]_{\rm odd} \! &\supset&\! 
2\mathcal{R}e \Bigg[\left(\mathcal{M}^{\left[\,\overline{l;{\rm i}}\,\right]}_{\gamma\gamma^*} 
+ \mathcal{M}^{\left[\,\overline{l;{\rm f}}\,\right]}_{\gamma\gamma^*}\right)^\dagger 
\bigg(\mathcal{M}^{\left[\,\overline{p(1);f}\,\right]}_{\gamma\gamma^*} 
+ \mathcal{M}^{\left[\,\widetilde{p(0);f}\,\right]}_{\gamma\gamma^*} 
+ \mathcal{M}^{\left[p(2);f\right]}_{\gamma\gamma^*} 
+ \mathcal{M}^{\left[p(0);f\boxtimes\right]}_{\gamma\gamma^*} 
\nonumber\\
&& \hspace{0.8cm} +\, \mathcal{M}^{\left[\rm \pi (c)\right]}_{\gamma\gamma^*} 
+\mathcal{M}^{\left[\rm \pi (d)\right]}_{\gamma\gamma^*}\bigg) 
+ \bigg(\mathcal{M}^{\left[\,p(1);{\rm f}\,\right]}_{\gamma\gamma^*} 
+\mathcal{M}^{\left[\,\overline{p(0);{\rm i}}\,\right]}_{\gamma\gamma^*} 
+ \mathcal{M}^{\left[\,\overline{p(0);{\rm f}}\,\right]}_{\gamma\gamma^*} 
+\mathcal{M}^{\left[\,p(0);{\rm i}\otimes \,\right]}_{\gamma\gamma^*} 
\nonumber\\
&& \hspace{0.8cm} +\, \mathcal{M}^{\left[\,p(1);{\rm v}\,\right]}_{\gamma\gamma^*}\bigg)^\dagger
\bigg(\mathcal{M}^{\left[\widetilde{\,l;{\rm i}\,}\right]}_{\gamma\gamma^*}  
+ \mathcal{M}^{\left[\widetilde{\,l;{\rm f}\,} \right]}_{\gamma\gamma^*} 
+ \mathcal{M}^{\left[\rm \pi (a)\right]}_{\gamma\gamma^*} 
+\mathcal{M}^{\left[\rm \pi (b)\right]}_{\gamma\gamma^*}\bigg)\Bigg] 
\sim \mathcal{O}\left(\frac{1}{M^2}\right)\,, \qquad\,
\label{eq:So_NNNLO_dropped} 
\end{eqnarray}
makes it evident that they yield contributions at ${\mathcal O}(1/M^2)$, which are beyond the required
accuracy of our NLO cross section. Thus, the above combination of amplitudes is excluded from our 
treatment of the charge-odd component.

\subsection{N$^4$LO [i.e. ${\mathcal O}(e^3/M^4)$] amplitudes} 
Finally, there are only two possible amplitudes at this order, which can contribute to the radiative 
recoil corrections to the cross section up-to-and-including NLO [i.e., ${\mathcal O}(\alpha^3/M)$] in 
HB$\chi$PT. They are given as follows (cf. Fig.~\ref{fig:N4LO_extra}):
\begin{eqnarray} 
\label{eq:N4LO_49}
i \mathcal{M}^{[p(3);{\rm f}\otimes]}_{\gamma\gamma^*} \!&=&\!
\frac{i e^3}{16 M^4 Q^2}\left(1-\frac{ (p_p^\prime+k)^2}{\left[v \cdot( p_p^\prime+k+i0)\right]^2} \right)
\left[\Bar{u}(p^\prime) \gamma^\mu u(p) \right] 
\bigg[\chi^\dagger(p_p^\prime) \Big\{(v\cdot k)k^2-(v\cdot k)^3\Big\} (v\cdot\varepsilon^*) v_\mu\chi(p_p)\bigg] \,,\qquad\,
\end{eqnarray}
and
\begin{eqnarray}
i \mathcal{M}^{\left[\widehat{p(0)\,;{\rm f}\otimes}\right]}_{\gamma\gamma^*} \!\!&=&\!\!
\frac{ie^3}{16 M^4 Q^2}\!\left(1-\frac{ (p_p^\prime+k)^2}{\left[v\cdot( p_p^\prime+k+i0)\right]^2}\right)\!
\left[\Bar{u}(p^\prime) \gamma^\mu u(p)\right]\!
\bigg[\chi^\dagger(p^\prime_p) (v\cdot\varepsilon^*) \Big\{(v\cdot Q)Q^2 - (v\cdot Q)^3\Big\} v_\mu \chi(p_p) \bigg]\,.\qquad\,
\label{eq:N4LO_50}
\end{eqnarray} 
\end{widetext}
Both the above amplitudes involve the NLO component of the final-state proton propagator and are 
therefore chirally \underline{enhanced} in the {\it lab.}-frame in SPL. However, the first amplitude
leads to a scaleless phase-space integral which vanishes under DR, while the second amplitude yields
cross section terms of $\mathcal{O}(1/M^2)$ which lie beyond our working precision. Hence, both 
amplitudes are discarded from our analysis in the main text.
%

\section{Bremsstrahlung phase-space integration} 
\label{sec:integrals}

\subsection{$S$-frame kinematics}
The determination of the {\it lab.}-frame soft-photon bremsstrahlung cross section entails the phase-space integration
over the so-called {\it radiative tail} differential distribution, namely,
\begin{eqnarray*}
\left[{\rm d}\widetilde{\sigma}_{br}\right]_{\gamma\gamma^*} \!&=&\!\frac{(2\pi)^4\delta^{lab}_k}{8M E^\prime_p |{\bf p}|} 
\frac{{\rm d}^3{\bf p}^{\prime}}{(2\pi)^3 2E^\prime}\frac{{\rm d}^3{\bf k}^{\prime}}{(2\pi)^3 2E_{\gamma^*}}
\nonumber\\
&& \hspace{3cm} \times\,\frac{1}{4}\sum_{\rm spins}|\mathcal{M}_{\gamma\gamma^*}|^2\,,
\end{eqnarray*}   
with energy-conserving $\delta$-function given by
\begin{equation}
\delta^{lab}_k \equiv \delta\left(E + M + - E^{\prime} - \sqrt{({\bf Q}- {\bf k})^2 + M^2} - E_{\gamma^*}\right)\,,
\label{eq:delta_lab}
\end{equation}
as obtained after partially integrating the final-state proton three-momentum. The resulting integration is complicated 
by the dependence of the $\delta$-function on the photon emission angles, which renders the radiation spectrum 
anisotropic. This confines the photon-energy integration domain to an ellipsoidal region that is analytically 
challenging to evaluate. An alternative practical method for evaluating the phase-space integrals employs a technique 
originally introduced by Tsai in Ref.~\cite{Tsai:1961zz}, later formalized in Refs.~\cite{Mo:1968cg,Maximon:1969nw}, and
subsequently applied in various studies of radiative 
corrections~\cite{Maximon:2000hm,Talukdar:2020aui,Vanderhaeghen:2000ws,Bucoveanu:2018soy}. This technique of evaluation 
of the integrals first entails boosting from the laboratory to the S-frame, namely, the center-of-mass frame of the 
recoiling proton and the emitted soft-photon, i.e.,
\begin{equation}
{\bf p}_p^\prime + {\bf k} = {\bf Q}  \quad \stackrel{\rm S-frame}{\Longrightarrow} \quad {\bf Q}^S = 0\,,
\label{eq:delta_S}
\end{equation}
where ${\bf p}_p^\prime$ and  ${\bf k}$ are the respective {\it lab.}-frame three-momenta of the recoil proton and the 
emitted soft-photon, and ${\bf Q}$ is the three-momentum transferred in the $\ell$-p elastic scattering process. In this
way, the integrals reduce to standard spherical forms, as the $\delta$-function above becomes independent of the photon 
angles in SPL. This effectively transforms the {\it lab.}-frame kinematics for the bremsstrahlung process, 
$\ell+{\rm p}\to\ell+{\rm p}+\gamma^*$, into the kinematics for the ``reverse bremsstrahlung”  process, namely,
$\ell+{\rm p}+\gamma^*\to\ell+{\rm p}$, in the S-frame with $(k^{S})^\mu\equiv (E^S_{\gamma^*},{\bf k}^S)\to 0$, subject
to the constraint:
\begin{equation}
\lim_{k\to 0}\delta^{lab}_k  \quad \stackrel{\rm S-frame}{\Longrightarrow} 
\quad \delta^S \equiv \delta\left(E^S + E_p^S - E^{\prime S} - E_p^{\prime S}\right)\,.
\end{equation}
With the {\it lab.}-frame detector resolution defined by the parameter $\Delta_{\gamma^*}$, the following relationships 
can be justified:
\begin{eqnarray}
&& \hspace{-0.5cm} \text{(i)} \,  E_p^{\prime S} \approx M\,, \quad \,\, \text{(ii)} \,  E^S \approx E^\prime = \frac{E}{\eta}\,, \quad  
\text{(iii)} \,\,  E^{\prime S} \approx E \,,   
\nonumber\\
&& \hspace{-0.5cm}  \text{(iv)} \,  E_p^S \approx E_p^\prime\,, \quad   \text{(v)} \,  \cos\theta_S \approx \cos\theta\,, \quad 
\text{(vi)} \, \Delta_S \approx \eta \Delta_{\gamma^*}\,, \quad\,\,
\nonumber\\
\label{eq:LS_transform}
\end{eqnarray}
where $\theta$ and $\theta_S$ denote the {\it lab.}-frame and S-frame scattering angle for the outgoing lepton, 
respectively. It is also notable that $\Delta_S/\eta$ is the upper limit of the soft-photon energy integral 
corresponding to the detector resolution $\Delta_S$ in the S-frame. A comprehensive treatment of the S-frame 
kinematics is provided in the appendix of the work by Talukdar \textit{et al.}~\cite{Talukdar:2020aui}.
 
\subsection{Integrations of the charge-even component}
The charge-even soft-photon bremsstrahlung differential cross section up-to-and-including NLO in HB$\chi$PT
is obtained by integrating over the elastic radiative tail distribution. This involves the following integrals,
expressed in terms of {\it lab.}-frame variables:
\begin{eqnarray*}
\int \frac{{\rm d}^3{\bf k}}{k} \frac{1}{(p \cdot k)^2} \delta^{lab}_k &,& 
\int \frac{{\rm d}^3{\bf k}}{k} \frac{1}{(p^\prime \cdot k)^2} \delta^{lab}_k \,\,\,, 
\\
\int \frac{{\rm d}^3{\bf k}}{k} \frac{(p \cdot p^\prime)}{(p \cdot k)(p^\prime \cdot k)} \delta^{lab}_k  &,&
\int \frac{{\rm d}^3{\bf k}}{k} \frac{(v \cdot k)}{(p \cdot k)^2} \delta^{lab}_k \,\,\,,
\\
\int \frac{{\rm d}^3{\bf k}}{k} \frac{(v \cdot k)}{(p^\prime \cdot k)^2} \delta^{lab}_k &,&
\int \frac{{\rm d}^3{\bf k}}{k} \frac{(p \cdot p^\prime) (v \cdot k)}{(p \cdot k)(p^\prime \cdot k)} \delta^{lab}_k\,\,\,,
\end{eqnarray*}
where $E_\gamma^*=k=|{\bf k}|\to 0$ is the  soft-photon energy/three-momentum. To simplify the treatment of 
the ``inelastic'' $\delta^{lab}_k$-functions, Eq.~\eqref{eq:delta_lab}, Tsai's technique~\cite{Tsai:1961zz} 
is employed, which involves boosting to the S-frame. The transformation renders $\delta^{lab}_k$ effectively 
``elastic", resulting in the simplified $\delta^S$-function of Eq.~\eqref{eq:delta_S}. With the 
$\delta$-function becoming independent of the photon angles in the SPL, the integrals are readily evaluated
in the S-frame. As discussed in the main text, some of these integrals exhibit IR 
divergences, which are isolated using DR by analytical continuation to $D-1=3-2\varepsilon_{\rm IR}$ spatial
dimensions, with $\varepsilon_{\rm IR}<0$. The resulting S-frame integrals, re-expressed in terms of 
{\it lab.}-frame kinematic variables {\it via} the kinematic transformations~\eqref{eq:LS_transform},
are presented below:
\begin{eqnarray}
L^{(S)}_{\rm ii} \!&\stackrel{\rm DR}{\longmapsto}&\!
\frac{m^2_l}{2}\int\frac{{\rm d}^{D-1}{k^S}}{k^S}\frac{1}{(p^{S}\cdot k^S)^2}
\nonumber\\
&=&\!\! \pi\left[\frac{1}{|\varepsilon_{\rm IR}|}+\gamma_E-\ln\left(\frac{4\pi\mu^2}{m_l^2}\right)\right]
+2\pi\widetilde{L}^{(S)}_{\rm ii}\,,
\nonumber\\
&& \hspace{-1.6cm} \text{where} 
\nonumber\\
\widetilde{L}^{(S)}_{\rm ii} \!\!&=&\!\! \frac{1}{2}\ln\left(\frac{4\Delta^2_{\gamma^*}}{m_l^2}\right)
-\frac{1}{\beta^\prime}\ln\sqrt{\frac{1+\beta^\prime}{1-\beta^\prime}}\,,
\\
\nonumber\\
L^{(S)}_{\rm ff} \!&\stackrel{\rm DR}{\longmapsto}&\!
\frac{m^2_l}{2}\int\frac{{\rm d}^{D-1}{k^S}}{k^S}\frac{1}{(p^{\prime S}\cdot k^S)^2}
\nonumber\\
&=&\!\! \pi\left[\frac{1}{|\varepsilon_{\rm IR}|}+\gamma_E-\ln\left(\frac{4\pi\mu^2}{m_l^2}\right)\right]
+2\pi\widetilde{L}^{(S)}_{\rm ff}\,,
\nonumber\\
&& \hspace{-1.6cm} \text{where} 
\nonumber\\
\widetilde{L}^{(S)}_{\rm ff} \!\!&=&\!\! \frac{1}{2}\ln\left(\frac{4\Delta^2_{\gamma^*}}{m_l^2}\right)
-\frac{1}{\beta}\ln\sqrt{\frac{1+\beta}{1-\beta}}\,,
\\
\nonumber\\
L^{(S)}_{\rm if} \!&\stackrel{\rm DR}{\longmapsto}&\!
\int\frac{{\rm d}^{D-1}{k^S}}{k^S}\frac{(p^S \cdot p^{\prime S})}{(p^S\cdot k^S)(p^{\prime S}\cdot k^S)}
\nonumber\\
&=&\!\! \pi\left[\frac{1}{|\varepsilon_{\rm IR}|}+\gamma_E-\ln\left(\frac{4\pi\mu^2}{m_l^2}\right)\right] \frac{\nu^2+1}{\nu}
\ln\!\left[\frac{\nu+1}{\nu-1}\right] 
\nonumber\\
&& +\,2\pi\widetilde{L}^{(S)}_{\rm if}\,,
\nonumber\\
&& \hspace{-1.6cm} \text{where} 
\nonumber\\
\widetilde{L}^{(S)}_{\rm if} \!\!&=&\!\! \frac{\nu^2_l+1}{2\nu_l}\Bigg[\ln\left(\frac{4\Delta^2_{\gamma^*}}{m_l^2}\right)
\ln\left[\frac{\nu_l+1}{\nu_l-1}\right]-\ln^2\!\!\sqrt{\frac{1+\beta^\prime}{1-\beta^\prime}} 
\nonumber\\
&& \hspace{1.1cm} + \ln^2\sqrt{\frac{1+\beta}{1-\beta}} 
+ \text{Li}_2\left(1-\frac{\lambda_\nu E-E^{\prime}}{(1-\beta)E\xi_\nu}\right)
\nonumber
\end{eqnarray}
\begin{eqnarray}
&&  +\,\text{Li}_2\left(1-\frac{\lambda_\nu E-E^{\prime}}{(1+\beta)E\xi_\nu}\right)
\nonumber\\
&& -\,\text{Li}_2\left(1-\frac{\lambda_\nu E-E^{\prime}}{(1-\beta^\prime)E^\prime \lambda_\nu\xi_\nu}\right)
\nonumber\\
&& -\,\text{Li}_2\left(1-\frac{\lambda_\nu E-E^{\prime}}{(1+\beta^\prime)E^\prime\lambda_\nu\xi_\nu}\right)\Bigg]\,, 
\\
\nonumber\\
LL^{(S)}_{\rm ii} \!&\stackrel{\rm DR}{\longmapsto}&\! 
\int\frac{{\rm d}^{d-1}{k^S}}{k^S}\frac{(v\cdot k^S)}{(p^S\cdot k^S)^2}
= 2\pi\Delta_{\gamma^*} \,,
\\
\nonumber\\
LL^{(S)}_{\rm ff} \!&\stackrel{\rm DR}{\longmapsto}&\!
\int\frac{{\rm d}^{d-1}{k^S}}{k^S}\frac{(v\cdot k^S)}{(p^{\prime S}\cdot k^S)^2}
= 2\pi\Delta_{\gamma^*} \,,
\\
\nonumber\\
LL^{(S)}_{\rm if} \!&\stackrel{\rm DR}{\longmapsto}&\!
\int\frac{{\rm d}^{d-1}{k^S}}{k^S}\frac{(p^S \cdot p^{\prime S})(v\cdot k^S)}{(p^S\cdot k^S)(p^{\prime S}\cdot k^S)}
\nonumber\\
&=&\!\! 4\pi \Delta_{\gamma^*} \left(\frac{1+\nu_l^2}{2 \nu_l}\right)\ln{\left[\frac{\nu_l+1}{\nu_l-1}\right]}\,,
\end{eqnarray}
Here, $\mu$ denotes the renormalization scale and $\gamma_E=0.577216...$ is the Euler-Mascheroni constant. 
After the phase-space integrations are carried out, the resulting expressions are boosted back into the 
{\it lab.}-frame using the same transformation relations~\eqref{eq:LS_transform}. The corresponding 
{\it lab.}-frame expressions for these integrals have been presented in the main text, as given in 
Eqs.~\eqref{eq:Lii}, \eqref{eq:Lff}, \eqref{eq:Lif}, \eqref{eq:LLii}, and \eqref{eq:LLif}.

\subsection{Integrations of the charge-odd component}
For the evaluation of our charge-odd soft-photon bremsstrahlung differential cross section 
up-to-and-including NLO in HB$\chi$PT, we encounter the following types of phase-space integrals, as 
expressed in terms of the {\it lab.}-frame kinematical variables:
\begin{eqnarray*}
\int \frac{{\rm d}^3{\bf k}}{k} \frac{1}{(p \cdot k)} \delta^{lab}_k &,& 
\int \frac{{\rm d}^3{\bf k}}{k} \frac{1}{(p^\prime \cdot k)} \delta^{lab}_k \,\,\,, 
\\
\int \frac{{\rm d}^3{\bf k}}{k} \frac{1}{(p \cdot k)(v \cdot k)} \delta^{lab}_k  &,&
\quad \int \frac{{\rm d}^3{\bf k}}{k} \frac{1}{(p^\prime \cdot k) (v \cdot k)} \delta^{lab}_k\,\,\,,
\\
\int \frac{{\rm d}^3{\bf k}}{k} \frac{(v \cdot k)}{(p \cdot k)} \delta^{lab}_k  &,& 
\int \frac{{\rm d}^3{\bf k}}{k} \frac{(v \cdot k)}{(p^\prime \cdot k)} \delta^{lab}_k \,\,\,, 
\\
\int \frac{{\rm d}^3{\bf k}}{k} \frac{(v \cdot k)^2}{(p \cdot k)} \delta^{lab}_k  &,& 
\int \frac{{\rm d}^3{\bf k}}{k} \frac{(v \cdot k)^2}{(p^\prime \cdot k)} \delta^{lab}_k \,\,\,, 
\\
\int \frac{{\rm d}^3{\bf k}}{k} \frac{(k \cdot Q)}{(p \cdot k)} \delta^{lab}_k  &,& 
\int \frac{{\rm d}^3{\bf k}}{k} \frac{(k \cdot Q)}{(p^\prime \cdot k)} \delta^{lab}_k \,\,\,. 
\end{eqnarray*}
The integrals are initially evaluated in the S-frame and then expressed in terms of {\it lab.}-frame 
kinematic variables using the transformation rules~\eqref{eq:LS_transform} outlined earlier.
The following are their expressions: 
\begin{eqnarray}
I_1^{(S)} \!&=&\! \int \frac{\mathrm{d}^3 {\bf k}^S}{k^S} \frac{1}{(p^S \cdot k^S)} 
= \frac{4\pi \Delta_{\gamma^*}}{\beta^\prime E^\prime} \ln\sqrt{\frac{1+\beta^\prime}{1-\beta^\prime}}\,, \qquad\,
\end{eqnarray}
\begin{eqnarray}
I_2^{(S)} \!\!&=&\!\!\! \int \frac{\mathrm{d}^3 {\bf k}^S}{k^S} \frac{1}{(p^{\prime S} \cdot k^S)}
=\frac{4\pi \Delta_{\gamma^*}}{\beta E} \!\ln\!\sqrt{\frac{1+\beta}{1-\beta}}, 
\\
\nonumber\\
I_3^{(S)} \!&\stackrel{\rm DR}{\longmapsto}&\! \int \frac{d^{D-1} {\bf k}^S}{k^S} \frac{1}{(p^S \cdot k^S)(v \cdot k^S)}
\nonumber\\
&=&\!\!\! \frac{2 \pi}{\beta^\prime E^\prime} \!\left[ \frac{1}{|\varepsilon_{\rm IR}|} + \gamma_E 
- \ln\left(\frac{4\pi\mu^2}{m_l^2}\right) \right]\!\ln\!\sqrt{\frac{1+\beta^\prime}{1-\beta^\prime}} 
\nonumber\\
&& +\, 2 \pi \widetilde{I}^{(S)}_{3}\,, 
\nonumber\\
&& \hspace{-1.5cm} \text{where} 
\nonumber\\
\widetilde{I}^{(S)}_{3} \!\!&=&\!\!
\frac{1}{\beta^\prime E^\prime} \bigg[\ln\left(\frac{4 \Delta_{\gamma^*}^2 }{ m_l^2} \right) 
\ln\sqrt{\frac{1+\beta^\prime}{1-\beta^\prime}} 
\nonumber\\
&& \hspace{0.7cm}  +\, \frac{1}{2} \operatorname{Li}_2\left( \frac{2\beta^\prime}{\beta^\prime - 1} \right) 
- \frac{1}{2} \operatorname{Li}_2\left( \frac{2\beta^\prime}{\beta^\prime + 1} \right) \!\bigg], 
\nonumber\\
\\
I_4^{(S)} \!&\stackrel{\rm DR}{\longmapsto}&\! \int \frac{d^{D-1} {\bf k}^S}{k^S} \frac{1}{(p^{\prime S} \cdot k^S)(v \cdot k^S)}
\nonumber\\
&=&\!\! \frac{2 \pi}{\beta E} \left[ \frac{1}{|\varepsilon_{\rm IR}|} + \gamma_E 
- \ln\left(\frac{4\pi\mu^2}{m_l^2}\right) \right] \ln\sqrt{\frac{1+\beta}{1-\beta}}
\nonumber\\
&& +\, 2 \pi \widetilde{I}^{(S)}_{4}\,,
\nonumber\\
&& \hspace{-1.5cm} \text{where} 
\nonumber\\
\widetilde{I}^{(S)}_{4} \!\!&=&\!\!  \frac{1}{\beta E} \bigg[\ln\left(\frac{4 \Delta_{\gamma^*}^2 }{ m_l^2} \right) 
\ln\sqrt{\frac{1+\beta}{1-\beta}} 
\nonumber\\
&& \hspace{0.6cm} +\, \frac{1}{2} \operatorname{Li}_2\left( \frac{2\beta}{\beta - 1} \right) 
- \frac{1}{2} \operatorname{Li}_2\left( \frac{2\beta}{\beta + 1} \right) \bigg]\,, 
\nonumber\\
\\
I_5^{(S)} \!\!\!\!&=&\!\!\!\! \int \frac{\mathrm{d}^3 {\bf k}^S}{k^S} \frac{(v \cdot k^S)}{(p^S \cdot k^S)} 
= \frac{2\pi \Delta_{\gamma^*}^2}{\beta^\prime E^\prime} \ln\sqrt{\frac{1+\beta^\prime}{1-\beta^\prime}}\,, \quad\,
\\
\nonumber\\
I_6^{(S)} \!\!\!\!&=&\!\!\!\! \int \frac{\mathrm{d}^3 {\bf k}^S}{k^S} \frac{(v \cdot k^S)}{(p^{\prime S} \cdot k^S)} 
= \frac{2 \pi \Delta_{\gamma^*}^2}{\beta E} \ln\sqrt{\frac{1+\beta}{1-\beta}}\,, \quad\,
\\
\nonumber\\
I_7^{(S)} \!\!\!\!&=&\!\!\!\! \int \frac{\mathrm{d}^3 {\bf k}^S}{k^S} \frac{(v \cdot k^S)^2}{(p^S \cdot k^S)} 
=\frac{4\pi\Delta_{\gamma^*}^3}{3\beta^\prime E^\prime} \ln\sqrt{\frac{1+\beta^\prime}{1-\beta^\prime}}\,, \quad\,
\\
\nonumber\\
I_8^{(S)} \!\!\!\!&=&\!\!\!\! \int \frac{\mathrm{d}^3 {\bf k}^S}{k^S} \frac{(v \cdot k^S)^2}{(p^{\prime S} \cdot k^S)} 
= \frac{4\pi\Delta_{\gamma^*}^3}{3\beta E} \ln\sqrt{\frac{1+\beta}{1-\beta}}, \qquad\,  
\\
\nonumber\\
I_9^{(S)} \!\!\!\!&=&\!\!\!\! \int \frac{\mathrm{d}^3 {\bf k}^S}{k^S} \frac{(k^S \cdot Q^S)}{(p^S \cdot k^S)} 
\nonumber\\
&=&\! \frac{-2 \pi E\Delta_{\gamma^*}^2}{E^{\prime} \beta^\prime}\left[\left(1-\frac{\beta}{\beta^\prime} \right) 
\ln\sqrt{\frac{1+\beta^\prime}{1-\beta^\prime}}-\beta\right], \qquad\,
\\
\nonumber\\
I_{10}^{(S)}\!\!\!\!&=&\!\!\!\! \int \frac{\mathrm{d}^3 {\bf k}^S}{k^S} \frac{(k^S \cdot Q^S)}{(p^{\prime S} \cdot k^S)}
\nonumber\\
&=&\! \frac{2 \pi E^{\prime}\Delta_{\gamma^*}^2}{E \beta} \left[ \left(1-\frac{\beta^\prime}{\beta} \right) 
\ln\sqrt{\frac{1+\beta}{1-\beta}}-\beta^\prime\right]\,. \quad\,
\end{eqnarray}
The corresponding expressions for these integrals, after being boosted back to the {\it lab.}-frame, have been
presented in the main text, as given in Eqs.~\ref{eq:I1},~\ref{eq:I2},~\ref{eq:I3},~\ref{eq:I4},~\ref{eq:I5},
~\ref{eq:I6},~\ref{eq:I7},~\ref{eq:I8},~\ref{eq:I9}, and \ref{eq:I10}.


\bibliographystyle{apsrev}

\end{document}